\documentclass[a4paper,10pt]{article}
\usepackage{jheppub,color,graphicx,slashed,multirow,hyperref,amssymb,amsmath,cleveref,colortbl}
\usepackage[utf8]{inputenc}
\usepackage[T1]{fontenc}
\usepackage[section]{placeins}
\usepackage{fancyvrb}
\usepackage[formats]{listings}
\usepackage{textcomp}
\usepackage{tcolorbox,enumitem,verbatim,titlesec}
\titleformat{\paragraph}{\normalfont\normalsize\bfseries}{\theparagraph}{1em}{}
\titlespacing*{\paragraph}{0pt}{3.25ex plus 1ex minus .2ex}{1.5ex plus .2ex}

\newcommand{\GeV}{{\rm\ GeV}}

\newcommand{\TeV}{{\rm\ TeV}}
\newcommand{\fb}{{\rm\ fb}}

\definecolor{coolblack}{rgb}{0.0, 0.18, 0.39}
\newcommand\cbc[1]{{\color{coolblack}{#1}}}

\def\ie{{\it i.e.}}
\def\eg{{\it e.g.}}

\newcommand{\fr}{{\sc\small FeynRules}}
\newcommand{\lhapdf}{{\sc\small LHAPDF6}}
\newcommand{\mg}{{\sc\small MG5\_aMC}}
\newcommand{\ms}{{\sc\small MadSpin}}
\newcommand{\mw}{{\sc\small MadWidth}}

\newcommand{\ma}{{\sc MadAnalysis}~5}

\newcommand{\be}{\begin{equation}}
\newcommand{\ee}{\end{equation}}
\def\bsp#1\esp{\begin{split}#1\end{split}}
\def\bpm{\begin{pmatrix}}
\def\epm{\end{pmatrix}}

\title{Single production of vector-like quarks: the effects of large width, interference and NLO corrections}

\author[a]{Aldo Deandrea}
\author[b]{\!\!, Thomas Flacke}
\author[c,d]{\!\!, Benjamin~Fuks}
\author[e,f]{\!\!, Luca Panizzi}
\author[c]{\! and Hua-Sheng Shao}

\emailAdd{deandrea@ipnl.in2p3.fr}
\emailAdd{flacke@ibs.re.kr}
\emailAdd{fuks@lpthe.jussieu.fr}
\emailAdd{luca.panizzi@physics.uu.se}
\emailAdd{huasheng.shao@lpthe.jussieu.fr}

\affiliation[a]{Univ. Lyon, Universit{\' e} Claude Bernard Lyon 1, CNRS/IN2P3,
 IP2I UMR5822, F-69622, Villeurbanne, France}
  \affiliation[b]{Center for Theoretical Physics of the Universe, Institute for Basic Science (IBS), Daejeon 34126,
	Korea}
\affiliation[c]{Laboratoire de Physique Th\'eorique et Hautes Energies (LPTHE),
  UMR 7589, Sorbonne Universit\'e et CNRS, 4 place Jussieu, 75252 Paris Cedex 05, France}
  \affiliation[d]{Institut Universitaire de France, 103 boulevard Saint-Michel,
  75005 Paris, France}
\affiliation[e]{Department of Physics and Astronomy, Uppsala University, 
  Box 516, SE-751 20 Uppsala, Sweden}
\affiliation[f]{School of Physics and Astronomy, University of Southampton, Highfield, Southampton SO17 1BJ, UK}

\abstract{We provide a comprehensive discussion, together with a complete setup for simulations, relevant for the production of a single vector-like quark at hadron colliders. Our predictions include finite width effects, signal-background interference effects and next-to-leading order QCD corrections. We explicitly apply the framework to study the single production of a vector-like quark $T$ with charge 2/3, but the same procedure can be used to analyse the single production of vector-like quarks with charge $-4/3$, $-1/3$, 2/3 and 5/3, when the vector-like quark interacts with the Standard Model quarks and electroweak bosons. Moreover, this procedure can be straightforwardly extended to include additional interactions with exotic particles. We provide quantitative results for representative benchmark scenarios characterised by the $T$ mass and width, and we determine the role of the interference terms for a range of masses and widths of phenomenological significance. We additionally describe in detail, both analytically and numerically, a striking feature in the invariant mass distribution appearing only in the $T\rightarrow th$ channel.}

\begin{document}
\preprint{ CTPU-PTC-21-18}

%\setpagewiselinenumbers
%\linenumbers

\maketitle
\flushbottom

%%%%%%%%%%%%%%%%%%%%%%%%%%%%%%%%%%%%%%%%%%%%%%%%%%%%%%%%%%%%%%%%%%%%%%%%%%%%%%
\section{Introduction}
\label{sec:intro}
Heavy spin-1/2 particles transforming as triplets under the QCD gauge group and with the same left-handed and right-handed couplings to gauge bosons, known as Vector-Like Quarks (VLQ), play a central role in many Standard Model (SM) extensions which address the hierarchy problem. Those include composite Higgs models \cite{Kaplan:1983fs,Kaplan:1991dc,Agashe:2004rs}, extra-dimensional models \cite{Randall:1999ee,Chang:1999nh,Gherghetta:2000qt,Agashe:2004rs}, Little-Higgs models~\cite{ArkaniHamed:2002qx,Perelstein:2003wd,Schmaltz:2005ky} and a sub-class of supersymmetric models \cite{Bratchikov:2005vp,Martin:2009bg,Abdullah:2015zta,Abdullah:2016avr,Aguilar-Saavedra:2017giu,Araz:2018uyi,Zheng:2019kqu}. The origin of these states, often simply discussed at the effective level, can be in many cases traced back to a theory in which the VLQs are composite states of a new strong dynamics \cite{Barnard:2013zea,Ferretti:2013kya}. A composite origin of these states indicates that they may have both a large width with respect to their mass (see refs.~\cite{Moretti:2016gkr,Carvalho:2018jkq} for earlier studies on this matter) and non-standard decay modes \cite{Serra:2015xfa,Aguilar-Saavedra:2017giu,Chala:2017xgc,Bizot:2018tds,Han:2018hcu,Xie:2019gya,Benbrik:2019zdp,Cacciapaglia:2019zmj,Aguilar-Saavedra:2019ghg}. Most of the existing studies parameterise VLQ production and decay using the Narrow Width Approximation (NWA) and standard VLQ decay modes to $W$, $Z$ and Higgs bosons. Establishing the relevance and limitations of the standard approximations, as well as studying the effect of going beyond these approximations, is therefore relevant and may have in some cases a strong impact on the VLQ searches performed at the LHC and at future colliders.

Current collider searches mainly focus on the QCD production of a pair of VLQs~\cite{Aaboud:2017zfn,Aaboud:2017qpr,Aaboud:2018xuw,Aaboud:2018saj,Aaboud:2018xpj,Aaboud:2018wxv,Aaboud:2018pii,Sirunyan:2017pks,Sirunyan:2018qau,Sirunyan:2018omb,Sirunyan:2019sza,Sirunyan:2018yun,Sirunyan:2020qvb}, although the electroweak single production of a vector-like quark of narrow or moderate width has been recently addressed for VLQs of charge $-4/3$ ($Y$), $-1/3$ ($B$), 2/3 ($T$) and 5/3 ($X$) in the case where the VLQ decays exclusively into a third generation quark and an electroweak boson $(W,Z,h)$~\cite{Aaboud:2018saj,Aaboud:2018ifs,Sirunyan:2017ynj,Sirunyan:2018fjh,Sirunyan:2018ncp,Sirunyan:2019xeh}. QCD pair production searches impose that the VLQ masses satisfy $M_Y\gtrsim 1.3$~TeV,  $M_T\gtrsim 1.3 - 1.4$~TeV, $M_B\gtrsim 1.0 - 1.4$~TeV  and $M_X\gtrsim 1.3$~TeV, the exact bounds depending on the specific assumptions about the VLQ decays. The run 3 of LHC and the high-luminosity LHC (HL-LHC) operations are expected to extend the discovery (and exclusion) potential to higher masses in the coming years (see refs.~\cite{Matsedonskyi:2014mna,CMS:2013xfa,Barducci:2017xtw,CidVidal:2018eel} for projections), this increase being however modest because the QCD pair production cross section rapidly decreases with increasing VLQ mass due to phase-space suppression.

In contrast, electroweak VLQ single production is less phase-space suppressed as only one heavy particle is produced. It is thus very attractive to tackle the VLQ high-mass regime. However, single production channels depend on new physics coupling strengths, so that experimental analyses do not yield a direct bound on the VLQ mass. Instead, the searches result in bounds on the VLQ production cross sections (times branching ratios into the targeted final state). The VLQ partial widths depending on the same couplings, a given production cross section (at a given $M_Q$) thus leads to a finite width of the VLQ, and a large production cross section additionally leads to a large VLQ width. These considerations therefore imply that a consistent treatment of finite width effects in VLQ single production is mandatory.

We give in the following a comprehensive discussion of these issues and provide a setup for the simulation of VLQ single production at the next-to-leading order (NLO) in QCD, our framework allowing one to include finite width effects. In addition we highlight striking differences in the $T\rightarrow th$ decay channel, as compared to results in the narrow-width approximation. In section \ref{sec:model} we review the simplified VLQ model description which is used in usual VLQ searches at the LHC~\cite{Fuks:2016ftf}, and we introduce its extension beyond the narrow-width approximation.

In section \ref{sec:lo}, we present an extensive  analysis of the single production of a vector-like quark $T$ of charge 2/3 that is produced via $W$-boson exchanges. Our predictions are accurate at the leading-order (LO) accuracy, and include finite width effects. Moreover, we consider all possible VLQ decay modes into a pair of SM particles, assuming the VLQ solely interacts with the SM third generation. We thus focus on the $p p\rightarrow ht+X$, $p p\rightarrow Zt+X$ and $p p\rightarrow Wt+X$ processes. We compare large width effects in different schemes in section \ref{subsec:schemecomparison}, provide a parton-level analysis of the signals in section~\ref{sec:signal}, as well as an analysis including the signal-background interference in section~\ref{subsec:plinterference}. We finally determine a cross section parameterisation and show reference plots to pin down the vector-like quark width in the single-production channel in section \ref{sec:lo_2d_scans}. In the $p p\rightarrow ht+X$ channel, we find a feature which is already present for a VLQ with a narrow width, and which becomes crucial in the case of a VLQ with a large width: the $ht$ invariant-mass distribution is \emph{not} well described by a Breit-Wigner distribution, and instead exhibits an enhancement at low partonic centre-of-mass energies. This fact can be understood analytically (at least at LO), as shown in section \ref{subsec:parton_ME}, and has profound consequences for the interpretation of the single-production VLQ search results in the $p p\rightarrow ht+X$ channel. It indeed affects the $T\rightarrow ht$ signal rate as well as the $ht$ invariant mass distribution,  which is a currently used signal discriminant.

QCD NLO effects have been shown to be relevant in VLQ single production analyses, when the VLQ is a narrow object~\cite{Cacciapaglia:2018qep}. In section \ref{sec:nlo} we provide a quantitative discussion of QCD NLO corrections when considering a VLQ featuring a large width for several distributions which are potentially relevant in experimental searches. 

 Our work is summarised in section~\ref{sec:conclusion}. Whereas we focus on single $T$  production through $Wb$ fusion in the main part of this paper and in appendix~\cref{app:TviaW}, we provide all analogous key results and figures for single $T$ production through $Zt$ fusion in appendix~\ref{app:TviaZ}. Finally, in~\cref{app:Tech} we provide the simulation syntax that needs to be used to reproduce all the results of this paper. This syntax can be easily generalised for studying other processes involving VLQs with different charges.

%%%%%%%%%%%%%%%%%%%%%%%%%%%%%%%%%%%%%%%%%%%%%%%%%%%%%%%%%%%%%%%%%%%%%%%%%%%%%%

\section{Theoretical framework}
\label{sec:model}
\subsection{Model description}
For phenomenological purposes, we consider a simplified extension of the
Standard Model in which the SM field content is extended by a single
species of vector-like quarks $T$ of mass $M_T$. The latter is chosen to carry
an electric charge of 2/3 (for the sake of the example) and to couple to all
SM gauge and Higgs bosons. Working in the mass eigenbasis so that the mixing between the vector-like quark $T$ and the SM quarks is encoded in the masses and couplings, we consider that the VLQ dominantly couples to the top quark. After imposing an $SU(3)_c \times U(1)_Q$ gauge
symmetry, the Lagrangian of the considered simplified model
therefore reads~\cite{Fuks:2016ftf}
\be\label{eq:lag}\bsp
  \mathcal{L} = &\ {\cal L}_{\rm SM} +
    i \bar{T} \slashed{D} T - M_T \bar{T} T
    + \bigg[ h\bar T \big(\hat\kappa_L P_L+\hat\kappa_R P_R\big) u_q
    + \frac{g}{2 c_W} \bar T \slashed{Z}
          \big( \tilde{\kappa}_L P_L + \tilde{\kappa}_R P_R \big) u_q\\
  &\hspace{2cm}
    + \frac{g}{\sqrt{2}} \bar{T} \slashed{W}
       \big( \kappa_L P_L + \kappa_R P_R \big) d_q
    + \mathrm{h.c.}\bigg]\ .
\esp\ee
In our notation, $g$ denotes the weak coupling constant, $c_W$ is the cosine of
the electroweak mixing angle and $\kappa$, $\tilde\kappa$ and $\hat\kappa$
represent the electroweak couplings of the vector-like quark $T$ (as vectors in
the flavour space). Whilst those are well-defined in UV-complete models where
the representation of the vector-like quark is fixed, they are taken as free
parameters in our simplified model parametrisation. Moreover, $u_q$ and $d_q$
denote the Standard Model up-type and down-type quark fields, $Z_\mu$,
$W_\mu$ and $h$ stand for the weak and Higgs boson fields, and $P_L$ and $P_R$
are the usual left-handed and right-handed chirality projectors.

The last three terms in the above Lagrangian open the door to single
vector-like quark production at hadron colliders, whereas the usual
pair production mechanism is embedded in the QCD component of the covariant
kinetic term (that only includes VLQ couplings to gluons and photons as we work in the context of an $SU(3)_c \times U(1)_Q$ gauge
symmetry). In general, the $\kappa$ parameters are taken small so that the
vector-like quark stays narrow. This is in particular the case in many searches
for vector-like quarks at the LHC (see \eg\ refs.~\cite{Aaboud:2017zfn,
Aaboud:2017qpr,Aaboud:2018xuw,Aaboud:2018saj,Aaboud:2018xpj,Aaboud:2018wxv,
Aaboud:2018pii,Aaboud:2018ifs,Sirunyan:2017pks,Sirunyan:2018qau,
Sirunyan:2018omb,Sirunyan:2019sza} for recent results of the ATLAS and CMS
collaborations), but it is not generally valid in many theoretical scenarios. 
% This is, however, not necessarily the case. 
The UV embedding
of the simplified model introduced above may imply the existence of exotic decay
modes~\cite{Bizot:2018tds,Cacciapaglia:2019zmj}, so the $T$ quark could become
wide. Several experimental searches have consequently started to explore such a
configuration, at least for moderately broad vector-like quarks with a
width-to-mass ratio ranging up to 30\%~\cite{Sirunyan:2017ynj,Sirunyan:2018fjh,
Sirunyan:2018ncp,Sirunyan:2019xeh}.

In this case, the traditional approach to simulate a vector-like quark signal at
colliders in which the production and the decay sub-processes are factorised is
not valid anymore. On the one hand, the two sub-processes must be considered
together as a single, not factorisable, process. On the other hand, the
vector-like-quark propagator must be treated in a special manner.

The vector-like quark signals relevant for this work are
simulated using the \mg\ Monte Carlo generator~\cite{Alwall:2014hca},
which allows for the simulation of SM processes up to the next-to-leading-order
(NLO) accuracy in QCD, by relying on the implementation~\cite{Fuks:2016ftf} of the
model described above in the form of an NLO \fr/UFO library~\cite{
Alloul:2013bka,Christensen:2009jx,Degrande:2011ua,Degrande:2014vpa}. The simulation syntax relevant for the processes considered in this work is detailed in \cref{app:Tech}.

\subsection{Scattering processes with unstable particles\label{sec:width}}
It is well known that the asymptotic external states in scattering amplitudes must be stable in order to guarantee the unitarity of the $S$-matrix in quantum field theory. A proper treatment of unstable particles in perturbative scattering amplitudes requires a (Dyson) summation of two-point one-particle irreducible Feynman diagrams, which leads to propagators of the form
\begin{eqnarray}
G_D(p^2)&=&-\frac{i}{p^2-M_0^2+\Sigma(p^2)}\ .\label{eq:propogator}
\end{eqnarray}
In this expression, $M_0$ ($p^2$) is the bare mass (virtuality) of the unstable particle, and $\Sigma(p^2)$ is the amputated two-point Green's function. 
The introduction of such a Dyson summation amounts to reorganising the perturbative expansion of coupling constants entering in the numerators and denominators of the scattering amplitudes, where the latter are related to $\Sigma(p^2)$ in $G_D(p^2)$. However, one should bear in mind that the naive replacements of the propagators of unstable particles appearing in the amplitudes might spoil many important properties of the $S$-matrix, like gauge invariance and perturbative unitarity. A widespread proposal that addresses those issues relies on the complex pole of the propagator~(\ref{eq:propogator}), that is located in  $\bar{p}^2=\bar{M}^2-i\bar{\Gamma}\bar{M}$ and that is defined by
\begin{eqnarray}
\bar{p}^2-M_0^2+\Sigma(\bar{p}^2)&=&0\ . 
\end{eqnarray}
Such a scheme is known as the \textit{complex mass scheme}~\cite{Denner:1999gp,Denner:2005fg}. After carefully assessing several non-trivial theoretical issues~\cite{Frederix:2018nkq}, such an approach is in principle valid up to next-to-leading order in perturbation theory. In practice, the masses (as well as all parameters derived from those masses) of the unstable particle fields in the original Lagrangian should be redefined and renormalised in terms of these complex poles
\begin{eqnarray}
M_0^2 \to \tilde M^2&=&\bar{M}^2-i\bar{\Gamma}\bar{M}\ .
\end{eqnarray}
We refer the interested reader to section 5 of ref.~\cite{Frederix:2018nkq} for details.

Alternatively, eq.~(\ref{eq:propogator}) also suggests another way to regularise the propagator of an unstable particle by using
\begin{eqnarray}
G_R(p^2)&=&-\frac{i}{p^2-M^2+i\Gamma M},
\end{eqnarray}
where $M$ is the (renormalised) on-shell mass, and $\Gamma$ is the total decay width given by 
\begin{eqnarray}
\Im{\left(\Sigma(p^2=M^2)\right)}&=&\Gamma M\ ,
\end{eqnarray}
as stemming from the optical theorem. This approach amounts to (Dyson) summing only the first term in the Taylor series of the self-energy function around $p^2=M^2$. Such a new propagator leads to Breit-Wigner (BW) forms after squaring the amplitudes
\begin{eqnarray}
{\rm BW}(p^2)&=&\frac{1}{\left(p^2-M^2\right)^2+\Gamma^2 M^2}\ ,
\end{eqnarray}
which prevents the appearance of divergences in cross sections in the vicinity of $p^2=M^2$. A BW form admits the expansion in terms of distributions~\cite{Frederix:2018nkq},
\begin{eqnarray}
{\rm BW}(p^2)&=&\frac{\pi}{\Gamma M}\delta(p^2-M^2)+\mathcal{P}\left(\frac{1}{\left(p^2-M^2\right)^2}\right)+\mathcal{O}\left(\frac{\Gamma}{M}\right)\ ,\label{eq:BWexp}
\end{eqnarray}
where the first term introduces a Dirac delta function, and the $\mathcal{P}$ operator in the second term is the principal-value operator. A convenient strategy for calculating cross sections is to keep the first term only in eq.~(\ref{eq:BWexp}), which is usually referred to as the \textit{narrow width approximation} (NWA). Such an approximation is gauge invariant, and should work well at higher orders. The advantage of using the NWA is that a process which undergoes a long decay chain can be factorised into several shorter sub-processes, because in the limit of narrow width, the time scales related to the different sub-processes are quite distinct. Schematically, let us consider a process $ab\to c\ e\ f$ with the two particles $e$ and $f$ originating from the decay of a resonance $X$. Its cross section can be effectively rewritten as
\begin{eqnarray}
 \sigma_{a b\to c e f}& \simeq&  \sigma_{a b  \to c X} \times {\rm BR}_{X\to e f}\ ,
\end{eqnarray}
where ${\rm Br}_{X\to ef}$ is the branching fraction associated with the $X\to ef$ decay. The NWA is expected to hold if several conditions are fulfilled~\cite{Berdine:2007uv}. First, the resonance must yield a narrow mass peak. Second,
kinematical conditions implying that both the production and decay sub-processes
are allowed and occur far from threshold must be realised. Finally, the
internal propagator has to be separable from the matrix element (which is not
generally possible at the loop level), and interferences with any other
resonant or non-resonant diagram contribution must be negligible.

In all of the above approaches, we have kept the self-energy function $\Sigma(p^2)$ at constant values of the virtuality $p^2$ in the propagators. The non-trivial virtuality dependence in $\Sigma(p^2)$ could certainly lead to some numerical significance, in particular when widths are large enough relatively to masses. A classical example is the $Z$-boson line-shape analysis with an energy-dependent width~\cite{Bardin:1988xt}, in which one only maintains the $p^2$ dependence in the imaginary part of $\Sigma(p^2)$, while its real part is still evaluated at $p^2=M^2$. It corresponds to introduce a running width $\Gamma(p^2)$ via
\begin{eqnarray}
\label{eq:runningwidth}
\Im{\left(\Sigma(p^2)\right)}&=&\Gamma(p^2)M\ .
\end{eqnarray}
Although it is still unclear how such a \textit{running width scheme} works out beyond lowest order, the implementation at leading order is straightforward, albeit with potential gauge violation issues. In a full theory, $\Sigma(p^2)$ is calculable from first principles. However, in our simplified model case, we will assume an ansatz for $\Gamma(p^2)$ which follows the $Z$-boson lineshape~\cite{Bardin:1988xt}, {\it i.e.}
\begin{eqnarray}
\Im{\left(\Sigma(p^2)\right)}&\simeq& \frac{p^2}{M^2}\Gamma M.
\end{eqnarray}
For the purpose of assessing theoretical uncertainties inherent to the finite width effects, we believe it is sufficient to use the linear-$p^2$ dependent form of the running width $\Gamma(p^2)$.

In this work, we consider the single production of a vector-like quark $T$ whose
width-over-mass ratio can be large. We investigate the corresponding
phenomenology by relying not only on the NWA, but also on the complex mass
scheme together with the running width scheme in order to account for the impact of the finite width of the unstable particle
$T$~\cite{Denner:1999gp,Denner:2005fg}. We moreover design a strategy (see
section~\ref{sec:nlo} for more details)
to obtain results that are accurate at the next-to-leading order in the strong
coupling $\alpha_s$ regardless of the width of the particle, extending an earlier
study focusing on the narrow resonance case~\cite{Cacciapaglia:2018qep}.

\subsection{Setting the range of the vector-like quark width-over-mass ratio}
\label{sec:widthrange}

The width of a VLQ can become large, but how much can that be? From a theoretical point of view the width of a particle is not limited from above in any precise way. It can reach very large values (including values larger than the particle mass) depending on the values of the couplings and on the number of interactions of the particle. However at very large values of the width the description in term of a particle can become questionable for various reasons. 

The total decay width of a particle, and specifically of a VLQ for the purposes of this analysis, can become large in two ways: 1) assuming its decay channels are exclusively the SM ones, larger couplings then lead to larger widths; 2) the VLQ can decay to other final states besides the SM ones, implying the presence of new processes.

In the first case the main limitation for a phenomenological analysis is the requirement that couplings stay within the perturbative limit, so that our perturbative treatment remains valid when truncating away higher-order contributions. However, the main constraints in the large coupling scenarios come from different observables. In the case of a simplified model where the SM is augmented only with one VLQ, its width is strongly limited by electroweak precision data and flavour observables, regardless of its mixing with the SM quarks~\cite{Moretti:2016gkr,Chen:2017hak}. It would be possible to evade such constraints without introducing further decay channels for the VLQ if other, potentially heavier, VLQs are introduced and allowed to modify the mixing patterns. This option is justified from a theoretical point of view, as for example in realistic composite models, the number of VLQ multiplets is not necessarily limited to one. The presence of further multiplets can induce cancellations of effects which can potentially relax the constraints in regions where the VLQ couplings can be large enough to lead to non-narrow widths~\cite{Cacciapaglia:2015ixa,Cacciapaglia:2018lld}.

In the second case, the strongest assumption to make is that the further (unspecified) decay channels are not contributing to the signal in the SM final states through chain decays of the new particles. This assumption becomes stronger and stronger as the number of decay channels increase.

For these reasons, the approach we follow in this analysis is to limit the width-over-mass ratio to 50\%. Larger values would indeed be allowed, but the reliability of the interpretation would probably become questionable. Furthermore, the treatment of states with large width requires to assess the dependence of the results on the schemes described in the previous section, and such dependence is likely stronger as the width of the particle increases.

%%%%%%%%%%%%%%%%%%%%%%%%%%%%%%%%%%%%%%%%%%%%%%%%%%%%%%%%%%%%%%%%%%%%%%%%%%%%%%

\section{Predictions at the leading-order accuracy in QCD}
\label{sec:lo}
\begin{figure}[t!]
  \centering
  \includegraphics[width=.6\textwidth]{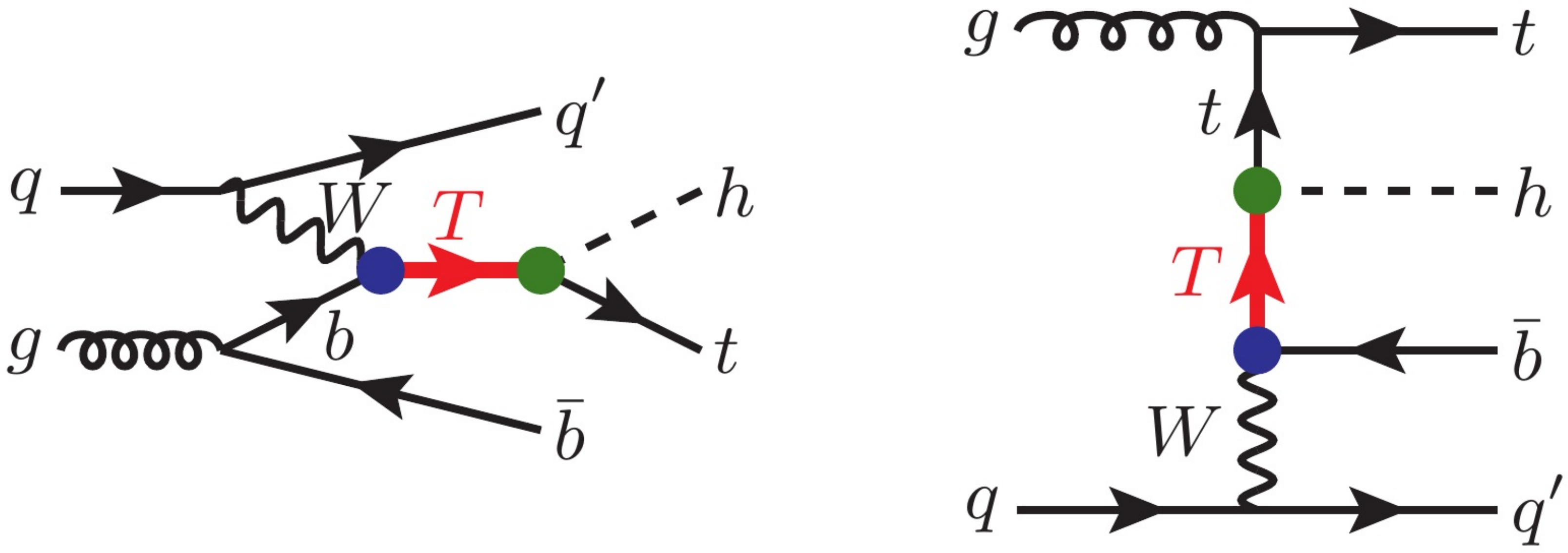}\vspace*{-0.2cm}
  \caption{\label{fig:Ht_res_tch_topologies} Representative Feynman diagrams for
    the $pp\to ht + X$ process in the four-flavour-number scheme. We illustrate
    both resonant $s$-channel (left) and non-resonant $t$-channel (right)
    exchange topologies, when induced by a heavy vector-like quark $T$.}
\end{figure}

We focus on the associated production of a single top quark $t$ or antiquark
$\bar t$ with a neutral SM Higgs or $Z$-boson,
\be
  p p \to ht + X \ (pp\to h \bar t + X) \qquad\text{and}\qquad 
  p p \to Zt + X \ (pp\to Z \bar t + X)\ ,
\label{eq:sgl1}\ee
as well as on the corresponding channel in which a $W$-boson is produced,
\be
  p p \to W^+b + X \ (pp\to W^- \bar b + X)\ .
\label{eq:sgl2}\ee
To stress which particles are propagating in a specific process, we label the processes by fully specifying the final state, the VLQ propagating in the topology and the SM gauge boson it interacts with to be produced. More precisely, we consider the following processes.
\begin{itemize}
\item $pp\to\{W,T\}\to Wbbj, Ztbj, htbj$ correspond to processes where the $T$ quark is singly produced in association with a jet, via its interaction with the $W$-boson and the bottom quark. The specification of the entire final state then allows for the explicit identification of the relevant VLQ decay channel. The propagation of the $W$-boson is also reflected by the presence of the bottom quark in the final state, arising from gluon splitting. 
\item $pp\to\{Z,T\}\to Wbtj, Zttj, httj$ correspond to processes where the $T$ quark is produced via its interaction with the $Z$-boson and the top quark, which is analogously reflected by the presence of the final-state top quark. 
\end{itemize}
This notation is redundant as we treat all processes in the four-flavour-number scheme, \ie\ without any initial $b$ quarks. The set of final state particles indeed includes the VLQ decay products, so that it would be already uniquely determined by the considered VLQ interactions. However, we keep this too detailed labeling for clarity.

In the vector-like quark model of section~\ref{sec:model}, new physics
contributions to the considered processes arise both from the $s$-channel resonant production of a heavy
quark $T$ that further decays into a $ht$, $Zt$  or $Wb$ system (together with
jets), as well as from non-resonant $t$-channel exchanges of the heavy quark.
As an illustration, representative leading-order (LO) Feynman diagrams for the
$pp\to ht$ process are shown in figure~\ref{fig:Ht_res_tch_topologies} for the
two classes of contributions, assuming four active quark flavours. Similar diagrams can be obtained for the other processes under consideration, with the Higgs boson being replaced by the relevant boson, and all internal and final-state top quarks being replaced by bottom
quarks in the case of $Wb+X$ production and $W$-boson-mediated VLQ production.

\subsection{Comparison between different schemes to treat the vector-like quark (large) width}\label{subsec:schemecomparison}

\begin{figure}[t]
  \centering
  \includegraphics[width=.45\textwidth]{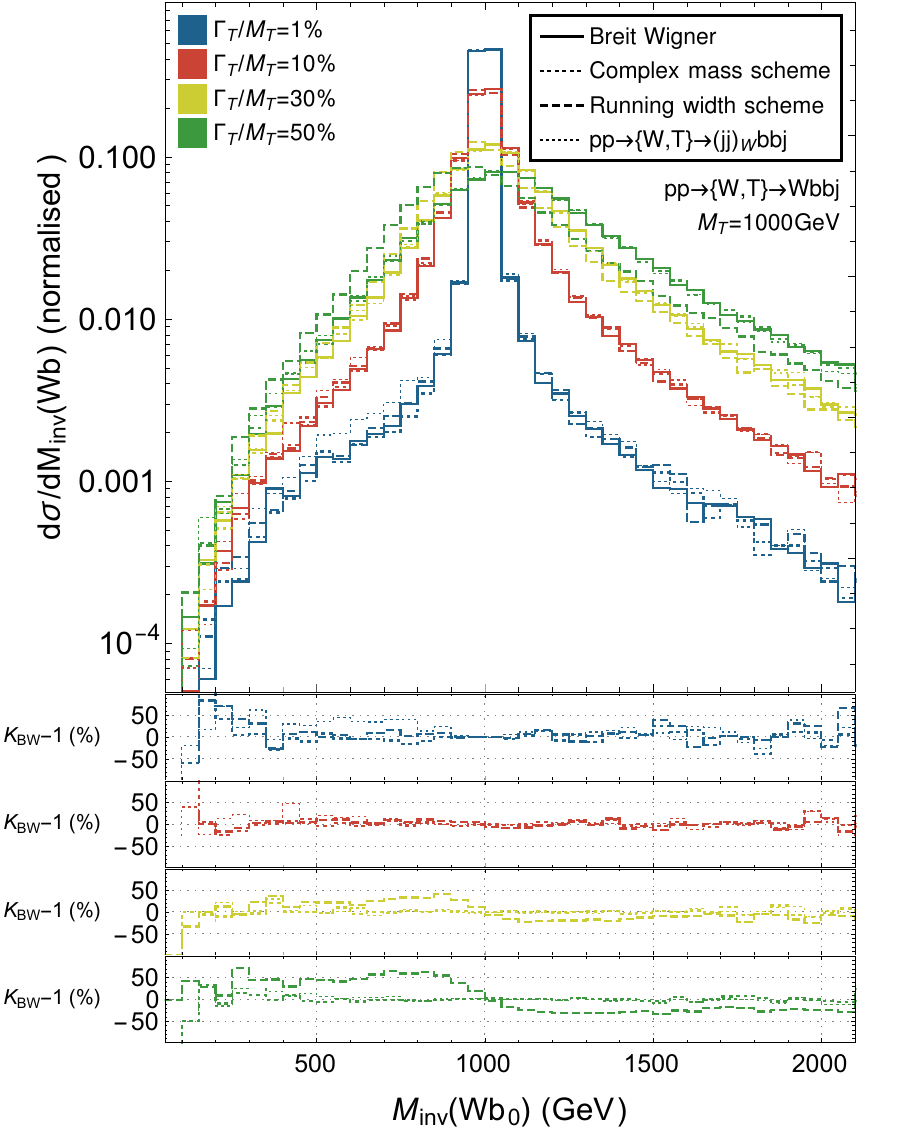}
  \includegraphics[width=.45\textwidth]{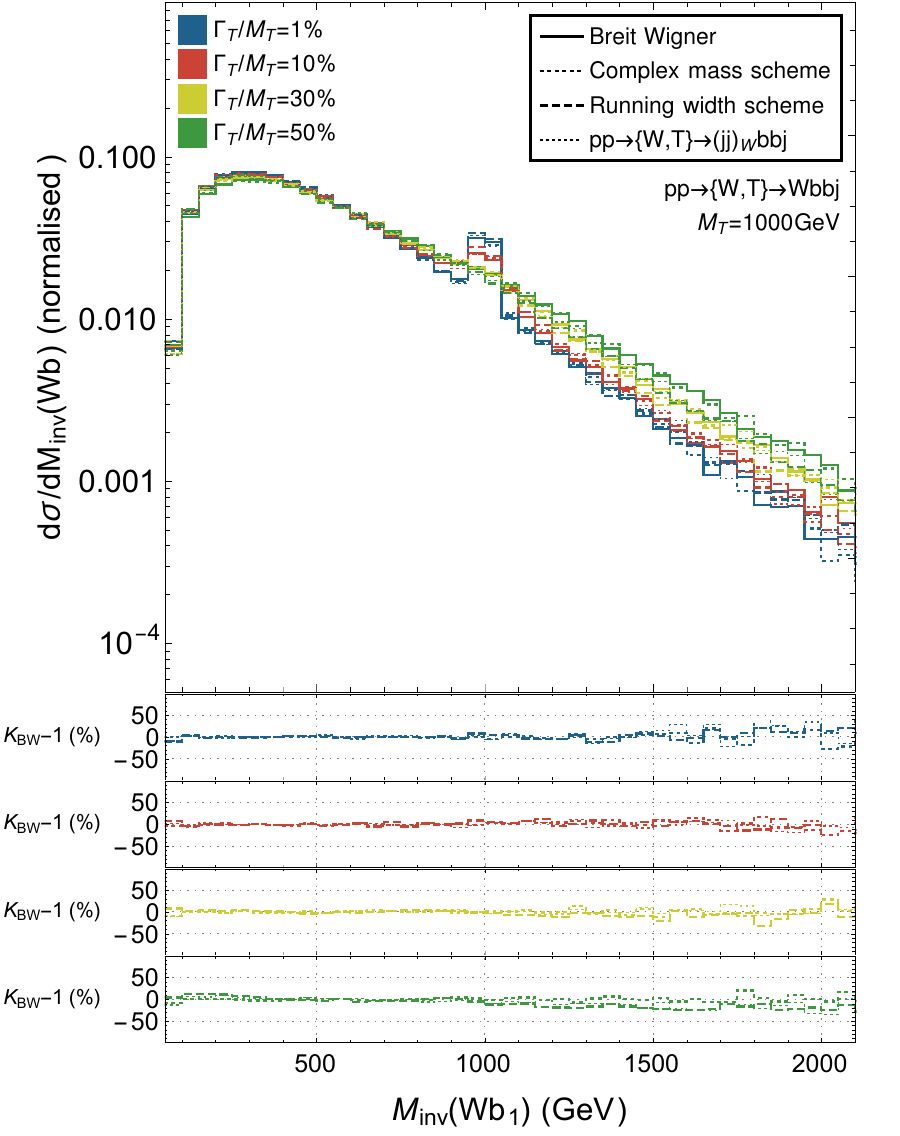}
  \\
  \includegraphics[width=.45\textwidth]{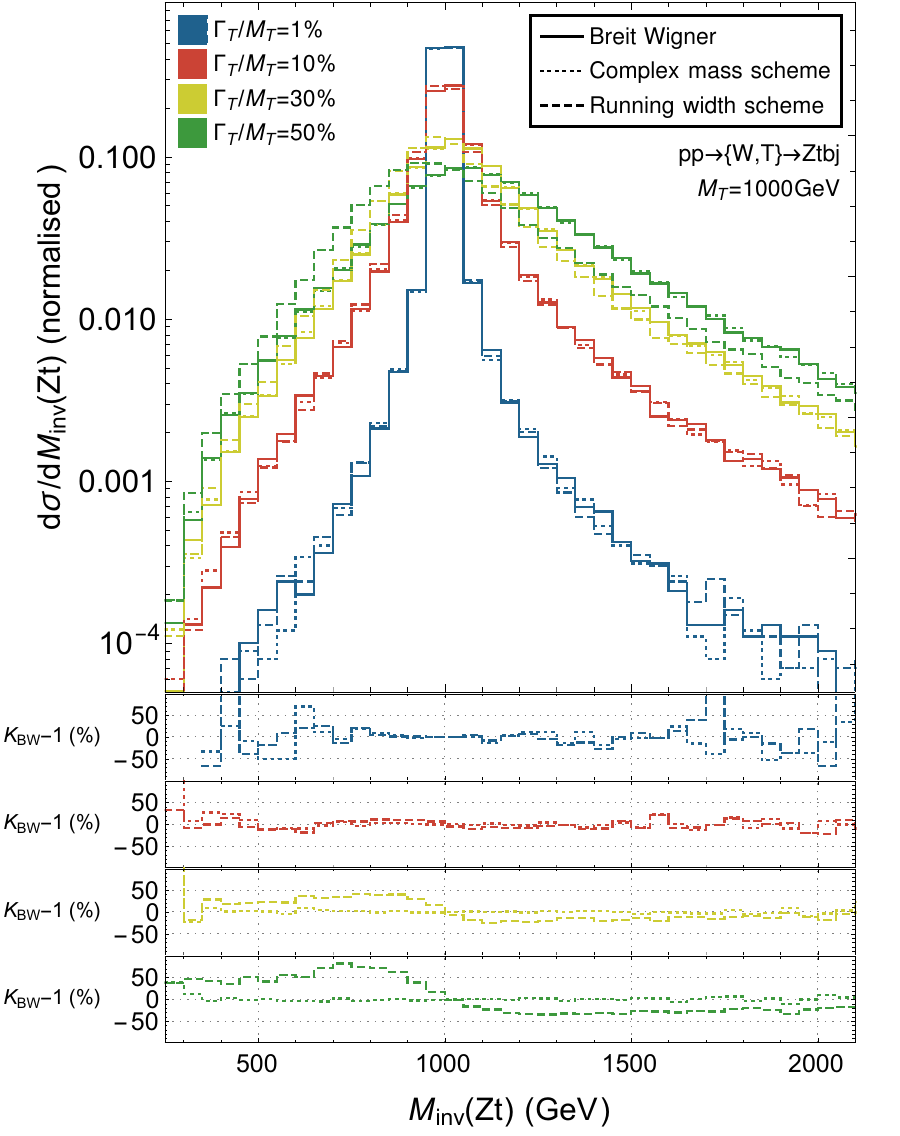}
  \includegraphics[width=.45\textwidth]{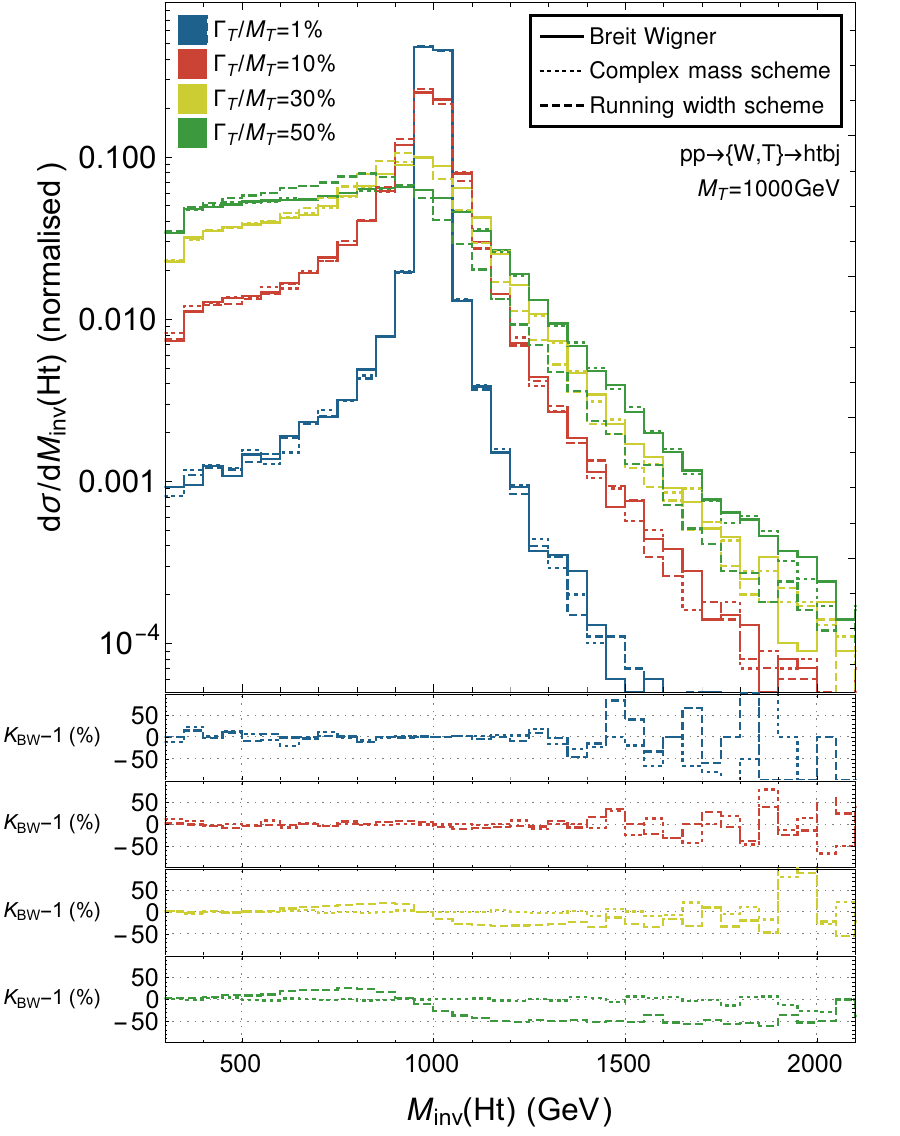}\vspace*{-0.2cm}
  \caption{\label{fig:schemecomparison} Comparison between different schemes to treat the VLQ width, for scenarios featuring $M_T=1000\GeV$ and for different processes: $pp\to\{W,T\}\to Wbbj$ (first row), $pp\to\{W,T\}\to Ztbj$ and $pp\to\{W,T\}\to htbj$ (second row). For each plot the upper panel shows normalised VLQ invariant-mass distributions at parton level, for different schemes and for different width-over-mass ratios. The lower panels provide the ratio between each scheme and the Breit-Wigner (BW) scheme, $K_\text{BW}-1={\text{scheme}-\text{BW}\over\text{BW}} (\%)$. For the process $pp\to\{W,T\}\to Wbbj$ two distributions are considered. They correspond to considering the leading $b$-jet (upper left) or the sub-leading $b$-jet (upper right) as part of the $T$ decay products. Still, only for the $Wbbj$ final state, the process in which the $W$-boson decays into jets is also considered (in the complex mass scheme), and the $W$-boson is there reconstructed from the pair of final-state light jets. For analogous results for $T$ production via $Zt$ interactions, we refer to figure~\ref{fig:schemecomparisonZT} in \cref{app:TviaZ}.}
\end{figure}

We are interested in scenarios featuring a vector-like quark $T$ with a large width. One of the leading systematic theoretical errors on the predictions could therefore stem from how we treat the unstable particle $T$ in the amplitudes, as discussed already in section \ref{sec:width}. The most obvious distribution useful to assess such an error is the invariant mass of the $T$ decay products. Examples of such comparisons can be found in figure~\ref{fig:schemecomparison} for the three considered processes. We do not include in the figures the invariant mass distribution in the NWA, as it consists of a pure Dirac delta function located at the pole mass $M_T$. We consider instead four different finite-width schemes at LO, that are summarised as follows.
\begin{itemize}
\item \textbf{Breit Wigner}. Within this scheme, we replace the $T$ propagator in the amplitude by
\begin{equation}
  \frac{i\left(\slashed{p}+M_T\right)}{p^2-M_T^2} \rightarrow \frac{i\left(\slashed{p}+M_T\right)}{p^2-M_T^2+i\Gamma_TM_T}.
\end{equation}
The introduction of the finite width in the denominator regulates the amplitude from any divergence at $p^2=M_T^2$, but violates gauge invariance. The results in this scheme are shown as solid curves in figure~\ref{fig:schemecomparison}.
\item \textbf{Complex mass scheme}. In this scheme, which has been described in section \ref{sec:width}, the corresponding distributions are displayed as thick dotted curves in figure~\ref{fig:schemecomparison}. We identify the mass and the width (that contribute to the complex mass) as in other schemes, using thus the pole mass and width.
\item \textbf{Complex mass scheme including the effects of the $W$-boson width}. The results are represented by thin dotted lines in figure~\ref{fig:schemecomparison}. Such a scheme has only been considered for the process $pp\to\{W,T\}\to (jj)_Wbbj$. Such a higher multiplicity in the final state allows us to account for the finite $W$-boson width, which has in contrast to be set to zero when the process is considered in the complex mass scheme with a final-state $W$-boson. 
\item \textbf{Running width scheme}. Such a scheme has been discussed in section \ref{sec:width}. The results are reported as dashed curves in figure~\ref{fig:schemecomparison}, for which only a linear $p^2$-dependence for the running width has been considered.
\end{itemize}

Figure~\ref{fig:schemecomparison} shows the heavy quark invariant mass distributions in the four considered width schemes, for $M_T=1$~TeV and for four different width-over-mass ratios of $\Gamma_T/M_T=1\%, 10\%, 30\%$ and $50\%$. The predictions are obtained at parton level with \mg, after implementing the non-standard propagators as detailed in ref.~\cite{Christensen:2013aua} and explicitly described in \cref{app:Tech}. All distributions have been normalised to unity. For the $pp\to\{W,T\}\to Wbbj$ process there is an ambiguity when reconstructing the invariant mass of the $T$ quark due to the presence of two bottom quarks in the final state. We consider both cases, using the leading-$p_T$ $b$-jet in the upper left figure and the sub-leading-$p_T$ $b$-jet in the upper right figure. We recall that what we refer to as $b$-jets are parton-level $b$-jets ({\it i.e.}\ bottom quarks and antiquarks), no realistic jet-reconstruction being performed. This allows us to focus on assessing the pure width scheme dependence of the results. For $Zt$ (lower left) and $ht$ (lower right) production, there is of course no such an ambiguity.

As anticipated, the differences between the schemes increase with $\Gamma_T/M_T$. While it is true that the results in the Breit Wigner scheme and in the complex mass scheme are quite close in general, running width effects become quite visible when $\Gamma_T/M_T=30\%$ and $50\%$. The energy-dependent width shifts the peaks of the $T$ lineshapes by a quantity $\Delta$ such that
\begin{equation}
    \Delta \sim M_T\left(1-\left(1+\Gamma_T^2/M_T^2\right)^{-1/2}\right).
    \end{equation}
This amounts to $106$ and $42$ GeV for the $50\%$ and $30\%$ cases, respectively. Such a shift is not surprising, as it has already been observed before in $Z$-lineshape studies~\cite{Bardin:1988xt}. Besides the shift, the running width contributions also distort the shapes of the invariant mass distributions, in particular in the $ht$ channel. This can be attributed to different behaviours in the low invariant mass region of the $ht$ system when compared to the $Zt$ and $Wb$ cases, as will be explained in detail in section~\ref{subsec:parton_ME}. On average, running width effects yield up to more than 50\% deviations with respect to the complex mass scheme when $\Gamma_T/M_T=30\%$ and $50\%$. They therefore represent a major source of systematic errors to account for when searching for large-width vector-like quarks at colliders.

The comparisons performed in figure~\ref{fig:schemecomparison} focus on production channels involving a $W$-boson, which dominate unless the $WTb$ coupling $\kappa$ is suppressed. An analogous comparison is performed when production involves $Z$-boson exchanges in figure~\ref{fig:schemecomparisonZT} in \cref{app:TviaZ}.

\subsection{Parton-level analysis of the signal}\label{sec:signal}

\subsubsection{Resonant and non-resonant signal contributions}\label{subsec:signal}

As mentioned above, both $s$-channel and $t$-channel diagrams contribute to the three 2-to-4 processes under consideration. The $t$-channel diagrams yield a subdominant contribution to the cross section in the case of a narrow $T$ resonance. Their contribution is, however, mildly dependent on the width of the vector-like quark $\Gamma_T$. Their relative
impact increases with increasing $\Gamma_T$ values, by virtue of the
strong dependence of the $s$-channel contributions on $\Gamma_T$.
For a small $T$-quark width, the $s$-channel component of the cross section
dominates, the NWA holds and the decay and production sub-processes can be
factorised. Moreover, the invariant mass distribution of the system of particles to which the
vector-like quark decays has a narrow peak at $M_T$. For increasing width, the peak widens, 
and the factorisation of the decay and production sub-processes
breaks down. The interference with other non-resonant diagrams becomes non-negligible, and needs to be included.

To quantify these statements, we simulate
the three signals of eqs.~\eqref{eq:sgl1} and \eqref{eq:sgl2} using \mg, and evaluate the
total production cross section for various vector-like quark masses $M_T$ and
width-over-mass ratios $\Gamma_T/M_T$. We focus on a light ($M_T= 1$~TeV) and
heavy ($M_T=2$~TeV) scenario, and adjust the couplings to recover both a
specific $\Gamma_T/M_T$ ratio (of 1\%, 10\% and 30\% respectively) and the
branching ratio relation
\be
 {\rm BR} (T\to W b) = 2\ {\rm BR} (T\to Z t) = 2\ {\rm BR} (T\to h t) = 0.5 \ .
\label{eq:brrelation}\ee
This choice of a $2\!:\!1\!:\!1$ relative magnitude for the three branching
ratios is motivated by the results obtained in the asymptotic limit $M_T\to
\infty$ for a vector-like quark lying in the ${\bf 1}_{2/3}$ representation of
the electroweak group, for which the Goldstone equivalence theorem dictates the
$2\!:\!1\!:\!1$ ratio.

\begin{table}
 \centering
 \resizebox{\columnwidth}{!}{
  \begin{tabular}{cc||ccc|ccc|ccc}
  \multirow{2}{*}{$M_T$}& \multirow{2}{*}{$\Gamma_T/M_T$} &
    \multicolumn{3}{c|}{$pp\to Wbbj$}& \multicolumn{3}{c|}{$pp\to Ztbj$}
     &\multicolumn{3}{c}{$pp\to htbj$}\\
    && $\sigma_{\rm tot}$ [fb] & $\sigma_s$ [fb] & $\sigma_t$ [fb]
     & $\sigma_{\rm tot}$ [fb] & $\sigma_s$ [fb] &$\sigma_t$ [fb]
     & $\sigma_{\rm tot}$ [fb] & $\sigma_s$ [fb] &$\sigma_t$ [fb]\\
  \hline\hline
   \multirow{3}{*}{1 TeV} & 1\% & 19.34 & 19.26 & 0.030
                                & 9.653 & 9.637 & 0.003
                                & 9.782 & 9.762 & 0.013 \\
                          &10\% & 187.1 & 184.0 & 3.004
                                & 92.33 & 91.82 & 0.305
                                & 104.5 & 103.6 & 1.294 \\
                          &30\% & 537.6 & 501.2 & 26.55
                                & 258.2 & 250.3 & 2.706
                                & 346.2 & 333.1 & 11.54 \\
   \hline
   \multirow{3}{*}{2 TeV} & 1\% & 0.316 & 0.316 & $\sim 0$
                                & 0.158 & 0.158 & $\sim 0$
                                & 0.169 & 0.169 & 0.001 \\
                          &10\% & 3.004 & 2.960 & 0.042
                                & 1.481 & 1.477 & 0.004
                                & 2.571 & 2.497 & 0.124 \\
                          &30\% & 8.189 & 7.691 & 0.365
                                & 3.930 & 3.823 & 0.039
                                & 12.71 & 11.90 & 1.094 \\
  \end{tabular}}\vspace*{-0.2cm}
  \caption{Total production cross sections for different vector-like quark
    masses and width-over-mass ratios in the complex mass scheme. The couplings are set to match the relation of
    eq.~\eqref{eq:brrelation}. We present results for proton-proton collisions
    at the LHC, at a centre-of-mass energy of 13~TeV and for the three processes
    of eqs.~\eqref{eq:sgl1} and \eqref{eq:sgl2}. For illustration purposes we separately show the individual resonant ($\sigma_s$) and
    non-resonant ($\sigma_t$) contributions, and the total physical
    rates ($\sigma_{\rm tot}$) which also include their interference.}
  \label{tab:totalrates}
\end{table}

We present in table~\ref{tab:totalrates} total cross section results for the
three processes under consideration in the case of LHC proton-proton collisions
at a centre-of-mass energy $\sqrt{s}=13$~TeV. We convolute the corresponding
hard-scattering matrix elements with the LO set of NNPDF 3.0 parton
densities~\cite{Ball:2014uwa}, handled  through the
\lhapdf\ package~\cite{Buckley:2014ana}. We additionally consider that the
bottom quark is massive ($m_b=4.7\GeV$), so that we rely on the four-flavour-number scheme
and include the bottom quark mass effects at the matrix-element level. Gauge
invariance is ensured (in particular in the large width case) through the use
of the complex mass scheme~\cite{Denner:1999gp,Denner:2005fg} described in \cref{sec:width}. The unphysical
factorisation and renormalisation scales have been set to half the sum of the
transverse masses of the final-state particles, on an event-by-event basis.

Mild kinematical cuts are used to limit the number of events populating regions of phase space which are likely to be excluded in experimental studies. Such cuts restrain the transverse momentum of the (parton-level) light and $b$-jets $(p_{Tj}$ and $p_{Tb}$) to be larger than 10 GeV, their pseudo-rapidity $\eta_j$ and $\eta_b$ to be below 5 (in absolute value) and that each jet is well separated in the transverse plane from each other by a $\Delta R$ distance of at least 0.1,
\begin{equation}
\label{eq:generationcuts}\begin{split}
  &p_{Tj}>10\GeV\ , \quad p_{Tb}>10\GeV\ ,\\[.2cm]
  &|\eta_j|<5\ ,\qquad\qquad |\eta_b|<5,\\[.2cm]
  &\Delta R_{jj}>0.1\ , \qquad \Delta R_{bb}>0.1\ , \qquad \Delta R_{jb}>0.1\;.
\end{split}\end{equation}
The cross section values are found to increase
with the width-over-mass ratio. This is driven by the $\kappa$, $\hat\kappa$ and
$\tilde\kappa$ couplings of the Lagrangian of eq.~\eqref{eq:lag} that need to be
larger to achieve a larger width without extending the model field content and
hence opening up new exotic decay channels.
More specifically, we have used
\be\bsp
  M_T = 1~{\rm TeV}, \ \ \Gamma_T/M_T = 1\%: \quad &
     \kappa_L = 0.123,\ \ \tilde\kappa_L = 0.129\ ,\ \ \hat\kappa_L = 0.510\ ,\\
  M_T = 1~{\rm TeV}, \ \ \Gamma_T/M_T = 10\%: \quad &
     \kappa_L = 0.390,\ \ \tilde\kappa_L = 0.408\ ,\ \ \hat\kappa_L = 1.612\ ,\\
  M_T = 1~{\rm TeV}, \ \ \Gamma_T/M_T = 30\%: \quad &
     \kappa_L = 0.676,\ \ \tilde\kappa_L = 0.707\ ,\ \ \hat\kappa_L = 2.792\ ,\\
  M_T = 2~{\rm TeV}, \ \ \Gamma_T/M_T = 1\%: \quad &
     \kappa_L = 0.062,\ \ \tilde\kappa_L = 0.062\ ,\ \ \hat\kappa_L = 0.503\ ,\\
  M_T = 2~{\rm TeV}, \ \ \Gamma_T/M_T = 10\%: \quad &
     \kappa_L = 0.195,\ \ \tilde\kappa_L = 0.197\ ,\ \ \hat\kappa_L = 1.592\ ,\\
  M_T = 2~{\rm TeV}, \ \ \Gamma_T/M_T = 30\%: \quad &
     \kappa_L = 0.338,\ \ \tilde\kappa_L = 0.342\ ,\ \ \hat\kappa_L = 2.757\ ,
\esp\label{eq:benchmarks}\ee
with all right-handed couplings $\kappa_R$, $\tilde\kappa_R$ and $\hat\kappa_R$
being fixed to 0. This configuration yields cross
sections that are roughly 10 and 30 times larger when the width-over-mass
ratio is fixed to 10\% and 30\% with respect to the narrow-width case (1\%) for
all three processes.

\begin{figure}
  \centering
  \includegraphics[width=.325\textwidth]{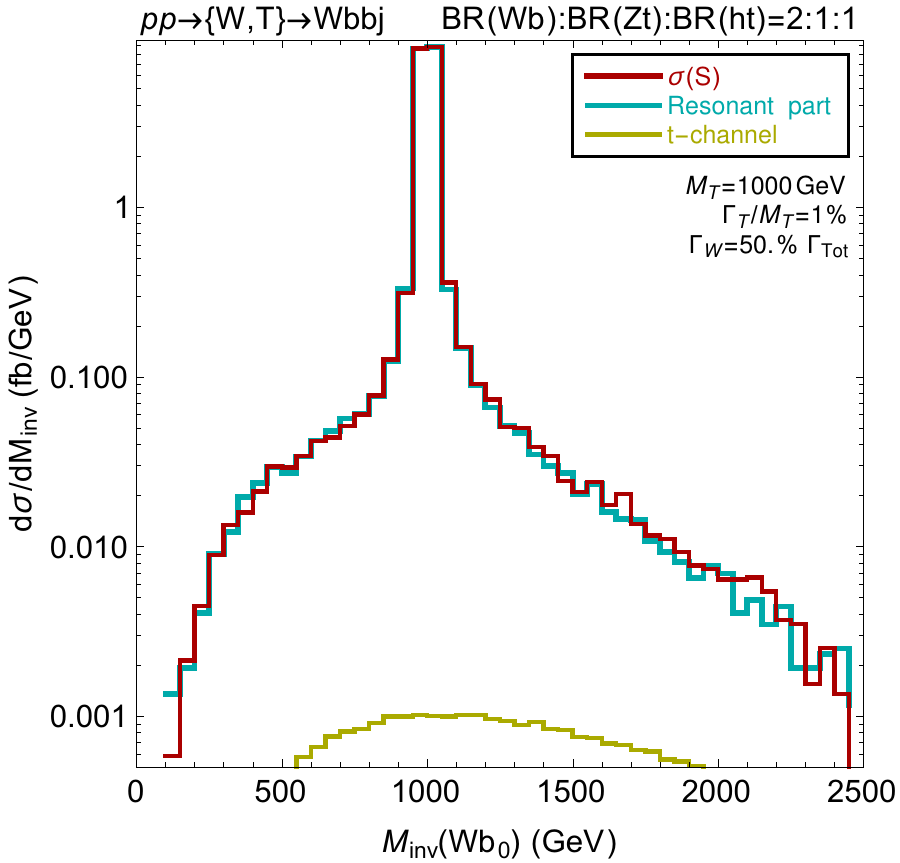}
  \includegraphics[width=.325\textwidth]{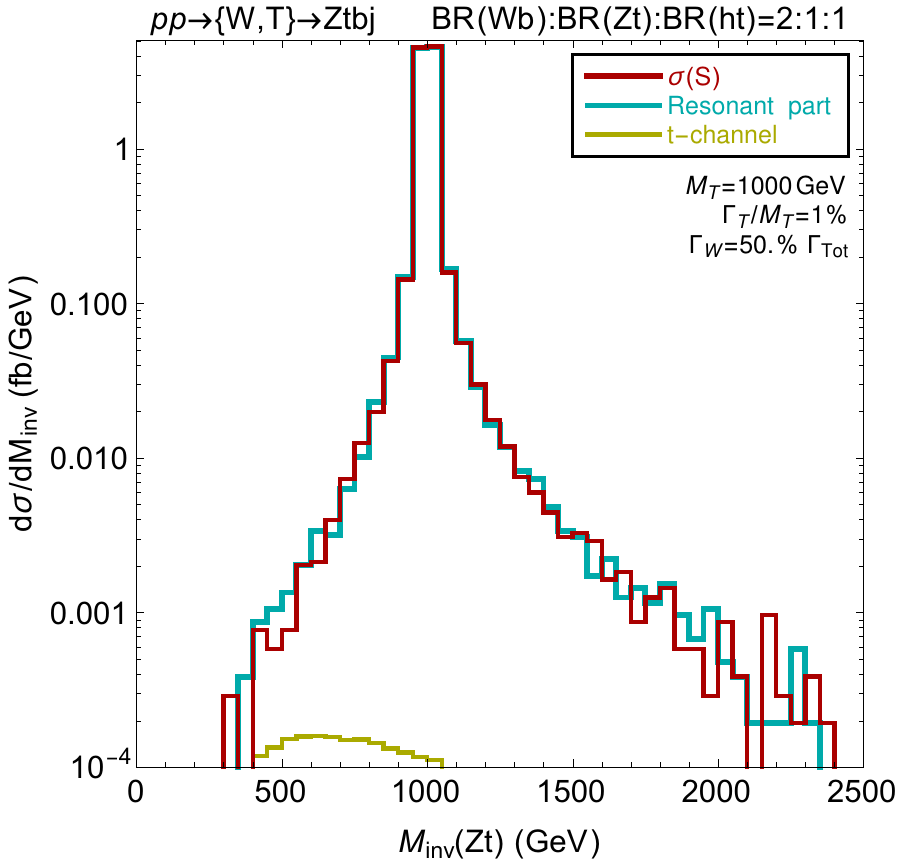}
  \includegraphics[width=.325\textwidth]{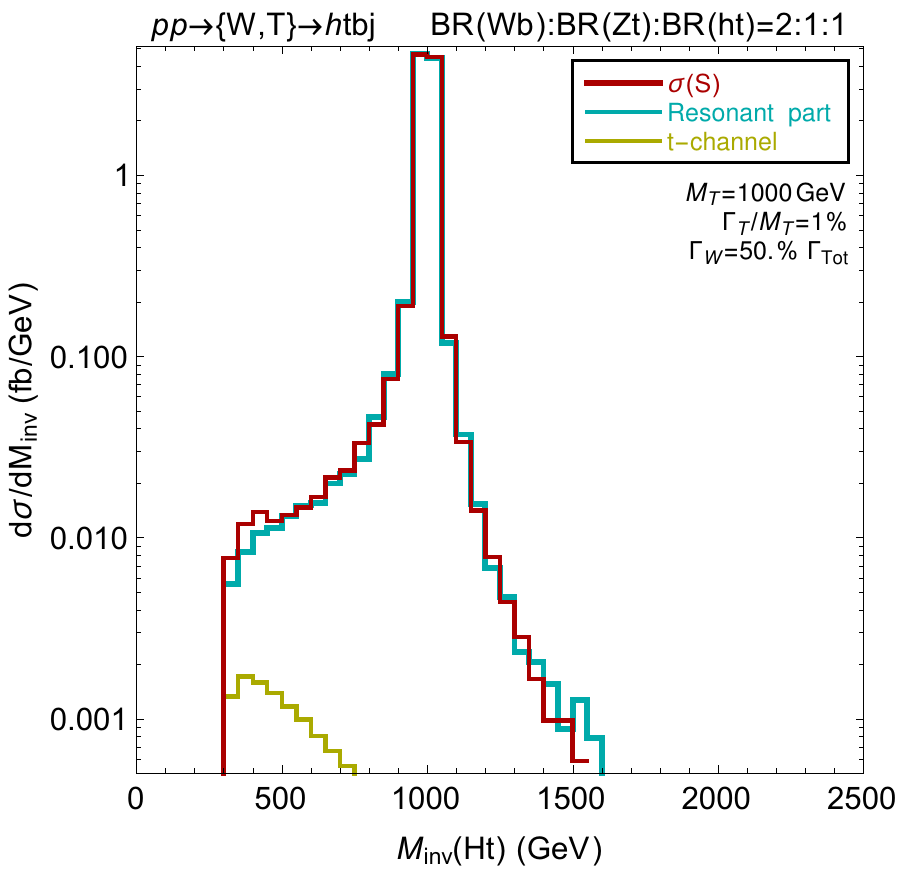}
  \\
  \includegraphics[width=.325\textwidth]{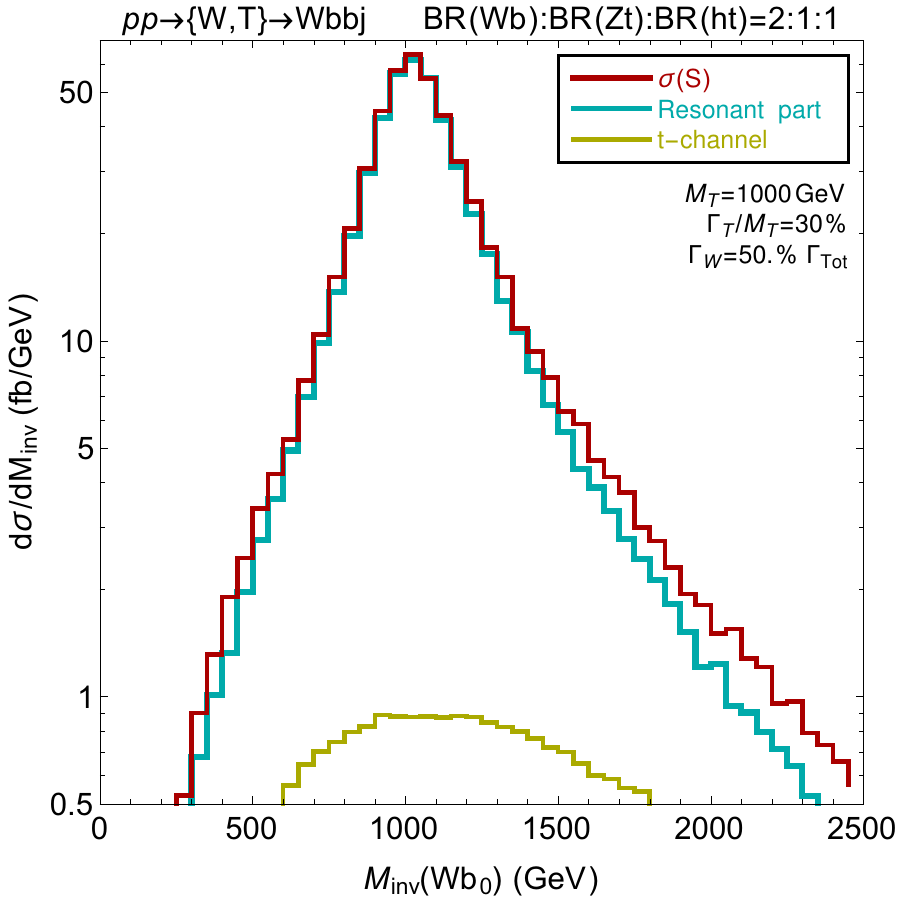}
  \includegraphics[width=.325\textwidth]{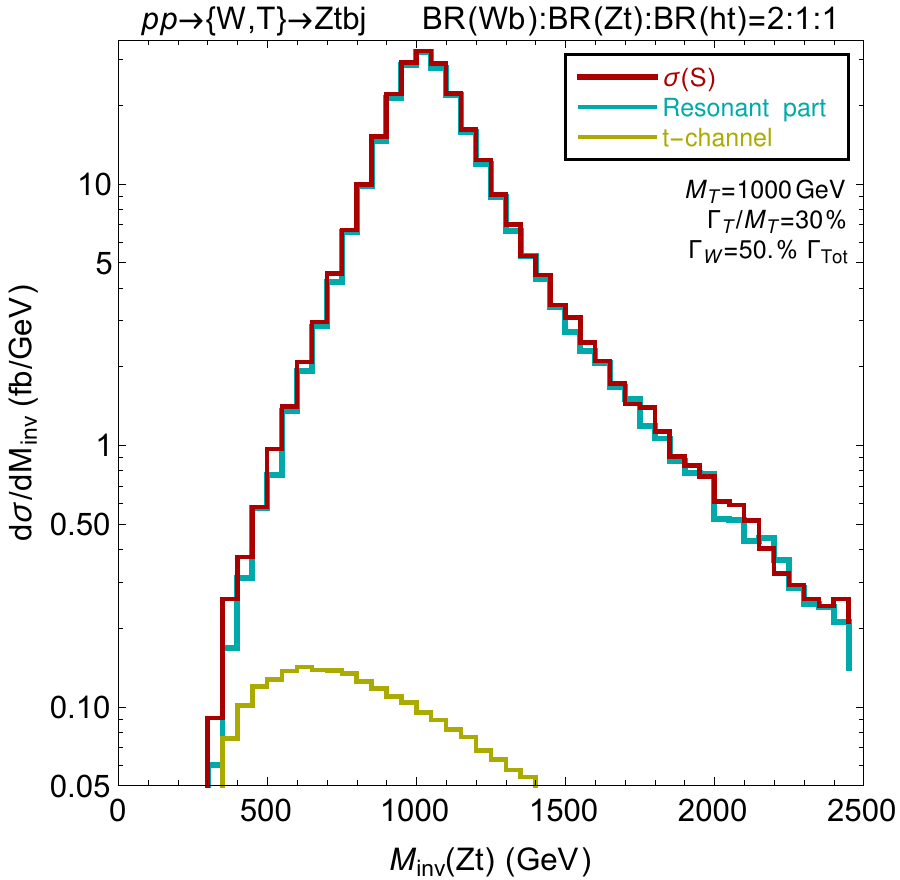}
  \includegraphics[width=.325\textwidth]{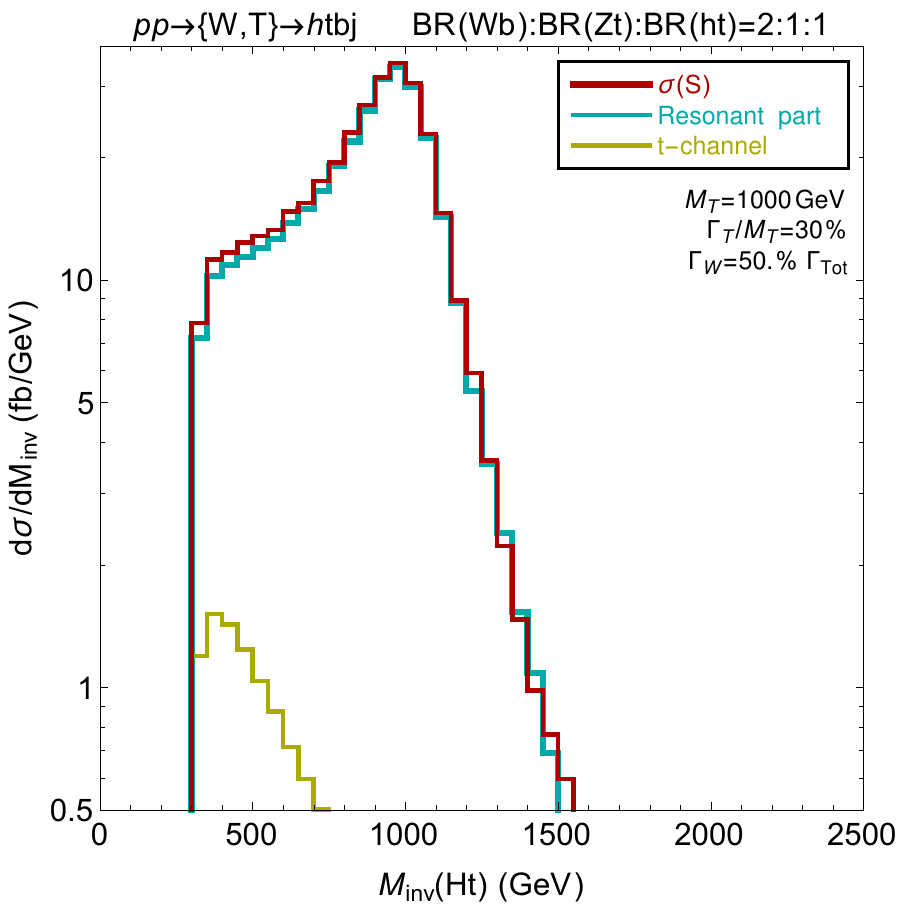}\vspace*{-0.2cm}
  \caption{\label{fig:res_tch_contributions} Individual contributions of the
    resonant
    $s$-channel (teal) and non-resonant $t$-channel (yellow) diagrams to
    $pp\to Wbbj$ (left), $pp\to Ztbj$ (centre) and $pp\to htbj$ (right)
    production, when mediated by the exchange of a vector-like quark $T$ of mass
    $M_T=1$~TeV in the narrow-width ($\Gamma_T/M_T=1\%$, top row) and
    large-width ($\Gamma_T/M_T=30\%$, bottom row) cases. We present
    distributions in the invariant mass of the $Wb$ (left), $Zt$ (centre) and
    $ht$ (right) systems, and indicate by red lines the full differential cross
    sections that include both the resonant and non-resonant components, as well
    as their interference. For all these results, the couplings are fixed
    as in eq.~\eqref{eq:benchmarks} in order to reproduce the branching ratio
    relation of eq.~\eqref{eq:brrelation}. For analogous results for a vector-like quark $T$ of mass
    $M_T=2$~TeV, we refer to figure~\ref{fig:res_tch_contributions_W_2000}, while figures~\ref{fig:res_tch_contributions_Z_1000} and \ref{fig:res_tch_contributions_Z_2000} focus on $T$ production through $Zt$ exchanges.}
\end{figure}

For a narrow-width configuration ($\Gamma_T/M_T = 1\%$), the largest cross
sections are associated with the $pp\to Wbbj$ process. The $\sigma_{\rm tot}(pp
\to Wbbj)$ cross section is found to be twice larger than for the $pp\to Ztbj$
and $pp\to htbj$ processes, the latter two cross sections being of a similar
size. This pattern directly stems from the branching ratio relation of
eq.~\eqref{eq:brrelation}, the coupling values of eq.~\eqref{eq:benchmarks} and
the available phase space that is reduced when a final-state top quark is
involved. Larger $\Gamma_T/M_T$ values impact this $2\!:\!1\!:\!1$ relation
between the cross sections. While the
relation $\sigma_{\rm tot}(pp \to Wbbj) = 2 \sigma_{\rm tot}(pp \to Ztbj)$ still
holds, the $\sigma_{\rm tot}(pp \to htbj)$ cross section is enhanced by a factor
of a few. This feature stems from the different Lorentz structures involved in
the dominant contribution to the corresponding amplitudes, which lead to a
different dependence of the partonic cross section on the mass and width of the
vector-like quark, as will be discussed in detail in section~\ref{subsec:parton_ME}. As can be seen from table~\ref{tab:totalrates}, the cross section is dominated by resonant $s$-channel diagram contributions $\sigma_s$.
While non-resonant contributions are not negligible, they only
account for a few percents of the cross section for
$\Gamma_T/M_T=10\%$ and 30\%. Restricting the analysis to the sole $s$-channel
pieces is therefore a good approximation to understand the leading large
width effects. This might not always be the case. For processes where the heavy $T$ quark is singly produced via its interaction with the $Z$-boson, the relevance of $t$-channel is indeed larger, as shown in \cref{app:TviaZ}.

Figure~\ref{fig:res_tch_contributions} further illustrates the relative
impact of the resonant $s$-channel and non-resonant $t$-channel contributions to
the vector-like-quark-induced production of a $Wb$, $Zt$ and $ht$ system. We
present three distributions in the corresponding invariant masses, namely the
$M_{\rm inv}(Wb)$, $M_{\rm inv}(Zt)$ and $M_{\rm inv}(ht)$ spectra in the left,
central and right panels of the figure. In this example, we consider
scenarios with $M_T=1$~TeV, and a width-over-mass ratio $\Gamma_T/M_T = 1\%$
(upper row) and 30\% (lower row).

Figure~\ref{fig:res_tch_contributions} also quantifies the expected dependence of the invariant mass distribution on $\Gamma_T/M_T$. We can firstly notice the direct impact of the
vector-like quark width on the broadness of the peak around $M_T$.
It is quite narrow for $\Gamma_T/M_T=1\%$ and much broader for $\Gamma_T/M_T =
30\%$. Secondly, as confirmed by
the total cross section results of table~\ref{tab:totalrates}, the $t$-channel
contribution is negligibly small when the width-over-mass ratio is small. In
this configuration, the NWA moreover holds, so that the full $2\to4$ cross
section can be approximated by
\be
  \sigma_{\rm tot}(p p \to Bq_3bj) \approx \sigma_s(p p \to Bq_3bj)
     \approx \sigma(p p \to Tbj) \times {\rm BR}(T\to Bq_3) \ ,
\label{eq:sigtotapprox}\ee
with $Bq_3 = Wb$, $Zt$ or $ht$. The results of the upper row in
figure~\ref{fig:res_tch_contributions} additionally demonstrate that for the
three considered processes, this approximation holds at the differential level
too, even when the $T$ quark is far off-shell, \ie\ for $|M_{\rm inv}(Bq_3)-M_T|
\gg \Gamma_T$. The bulk of the differential cross section is located
around $M_{\rm inv}(Bq_3)\sim M_T$, so that a standard parton-level simulation
making use of \mg\ for the hard process ($p p \to Tbj$), 
\ms~\cite{Artoisenet:2012st} and \mw~\cite{Alwall:2014bza} for the decay process
($T\to Bq_3$) so that both off-shell and spin correlation effects are retained
would be justified\footnote{We emphasise that in this section addressing LO predictions, such a factorised simulation chain is nowhere used. We always consider
the full $2\to 4$ process as the hard-scattering process, without any
approximation, and thus include both the $s$-channel and $t$-channel components
regardless the actual value of the vector-like quark width.}. For a
larger width-over-mass ratio $\Gamma_T/M_T = 30\%$, the slope of the $M_{\rm
inv}(Bq_3)$ distribution around $M_T$ is much milder, so that the entire
$M_{\rm inv}(Bq_3)$ range contributes significantly to the total cross section.
Therefore, although the first approximation in eq.~\eqref{eq:sigtotapprox} is
still valid at the level of a few percents both at the differential and total
cross section level (see also table~\ref{tab:totalrates}), the heavy quark
production and decay processes cannot be factorised anymore, \ie\ the last
approximation in eq.~\eqref{eq:sigtotapprox} does not hold anymore. The simulation of the full
$2\to4$ process should therefore be considered.

\subsubsection{Deciphering the resonant contribution to the signal}
\label{subsec:parton_ME}

Figure~\ref{fig:res_tch_contributions}  also demonstrates the qualitatively different behaviours of the invariant mass in the $Wb$ and $Zt$ system compared to the $ht$ system for both a narrow width ($\Gamma_T/M_T = 1\%$) and a broad width ($\Gamma_T/M_T = 30\%$). The $M_{\rm inv}(Wb)$ and $M_{\rm inv}(Zt)$ spectra are well-described by a Breit-Wigner distribution while the $M_{\rm inv}(ht)$ spectrum exhibits a larger asymmetry and in particular less of a decrease at invariant masses below the peak. To understand this maybe surprising feature in more detail, we study the resonant contribution to the signal at an analytical level. In this section (and in this section, only), we perform a number of approximations as detailed below.

Our previous findings show that it is reasonable to ignore the
$t$-channel contributions to the single-production cross section.\footnote{The only visible contribution of $t$-channel and its interference with the $s$-channel is in the tail of the $Wbbj$ process, which is however not likely to produce enough signal events to reconstruct a shape. This will be clarified in the next section after the SM background is included.} As suggested by the topology of the left diagram in
figure~\ref{fig:Ht_res_tch_topologies}, we will moreover rely, for the
computations in this section, on the effective $W$-boson approximation~\cite{
Dawson:1984gx,Kane:1984bb,Kunszt:1987tk} in which the $W$-boson is treated as a
constituent of the proton. Such an approximation is known to be sufficient to
build a succinct picture of the process dynamics when the relevant scales are
much larger than the $W$-boson mass. This allows us to focus on the $2\to 2$
partonic processes,
\be
  W(k_1) b(k_2) \to b(p_1) W(p_2)\ , \qquad
  W(k_1) b(k_2) \to t(p_1) Z(p_2)\ , \qquad
  W(k_1) b(k_2) \to t(p_1) h(p_2)\ ,
\ee
once the initial $g\to b\bar b$ splitting of the $2\to 4$ process has been
factorised out. We denote by $k_i$ and $p_i$ (with $i=1,2$) the initial-state
and final-state four-momenta respectively, and all processes proceed via a
$T$-exchange in the $s$-channel. The three amplitudes $i{\cal M}_{Wb}$,
$i{\cal M}_{Zt}$ and $i{\cal M}_{ht}$ are respectively given by
\be\bsp
  i{\cal M}_{Wb} = &\
     \frac{-ig^2}{2 s_T} \varepsilon_\mu(k_1)\varepsilon_\nu^*(p_2)
     \Big[\bar u(p_1) \gamma^\nu
     \big(\kappa_L P_L \!+\! \kappa_R P_R\big)
     \big(\slashed{k}_1\!+\!\slashed{k}_2\!+\!\tilde M_T\big)\gamma^\mu
    \big(\kappa_L P_L \!+\! \kappa_R P_R\big) v(k_2)\Big] \ ,\\
  i{\cal M}_{Zt} = &\
     \frac{-ig^2}{2\sqrt{2} c_W s_T} \varepsilon_\mu(k_1)\varepsilon_\nu^*(p_2)
     \Big[\bar u(p_1) \gamma^\nu
     \big(\tilde\kappa_L P_L \!+\! \tilde\kappa_R P_R\big)
     \big(\slashed{k}_1\!+\!\slashed{k}_2\!+\!\tilde M_T\big)\gamma^\mu
    \big(\kappa_L P_L \!+\! \kappa_R P_R\big) v(k_2)\Big] \ ,\\
  i{\cal M}_{ht} = &\
     \frac{-ig}{\sqrt{2} s_T} \varepsilon_\mu(k_1)
     \Big[\bar u(p_1) \big(\hat\kappa_R P_L \!+\! \hat\kappa_L P_R\big)
     \big(\slashed{k}_1\!+\!\slashed{k}_2\!+\!\tilde M_T\big)\gamma^\mu
    \big(\kappa_L P_L \!+\! \kappa_R P_R\big) v(k_2)\Big] \ ,
\esp\ee
after having introduced the reduced Mandelstam variable $s_T \equiv s -
\tilde M_T^2$. The couplings  $\kappa$, $\tilde\kappa$, $\hat\kappa$  refer respectively to the  
coupling of $Wb$, $Zt$ and $ht$ with the VLQ $T$.
The previous expressions are valid in the complex mass scheme,
in which the vector-like quark complex squared mass reads
\be \tilde M_T^2 = M_T^2 - i M_T \Gamma_T \ .\ee
The corresponding partonic cross sections are obtained by squaring those
amplitudes and integrating the results, multiplied by the flux factor, over the
phase space. After accounting for a summation over the final-state and an
average over the initial-state helicities and colour quantum numbers, we obtain
\be\bsp
  \sigma_{Wb} = &\ \frac{{\cal N}_{Wb}(s)}{|s_T|^2} \! \bigg[
    \big(\kappa_L^2\kappa_L^2\!+\!\kappa_R^2\kappa_R^2\big) \!+\!
    \big(\kappa_L^2\kappa_R^2\!+\!\kappa_R^2\kappa_L^2\big)
      \frac{\big|\tilde M_T^2\big|}{s} \!+\!
    \frac{12 M_t M_W^2 \Re\{\tilde M_T\}
        \kappa_L\kappa_R\big(\kappa_L^2+\kappa_R^2\big)}
        {s^2 \!+\! (M_W^2\!-\!2M_t^2) s \!+\! M_t^4 \!-\! 2 M_W^4 \!+\! 
        M_t^2 M_W^2}\bigg]\ ,\\
   \sigma_{Zt} = &\ \frac{{\cal N}_{Zt}(s)}{|s_T|^2}\bigg[
    \big(\tilde\kappa_L^2\kappa_L^2\!+\!\tilde\kappa_R^2\kappa_R^2\big) +
    \big(\tilde\kappa_L^2\kappa_R^2\!+\!\tilde\kappa_R^2\kappa_L^2\big)
      \frac{\big|\tilde M_T^2\big|}{s} +
    \frac{12 M_t M_Z^2 \Re\{\tilde M_T\}
        \tilde\kappa_L\tilde\kappa_R\big(\kappa_L^2+\kappa_R^2\big)}
        {s^2 \!+\! (M_Z^2\!-\!2M_t^2) s \!+\! M_t^4 \!-\! 2 M_Z^4 \!+\! 
        M_t^2 M_Z^2}\bigg]\ ,\\
  \sigma_{ht} = &\ \frac{{\cal N}_{ht}(s)}{|s_T|^2}\bigg[
    \big(\hat\kappa_L^2\kappa_L^2\!+\!\hat\kappa_R^2\kappa_R^2\big)
      \frac{\big|\tilde M_T^2\big|}{s} +
    \big(\hat\kappa_L^2\kappa_R^2\!+\!\hat\kappa_R^2\kappa_L^2\big) +
       \frac{4 M_t \Re\{\tilde M_T\}}{s-M_h^2+ M_t^2}
        \hat\kappa_L\hat\kappa_R \big(\kappa_L^2 \!+\! \kappa_R^2\big)\bigg] \ ,
\esp\ee
where the normalisation factors are given by
\be\bsp
  {\cal N}_{Wb}(s) = & \frac{g^4\big(s+2M_W^2\big)\big(s^2 \!+\! (M_W^2\!-\!
     2M_t^2) s \!+\! M_t^4\!-\! 2 M_W^4 \!+\!M_t^2 M_W^2\big)}
    {768\pi M_W^4 s }
    \sqrt{\lambda\big(s,M_t^2,M_W^2\big)}\ , \\
  {\cal N}_{Zt}(s) = & \frac{g^4\big(s+2M_W^2\big)\big(s^2 \!+\! (M_Z^2\!-\!
     2M_t^2) s \!+\! M_t^4\!-\! 2 M_Z^4 \!+\!M_t^2 M_Z^2\big)}
    {1536\pi c_W^2 M_W^2 M_Z^2 s }
    \sqrt{\lambda\big(s,M_t^2,M_Z^2\big)}\ , \\
  {\cal N}_{ht}(s) = & \frac{g^2\big(s+2M_W^2\big)\big(s-M_h^2+
     M_t^2\big)}{384\pi M_W^2 s }
     \sqrt{\lambda\big(s,M_t^2,M_h^2\big)}\ .
\esp\ee
In the expressions above, we have ignored the effects of the width of the top quark,
Higgs, $Z$- and $W$-bosons, whose respective masses $M_t$, $M_h$, $M_Z$ and $M_W$
are thus real. Moreover, the $b$-quark mass has been neglected all along the
calculations and $\lambda(x,y,z)$ denotes the usual K\"all\'en function. This
approximation is sufficient for the features which we aim to exhibit in this
section.

\begin{figure}
  \centering
  \includegraphics[width=.48\textwidth]{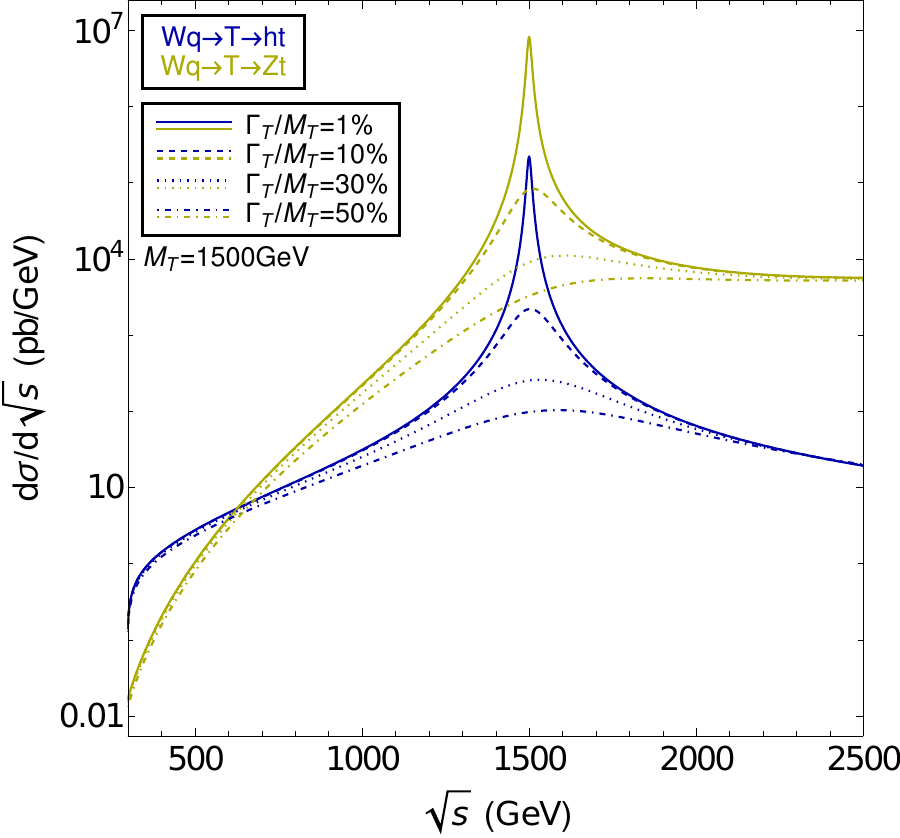}
  \includegraphics[width=.48\textwidth]{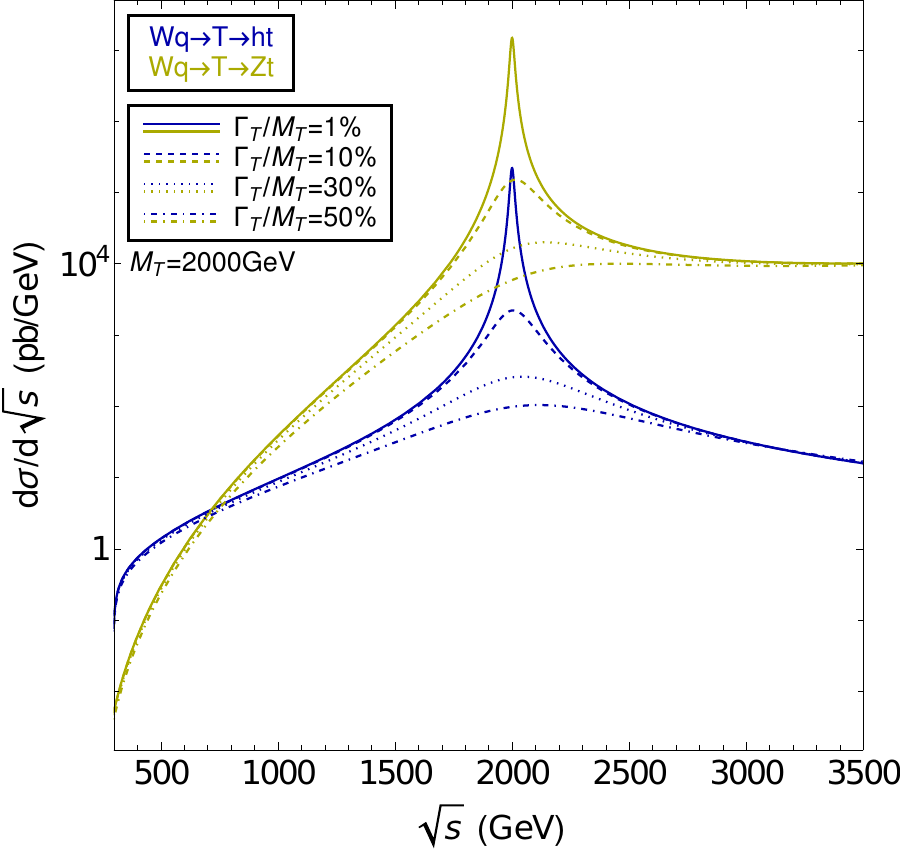}\vspace*{-0.2cm}
  \caption{\label{fig:feature_chiralities} Partonic cross section for the
    $Wq\to ht$ (blue) and $Wq \to Zt$ (yellow) processes, that both proceed via
    the $s$-channel exchange of a vector-like quark $T$. We depict the
    dependence of the cross section on the partonic centre-of-mass energy
    $\sqrt{s}$, for different values of the $\Gamma_T/M_T$ ratio. We
    consider vector-like quark setups with $M_T=1500\GeV$ (left panel) and
    2000~GeV (right panel), and chiralities of the $T$ couplings that are chosen
    left-handed and fixed as in eq.~\eqref{eq:benchmarks} to reproduce the
    branching ratio relation of eq.~\eqref{eq:brrelation}.}
\end{figure}
We observe a very different dependence of the cross sections on $\tilde M_T$
according to the spin quantum numbers of the final-state boson, arising from the
different Lorentz structures in the vector ($Wbbj$ or $Ztbj$ production) or
scalar ($htbj$ production) cases.  In concrete composite Higgs models, vector-like quarks in definite $SU(2)_L$ representations dominantly couple to the SM quarks through either the  left-handed ($\kappa_L$, $\tilde\kappa_L$, $\hat\kappa_L$) or right-handed ($\kappa_R$, $\tilde\kappa_R$, $\hat\kappa_R$) set of couplings, while the opposite-handed couplings are suppressed~\cite{delAguila:2000rc,Buchkremer:2013bha,Chen:2017hak}.
Realistic benchmark scenarios will thus feature large couplings of a given chirality, and
suppressed coupling of the other chirality. This yields
\be
   \sigma_{Wb} \sim \frac{{\cal N}_{Wb}(s)}{|s_T|^2} \ , \qquad
   \sigma_{Zt} \sim \frac{{\cal N}_{Zt}(s)}{|s_T|^2} \ , \qquad
   \sigma_{ht} \sim \frac{{\cal N}_{ht}(s)}{|s_T|^2}
      \frac{\big|\tilde M_T^2\big|}{s} \ ,
\label{eq:signal_behaviours}\ee
which shows that $ht$ production features a different dependence on the
vector-like quark mass and width. This is further illustrated in
\cref{fig:feature_chiralities}, in which we compare results for $Ztbj$
(yellow) and $htbj$ (blue) production, the $Wbbj$ channel that exhibits the same
behaviour as the $Ztbj$ channel being omitted. Following the example taken in
the beginning of this section, we set $\kappa_R=\tilde\kappa_R=\hat\kappa_R=0$
and present the dependence of the partonic cross section on the partonic
centre-of-mass energy $\sqrt{s}$ for two vector-like quark masses of $M_T =
1500$~GeV (left panel) and 2000~GeV (right panel). Four sets of curves are
included, for width-over-mass ratios of 1\% (solid), 10\% (dashed), 30\%
(dotted) and 50\% (dot-dashed).

We first recover the previous results, with the peak around $M_T$ becoming
broader with increasing values for the vector-like quark
width-over-mass ratio, merely flattening out in the extreme case of $\Gamma_T/M_T
= 50\%$. The especially interesting feature is, however, the impact of the
$\big|\tilde M_T^2\big|/s$ factor on the $htbj$ partonic cross section. Whereas
the $Ztbj$ production cross section is steeply falling for decreasing $\sqrt{s}$
values smaller than $M_T$, the $htbj$ production one plateaus until
the threshold value $\sqrt{s}\approx M_t+M_h$ is reached. In addition, for
$\sqrt{s}$ values larger than $M_T$, both cross sections present a smoothly
decreasing behaviour, the decrease being more pronounced in the $htbj$ case. For
both processes, the bulk of the integrated cross section is dominated by the
peak region in the narrow width case $\Gamma_T /M_T=1\%$ or quite-narrow-width
case $\Gamma_T /M_T=10\%$. Therefore, the phase space region defined by
$\sqrt{s}\not\in[M_T-n \Gamma_T, M_T+n \Gamma_T]$, with $n$ being equal to a
few, only yields sub-leading contributions for both processes. For broader widths, the situation changes, as the entire $\sqrt{s}$ range contributes for $htbj$, in contrast to $Ztbj$ where the partonic cross
section falls by several orders of magnitude for a decreasing $\sqrt{s}$-value. After
accounting for a convolution with the parton density functions, this results in
a dramatic relative increase of the $htbj$ cross section compared to the other
processes, as smaller $\sqrt{s}$ values are preferred in the parton density functions. This effect can yield, as shown in table~\ref{tab:totalrates},
$\sigma_{\rm tot}(pp\to htbj)$ being larger than $\sigma_{\rm tot}(pp\to Ztbj)$ by a factor of a few.

Even more importantly, this enhancement of the signal in the small $\sqrt{s}$
regime for a non-narrow vector-like quark plays a crucial role for the signal
and corresponding background modelling. Interference between the
new physics signal and the corresponding SM background (\ie\ the $pp\to Bq_3bj$
processes without any internal vector-like quark exchange) must be
accounted for, and cannot be neglected as is usually done in searches for
broad vector-like quarks at the LHC~\cite{Sirunyan:2017ynj,Sirunyan:2018fjh,
Sirunyan:2018ncp,Sirunyan:2019xeh}. Having large signal contributions in the
small $\sqrt{s}$ regime as shown in figure~\ref{fig:feature_chiralities}
significantly impacts the shape of distributions such as the $M_{\rm inv}(Bq_3)$
one, and can for instance lead to the apparition of a spurious secondary peak
driven by the parton densities. This is further detailed in the next
subsection.

\subsection{Parton-level analysis of the interfering signal and background}\label{subsec:plinterference}
As mentioned at the end of the previous section, vector-like quark contributions
to any of the considered processes cannot be taken independently of the
corresponding background contributions if the width of the vector-like quark is
large. Vector-like quark and SM-like diagrams can indeed interfere quite
substantially. In this section we consider again the processes of
eqs.~\eqref{eq:sgl1} and \eqref{eq:sgl2}, but this time by including both the
$t$-channel and $s$-channel new physics contributions studied in
section~\ref{subsec:signal}, and the SM diagrams yielding the same final state.
We present results for a subset of the benchmark scenarios introduced in
eq.~\eqref{eq:benchmarks}, namely those featuring a width-over-mass ratio
$\Gamma_T/M_T$ of 1\% (narrow configuration) and 30\% (broad configuration).

Our predictions are obtained by convoluting the full (in the signal plus
background sense) LO $2\to 4$ squared amplitudes for the $pp\to Wbbj$, $Ztbj$
and $htbj$ processes with the LO set of NNPDF~3.0 parton densities~\cite{
Ball:2014uwa} (handled through \lhapdf~\cite{Buckley:2014ana}), as performed by
the \mg\ event generator~\cite{Alwall:2014hca}. The same cuts of \cref{eq:generationcuts} are imposed at the generator level. 
Our calculations are achieved in the
complex mass scheme and include three components, namely an SM piece independent
of the presence of a vector-like quark, a pure vector-like quark piece (that has
been studied in details in section~\ref{subsec:signal}), and the interference
between the SM and the new physics diagrams. The latter is expected to be
negligible in the narrow width case, but to contribute to a significant extent
for broad-width scenarios.

\begin{figure}
  \centering
  \includegraphics[width=.30\textwidth]{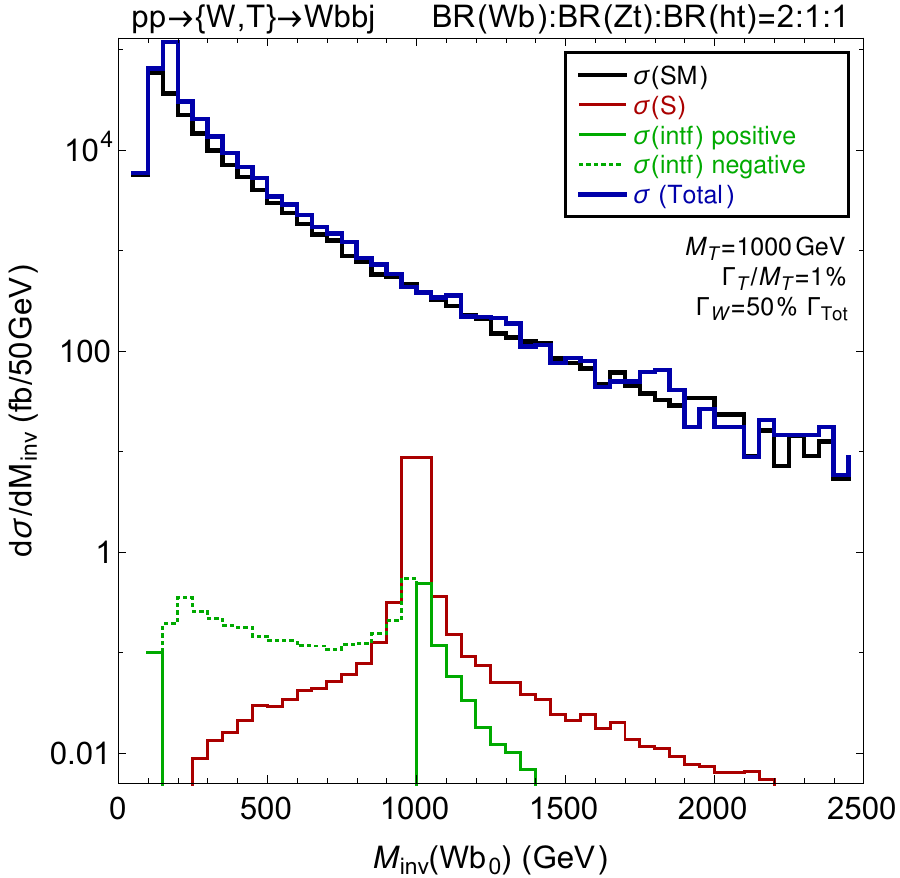}
  \includegraphics[width=.30\textwidth]{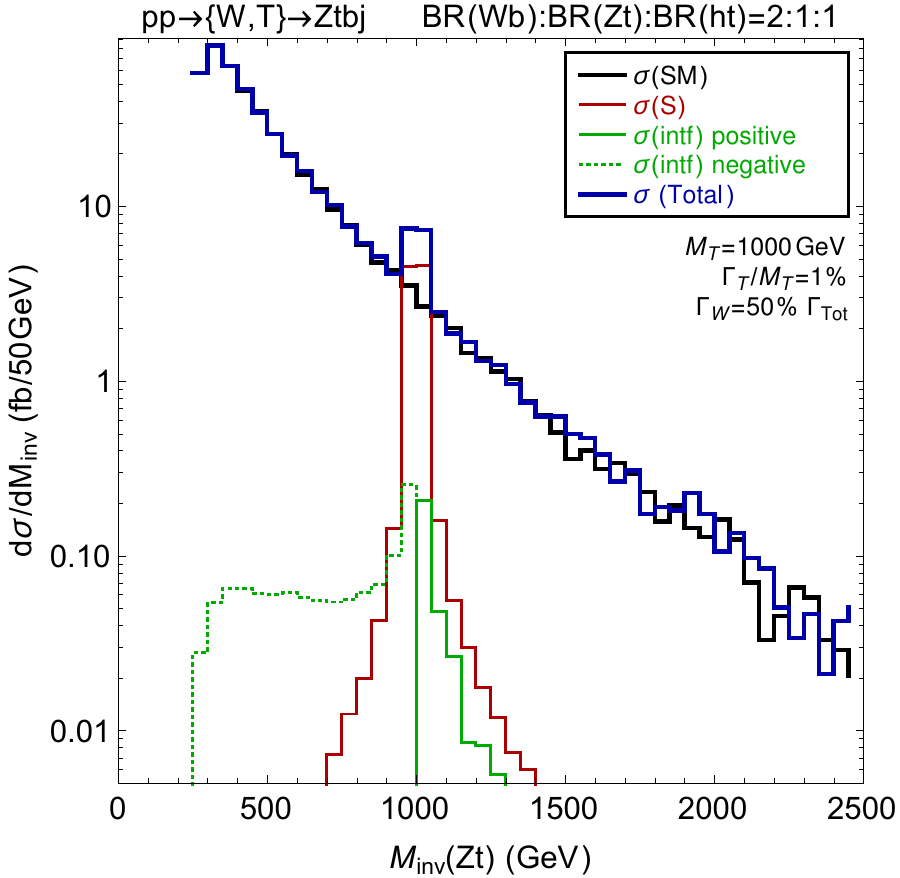}
  \includegraphics[width=.30\textwidth]{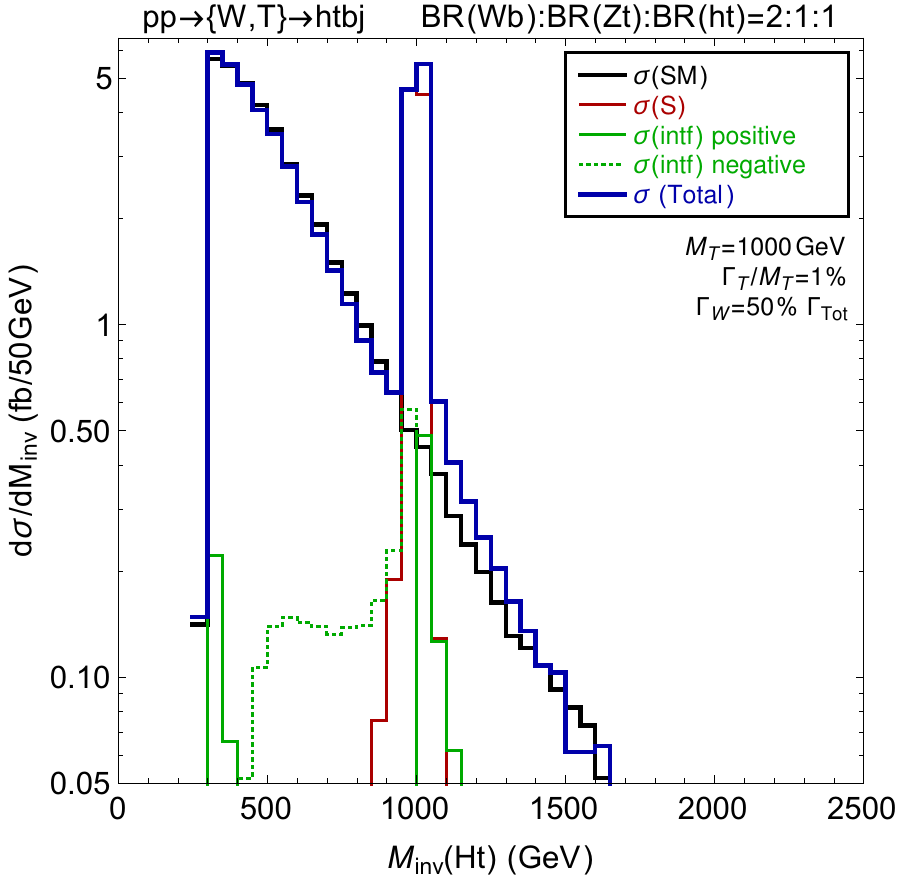}\\
  \includegraphics[width=.30\textwidth]{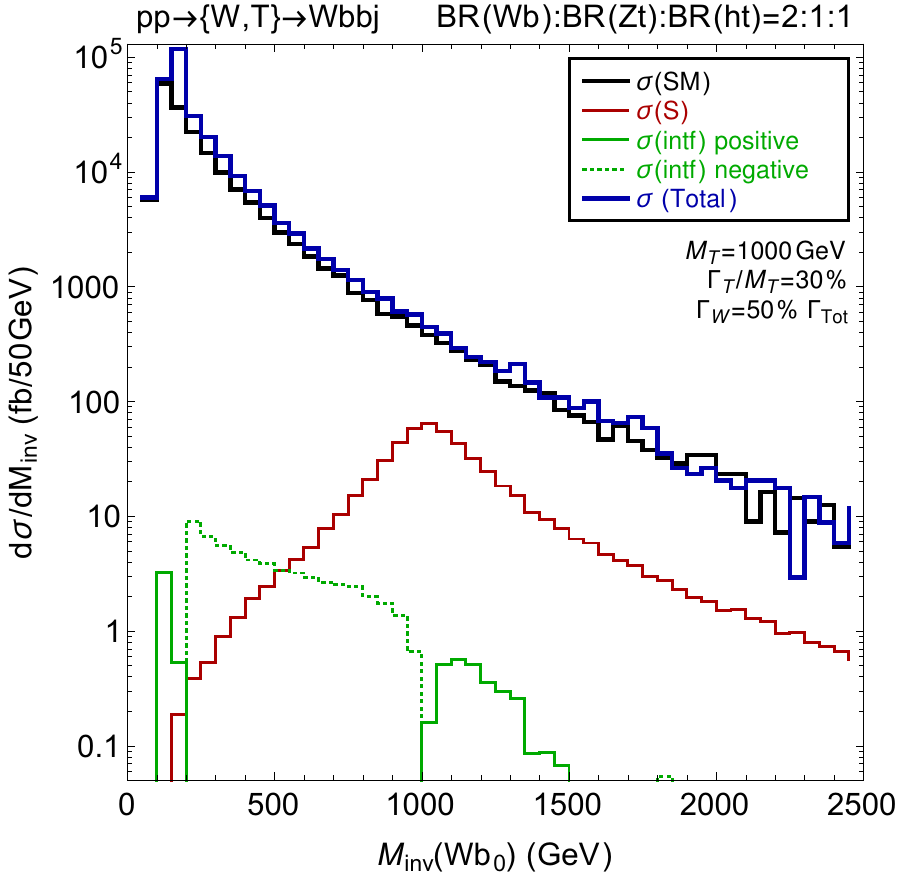}
  \includegraphics[width=.30\textwidth]{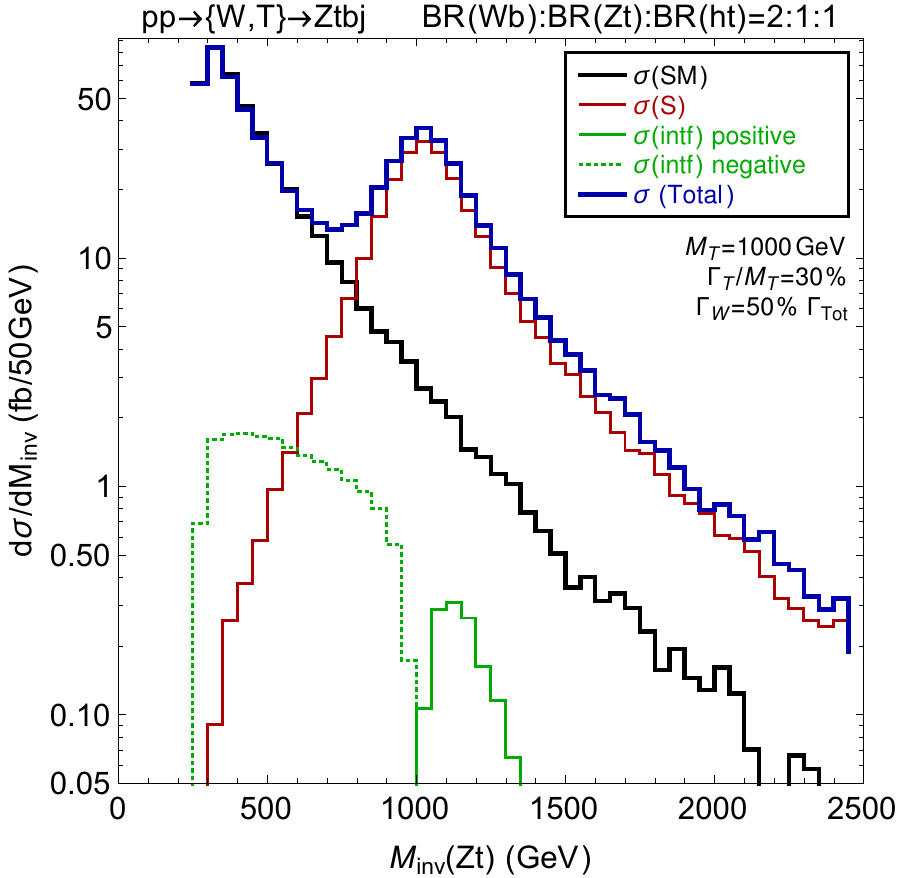}
  \includegraphics[width=.30\textwidth]{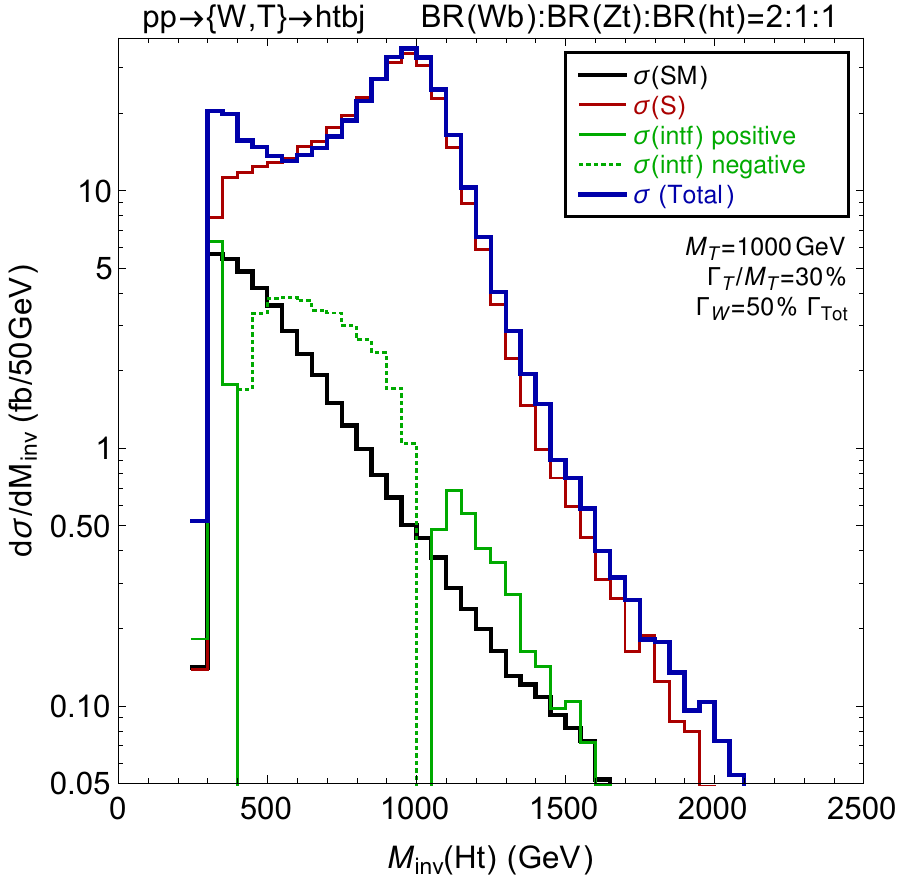}\vspace*{-.2cm}
  \caption{\label{fig:SB_int_sums_1000} Parton-level distributions in the
    invariant mass of the $Wb$ (left), $Zt$ (centre) and $ht$ (right) systems
    for the $pp\to Wbbj$, $pp\to Ztbj$ and $pp\to htbj$ processes respectively.
    We include the SM contribution (black), the new physics contributions
    stemming from a vector-like quark of mass $M_T=1\TeV$ (red) and the absolute
    value of their interference (green). We indicate by a dashed line the region
    in which the interference is negative. We consider a narrow
    width scenario ($\Gamma_T/M_T=1\%$, top row) and a large width one
    ($\Gamma_T/M_T=30\%$, bottom row), and the $T$ couplings are fixed as in
    eq.~\eqref{eq:benchmarks}. For analogous results for $T$ production through $Zt$ exchanges, see figure~\ref{fig:SBZ_int_sums_1000} and \ref{fig:SBZ_int_sums_2000}.}
\end{figure}

\begin{figure}
	\centering %\vspace*{.5cm}
  \includegraphics[width=.30\textwidth]{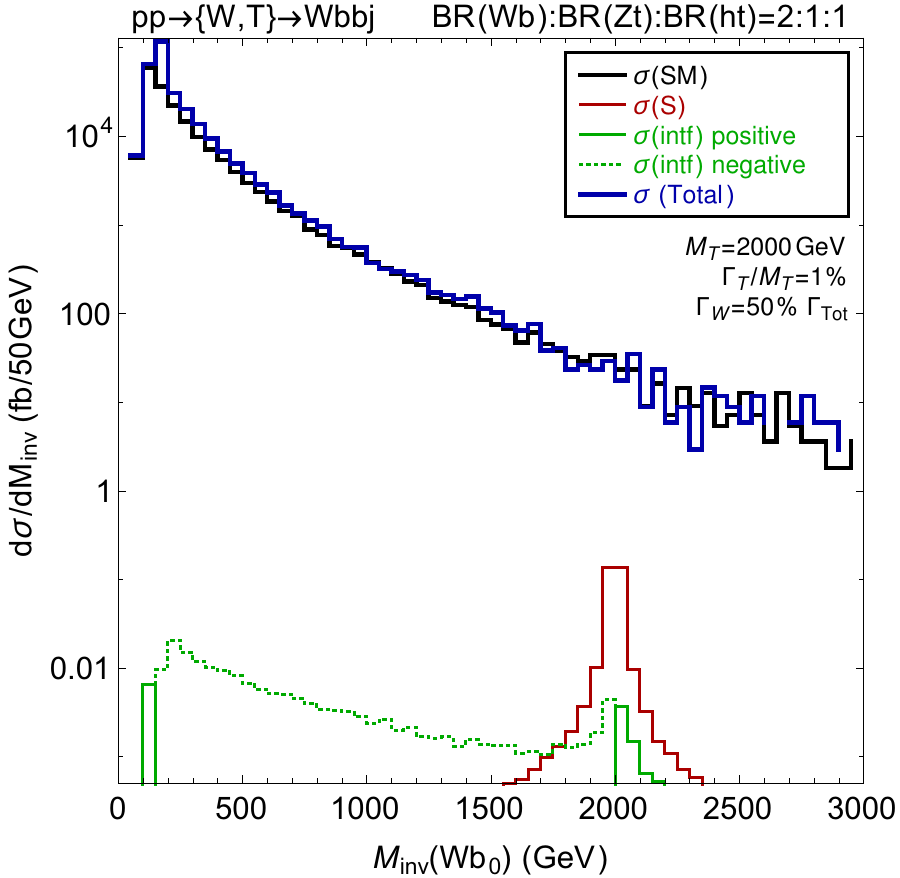}
  \includegraphics[width=.30\textwidth]{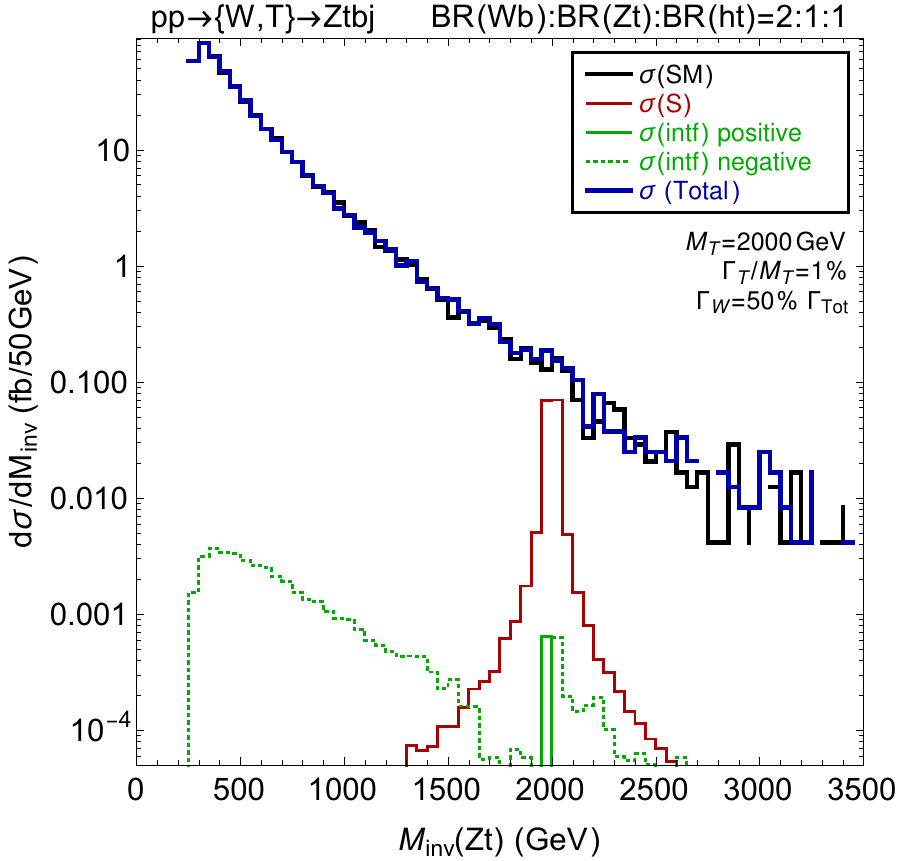}
  \includegraphics[width=.30\textwidth]{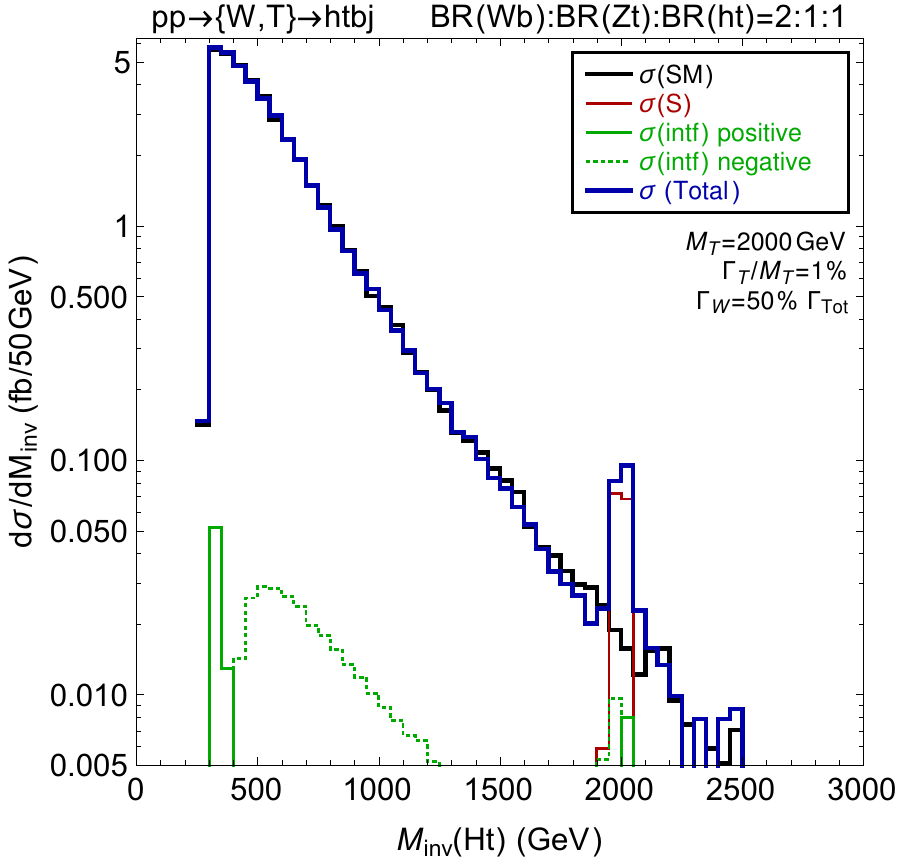}\\
  \includegraphics[width=.30\textwidth]{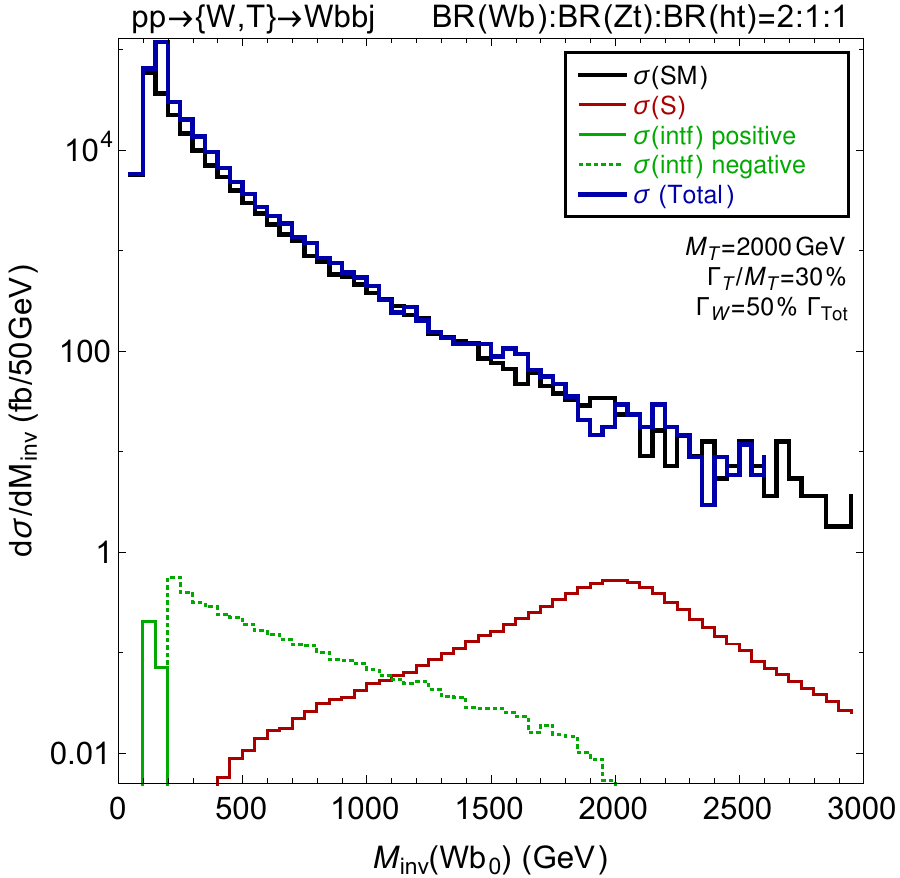}
  \includegraphics[width=.30\textwidth]{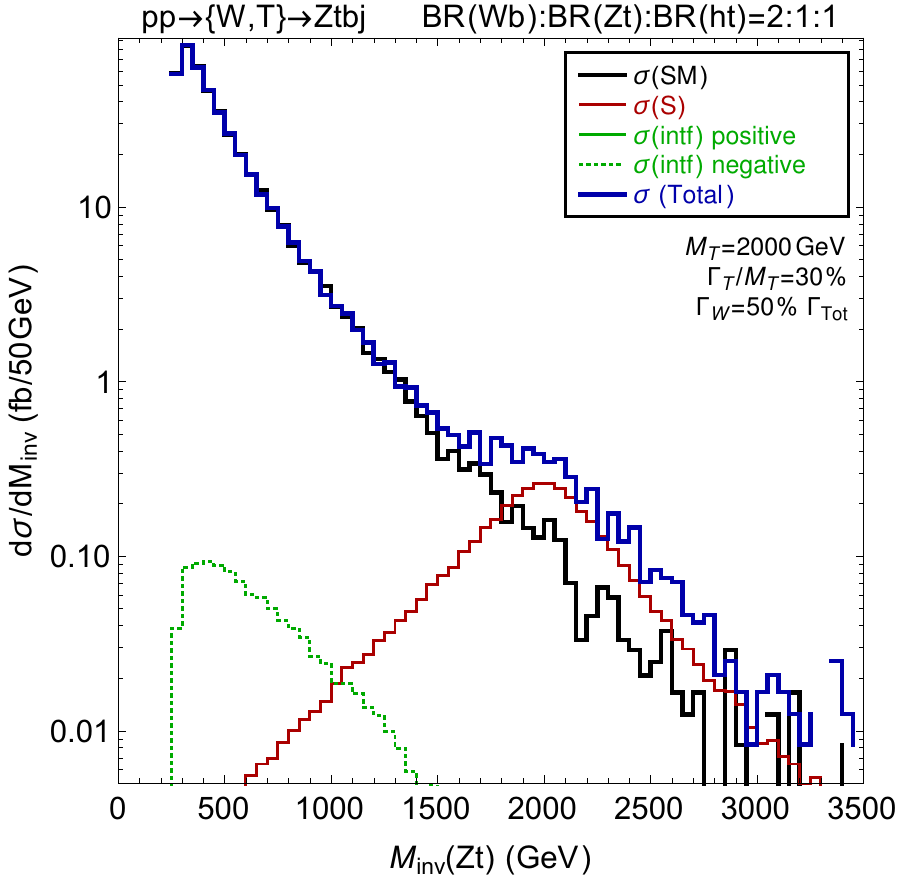}
  \includegraphics[width=.30\textwidth]{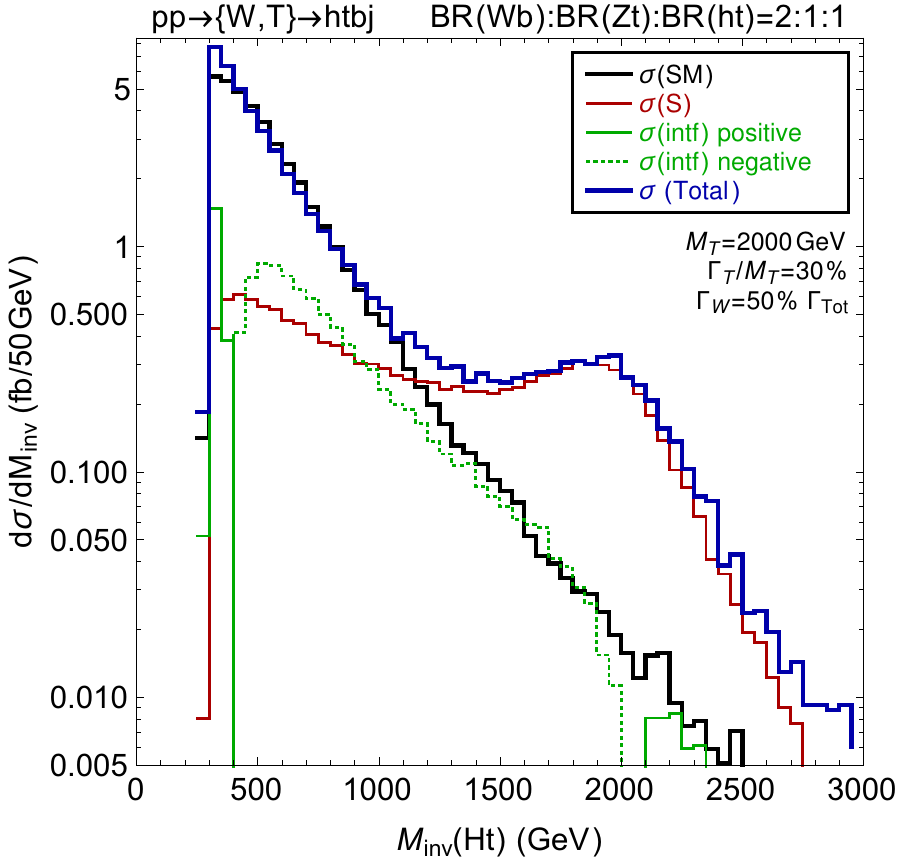}\vspace*{-.2cm}
  \caption{\label{fig:SB_int_sums_2000}
    Same as in figure~\ref{fig:SB_int_sums_1000} for $M_T=2\TeV$.}
\end{figure}

Our results are shown in figures~\ref{fig:SB_int_sums_1000} and
\ref{fig:SB_int_sums_2000} for a vector-like quark mass of $M_T = 1$ and 2~TeV
respectively. We show, in those figures, distributions in the invariant masses
$M_{\rm inv}(Wb)$, $M_{\rm inv}(Zt)$ and $M_{\rm inv}(ht)$ of the $Wb$ (left),
$Zt$ (centre) and $ht$ (right) systems, when produced through the
$pp \to Wbbj$, $pp\to Ztbj$ and $pp\to htbj$ processes
respectively. Like in the previous section, we consider scenarios featuring a
width-over-mass of 1\% (top row) and 30\% (bottom row), as stemming from the
benchmarks defined in eq.~\eqref{eq:benchmarks}. The invariant-mass spectra
resulting from including all SM and new physics contributions are given by solid
blue lines.

\subsubsection{$Wbbj$ production}\label{sec:wbbj}

The $pp\to Wbbj$ process features the largest (differential and total) cross
section, that is found to be 3--5 orders of magnitude greater than for the other
processes under consideration. This is due to ${\cal O}(g_s^3 g)$ diagrams in
which a $b\bar b$ pair originates from gluon splitting. Those diagrams dominate
the total SM contribution to $\sigma_{\rm tot}(pp\to Wbbj)$, and have moreover
no counterpart for the other two
processes. The $M_{\rm inv}(Wb)$ spectra
(blue lines) are therefore well approximated by the sole SM contributions (black
lines). The vector-like contributions (see also table~\ref{tab:totalrates})
are much smaller and therefore irrelevant, regardless of the actual values of
the vector-like quark mass and width (four left sub-figures). As a consequence,
relying on this channel to potentially probe vector-like quark single production
would require advanced analysis strategies to unravel the signal from the
background, both in the narrow and broad vector-like quark cases. The
gluon-splitting origin of the dominant background contribution could, for
instance, be used to design dedicated analysis selection cuts.

Focusing on the signal only (red lines), we recover the results of
figure~\ref{fig:res_tch_contributions}, the peak being well centred on $M_T$ and
of a Gaussian shape in the narrow case, and much broader and distorted (due to
parton density effects as a large range in Bjorken-$x$ is probed) in the
large-width case. The impact of the interference between the SM and the new
physics contributions is also indicated in the figures, destructive
interferences being shown through dashed green lines and constructive ones
through solid green lines.

In the narrow width case, the interference terms yield a 10\%-level effect on
the signal in the region of the peak defined by $M_{\rm inv}(Wb)\in [M_T-n
\Gamma_T, M_T +n \Gamma_T]$, with $n$ being equal to a few. It is destructive
for invariant masses smaller than $M_T$ and larger than $M_W$, and constructive
otherwise. Outside the peak, the interference is much larger (in
absolute value) than the pure new physics contribution, although both of them
are negligible relatively to the SM background component. They are indeed at
least 4 orders of magnitude weaker.

In the broad vector-like case, the pure new physics contribution to $pp\to Wbbj$
production is this time only 2 orders of magnitude smaller than the SM
background at the level of the peak, \ie\ for $M_{\rm inv}(Wb)\sim M_T$. This
results from the larger vector-like quark coupling values that are needed to
accommodate a larger width without invoking exotic vector-like quark decay
modes. Such $\kappa$, $\tilde\kappa$ and $\hat\kappa$ values indeed additionally
enhance the signal cross section with respect to the narrow-width case (see
also discussions and results in \cref{subsec:signal}). The peak
is, as expected and as already mentioned above, much broader. The signal
therefore significantly contribute to almost the entire considered
$M_{\rm inv}(Wb)$ range. It stays nevertheless orders magnitude smaller than the
background, in particular as soon as $|M_{\rm inv}(Wb)-M_T|$ is larger than
$n\Gamma_T$. The relative difference with the background is more pronounced in
the small invariant-mass regime, where the SM contributions are drastically
enhanced (close to $M_{\rm inv}(Wb)\sim M_W$).
Turning to the interference, the situation is different from the narrow-width
case. The interference between the new physics and SM contributions is
sub-leading compared with the signal, except for the small invariant-mass regime
where the SM contribution to the amplitude is huge. As in the narrow
case, it is destructive for $M_{\rm inv}(Wb)$ smaller than $M_T$ and larger than
$M_W$, and constructive otherwise.

The strong enhancement of the background contributions by virtue of diagrams
featuring gluon splittings into $b\bar b$ pairs nevertheless makes any potential
new physics
observation through this process challenging, regardless of the value of the
vector-like quark width. This may require the design of a dedicated analysis,
which goes beyond the scope of this work in which we only aim to depict
the importance of a correct treatment of the vector-like quark width in the
modelling of single vector-like quark production signals.

\subsubsection{$Ztbj$ production}

We expect that the $pp\to Ztbj$ process leads to a {\it signal} behaviour
that is similar to the $pp\to Wbbj$ case, as demonstrated in
eq.~\eqref{eq:signal_behaviours}. A notable difference nevertheless comes from
the associated SM background, and thus its interference with the new physics
contributions. SM $Wbbj$ production has been found to be dominated by diagrams
featuring a $g\to b\bar b$ splitting. Such diagrams have no equivalent in the
$Ztbj$ case, as $g\to t\bar t$ splittings yield not only different kinematics,
but also a suppression stemming from the heavy mass of the top
quark. This can be seen in the middle panels of
figures~\ref{fig:SB_int_sums_1000} and \ref{fig:SB_int_sums_2000}, where the SM
spectrum in the invariant mass of the $Zt$ system (black lines) is reduced by
about 2 orders of magnitude relatively to the $Wbbj$ case. Consequently, the
$Ztbj$ signal, that is only a factor of 2 smaller than the corresponding $Wbbj$
signal, has a chance to leave potentially observable effects in distributions
such as the $M_{\rm inv}(Zt)$ invariant-mass one considered in this section.
This is illustrated on all the four sub-figures relevant for the $pp\to Ztbj$
process. In those sub-figures, we directly compare
predictions for the full (SM plus new physics) process (blue lines) with
predictions for the pure SM (black lines) or pure new physics (red lines) cases.

In the narrow-width scenarios (top lines of the two figures), a clear
Breit-Wigner signal peak is observed at $M_{\rm inv}(Zt)\sim M_T$. At the
level of the peak, the interference of the signal with the SM background is
sub-leading and of a few percent. For $M_{\rm inv}(Zt)<M_T - n \Gamma_T$ (with
$n$ being
equal to a few), the interference is largely dominating over the pure signal
contributions. However, the background is also much larger, so that we do not
obtain any noticeable net effect on the full invariant mass spectrum. We thus
recover the usual configuration in which the signal and the background can be
treated independently to a good approximation, as implemented in all searches
for single vector-like quark production so far.

The situation is different in the large-width case. First, the
shape of the full invariant mass spectrum (that includes both its SM and new
physics components) is distorted and shifted for $M_{\rm inv}(Zt) \gtrsim M_T -
\Gamma_T$ for the two considered benchmark scenarios. This effect solely comes
from the sum of the individual background and signal contributions, their
interference being found to be sub-leading as in the narrow-width case. Once
again, the interference of the signal with the background can thus be safely
neglected. By comparing the full result to its components, we find that the SM
prediction is equal to the full one for $M_{\rm inv}(Zt) \lesssim M_T-\Gamma_T$.
We then observe a smooth departure from the SM in the $M_{\rm inv}(Zt)\in [M_T -
\Gamma_T, M_T]$ region, and an increase of the
distribution by about an order of magnitude for $M_{\rm inv}(Zt) > M_T$
regardless of the $M_{\rm inv}(Zt)$ value. This increase directly originates
from the dependence of the new physics partonic cross section on the partonic
centre-of-mass energy $\sqrt{s}$. This is demonstrated in
figure~\ref{fig:feature_chiralities} (yellow lines), in which we show that the
partonic cross section for $Ztbj$ production is constant for $\sqrt{s} > M_T +
\Gamma_T$, and not suppressed by the internal propagators. We thus recover the
signal shapes shown in figures~\ref{fig:SB_int_sums_1000} and
\ref{fig:SB_int_sums_2000}, the decrease with $M_{\rm inv}(Zt)$ being driven by
the parton densities that involve larger Bjorken-$x$ values.

Those results motivate a usage of more inclusive signal regions in new
physics experimental searches for vector-like quark single production and decay
into a $Zt$ system, instead of only targeting the reconstruction of a
Breit-Wigner peak. The latter option stays powerful for searches for a narrow
vector-like quark. However, the former option is in contrast very promising in
the large width case. The differential cross section is indeed enhanced for
invariant-mass values much larger than the vector-like quark mass $M_T$ or even
than $M_T+\Gamma_T$. One must however bear in mind that the signal distributions
could be suppressed relatively to the optimistic case presented in this section.
For instance, if the large width arises from the existence of exotic decay
channels, then the $\kappa$ and $\tilde\kappa$ couplings could be smaller,
accordingly reducing the signal cross section by a global factor. This is
addressed in section~\ref{sec:lo_2d_scans}.

\subsubsection{$htbj$ production}
This subsection is dedicated to the last of the considered processes, namely
$pp\to htbj$. The corresponding distributions in the invariant mass of the $ht$
system $M_{\rm inv}(ht)$ are expected to feature a behaviour that is different
from the case of the other two processes, as predicted by
eq.~\eqref{eq:signal_behaviours}. The distributions shown in
figure~\ref{fig:feature_chiralities} (blue lines) indeed exhibit a decrease of
the partonic cross section at large invariant masses $M_{\rm inv}(ht) > M_T$,
and the cross section is in addition approximately constant for
$M_{\rm inv}(ht) < M_T - n \Gamma_T$, with $n$ being equal to a few and with an
exact value that depends on the width-over-mass ratio. This situation contrasts
with the $pp\to Ztbj$ and $pp\to Wbbj$ cases where an opposite behaviour is
found. Consequently, we expect, at least for broad vector-like quarks that yield
a wider and less
pronounced invariant-mass peak, a large signal contribution for invariant masses
much lower than the peak value $M_T$, and a potentially important role to be
played by the interference with the SM background. This is confirmed in the
sub-figures shown in the last column of
figures~\ref{fig:SB_int_sums_1000} and \ref{fig:SB_int_sums_2000}.

We focus first on predictions for narrow-width scenarios (upper right panel in
the two figures). Here, as for the other
processes, the signal contribution corresponds to a wide peak centred on $M_T$.
As in the $pp\to Ztbj$ case, there is no enhancement of the SM background as
gluon splittings into a top-antitop pair are kinematically suppressed. The SM
spectrum (black line) is thus relatively small enough to make the Breit-Wigner
peak visible in the two considered scenarios. However,
the interference between the signal and the SM background is this time not
systematically negligible. The signal amplitude significantly contributes in the
$M_{\rm inv}(ht) < M_T - n \Gamma_T$ regime, so that its interference with the
SM amplitude can get enhanced in this kinematical regime too. This is the case
for the light vector-like quark
scenario with $M_T=1$~TeV, where the full (SM plus new physics) distribution
(blue line) slightly deviates from the SM one by about 10\%. Those effects (\ie\
a constant signal partonic cross section that can benefit from an extra
parton density enhancement due to intermediate and lower probed
Bjorken-$x$ values)  are nevertheless suppressed by the heavy
quark mass. There are in particular found negligible for the $M_T=2$~TeV
scenario.

For broad width scenarios (lower right panels in
figures~\ref{fig:SB_int_sums_1000} and \ref{fig:SB_int_sums_2000}), we observe
a dramatic increase of the signal (and therefore its interference with the
background too) for all invariant-mass values. This results directly from the
dependence of the partonic cross
section on the vector-like quark mass and width (or the vector-like quark
complex mass) detailed in section~\ref{subsec:parton_ME}. For $M_T=1$~TeV, the
signal distribution is a factor of a few larger than the background one for all
considered $M_{\rm inv}(ht)$ values, and thus is probably impossible to miss in the
context of a search. Close to the production threshold of the $ht$ system or at
invariant masses higher than the peak region, both the SM and signal components
contribute equally to the full rate, and interfere. As a result, the full
distribution exhibits a two-peak structure, with a first peak around $M_T$ and a
second one around the $ht$ threshold. The impact of the interference is of
about 10\%.

The situation is slightly different for heavier vector-like quarks. Here, the
low invariant-mass part of the spectrum is dominated by its SM component, the
new physics contributions (and their interference with
the SM) being not significant enough as one is too far from the peak. They
hence only leads to a distortion of the spectrum shape of about 10\%. In
particular,
the differences are larger close to threshold, where the signal distribution
exhibits a spurious peak resulting from the convolution of the parton densities
at small and intermediate Bjorken-$x$ values and a constant partonic cross
section. Around the (broad) peak, the signal is one order of magnitude larger than
the SM background, and thus almost equal to the full contribution.

As for the case of $Ztbj$ production, we recall that the entire signal
contributions can be globally reduced by introducing exotic vector-like quark
decay modes. This would indeed allow for smaller coupling values to accommodate
the large width, so that the corresponding new physics effects at the level of
the considered invariant-mass spectra could be rendered potentially not so
visible on top of the SM background. However, the shape of the new physics
contributions at the partonic level implies that the signal cannot be considered
independently of the background as soon as the width gets large. Interferences
indeed matter. This is an important difference with respect to a setup in which
the vector-like quark decays into a vector boson, where the interference
between the signal and the background is always sub-leading.

\subsection{Pinning down the vector-like quark width at the LHC}
\label{sec:lo_2d_scans}

\subsubsection{Large width effects on total cross sections}

In the previous section, we have described collider features that emerge
from the single production of an up-type vector-like quark $T$ of a given
chirality and with a
potentially large width. We have assumed, however, that all decay modes
consisted of decays into a Standard Model quark and a weak or a Higgs boson. The
size of the vector-like quark couplings to the $W$, $Z$ and Higgs boson
$\kappa$, $\tilde\kappa$ and $\hat\kappa$ is therefore related to the
vector-like quark
width $\Gamma_T$ once the relative contributions of the three decay modes are
fixed. In concrete models, exotic decay modes could nevertheless exist~\cite{
Bizot:2018tds,Cacciapaglia:2019zmj}, modifying the connection between the width
and the above couplings. The only requirement that survives implies that the sum
of the three partial widths cannot exceed the vector-like-quark total width.

In this section, we investigate the impact of the existence of extra decay
channels on the rates associated with the three processes of
eqs.~\eqref{eq:sgl1} and \eqref{eq:sgl2}, the vector-like quark coupling chirality being
fixed, {\it i.e.} taking only one of the $\{\kappa_L,
\tilde\kappa_L, \hat\kappa_L\}$ and $\{\kappa_R, \tilde\kappa_R, \hat\kappa_R\}$
sets of couplings non-zero. As above-mentioned, such a choice is motivated by the fact that in
concrete composite models, the vector-like quark coupling chirality is related to its
representation under $SU(2)_L$.
In the following, the considered set of dominant couplings ($L$ or $R$) is generically
denoted by $\{\kappa, \tilde\kappa, \hat\kappa\}$ while the other set of couplings is taken as vanishing.

In order to provide model-independent results, we consider the vector-like quark
width $\Gamma_T$ and mass $M_T$ as free parameters, and factor out of the
vector-like quark couplings appearing in the cross sections from the functional
form of the latter. These cross sections can hence be written, for each of the
considered processes and after including both the SM contributions, the
vector-like quark contributions and their interference, as
\be\bsp
 \sigma_{\rm tot}(pp\to Wbbj) =&\
    \sigma_{Wb}^{\rm SM} 
      + \kappa^4 \ \hat\sigma_{Wb}^{\rm VLQ}(M_T, \Gamma_T)
      + \kappa^2\ \hat\sigma_{Wb}^{\rm int}(M_T, \Gamma_T) \ ,\\
 \sigma_{\rm tot}(pp\to Ztbj) =&\
    \sigma_{Zt}^{\rm SM}
      + \kappa^2\tilde\kappa^2 \ \hat\sigma_{Zt}^{\rm VLQ}(M_T, \Gamma_T)
      + \kappa\tilde\kappa\ \hat\sigma_{Zt}^{\rm int}(M_T, \Gamma_T) \ ,\\
 \sigma_{\rm tot}(pp\to htbj) =&\
    \sigma_{ht}^{\rm SM}
      + \kappa^2\hat\kappa^2 \ \hat\sigma_{ht}^{\rm VLQ}(M_T, \Gamma_T)
      + \kappa\hat\kappa\ \hat\sigma_{ht}^{\rm int}(M_T, \Gamma_T) \ .
\esp\label{eq:xsec_param}\ee
These expressions involve `bare' components $\hat\sigma$
that are independent of the $\kappa$, $\tilde\kappa$ and $\hat\kappa$ parameters
and that solely depend on the vector-like quark mass and width. In our notation,
$\hat\sigma^{\rm VLQ}$ and $\hat\sigma^{\rm int}$ represent the bare
contributions to the pure vector-like component and to its interference with the
SM piece respectively. We however assume here that at least one
of the vector-like quark coupling appearing in the signal diagrams is a coupling
to the $W$-boson (as shown in the diagrams of
figure~\ref{fig:Ht_res_tch_topologies}). Generalisations are treated in
\cref{app:TviaZ}.

From tabulated
values of the bare cross sections as a function of $M_T$ and $\Gamma_T$, it is
then possible to evaluate the corresponding total rate for any specific set of
$\kappa$, $\tilde \kappa$ and $\hat\kappa$ coupling values. Moreover, the
formul\ae\ in eq.~\eqref{eq:xsec_param} can be generalised to scenarios featuring
several vector-like quark species. The number of bare cross sections increases
in this case, as one needs to also account for `signal-signal' interferences.
Whereas the different pieces could be dependent of the gauge choice (as
$W$-boson and $Z$-boson are both involved), the sum will always be
gauge-invariant.

Nevertheless, we focus in this work on the simplest scenarios where only one
single vector-like quark species is added to the SM field content.
The expressions of eq.~\eqref{eq:xsec_param} can then be used to derive the
total rates associated with the $pp\to Wbbj$, $pp\to Ztbj$ and $pp\to htbj$
processes. The first term is the pure SM component in which we only
consider diagrams free from any vector-like quark propagators. These rates are
given, for LHC proton-proton collisions at a centre-of-mass energy of 14~TeV\footnote{For the calculation of the SM total cross sections, and for the results shown in figure~\ref{fig:sigmahat_W_T_14TeV} and in the appendices, simulations only include a minimal cut on the transverse momentum of the final state light and $b$-jets,  $p_{Tj,b}>1\GeV$, at the generator level.}, by
\be
  \sigma_{Wb}^{\rm SM} = 990.0~{\rm pb}, \qquad
  \sigma_{Zt}^{\rm SM} = 660.8~{\rm fb} \qquad\text{and}\qquad
  \sigma_{ht}^{\rm SM} = 65.7~{\rm fb}.
  \label{eq:SMxsecs}
\ee
As already detailed in section~\ref{sec:wbbj}, the SM rate for $Wbbj$ production
is of about 3 orders of magnitude larger than for the other processes due to the
dominant contributions of diagrams featuring $g\to b\bar b$ splittings.
Unraveling the vector-like quark signal (and its interference with the SM
background) may in this case be thus more challenging and require a more
sophisticated analysis.

\begin{figure}
  \centering
  \includegraphics[width=.325\textwidth]{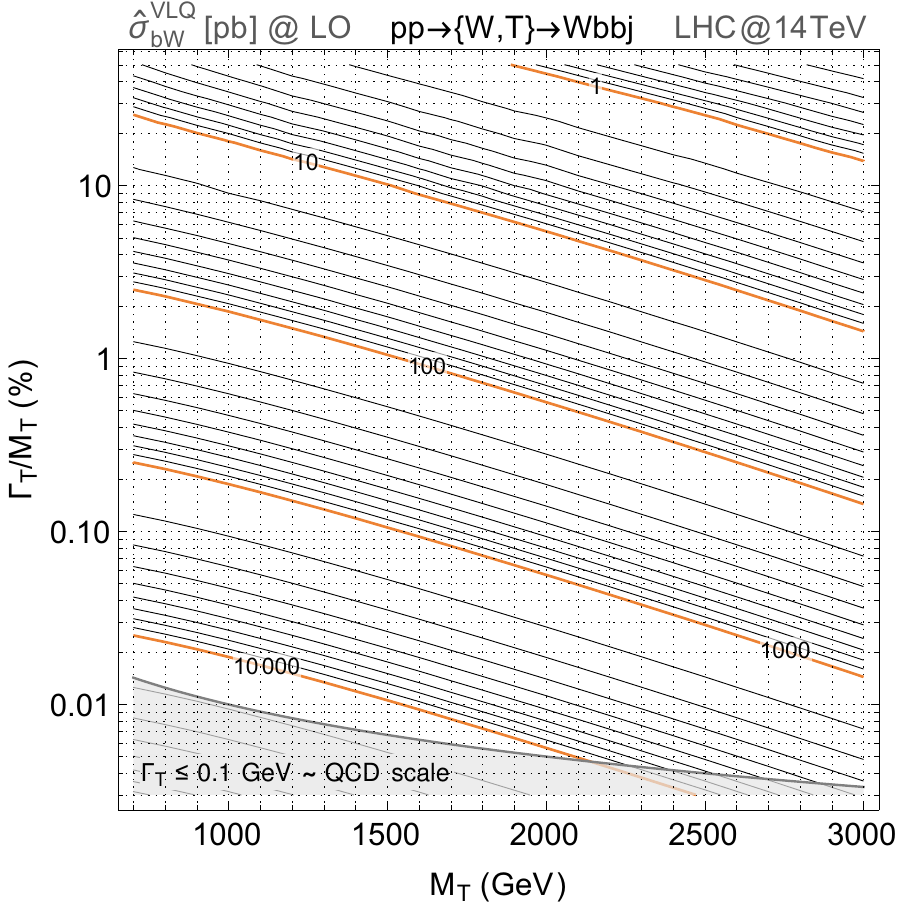}
  \includegraphics[width=.325\textwidth]{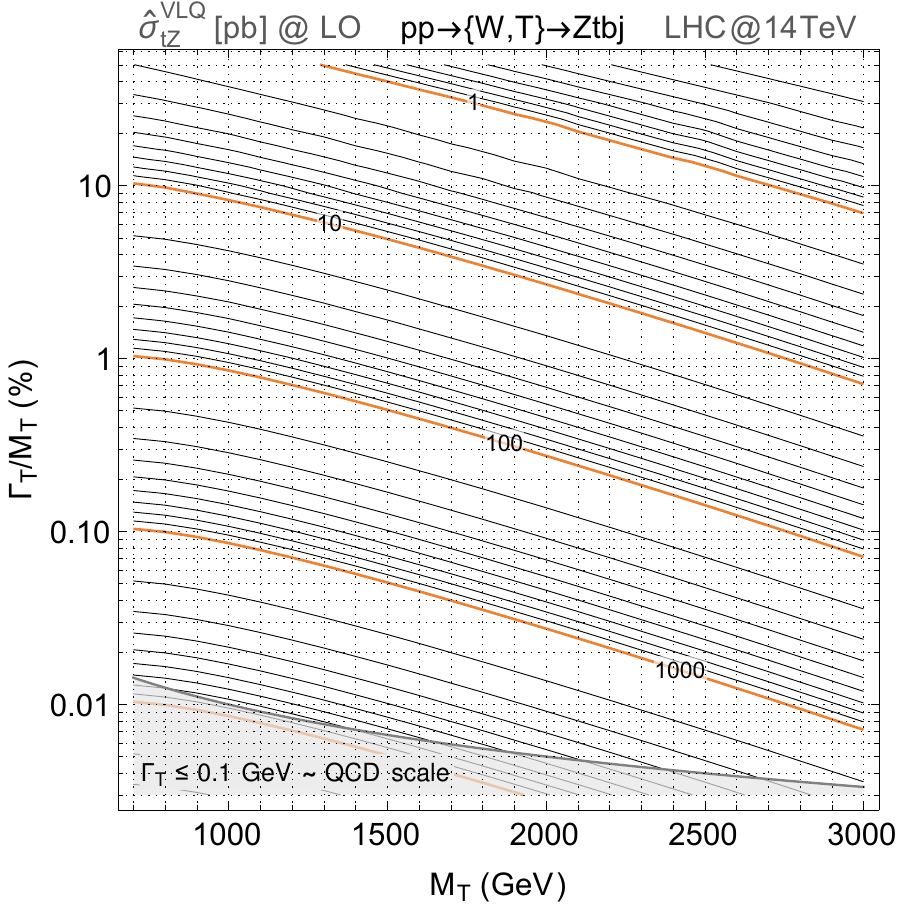}
  \includegraphics[width=.325\textwidth]{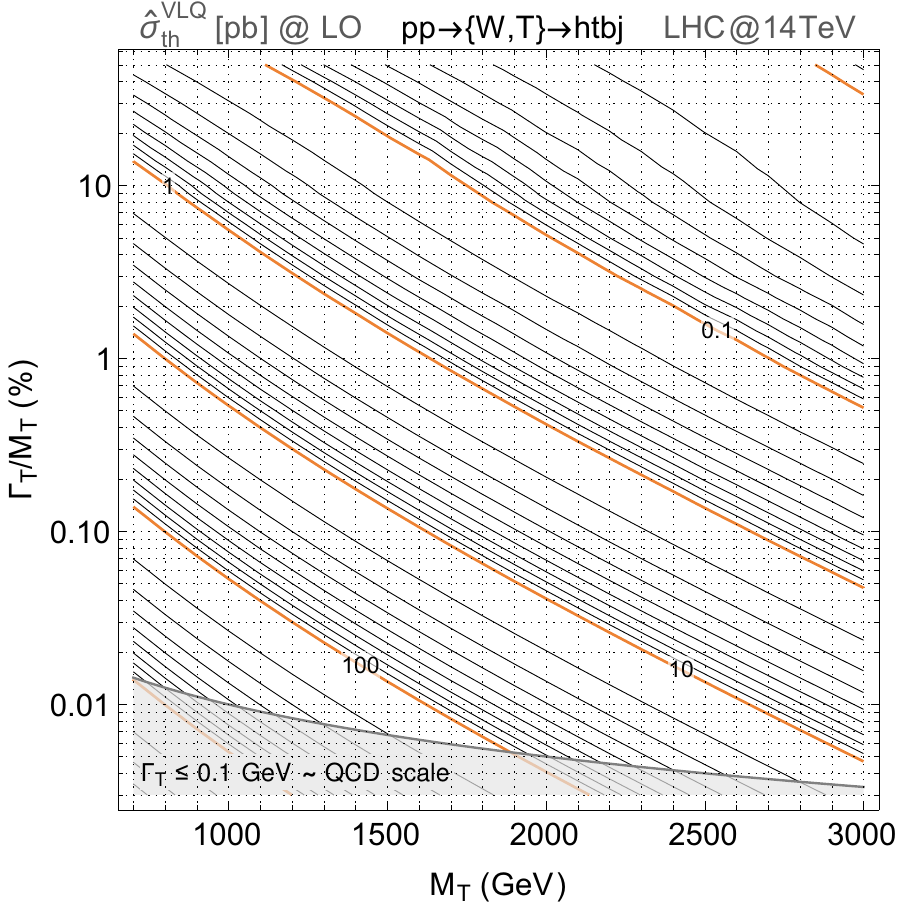}
  \\
  \includegraphics[width=.325\textwidth]{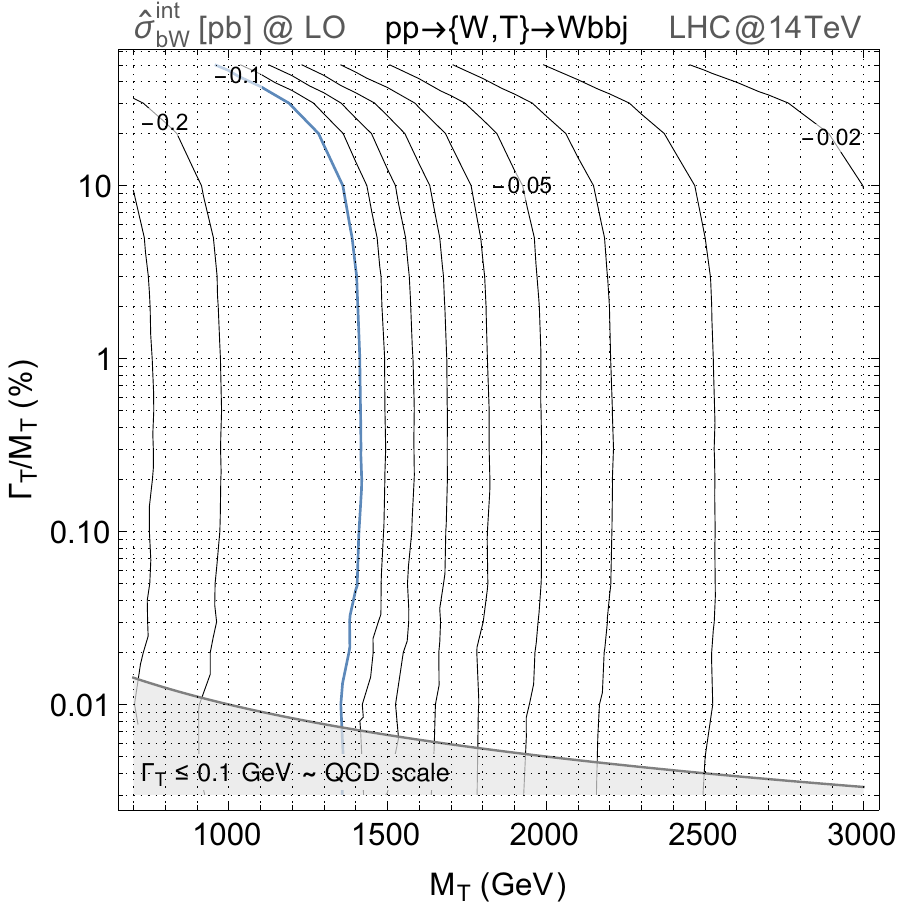}
  \includegraphics[width=.325\textwidth]{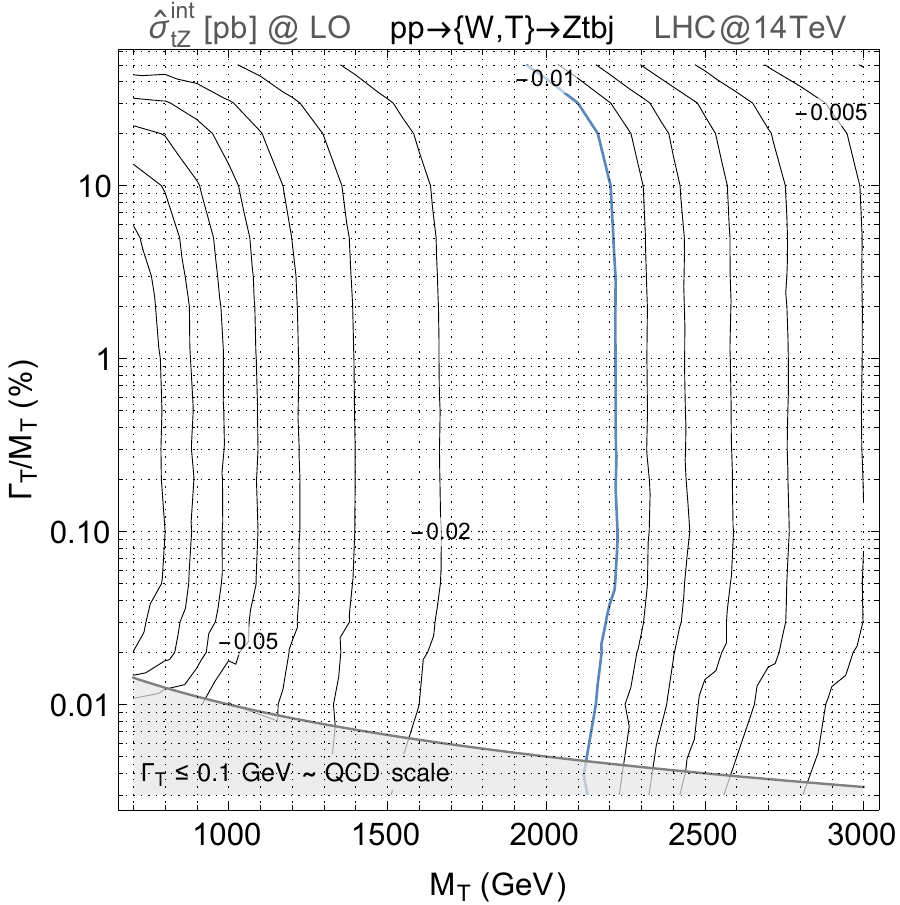}
  \includegraphics[width=.325\textwidth]{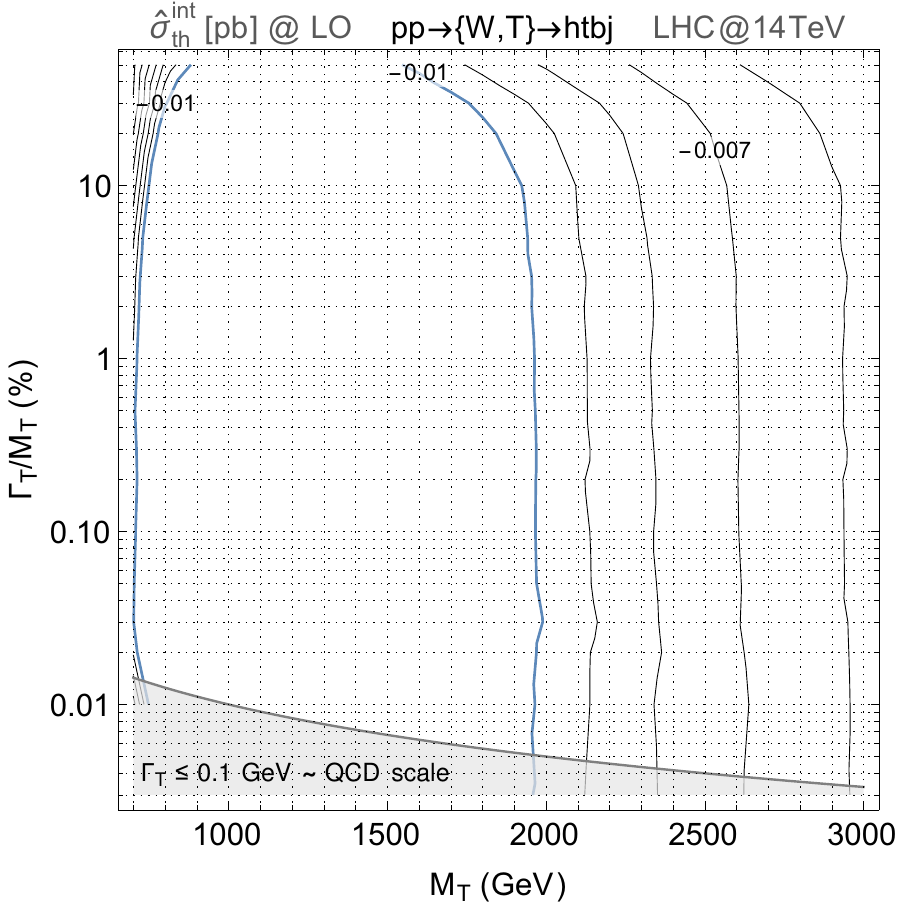}\vspace*{-0.2cm}
  \caption{\label{fig:sigmahat_W_T_14TeV} Values of the $\hat\sigma^{\rm VLQ}$
    (top row) and $\hat\sigma^{\rm int}$ (bottom row) bare cross sections, in pb,
    for the $pp\to Wbbj$ (left), $Ztbj$ (centre) and $Htbj$ (right) processes,
    for 14 TeV LHC proton-proton collisions. The bare cross sections are shown
    in $(M_T, \Gamma_T/M_T)$ planes. Analogous results for 13 TeV LHC proton-proton collisions are shown in figure~\ref{fig:sigmahat_W_T_13TeV}, and the case of $T$ production through $Zt$ exchanges is treated in figures~\ref{fig:sigmahat_Z_T_13TeV} and \ref{fig:sigmahat_Z_T_14TeV}.} 
\end{figure}

The bare vector-like quark cross sections $\hat\sigma_{Wb}^{\rm VLQ}$,
$\hat\sigma_{Zt}^{\rm VLQ}$ and $\hat\sigma_{ht}^{\rm VLQ}$ depend all on the
vector-like quark mass and width. This dependence is illustrated in the left,
central and right panels of the top row of figure~\ref{fig:sigmahat_W_T_14TeV}
for the $pp\to Wbbj$, $pp\to Ztbj$ and $pp\to htbj$ processes respectively. We
observe that the bare cross sections steeply
fall with the vector-like quark mass $M_T$, regardless of the width, and that
the decrease is more pronounced when the produced SM boson is a Higgs boson than
when it is a charged or neutral gauge boson. For instance, the
$\hat\sigma_{Wb}^{\rm VLQ}$ and $\hat\sigma_{Zt}^{\rm VLQ}$ bare cross
sections involving a vector-like quark of about 1~TeV are one order of magnitude
larger than those involving a vector-like quark of 3~TeV. In contrast, this
relative difference increases to two orders of magnitude (or slightly less) for
the $\hat\sigma_{ht}^{\rm VLQ}$ bare cross section. The latter indeed drops much
more quickly with $M_T$ than its counterparts for the two other processes. Such a
behaviour is, nevertheless, not so unexpected, as the production of a scalar
bosons involves a different chirality structure at the level of the matrix
element. On the contrary, the width dependence of the bare cross section is
similar for all three cases when $\Gamma_T/M_T$ is not too large. Increasing the
width-over-mass ratio by a factor of 10 indeed leads to a reduction of the cross
section by roughly the same factor of 10, all other parameters being fixed.
However, for larger width-over-mass ratios, the different mass dependence of the
$pp\to htbj$ partonic cross section that we discussed in the previous subsections
kicks in, and changes the picture. This can be seen in particular in the upper
part of the top-right panel of figure~\ref{fig:sigmahat_W_T_14TeV}, the
separation between equally spaced cross section iso-contours being larger and
larger.

In the lower row of figure~\ref{fig:sigmahat_W_T_14TeV}, we present the
dependence of the $\hat\sigma_{Wb}^{\rm int}$, $\hat\sigma_{Zt}^{\rm int}$
and $\hat\sigma_{ht}^{\rm int}$ bare cross sections including the interference
between diagrams featuring vector-like quarks and diagrams of a purely SM
nature. We can first observe that in all cases, the width only plays a role when
the width-over-mass ratio is larger than 10\%. The mass dependence is,
moreover, quite mild, the magnitude of the bare interference changing solely by
a few when $M_T$ varies from 1 to 3~TeV. For small masses, the interference
however increases with the mass, to reach a maximum (in blue on the figures) for
$M_T\sim 1.5$~TeV for the $pp\to Wbjj$ process and for $M_T\sim 2$~TeV for the
other two processes, the maximum value being reduced in large width cases. This
is of upmost importance for the LHC, as this mass range lies within (or close to)
the expected reach of the future LHC operations. The
interference then slowly decreases with increasing mass values. Whilst in
general much smaller (by at least two orders of magnitude or more) than the pure
vector-like quark contributions $\hat\sigma^{\rm VLQ}$, one must keep in mind
than the physical cross section includes a product of bare interference
contributions and two powers of the couplings. In contrast, the pure
vector-like quark component involves a quartic dependence on the couplings, so
that for large widths, the $\hat\sigma^{\rm VLQ}$ and $\hat\sigma^{\rm int}$
contributions to the signal are both important (as already observed in the
previous subsection).

From the maps in figure~\ref{fig:sigmahat_W_T_14TeV}, it is now straightforward to derive the full cross section for
the $pp\to Wbbj$, $Ztbj$ and $Htbj$ processes, that are relevant for single vector-like
quark production at the LHC.

\begin{figure}
	\centering
	\includegraphics[width=.325\textwidth]{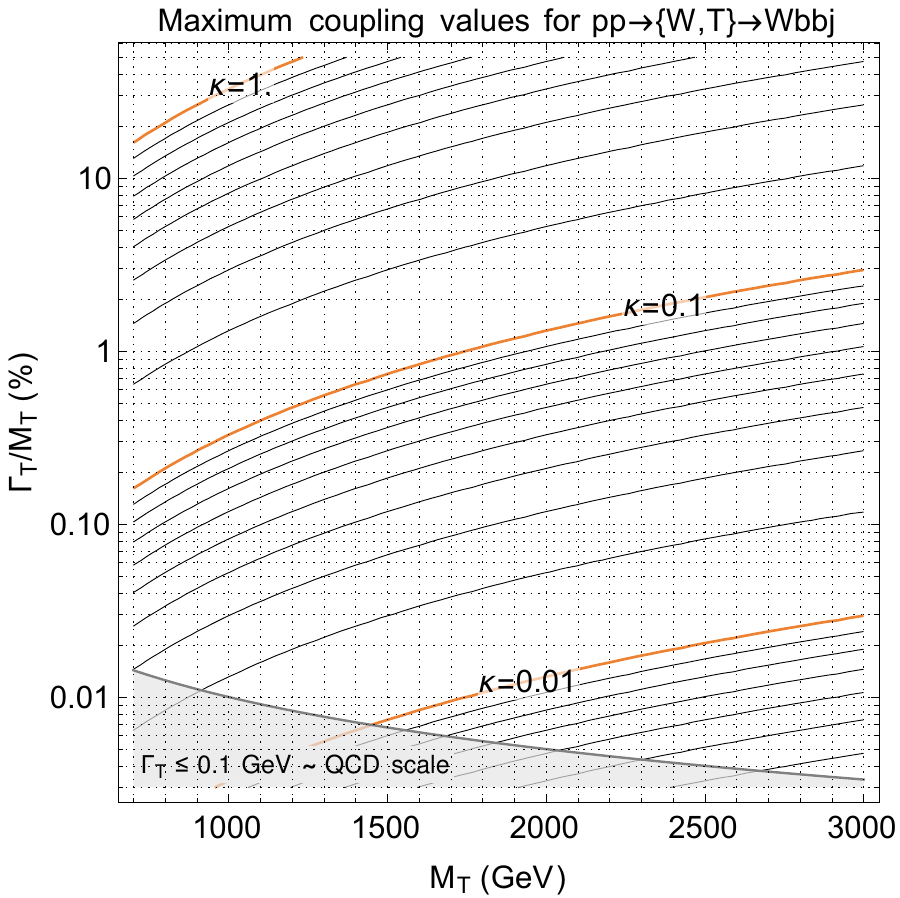}
	\\
	\includegraphics[width=.325\textwidth]{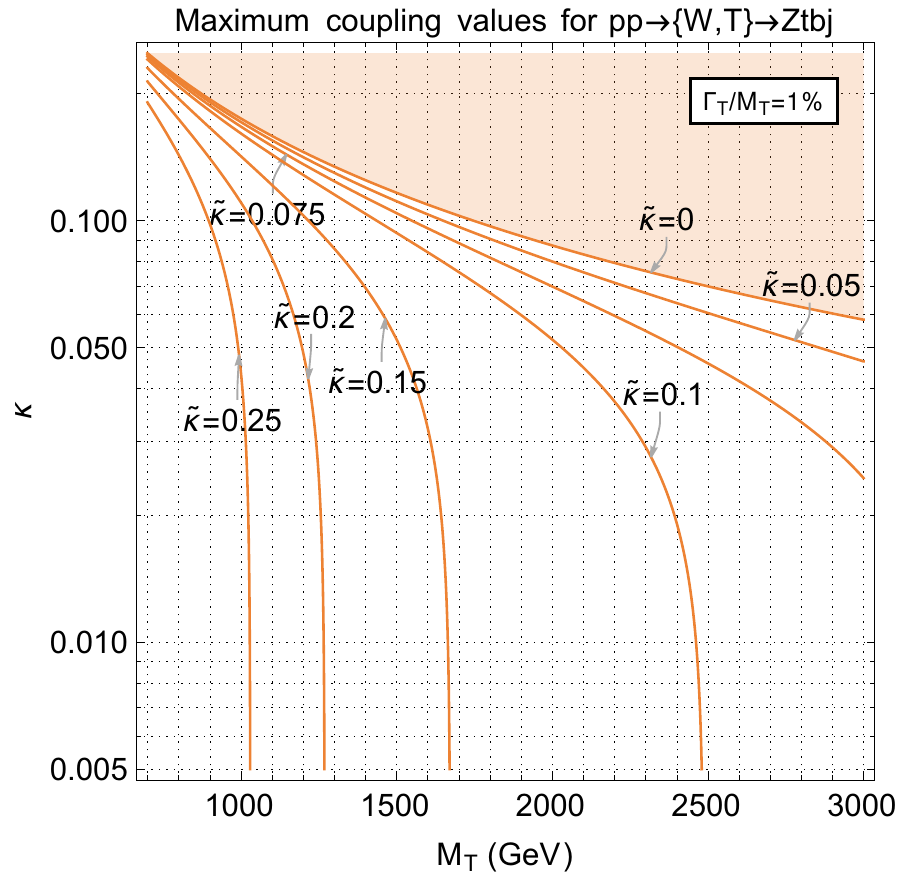}
	\includegraphics[width=.325\textwidth]{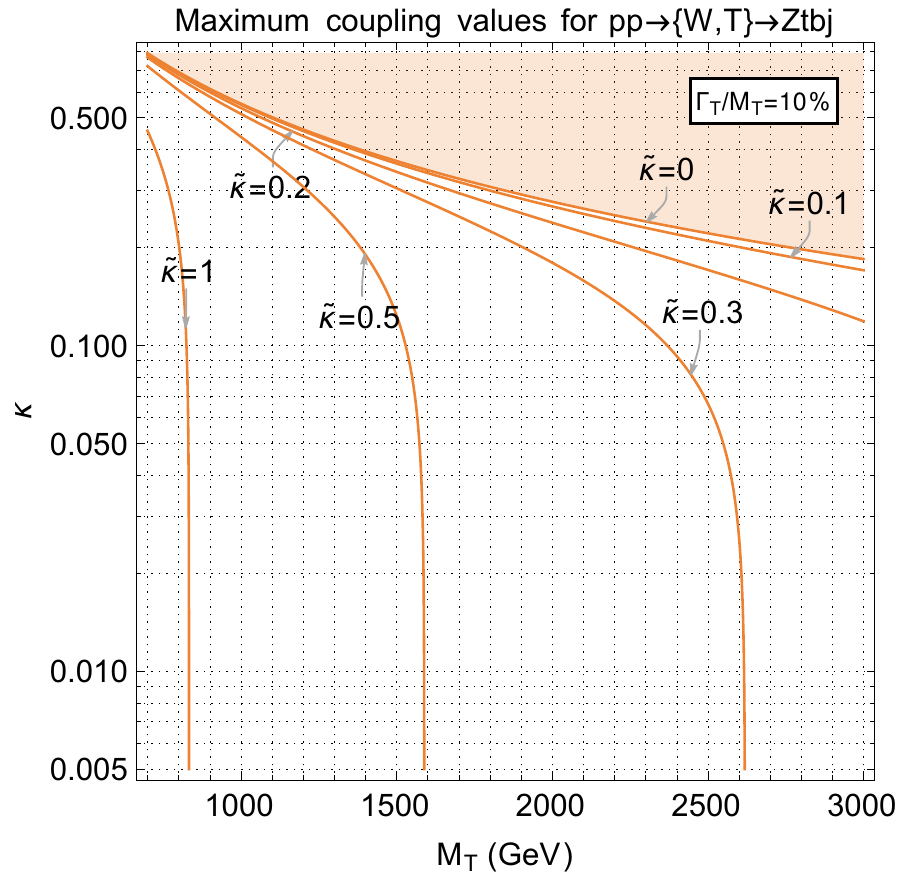}
	\includegraphics[width=.325\textwidth]{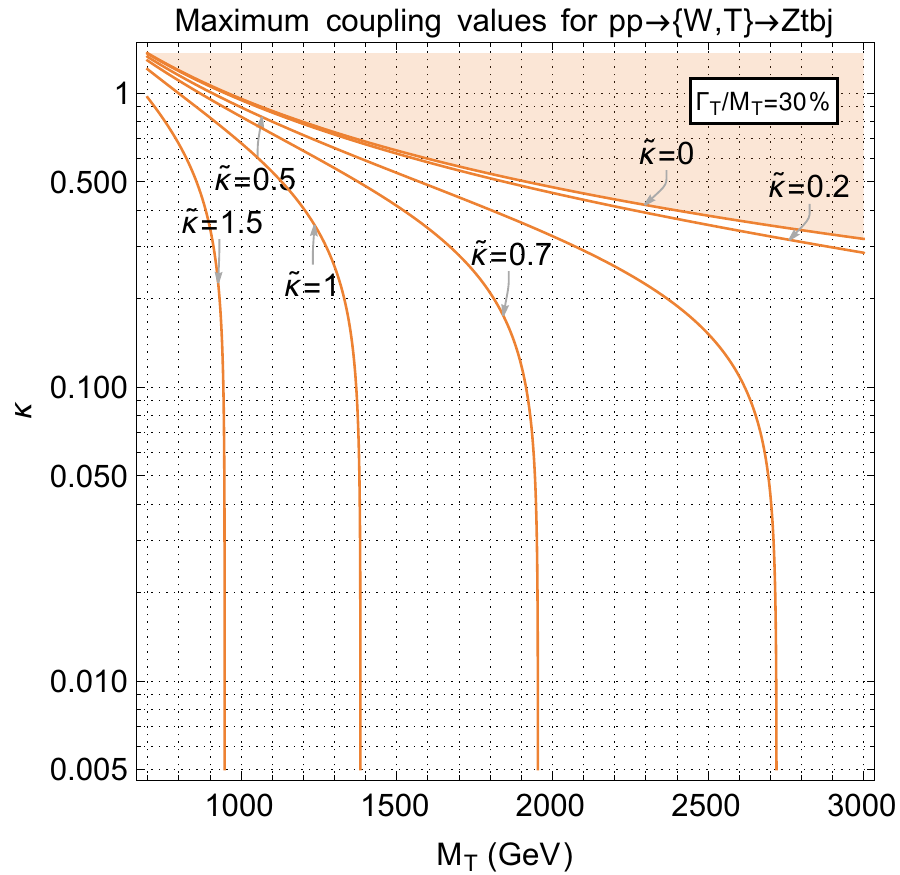}
	\\
	\includegraphics[width=.325\textwidth]{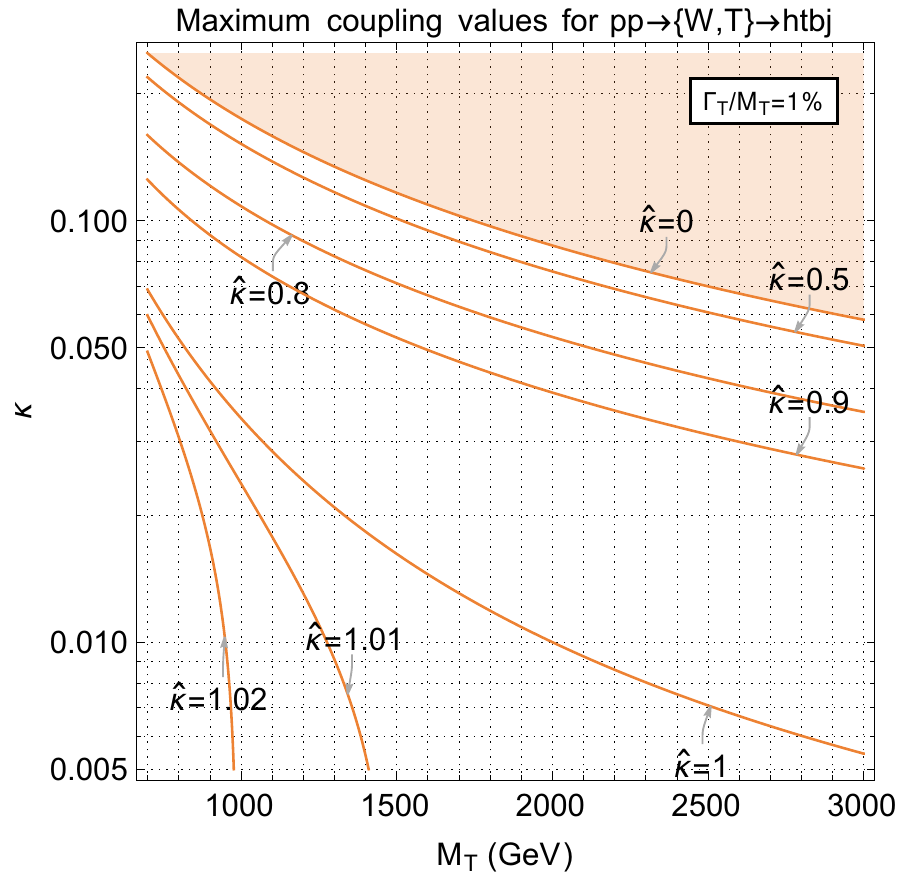}
	\includegraphics[width=.325\textwidth]{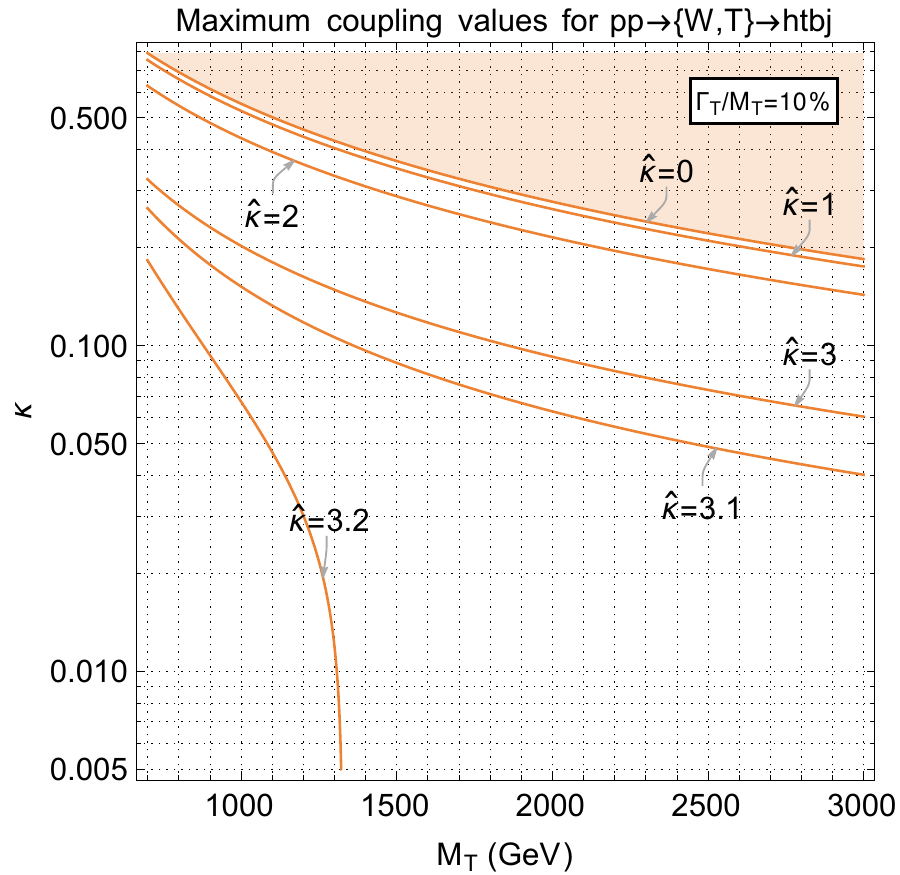}
	\includegraphics[width=.325\textwidth]{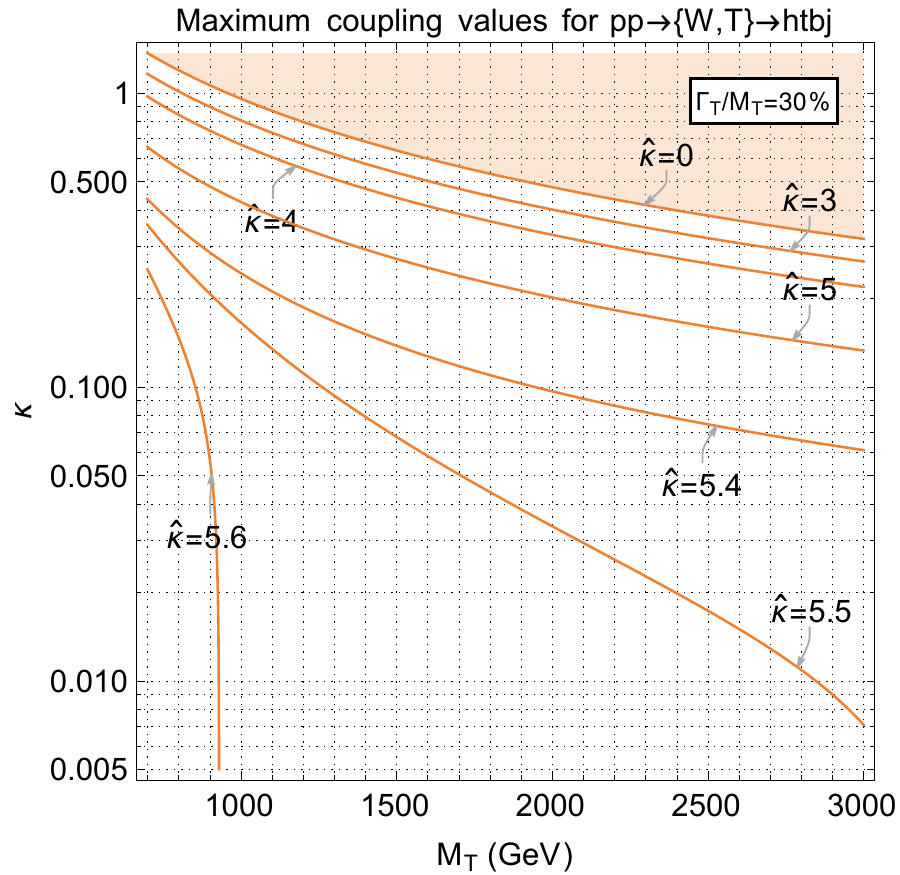}\vspace*{-0.2cm}
	\caption{\label{fig:maxkappa_T} Maximum values of the couplings which saturate the values of the total width of the $T$ VLQ for the $pp\to Wbbj$ (top), $Ztbj$ (centre) and $Htbj$ (bottom) processes and for different values of the $\Gamma_T/M_T$ ratio for the $Ztbj$ and $Htbj$ processes. The couplings  $\kappa$, $\tilde\kappa$, $\hat\kappa$  refer respectively to the coupling of $Wb$, $Zt$ and $ht$ with the VLQ $T$. As the partial widths depend on the couplings squared, the limits above apply to the absolute values of the couplings.}
\end{figure}

In addition to these cross section maps, we provide in figure~\ref{fig:maxkappa_T} the maximum values of the couplings $\kappa$, $\tilde{\kappa}$ and $\hat{\kappa}$ that are needed to saturate the total width $\Gamma_T$ for the three final states and for various width-over-mass ratios. For $pp\to Wbbj$, $T$ VLQ production and decay proceed through the same coupling $\kappa$, which is maximised (for a given mass and width) if all other $T$ couplings are zero. The maximal $\kappa$ value is thus determined as a function of $\Gamma_T/M_T$ and $M_T$, as shown in the top panel of figure~\ref{fig:maxkappa_T}. For $pp\to Ztbj$ (middle row of figure~\ref{fig:maxkappa_T}), $\kappa\neq 0$ is required for the production process, while the decay proceeds through a $\tilde{\kappa}$ coupling. This increases the number of relevant quantities to three. We therefore show contours of maximal $\tilde{\kappa}$ coupling strength in the $\{M_T,\kappa\}$ plane for fixed $\Gamma_T/M_T$ values of $1\%,\, 10\%$ and $30\%$, assuming no further non-zero $T$ couplings besides $\kappa$ and $\tilde{\kappa}$.  Analogously, for the $pp\to htbj$ process (bottom row of figure~\ref{fig:maxkappa_T}), we show contours of maximal $\hat{\kappa}$ coupling strength in the $\{M_T,\kappa\}$ plane, assuming no further $T$ couplings other than $\kappa$ and $\hat{\kappa}$. 

\begin{figure}
	\centering
	\includegraphics[width=.45\textwidth]{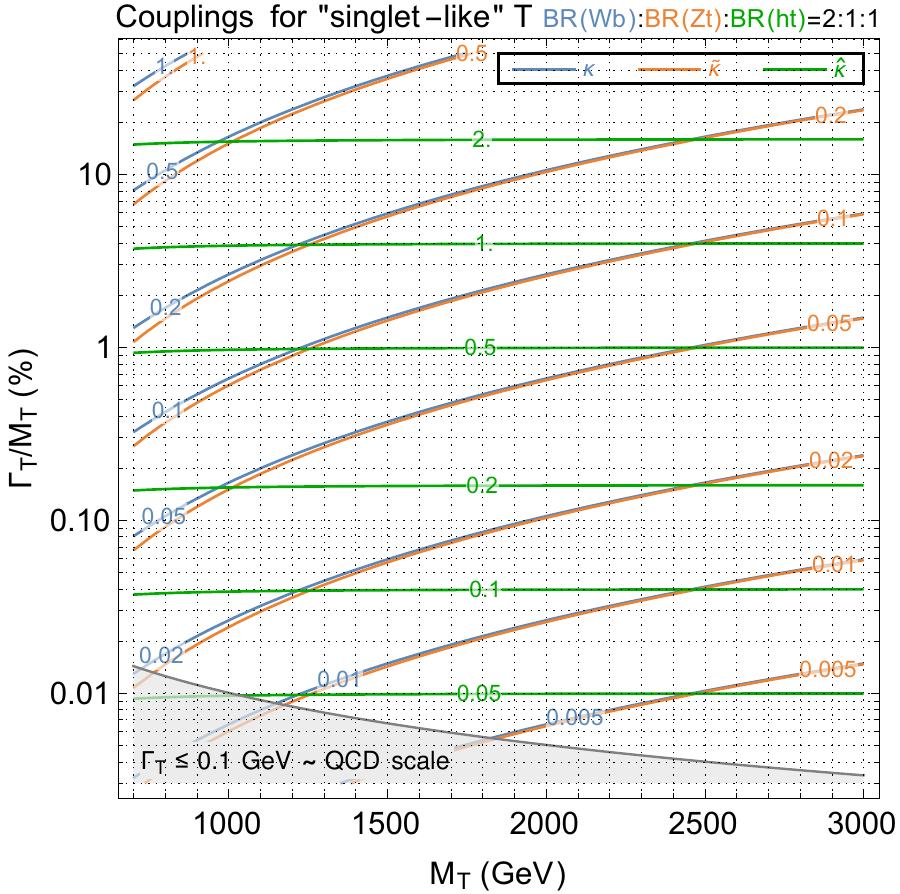}
	\includegraphics[width=.45\textwidth]{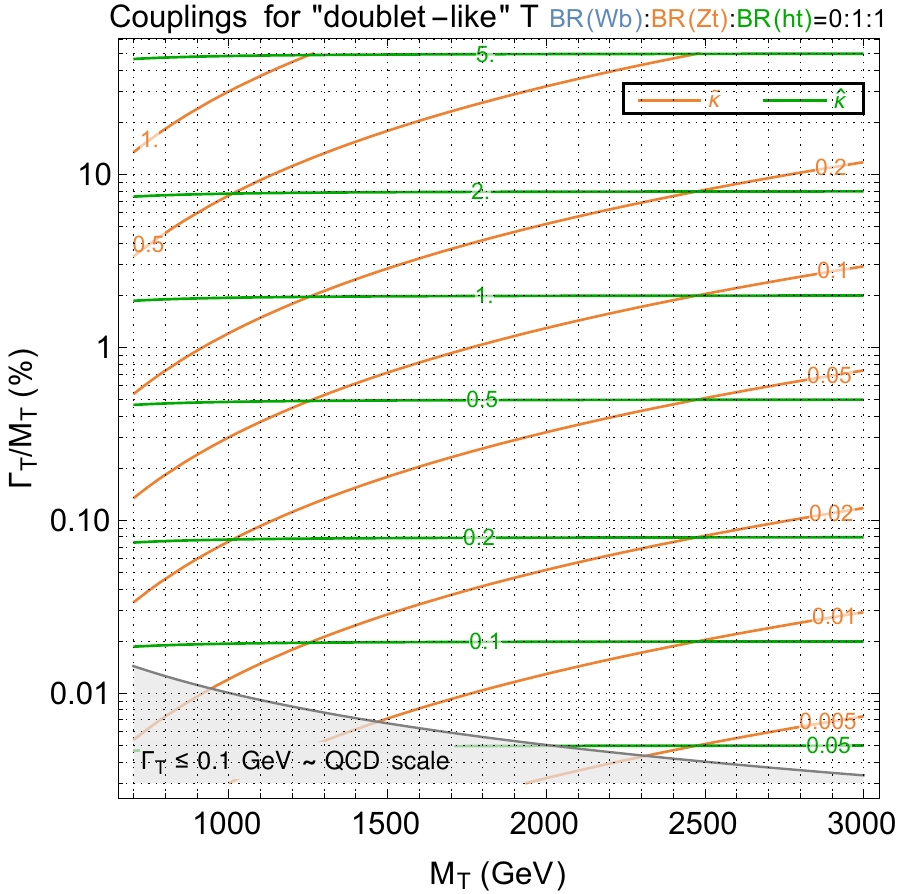}\vspace*{-0.2cm}
	\caption{\label{fig:singletdoubletT} Values of the couplings which saturate the values of the total width of the $T$ VLQ and reproduce the asymptotic relations satisfied by the VLQ branching ratios. We consider a ``singlet-like'' $T$ (left), {\it i.e.}\ BR$(Wb)$:BR$(Zt)$:BR$(ht)=2\!:\!1\!:\!1$, and a ``doublet-like'' $T$ (right), {\it i.e.}\ BR$(Wb)$:BR$(Zt)$:BR$(ht)=0\!:\!1\!:\!1$. The couplings  $\kappa$, $\tilde\kappa$, $\hat\kappa$  refer respectively to the coupling of $Wb$, $Zt$ and $ht$ with the VLQ $T$. As the partial widths depend on the couplings squared, the results above apply to the absolute values of the couplings.}
\end{figure}

In figure~\ref{fig:singletdoubletT} we also provide maps of maximal coupling values for a given relative width for two popular benchmark models: the ``singlet-like'' $T$ model, for which in the high mass limit the branching ratios are related by the Goldstone boson equivalence theorem as BR$(Wb)$:BR$(Zt)$:BR$(ht)=2\!:\!1\!:\!1$ (already mentioned in \cref{subsec:signal}), and the ``doublet-like'' $T$ model, for which the relations between the branching ratios are BR$(Wb)$:BR$(Zt)$:BR$(ht)=0\!:\!1\!:\!1$. In a consistent simplified model with a singlet or a doublet $T$ VLQ interacting with the SM quarks, these relations are valid to a good approximation in the NWA and for large $T$ masses. As the width increases, the validity range of these asymptotic relations solely concerns larger values of the $T$ mass. In our analysis, however, the branching ratio relations are imposed by hand, and the couplings necessary to obtain them are computed correspondingly, thus implicitly implying the presence of unidentified new physics which affects the relations between the $T$ couplings but does not impact the single production results otherwise. In other words, these benchmarks are purely phenomenological, and used only for the purpose of comparison with the corresponding NWA scenarios.

To illustrate the use of the cross section maps in figure~\ref{fig:sigmahat_W_T_14TeV} and the coupling maps of figures~\ref{fig:maxkappa_T} and \ref{fig:singletdoubletT}, we consider the example of the single production of a $T$ quark via $W$ exchanges, for a scenario in which $M_T=2\TeV$, $\Gamma_T/M_T = 30\%$ and ``singlet-like'' branching ratio relations. We focus on the $T\rightarrow th$ decay channel ({\it i.e.}\ on the process $pp\to\{W,T\}\to htbj$). From figure~\ref{fig:singletdoubletT} (left), the maximal couplings authorised for these parameter points are $\kappa = \tilde{\kappa} \simeq 0.34$ and $\tilde{\kappa}\simeq2.8$. From figure~\ref{fig:sigmahat_W_T_14TeV} one obtains $\hat{\sigma}^{\rm VLQ}_{ht}\simeq32\fb$ and $\hat{\sigma}^{\rm int}_{ht}\simeq-8.7\fb$ for the LHC at 14~TeV, and thus, via \cref{eq:xsec_param}, a VLQ contribution of $29\fb$ and a VLQ-SM interference contribution of $-8.3\fb$ to the full production cross section. This has to be compared to a SM background contribution of $65.7\fb$, as shown in \cref{eq:SMxsecs}. This illustrative example shows the relevance of the interference contribution for VLQs with broad width for high mass searches in VLQ single production.

A few remarks are however in order.
\begin{itemize}
	\item The signal cross section for a 30\% width-over-mass ratio and a $T$ mass of $M_T=2$~TeV is very large\footnote{We recall that the $htbj$ cross section is reduced  by the branching ratios of the top quark and that of the Higgs boson once a specific final-state signature, including the decay of all heavy SM particles, is targeted.} and possibly on the verge of exclusion.
	\item In the presence of cuts and selections, the splitting of the cross section into SM, interference and VLQ contributions as done in \cref{eq:xsec_param} can still be performed. New cross section maps, including the cuts, should however be produced, as figure~\ref{fig:sigmahat_W_T_14TeV} is only valid if no cuts are imposed)~\cite{Carvalho:2018jkq}. The relative contributions of the signal and the interference with the SM will indeed change. Nevertheless, an impact of interference is still to be expected, as by their very purpose, the cuts select events with signal-like kinematics. Background events surviving them hence occupy similar phase-space regions as signal events, which makes interference more likely.
\end{itemize} 

In the above example we discussed only one parameter point and one $T$ decay channel (with $T$ production from $Wb$). The same procedure can be applied to $T$ production from $Wb$ exchanges and $T$ decay into a $Wb$ or a $Zt$ system by relying instead on figure~\ref{fig:sigmahat_W_T_14TeV} for the LHC at 14 TeV, and on figure~\ref{fig:sigmahat_W_T_13TeV} for the LHC at 13 TeV. Production modes from $Zt$ exchanges could be addressed from the results presented in figures~\ref{fig:sigmahat_Z_T_13TeV} and \ref{fig:sigmahat_Z_T_14TeV}. The maximal couplings can here be read off from figure~\ref{fig:maxkappa_T} for branching ratios maximised for a specific decay channel, or from figure~\ref{fig:singletdoubletT} for singlet-like and doublet-like $T$ branching ratios. The coupling maps are only provided for the convenience of the reader. Maps for other branching ratios are easily generated from the analytical expressions of the partial widths.

\subsubsection{Large width effects on differential distributions}\label{sec:LOdifferential}
In the remainder of this section, we present examples where the large VLQ width and the interference of the new physics contributions with the SM ones impact differential distributions commonly used in searches for single VLQ production. We focus in particular on distributions which might allow to distinguish scenarios featuring different $M_T$ and/or different $\Gamma_T/M_T$ values.

The most obvious way to measure the mass and the width of a resonance would be to reconstruct its invariant mass distribution, if the targeted final state allows for it. This may however necessitate a sufficiently large excess of events, and may therefore not be the first sign of a hypothetical $T$ resonance. We thus focus on alternative but commonly used distributions as potential handles for the first detection of anomalies, and aim to illustrate the impact of broad widths and interference contributions on the determination of the cuts that could be imposed on the corresponding observables.

The distributions presented in this section are generated at LO (NLO QCD corrections being discussed in \cref{sec:nlo}), using the generator-level cuts of \cref{eq:generationcuts} for the $2\to4$ processes of eqs.~\eqref{eq:sgl1} and \eqref{eq:sgl2}. The cross sections reported in the legends of the plots reflect this setup. Moreover, all unstable SM particles but the Higgs boson are decayed (inclusively) by means of \mg, the Higgs boson being dealt with through {\sc Pythia 8}~\cite{Sjostrand:2014zea} as spin correlations are not relevant. {\sc Pythia 8} is also used to simulate parton showering, as well as hadronisation. The reconstruction of the objects in the final state is done through \ma~\cite{Conte:2012fm,Conte:2014zja,Araz:2020lnp}, using the anti-$k_T$ algorithm~\cite{Cacciari:2008gp} with radius parameter $R = 0.4$ as implemented in {\sc FastJet}~\cite{Cacciari:2011ma}.

\begin{figure}
	\centering
	\includegraphics[width=.325\textwidth]{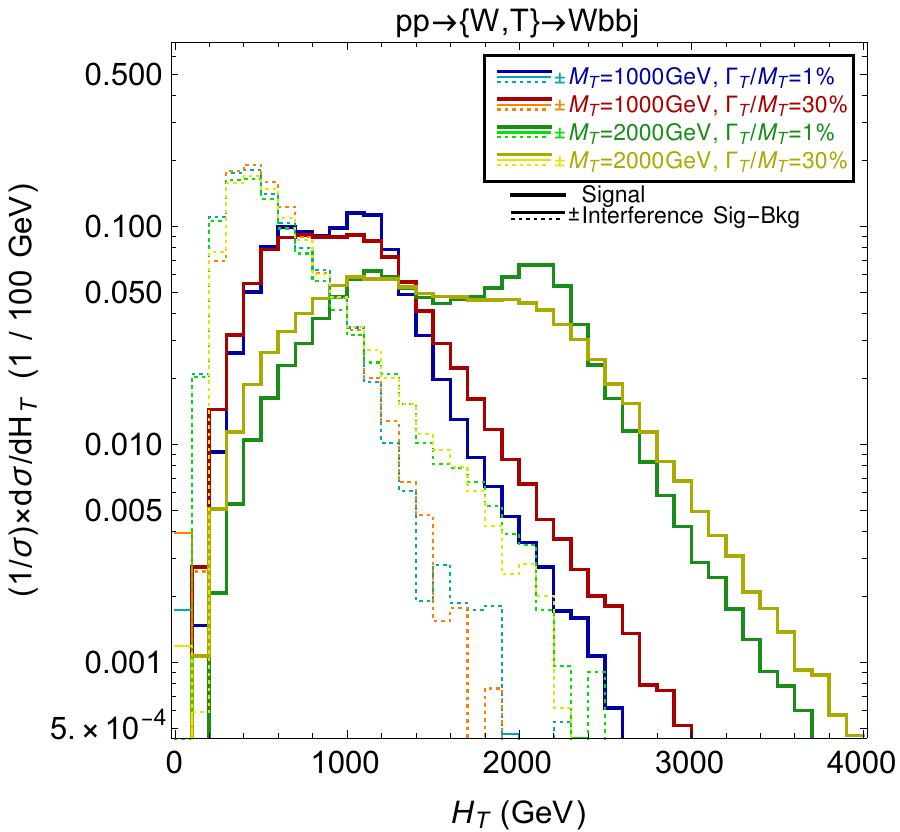}
	\includegraphics[width=.325\textwidth]{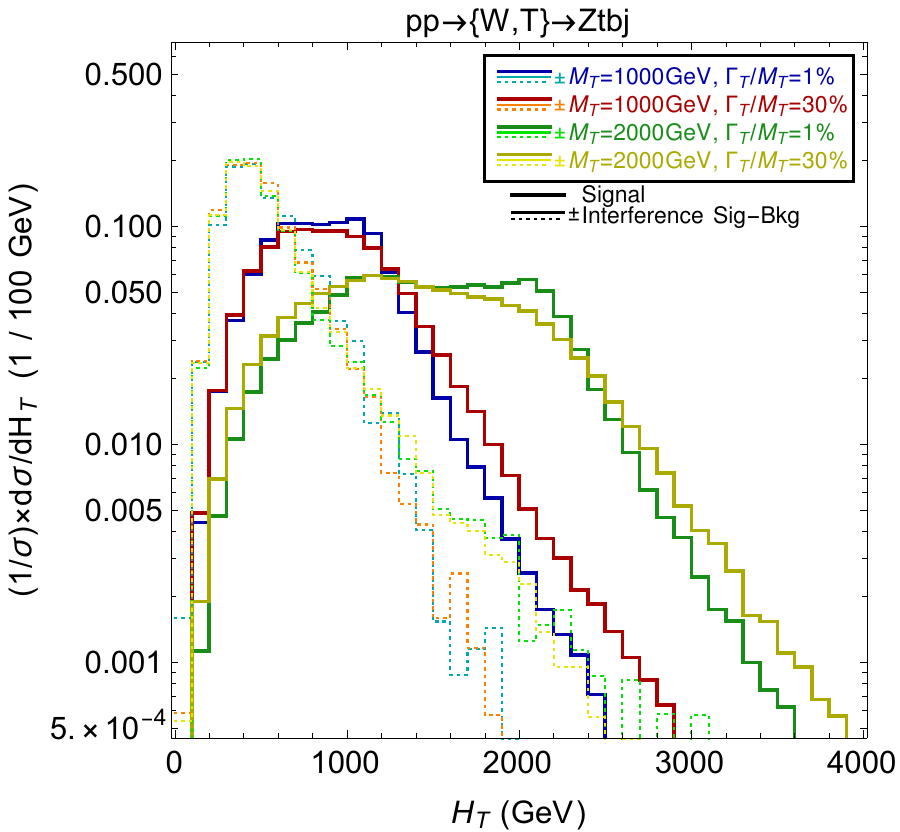}
	\includegraphics[width=.325\textwidth]{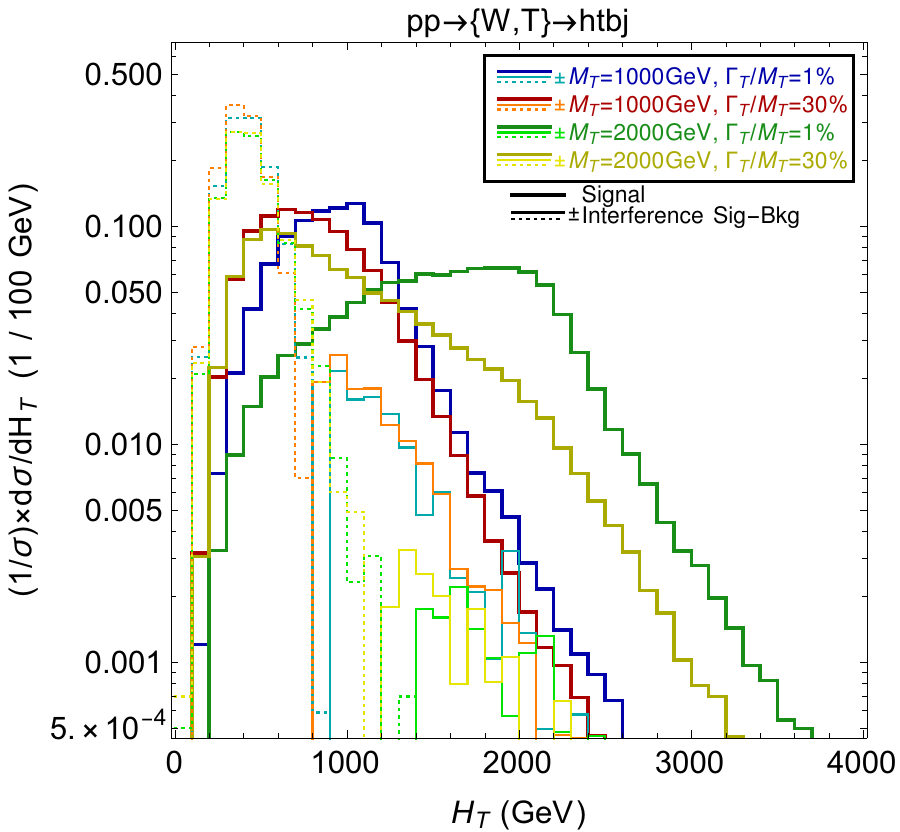}
	\\
	\includegraphics[width=.325\textwidth]{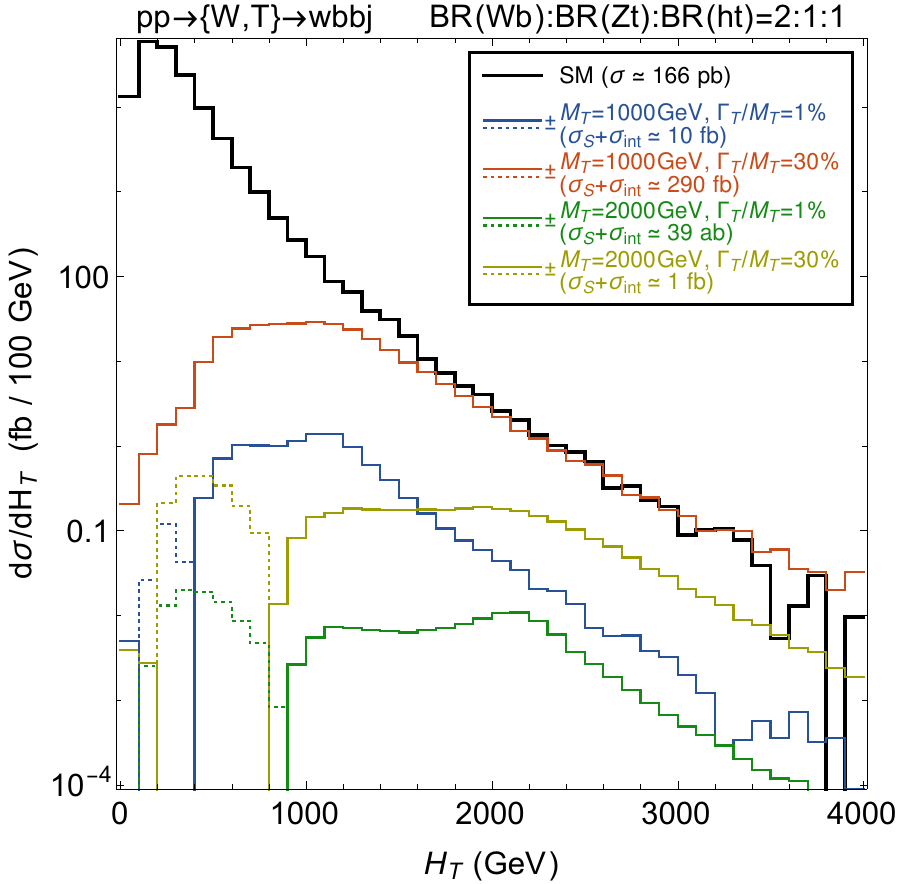}
	\includegraphics[width=.325\textwidth]{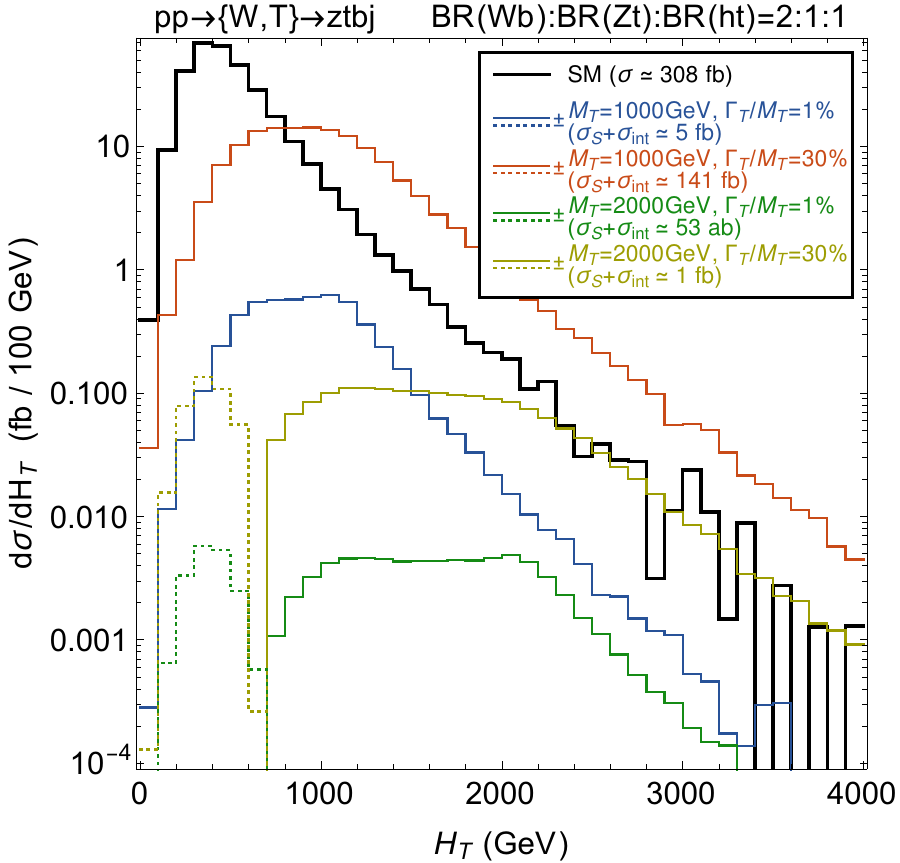}
	\includegraphics[width=.325\textwidth]{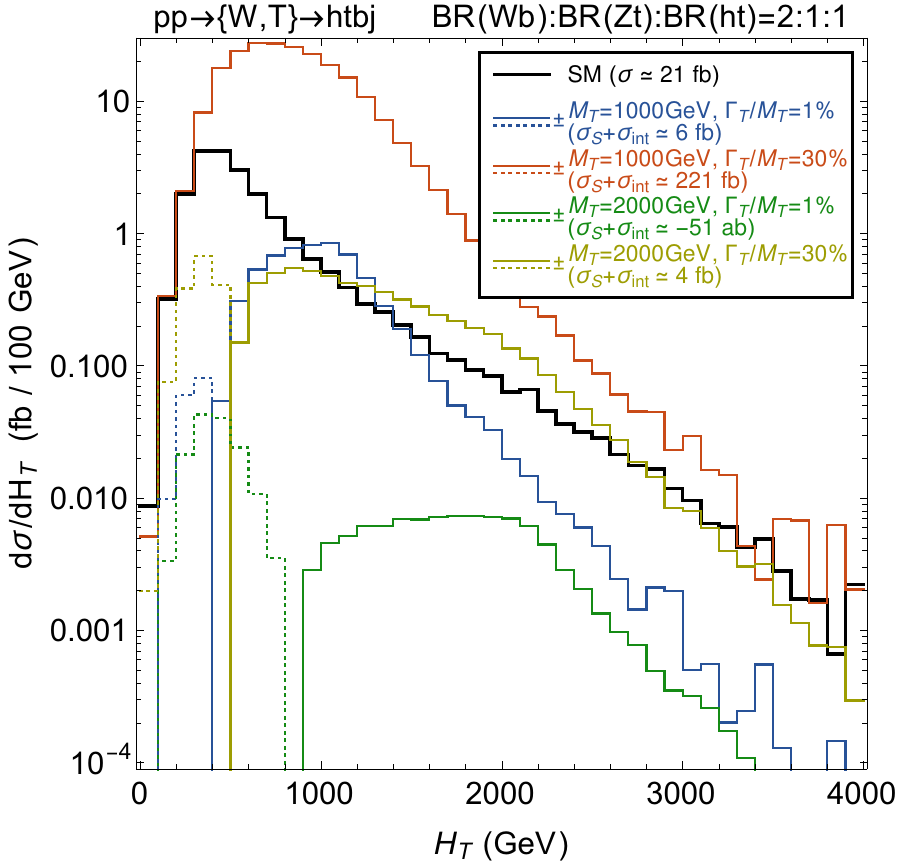}\vspace*{-0.2cm}
	\caption{\label{fig:recodistHT} $H_T$ distribution at the reconstructed level for the $pp\to Wbbj$ (left), $Ztbj$ (centre) and $htbj$ (right) processes. We consider two values of the $T$ mass ($1000\GeV$ and $2000\GeV$), and two values of the $\Gamma_T/M_T$ ratio (1\% and 30\%). In the top row, we focus on the shapes of the signal contributions as well as on those associated with the interference of the signal with the SM background. In the bottom row, the sum of the signal and interference contributions is performed following eq.~\eqref{eq:xsec_param}, the couplings are chosen according to the singlet case (see figure~\ref{fig:singletdoubletT}), and we compare the predictions with the SM expectation. Analogous results for $T$ production through $Zt$ exchanges are given in figure~\ref{fig:recodistHT_Z}.}
\end{figure}

\paragraph{Hadronic activity}

In figure~\ref{fig:recodistHT}, we present distributions for the scalar sum of the transverse momenta of all reconstructed light jets and $b$-jets ($H_T$), for the processes $pp\to\{W,T\}\to Wbbj$ (left), $Ztbj$ (centre) and $htbj$ (right). We consider two benchmark masses of $M_T=1\TeV$ and $2\TeV$, both for a narrow ($\Gamma_T/M_T=1\%$) and a large ($30\%$) width configuration. The top row of the figure focuses on the visualisation of the signal shape of the $H_T$ distributions, so that all spectra are normalised to 1. We consider both the pure VLQ signal component (thick lines), as well as its interference with the SM diagrams (thin solid lines for a positive interference and thin dotted lines for a negative one). The bottom row of the figure allows for the comparison of the $H_T$ SM background distribution (black) with the signal ones (coloured solid lines and coloured dotted lines when positive and negative respectively), after the pure signal component and its interference with the SM are combined. In this case, we assume maximal VLQ couplings and consider the singlet-like benchmark model. In this bottom row, the distributions are normalised to the cross sections reported in the legend for each curve. The signal cross sections are obtained by summing the pure signal and interference contributions, while the SM ones are lower than  in \cref{eq:SMxsecs} as they have been obtained using the kinematical cuts of \cref{eq:generationcuts}.

For the $Wbbj$ and $Ztbj$ channels the pure signal distributions exhibit mild differences when the VLQ width is increased for a given mass. The large width scenarios generally lead to spectra featuring a tail falling less rapidly, and the small width distributions exhibit two clear peaks, one at $M_T$ and another one at $M_T/2$, the second peak being much less evident in the large width case. The distributions originating from the interference do not present, on the other hand, any sizeable differences for different widths, but only depend on $M_T$. Those contributions are always negative. For the $Wbbj$ final state, we recall that the SM background rate lies two orders of magnitude above that in the other channels, and thus generally dominates the signal. The $M_T=1\TeV$ scenario with a large width consists of one exception in the high-$H_T$ regime, thanks to enhanced new physics couplings (that are required to yield a large width). Those large coupling values also explain why the total cross sections relevant for both large width scenarios are larger (as already pointed out in the previous sections). For the signal, the different role of the interference for different $M_T$ values is clearly visible: for $M_T=1\TeV$ the interference contribution plays a negligible role in the total signal shape when the width is large, but becomes important for small $H_T$ values if the width is small. For $M_T=2\TeV$ the interference is dominant and negative for $H_T\lesssim900\GeV$ in both the narrow and large width cases, and is negligible otherwise. For the $Ztbj$ final state, the background is less dominant, at least if as adopted here the coupling values are maximised to increase the VLQ width. The interplay between pure signal and interference contributions is stronger for higher $M_T$. This is in particular clearly visible at low $H_T$ for $M_T=2\TeV$ in the large width case, where the negative interference becomes dominant in the lower bins, as well as in the narrow case where the combined (signal plus interference) contributions are comparable and oscillate between positive and negative values. In the latter case, they nevertheless correspond to a very low (undetectable) cross section.

For the $htbj$ channel, the pure signal component of the spectrum for $M_T=1\TeV$ has a very similar distribution regardless of the width-over-mass ratio. The small differences are however more pronounced than for the two other processes, the peak of the distribution being more shifted towards lower $H_T$ values when $\Gamma_T/M_T$ increases. For $M_T=2\TeV$ the behaviour of the distributions in the small and large width scenarios is on the contrary very different: the peak at $H_T\sim M_T$ is visible in the narrow width case, although it is already quite broad, and is then completely lost in the large width case. Here, the distribution peaks instead at smaller $H_T$ values, closer to what the distributions feature for $M_T=1\TeV$ in the large-width case. The interference distributions are not as dependent on the total width, but the mass dependence is more pronounced than for the other processes. There is moreover a change of sign when $H_T\sim M_T$. The relative weight of the interference depends on $M_T$, this weight becoming more apparent for high $M_T$ values. In this way, the combined pure signal and interference distribution for $M_T=2\TeV$ and in the narrow width case is completely dominated by the negative interference contribution, the integrated rate being even negative. This contrasts with all other cases.

\paragraph{The transverse momentum spectrum of the leading light jet}
\begin{figure}
  \centering
  \includegraphics[width=.325\textwidth]{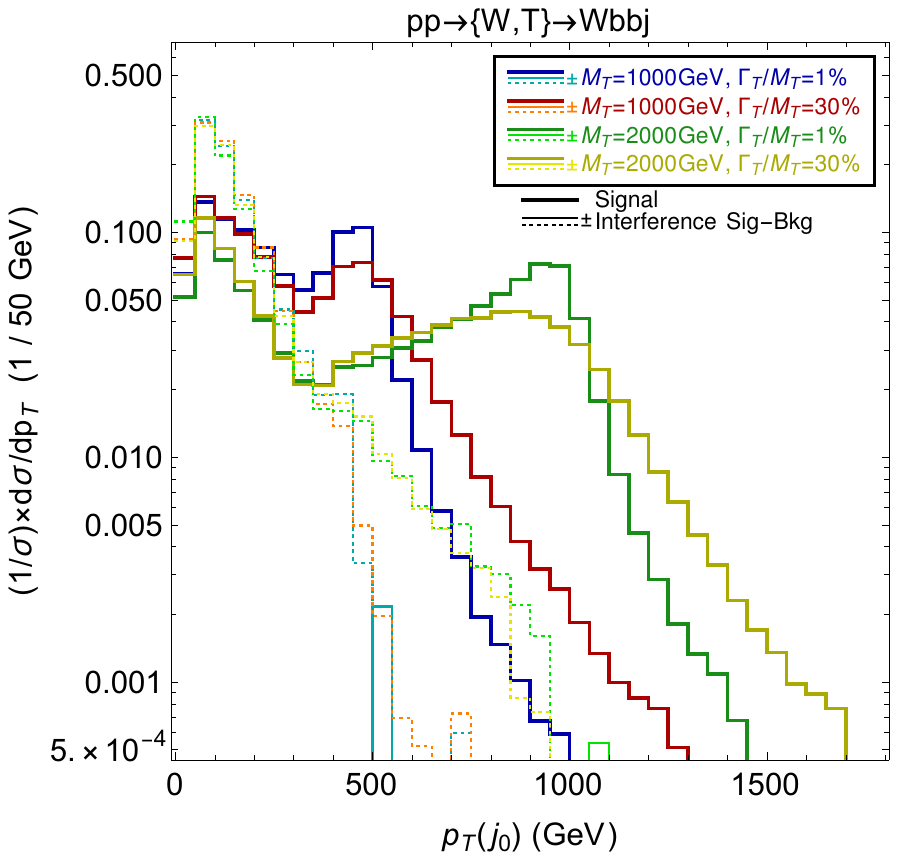}
  \includegraphics[width=.325\textwidth]{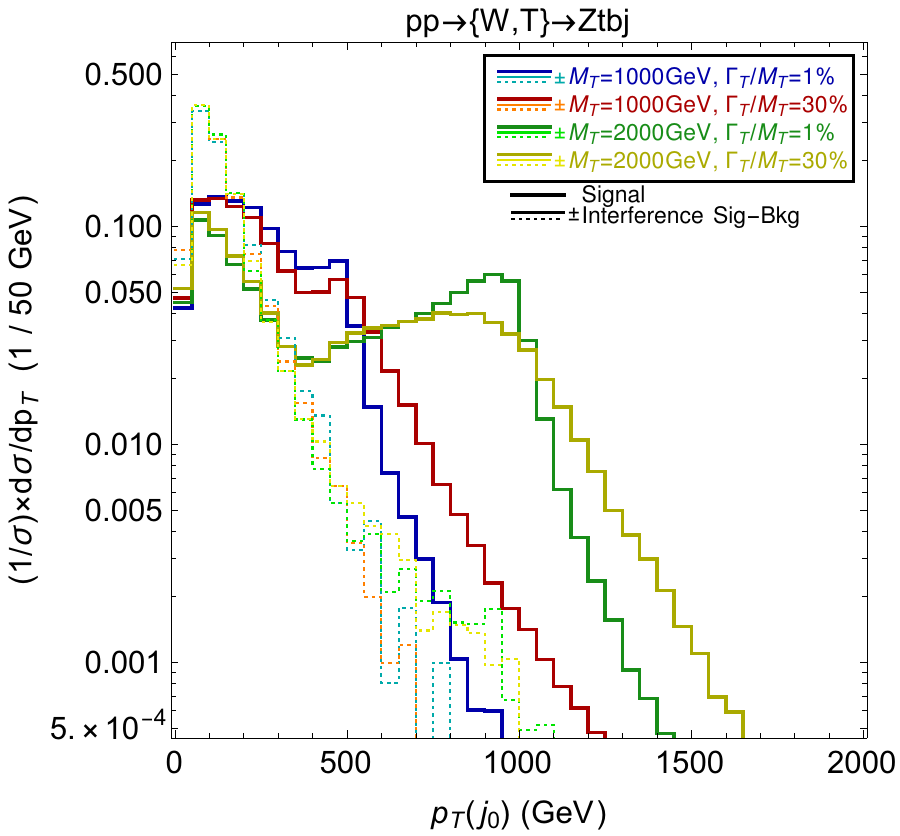}
  \includegraphics[width=.325\textwidth]{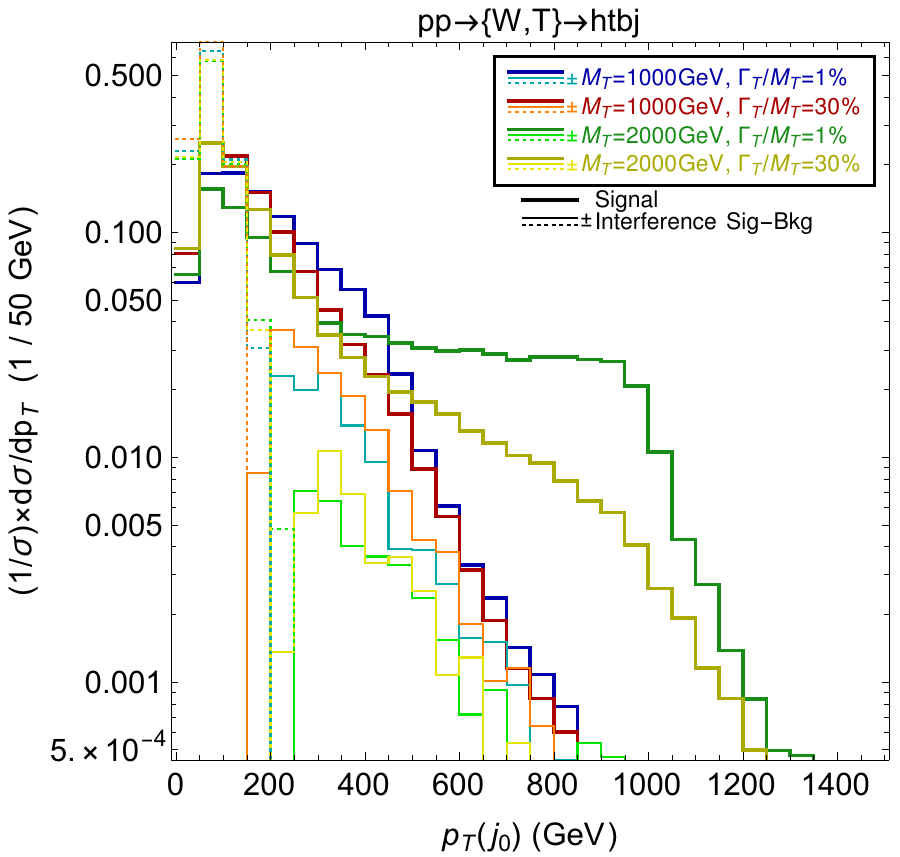}
  \\
  \includegraphics[width=.325\textwidth]{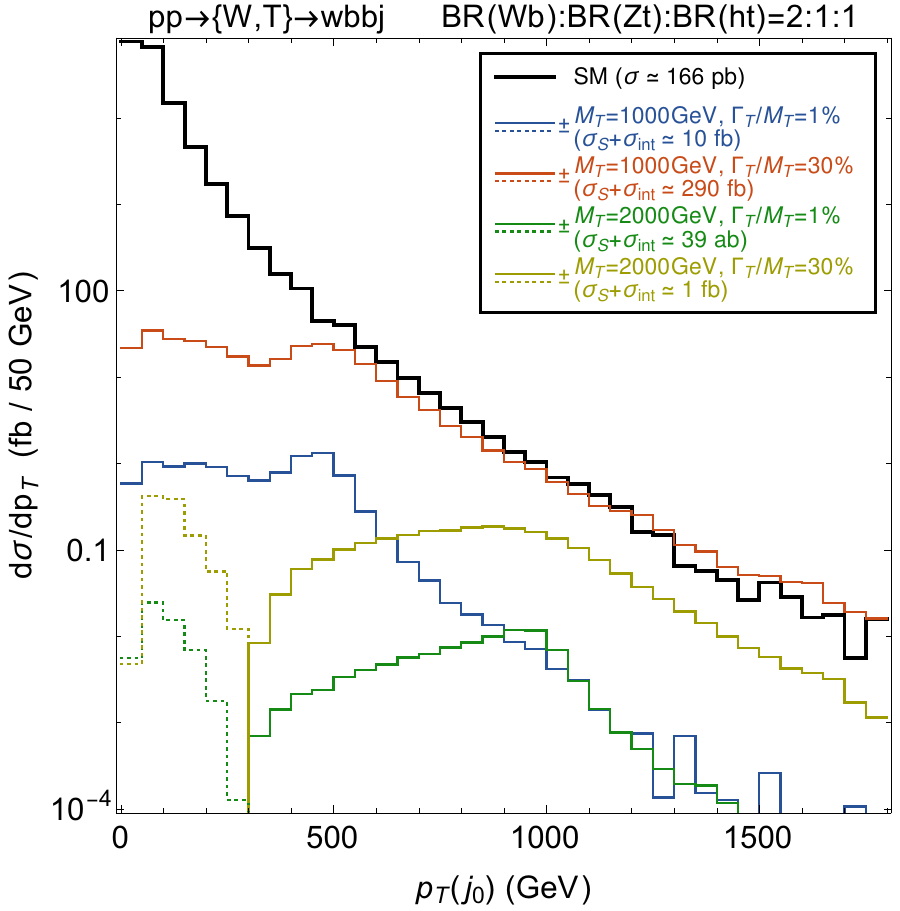}
  \includegraphics[width=.325\textwidth]{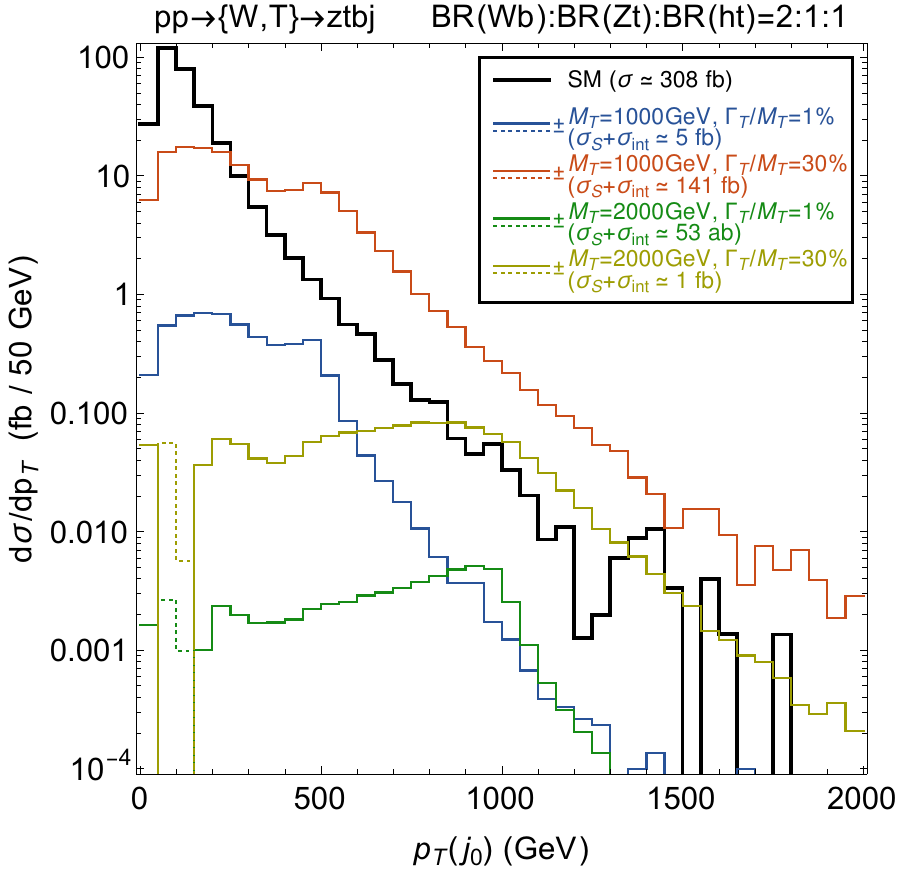}
  \includegraphics[width=.325\textwidth]{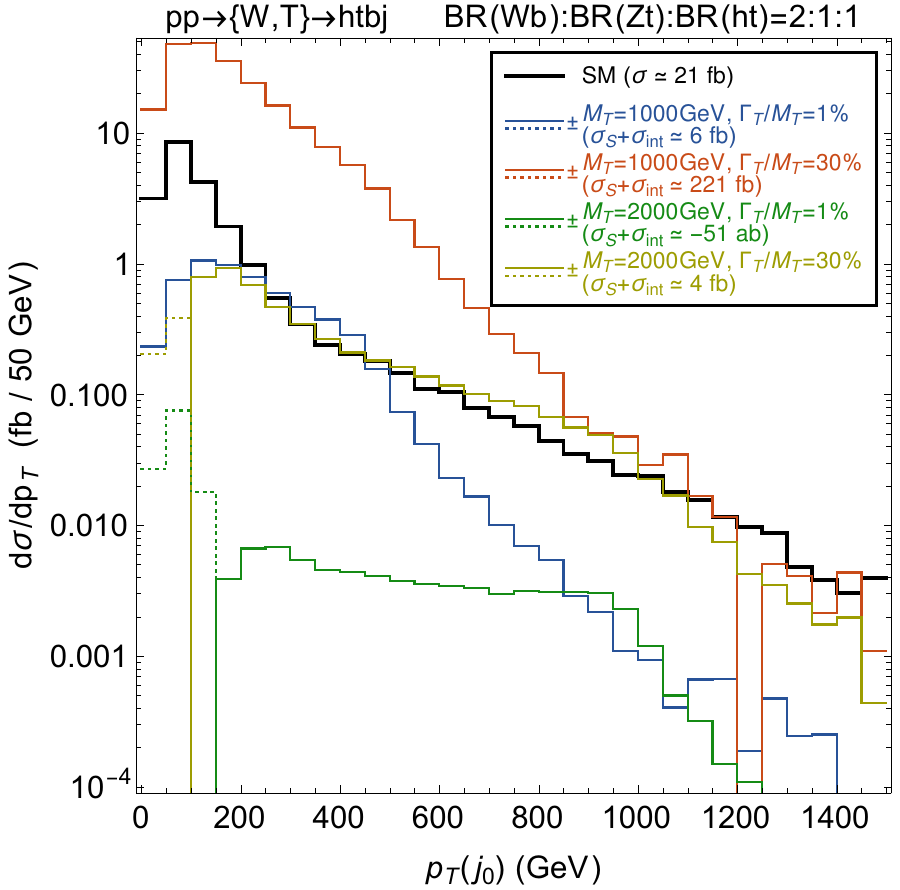}\vspace*{-0.2cm}
  \caption{\label{fig:recodistpTj0} Same as figure~\ref{fig:recodistHT} for the distribution in the $p_T$ of the leading light jet. Analogous results for $T$ production through $Zt$ exchanges are shown in figure~\ref{fig:recodistpTj0_Z}.}
\end{figure}

In this subsection, we consider another commonly-used observable in phenomenological analyses, and we study the distribution in the leading jet transverse momentum ($p_T(j_0)$). We investigate the same benchmark scenarios as in the previous section, the VLQ mass $M_T$ being set to either $1\TeV$ or $2\TeV$, and the width-over-mass ratio being fixed to either 1\% or 30\%. The results are shown in figure~\ref{fig:recodistpTj0} for a simulation chain in which we once again include the inclusive decays of all heavy SM particles. The top row of the figure shows the leading jet $p_T$ distributions for the  $pp\to Wbbj$, $Ztbj$ and $htbj$ final states, the results being normalised to one so that we could compare the shapes of the signal and interference distributions. The bottom row of the figure compares, as previously, the combined `pure signal and interference' predictions with the SM expectation, the cross sections for the SM background and for the sum of the pure VLQ signal and its interference with the SM are indicated in the legends of the subfigures.

In the $Wbbj$ and $Ztbj$ channels and for all benchmark points, the pure signal distributions display a peak at $\sim M_T/2$. Such a peak corresponds to a VLQ decay into a $b$-jet and a hadronically-decaying and boosted weak boson. The decay products of the latter are thus collimated so that they are reconstructed as a single jet with a $p_T$ equal to half the VLQ mass. Such a peak is not visible in the $htbj$ case, as the Higgs boson decay pattern is different. All signal spectra also feature a low energy peak, which results from the spectator jet produced in association with the $T$ quark. This peak is enhanced by events in which the SM bosons decay leptonically, and thus the spectator jet becomes the leading jet in terms of transverse momentum. Interference contributions play a role only at low  $p_T(j_0)$ and are negative, such that they reduce the potential impact of the low $p_T(j_0)$ peak as a handle on the signal. As can be expected, the peak at $M_T/2$ is broader for a larger $\Gamma_T/M_T$ value, so that the leading jet $p_T$ distribution is a good discriminator for both the $M_T$ and $\Gamma_T$ variables in the $Wbbj$ and $Ztbj$ channels. In \cref{sec:nlo}, we will quantify how NLO corrections affect those conclusions. For the $pp\to htbj$ process, a peak at low $p_T(j_0)$ is again visible. As for the other two processes, it originates from the spectator jet produced in association with the heavy $T$ quark. The corresponding enhancement of the contributions to the cross section at low invariant masses described in  \cref{subsec:parton_ME} is in addition clearly visible on top of the SM background.

\begin{figure}
  \centering
  \includegraphics[width=.325\textwidth]{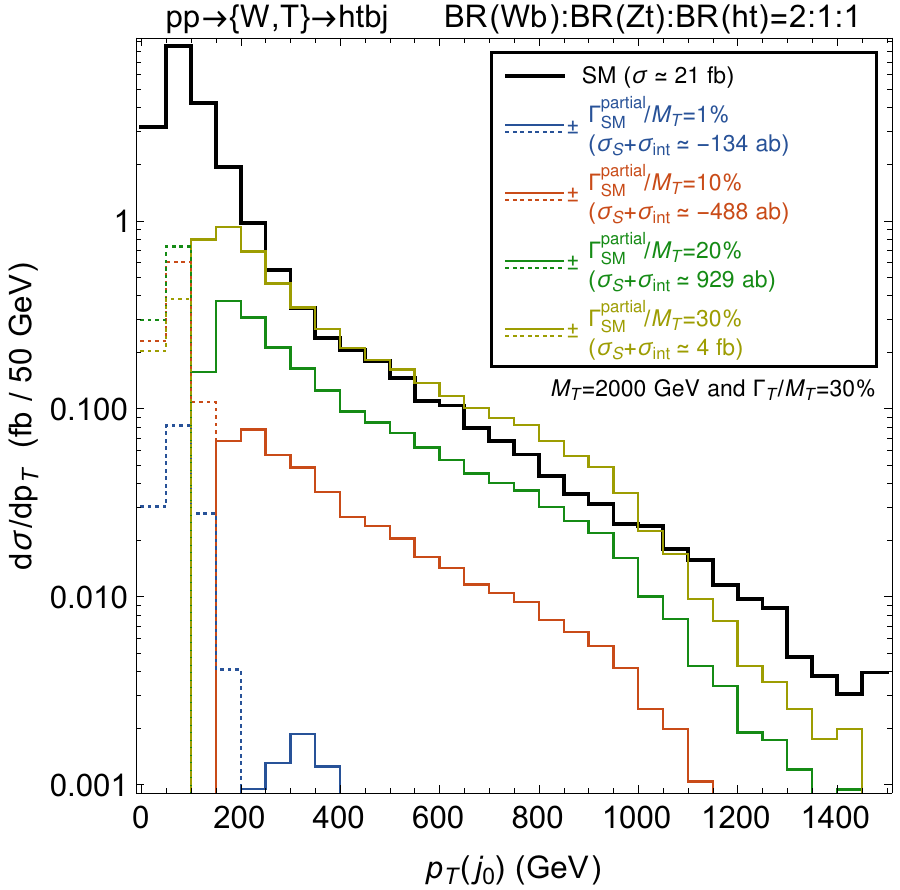}\hspace{.3cm}
  \includegraphics[width=.325\textwidth]{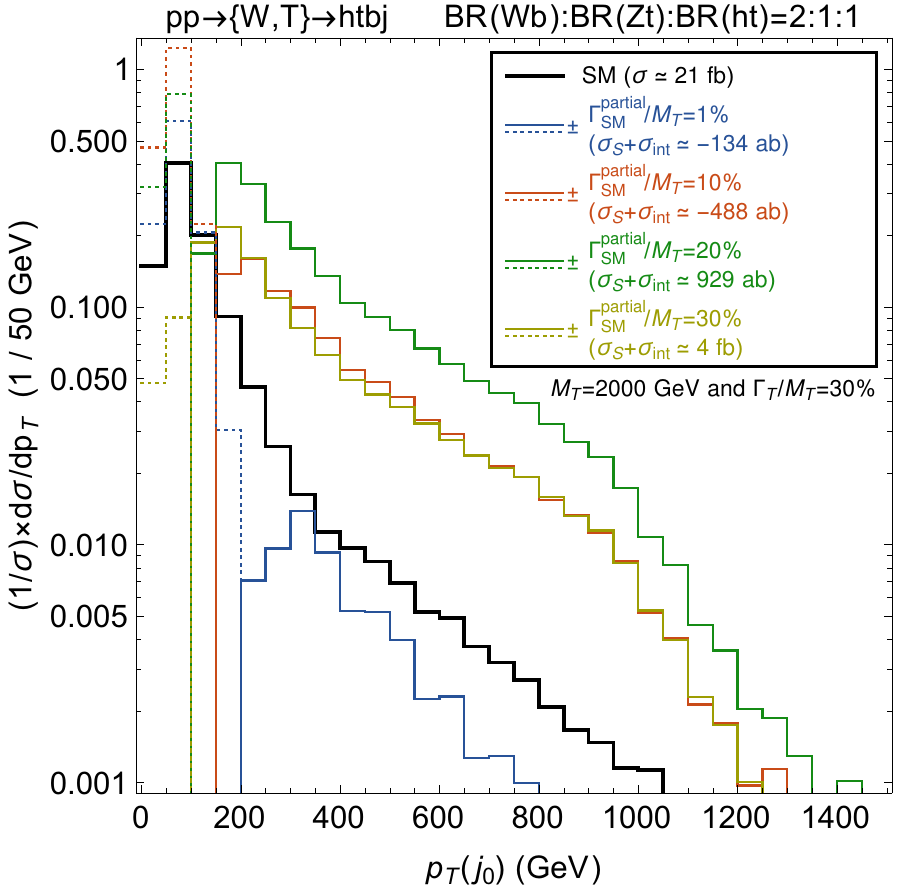}\vspace*{-0.2cm}
  \caption{\label{fig:recodistpTj0htbjpartial} Leading jet $p_T$ distribution for $pp\to htbj$. We fix $M_T=2000\GeV$ and $\Gamma_T/M_T=30\%$, and we set the VLQ couplings such that the branching ratios correspond to the singlet case and such that the sum of the partial widths for decays into SM final states corresponds to either $\Gamma_{\rm SM}^{\rm partial}/M_T=1\%$ (blue), $10\%$ (red) and $20\%$ (green), or saturates the total width (mustard). In the left panel the distributions are compared with the SM background and normalised to their respective cross section, while in the right panel we report the shapes of the distributions.}
\end{figure}

Before concluding this LO discussion, we provide a quantitative evaluation of how much the shapes of distributions would change by considering that the VLQ width is large due to exotic $T$ decay channels. In such a setup, we can build new physics scenarios where the VLQ has a large width and the $\kappa$, $\hat\kappa$ and $\tilde\kappa$ couplings are small. In figure~\ref{fig:recodistpTj0htbjpartial} we display the leading jet $p_T$ distribution resulting from the single production of a VLQ $T$ quark with a mass $M_T = 2\TeV$, and a width-over-mass ratio of 30\%. We focus on the $htbj$ channel, and we show shapes (right panel) and normalised-to-the-cross-section distributions (left panel) for the  SM background (black) and the combined `pure signal plus interference' VLQ signal (coloured). For reference, we first compute predictions where the $\kappa$, $\hat\kappa$ and $\tilde\kappa$ coupling values are maximised (and are thus the sole reason for a VLQ large width), in the ``singlet-like'' scenario. The corresponding results, already shown in the right panel of figure~\ref{fig:recodistpTj0}, are reported once again through the mustard curves in figure~\ref{fig:recodistpTj0htbjpartial}. In addition, we show distributions for the same VLQ mass and {\it total} width, for cases in which the SM \emph{partial} width is $\Gamma_{\rm SM}^{\rm partial}/M_T=1\%$ (blue), $10\%$ (red) and $20\%$ (green). As previously, all corresponding cross sections are indicated in the legend of the plots. Such a scenario could for example be realised if the $T$ quark (chain-)decays into a final state made of invisible objects. This is the case, among others, in scenarios with universal extra-dimensions in which VLQs belonging to $KK$-even tiers decay into a pair of $KK$-odd particles. The latter in turn decay to a final state involving dark matter and SM objects~\cite{Cacciapaglia:2009pa,Cacciapaglia:2013wha}, which potentially leads to signatures exhibiting large missing transverse momentum that are likely excluded by experimental cuts targeting visible VLQ decays. As said above, the presence of such an extra decay channel allows us to get a broad VLQ featuring smaller $\kappa$, $\hat\kappa$ and $\tilde\kappa$ couplings, and therefore a different signal (total and differential) cross section as compared to the predictions of the curves in mustard colour. 

The relative dependence on the $\kappa$, $\hat\kappa$ and $\tilde\kappa$ couplings being different for the interference and the pure signal component, we now study the full impact on the combined predictions. As the partial width into the SM final state becomes smaller with respect to the fixed total width, two effects can be seen. First, as said above, we observe an obvious reduction of the signal cross section, which becomes negative below $\Gamma_{SM}^{\rm partial}/M_T=20\%$. This translates the fact that the (destructive) interference between the VLQ signal and the SM diagrams becomes dominant. Second, the peak of the distribution is shifted towards higher values of $p_T(j_0)$, the shift being larger for smaller SM contributions to the total width. This is again entirely due to the interplay between the positive signal and the interference term which, on the contrary, does not have a definite sign. The shapes of the positive part of the distribution (the one which is likely to be targeted by any associated experimental cuts) are qualitatively similar, but with a positive rescaling only if the SM partial width accounts for  about 20\% of $M_T$ or more.

%%%%%%%%%%%%%%%%%%%%%%%%%%%%%%%%%%%%%%%%%%%%%%%%%%%%%%%%%%%%%%%%%%%%%%%%%%%%%%
\section{Predictions at the next-to-leading-order accuracy in QCD}
\label{sec:nlo}
When considering VLQ single production, the signal cross section and the shape of the differential distributions can receive non-trivial contributions from next-to-leading order corrections in the strong coupling $\alpha_s$, as VLQs carry colour charges. This could hence have a significant impact on the description of the processes, and be of crucial relevance for the optimisation of the analysis strategies. NLO QCD corrections have the advantage that as long as the strong sector of the SM is not modified, they do not depend on the details of the theory in which the VLQs appear, except for their masses via their propagation in the loops. In the present section an approximate treatment of QCD NLO corrections and their interplay with the VLQ finite width is presented, focusing on how kinematic distributions are modified, and under which limitations such a treatment is possible.

\subsection{Approximate treatment of the finite width at next-to-leading order (and its limitations)}

Consistent NLO QCD predictions for LHC processes involving the propagation of broad VLQs would require computations to be achieved in the complex mass scheme, as described in \cref{sec:width}. We can however only rely on an NLO UFO model~\cite{Fuks:2016ftf} that does not allow for complex masses. The necessary extension of such a UFO model would require the redefinition of all UV counter-terms~\cite{Frederix:2018nkq}. Such an effort consists of a laborious operation, as the automated framework allowing to do so has never been tested in a new physics context featuring new coloured particles, and would be meaningful only in cases of hints of anomalies in LHC data pointing towards the existence of VLQs. We therefore consider instead an approximate treatment that involves the combination of three ingredients for a given process and a given final state (as defined in eqs.~\eqref{eq:sgl1} and \eqref{eq:sgl2}). Those are based on the generation of different samples of events and are defined in the following.
\begin{enumerate}
 \item {\bf \{LO,LW\}}. We generate a first sample of events at leading order in QCD, in which we include a finite VLQ width value, we use the complex mass scheme and we consider the 2-to-4 processes of eqs.~\eqref{eq:sgl1} and \eqref{eq:sgl2}. We enforce diagrams to feature the propagation of a single VLQ by exploiting the \verb+VLQ+ coupling order labeling of the new physics interactions implemented in the UFO model, as described in \cref{app:Tech}. This allows us to include both off-shell contributions and topologies where the VLQ does not propagate resonantly, as well as their interference. 
 \item {\bf \{LO,NWA\}}. We generate a second sample of events at leading order in QCD, in which the VLQ is treated in the narrow-width approximation regardless of its width. In other words, its resonant production is factorised from its decay.
 \item {\bf \{NLO,NWA\}}. We generate a third sample of events at next-to-leading order in QCD, in which the VLQ is treated in the narrow-width approximation. Such computations are reliable from the UFO model designed in ref.~\cite{Fuks:2016ftf}.
\end{enumerate}

For each observable $\cal O$ of interest, we define a differential $K$-factor as the ratio of the predictions obtained with the \{NLO,NWA\} and \{LO,NWA\} samples. We then estimate the NLO differential cross section in the large width regime \{NLO,LW\} by rescaling the corresponding LO differential cross section \{LO,LW\} with such a $K$-factor,
\begin{equation}
 \label{eq:NLOLW}
 \left({d\sigma\over d\mathcal{O}}\right)_{\text{\{NLO,LW\}}} \simeq
 {\left({d\sigma\over d\mathcal{O}}\right)_{\text{\{NLO,NWA\}}} \over 
 \left({d\sigma \over d\mathcal{O}}\right)_{\text{\{LO,NWA\}}}} 
 \times 
 \left({d\sigma\over d\mathcal{O}}\right)_{\text{\{LO,LW\}}} \equiv 
 K_{\text{NWA}}
 \times 
 \left({d\sigma\over d\mathcal{O}}\right)_{\text{\{LO,LW\}}}\ .
\end{equation}
Such an approximation is based on the assumption that the differential $K$-factor does not significantly depend on the VLQ width-over-mass ratio. This corresponds to assuming that the contributions from the topologies where the VLQ can be resonantly produced dominate, while the contributions from any other topology is sub-leading. This assumption has been verified for the final states originating from the single production of a vector-like quark $T$ from $W$-boson exchanges (see \cref{subsec:signal}). Other final states and production modes have, however, to be treated individually. For example, not all sub-processes involving the single production of a $T$ quark via $Z$-boson exchanges satisfy the above assumptions (see \cref{app:TviaZ}). 
Furthermore, this procedure can only be applied to the pure signal component of the cross section, and not to its interference with the SM background. This comes from the fact that the structure of the interference is not compatible  with the factorisation of the VLQ production and decay processes like in the NWA. Therefore, if the interference contribution has a large impact on the cross section and on the kinematics of the final state, the approximate NLO treatment proposed above might not be accurate enough when the width of the VLQ is large.

The asymmetric differential uncertainties associated with the scale systematics ($\Delta_\pm^{\text{scale}} f$) and the parton distribution function (PDF) systematics ($\Delta_\pm^{\text{PDF}} f$) have been obtained from the above relation and the correlations between the samples. The errors are estimated according to the following expression, that is valid on a bin-by-bin basis,
\begin{eqnarray}
{\Delta_\pm^{\text{scale,PDF}} f_{\text{\{NLO,LW\}}} \over f_{\text{\{NLO,LW\}}}} &=& \pm \sqrt{ \sigma^{u,\text{scale,PDF}}_\pm + \sigma^{c,\text{scale,PDF}}_\pm }\ .
\end{eqnarray}
In this notation, $f$ represents any differential cross section $d\sigma\over d\mathcal{O}$, and $\sigma^u_\pm$ and $\sigma^c_\pm$ are the uncorrelated and correlated components of the uncertainties respectively. For both the scale variation and the PDF systematics, these quantities are defined as
\begin{eqnarray}
\sigma^u_\pm &=& \left(\Delta_\pm f_{\text{\{NLO,NWA\}}} \over  f_{\text{\{NLO,NWA\}}}\right)^2 + 
\left(\Delta_\pm f_{\text{\{LO,NWA\}}} \over f_{\text{\{LO,NWA\}}}\right)^2 + 
\left(\Delta_\pm f_{\text{\{LO,LW\}}} \over f_{\text{\{LO,LW\}}}\right)^2\ , \\
\sigma^c_\pm &=& 
2 \rho_{ f_{\text{\{NLO,NWA\}}},f_{\text{\{LO,LW\}}} }\left( {\Delta_\pm f_{\text{\{NLO,NWA\}}} \over  f_{\text{\{NLO,NWA\}}}} {\Delta_\pm f_{\text{\{LO,LW\}}} \over f_{\text{\{LO,LW\}}}} \right) \nonumber\\
&-& 2 \rho_{ f_{\text{\{NLO,NWA\}}},f_{\text{\{LO,NWA\}}} }\left( {\Delta_\pm f_{\text{\{NLO,NWA\}}} \over  f_{\text{\{NLO,NWA\}}}} {\Delta_\pm f_{\text{\{LO,NWA\}}} \over f_{\text{\{LO,NWA\}}}} \right) \nonumber\\
&-& 2 \rho_{ f_{\text{\{LO,LW\}}},f_{\text{\{LO,NWA\}}} }\left( {\Delta_\pm f_{\text{\{LO,LW\}}} \over  f_{\text{\{LO,LW\}}}} {\Delta_\pm f_{\text{\{LO,NWA\}}} \over f_{\text{\{LO,NWA\}}}} \right)\;,
\end{eqnarray}
where $\rho_{a,b}$ stands for the correlation between two quantities $a$ and $b$. It is defined as $\rho_{a,b}={V_{ab}\over\sigma_a\sigma_b}$, with $V_{ab}$ being the covariance between the quantities $a$ and $b$, and $\sigma_{a}$ and $\sigma_b$ their standard deviations. For the scale variation systematics, the uncertainties have been evaluated as the maximum relative difference (in the positive and negative ranges) between the value of the observable at each scale choice and the central value, while for the PDF uncertainties the conventions of ref.~\cite{Butterworth:2015oua} have been followed.

The total uncertainties on the samples are then computed by linearly combining the scale variation and the PDF uncertainties,
\begin{equation}
\Delta_\pm f_{\text{\{NLO,LW\}}} = \Delta_\pm^{\text{scale}} f_{\text{\{NLO,LW\}}} + \Delta_\pm^{\text{PDF}} f_{\text{\{NLO,LW\}}}\;.
\end{equation}
We nevertheless provide the individual contributions to the errors in separate panels in the results below.

An extra uncertainty should be associated with the choice of the scheme for the treatment of the VLQ finite width. This will not be explicitly added in the error bands shown on the plots below. We refer to the estimations from figure~\ref{fig:schemecomparison} for the invariant mass distribution. We have found that those uncertainties are larger for larger width-over-mass ratios, reaching values larger than about 50\% around the resonance peak. This contribution depends on the process and on the region in the phase space defined by the considered differential distribution.

\subsection{Phenomenology at the next-to-leading order in QCD}

The combined impact of NLO corrections and large width effects on the shape of the signal distributions can be sizeable. As will become clearer in the rest of this section, each of these two contributions can dominate over the other one, depending on the observable. As meaningful examples to illustrate their joint role, we consider various distributions relevant for single VLQ production processes involving of $W$-bosons exchanges, for a scenario in which the $T$ quark has a mass of $m_T=1000$~GeV. All presented distributions are evaluated at the reconstructed level without the inclusion of any detector effects, analogously to what has been done for figures~\ref{fig:recodistHT}, \ref{fig:recodistpTj0} and \ref{fig:recodistpTj0htbjpartial}.

\subsubsection{Leading jet and leading muon transverse momentum spectrum}

\begin{figure}[t]
\centering
\includegraphics[width=0.325\textwidth]{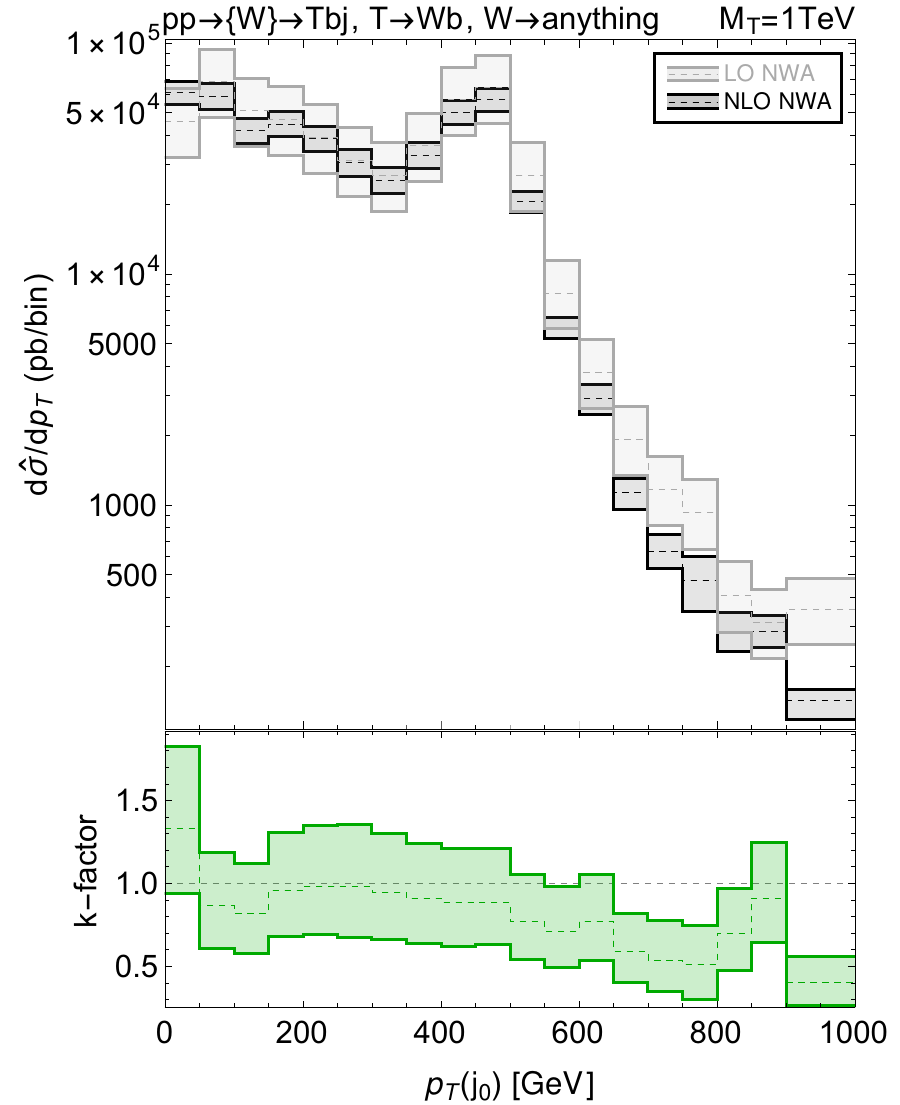}
\includegraphics[width=0.325\textwidth]{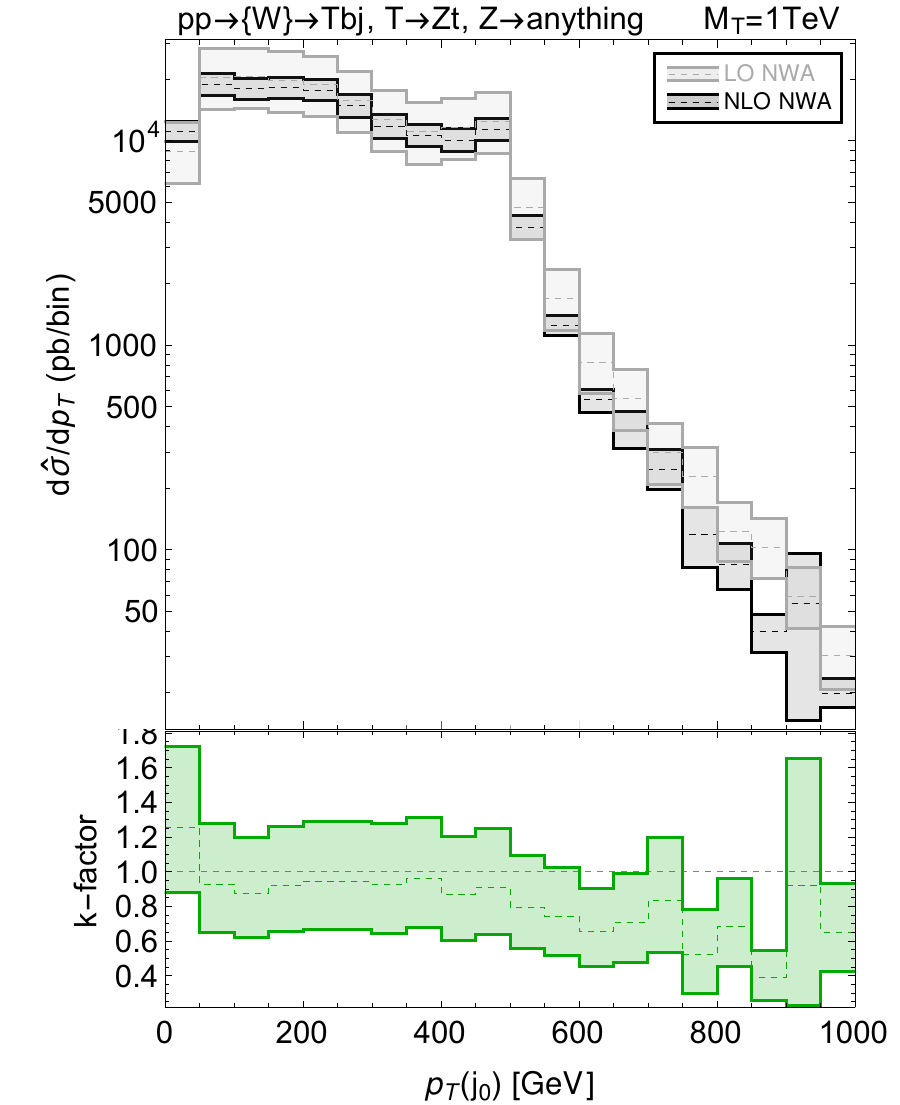}
\includegraphics[width=0.325\textwidth]{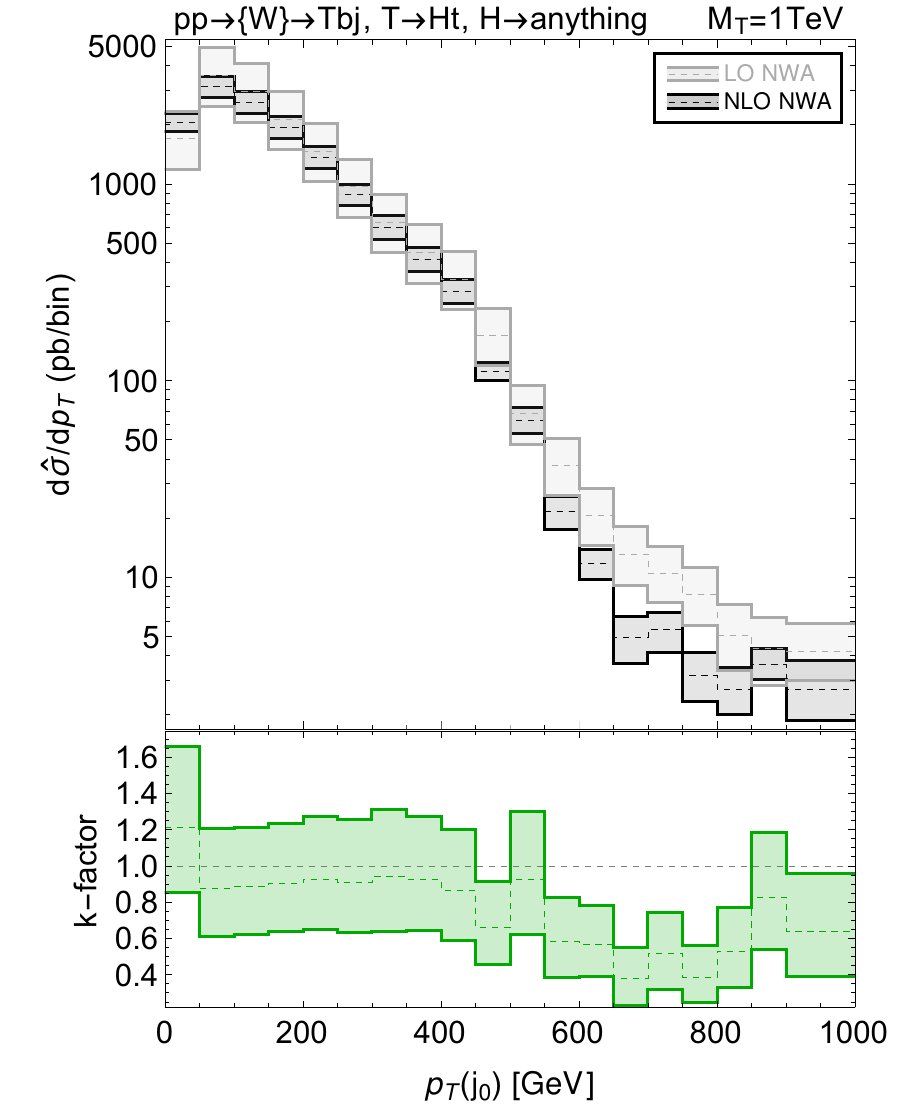}\\
\includegraphics[width=0.325\textwidth]{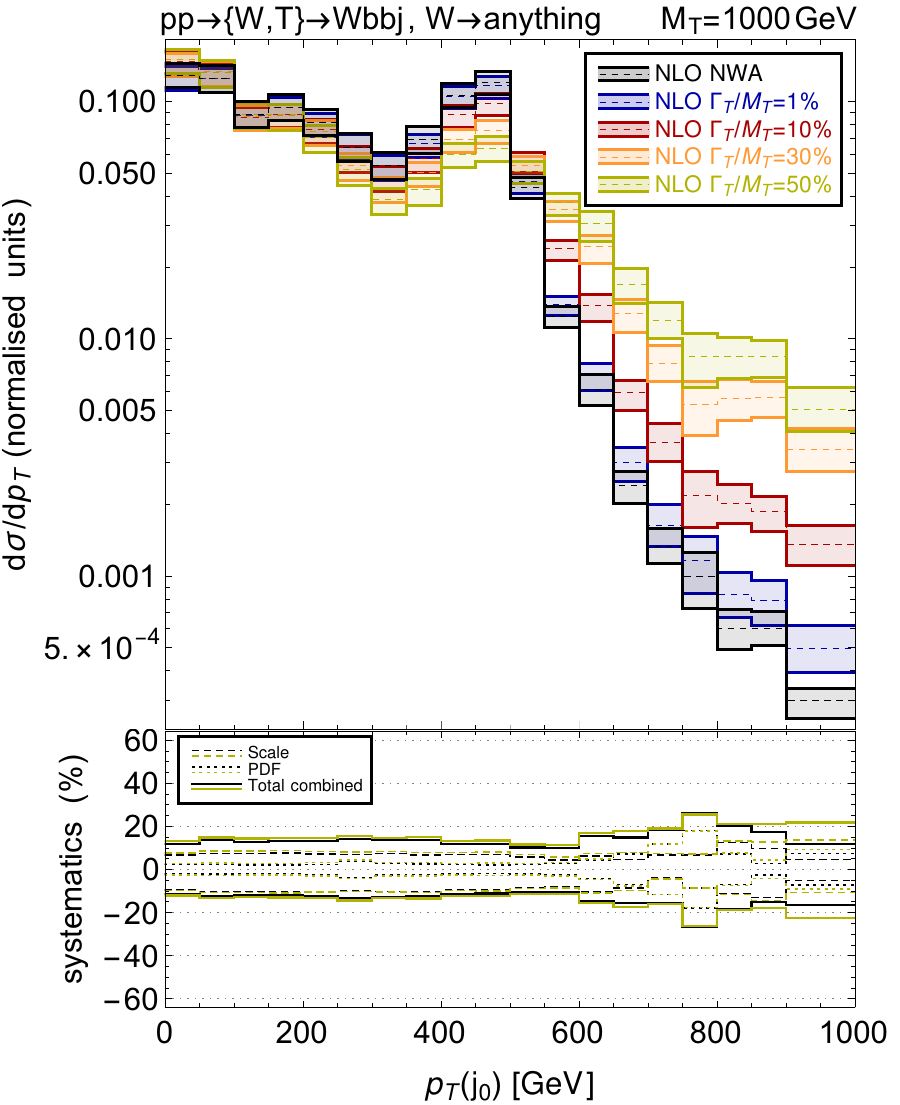}
\includegraphics[width=0.325\textwidth]{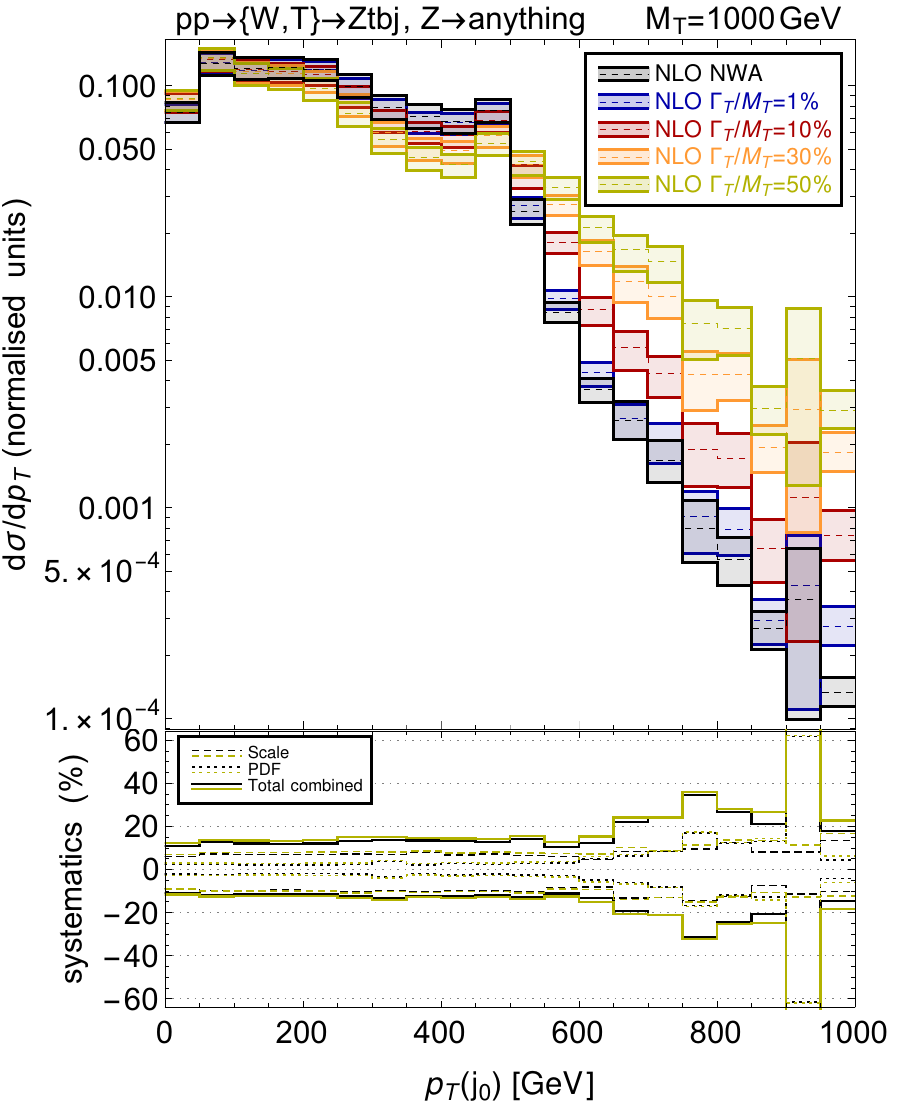}
\includegraphics[width=0.325\textwidth]{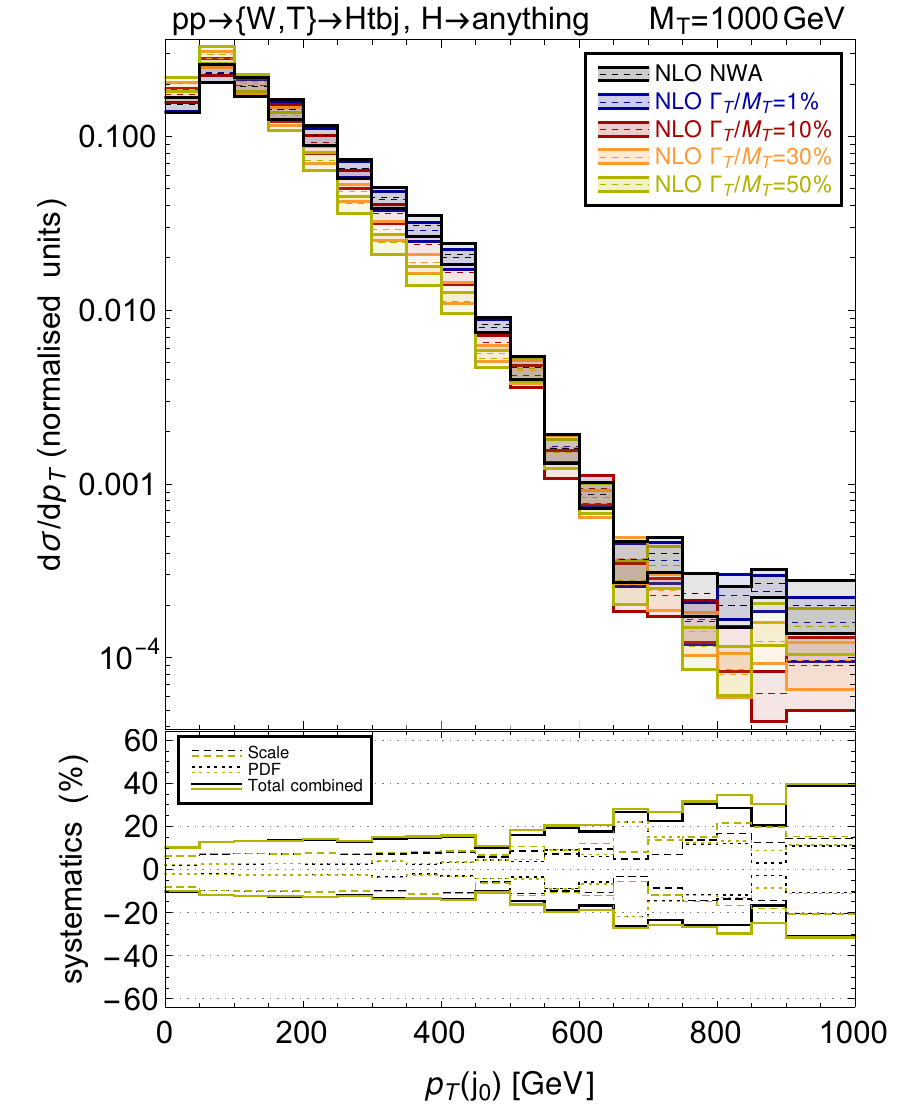}\vspace*{-0.2cm}
\caption{\label{fig:jet0pT} Distribution of the transverse momentum of the leading jet for single $T$ production in association with a jet. We fix $m_T=1000$~ GeV, and we consider the three channels $pp\to\{W,T\}\to Wbbj$ (left), $pp\to\{W,T\}\to Ztbj$ (centre) and $pp\to\{W,T\}\to htbj$ (right). The top row compares the \{LO,NWA\} and \{NLO,NWA\} distributions after factorising out the production coupling ({\it i.e.} the $y$-axis represents the values of the bare cross section $\hat\sigma^{\rm VLQ}_{bW,tZ,th}$ defined in eq.~\eqref{eq:xsec_param}), and it additionally includes the bin-by-bin $K$-factor resulting from the ratio of these two predictions. In the bottom row a comparison is made between the \{NLO,NWA\} and the \{NLO,LW\} distributions for $\Gamma_T/M_T =$ 1\% ({\it i.e.} equivalent to approximately the NWA), 10\%, 30\% and 50\%. The error bands in the upper panel combines scale and PDF uncertainties, while in the lower panel both their individual relative contributions are shown, together with their combination for the NWA and largest width (50\%) scenarios.}
\end{figure}

The first distribution that we consider is the $p_T$ of the leading jet, shown in figure~\ref{fig:jet0pT}. Independently of the value of the $\Gamma_T/M_T$ ratio, the distributions exhibit a falling shape for all three considered processes, with a peak around $M_T/2$ for the $Wbbj$ and $Ztbj$ processes. This peak corresponds to a configuration in which the leading jet arises from the decay of the SM boson emerging from the $T$ quark, this boson decaying into a boosted and collimated di-jet system that is reconstructed as a single jet. As at LO (see section~\ref{sec:LOdifferential}), a further peak is visible at lower energies and corresponds to the spectator jet produced in association with the $T$ quark. In the case of the $htbj$ channel the peak at $M_T/2$ is less pronounced (and actually mostly vanishes) due to the different decay pattern of the Higgs boson. Those results illustrate the combined role of the width and the NLO corrections in the determination of the shapes, that we will now discuss in detail.

The contribution of the \{NLO,NWA\} corrections to the \{LO,NWA\} $p_T(j_0)$ distribution is shown in the upper row of figure~\ref{fig:jet0pT}. The bulk of the effects lies in the low energy part and in the high-$p_T$ tail of the spectrum, where the differential $K$-factors can reach $\{+20, +30\}\%$ and $\{-40,-60\}\%$ respectively (barring larger statistical fluctuations in the tail of the distributions). In the central part of the distribution, corresponding to the peak around $M_T/2$, the \{NLO,NWA\} corrections are negligible, the $K$-factor lying around 1. The NLO contributions have therefore the overall effect of shifting the distribution towards the soft region. 

The role of the finite width and its interplay with the NLO corrections is presented in the bottom row of figure~\ref{fig:jet0pT}, where the \{NLO,LW\} $p_T(j_0)$ distributions (which all have the same differential $K$-factor according to eq.~\eqref{eq:NLOLW}) are shown for different values of the $\Gamma_T/M_T$ ratio (1\%, 10\%, 30\% and 50\%). Those are compared with the \{NLO,NWA\} predictions (which are shown both in the upper and lower rows of the figure as a reference). In this case we normalise all distributions such that their integral gives $1$. This allows for a shape comparison, and the actual values of the couplings would have the effect of rescaling the distributions without modifying their shapes. The overall normalisation has been deduced from eq.~\eqref{eq:xsec_param}. For the $Wbbj$ and $Ztbj$ processes, the shape of the distribution is mostly affected by the width for $p_T$ values around and above the peak at $M_T/2$. This can be understood from the fact that the kinematics of the boosted weak boson (reconstructed as the leading jet) is determined by the $T$ width: a larger width implies that it becomes more probable that the emitted boson has a larger momentum. From a quantitative point of view, the difference between the \{NLO,NWA\} distribution and the distribution corresponding to $\Gamma/M=50\%$ increases with $p_T$, the large width-over-mass ratio leading to an enhancement of the differential cross section by about one order of magnitude in the last bin. For the $htbj$ process the relative impact of the finite width of the $T$ quark is much lower, the spectrum being in fact mostly insensitive to the width value. This can be explained by the much larger impact of the phase space region with small partonic centre-of-mas energy in the determination of the kinematics of the process (see section~\ref{subsec:parton_ME}).

\begin{figure}[t]
\centering
\includegraphics[width=0.325\textwidth]{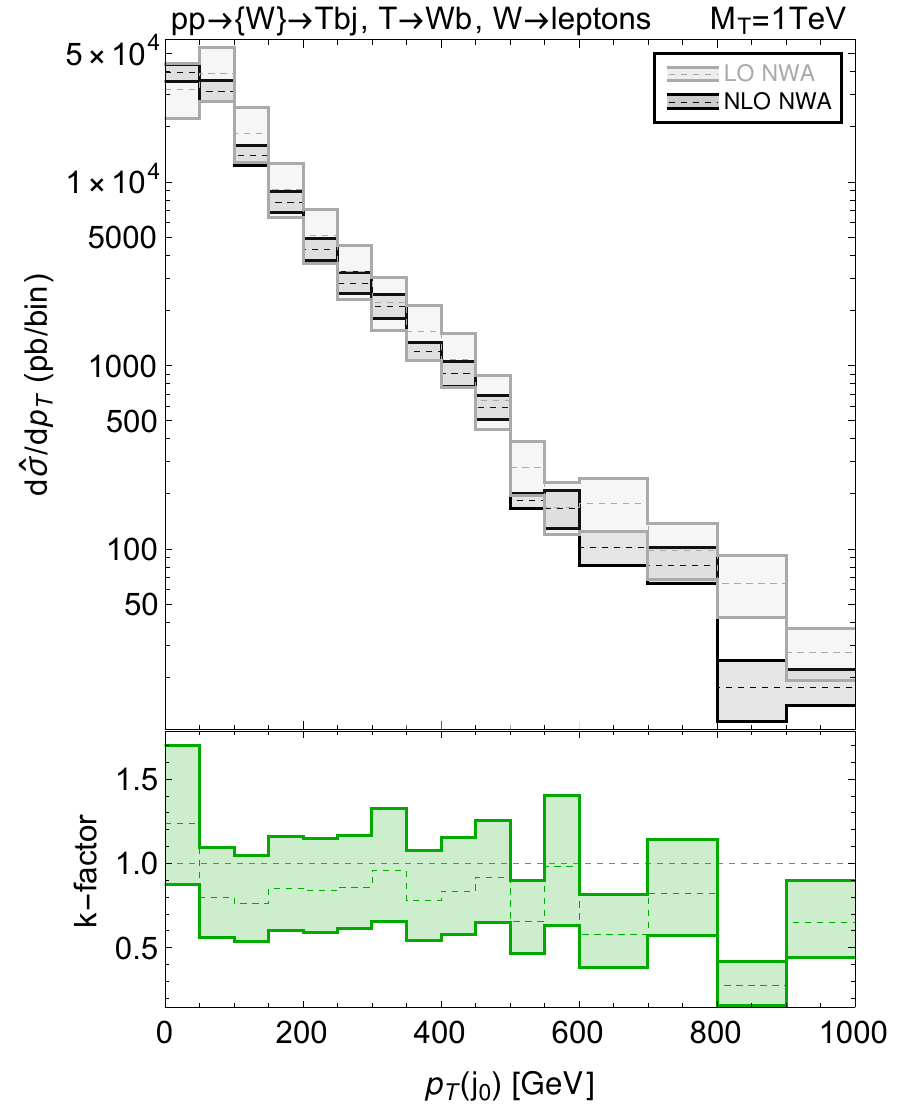}
\includegraphics[width=0.325\textwidth]{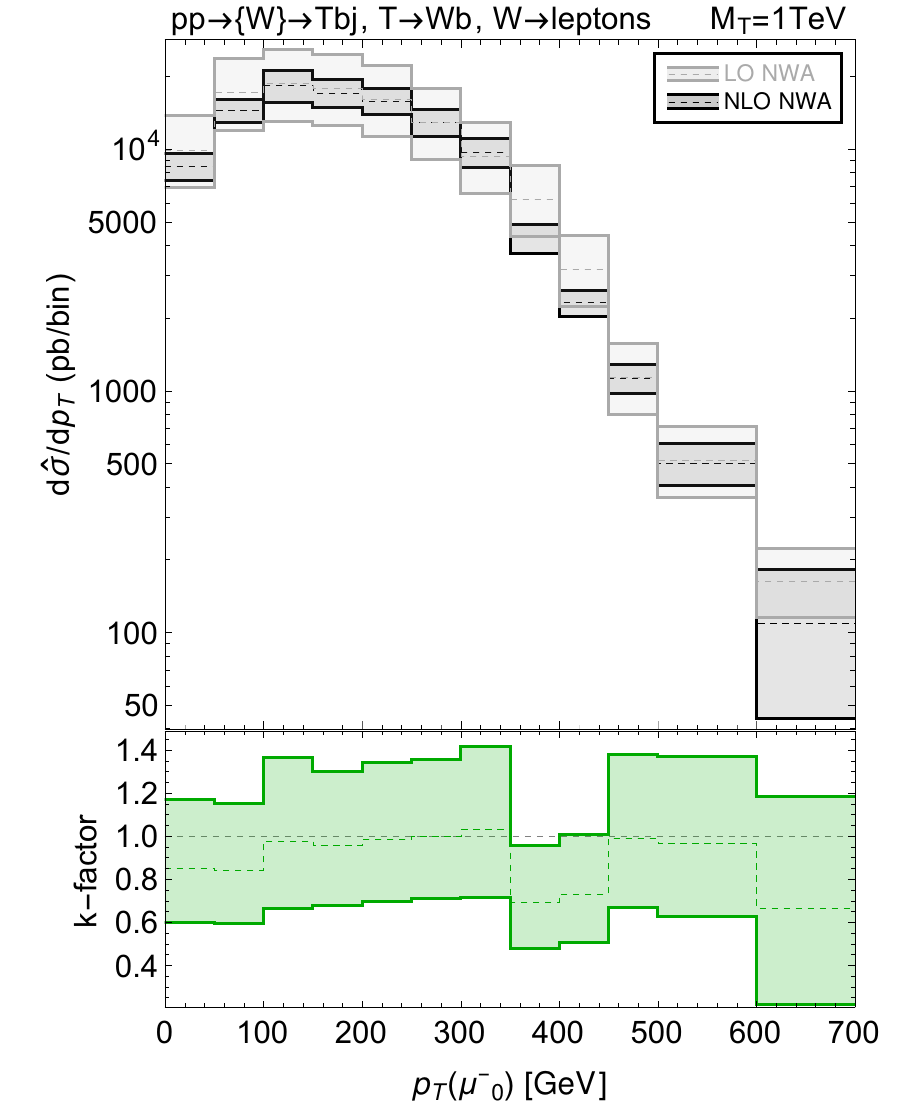}\\
\includegraphics[width=0.325\textwidth]{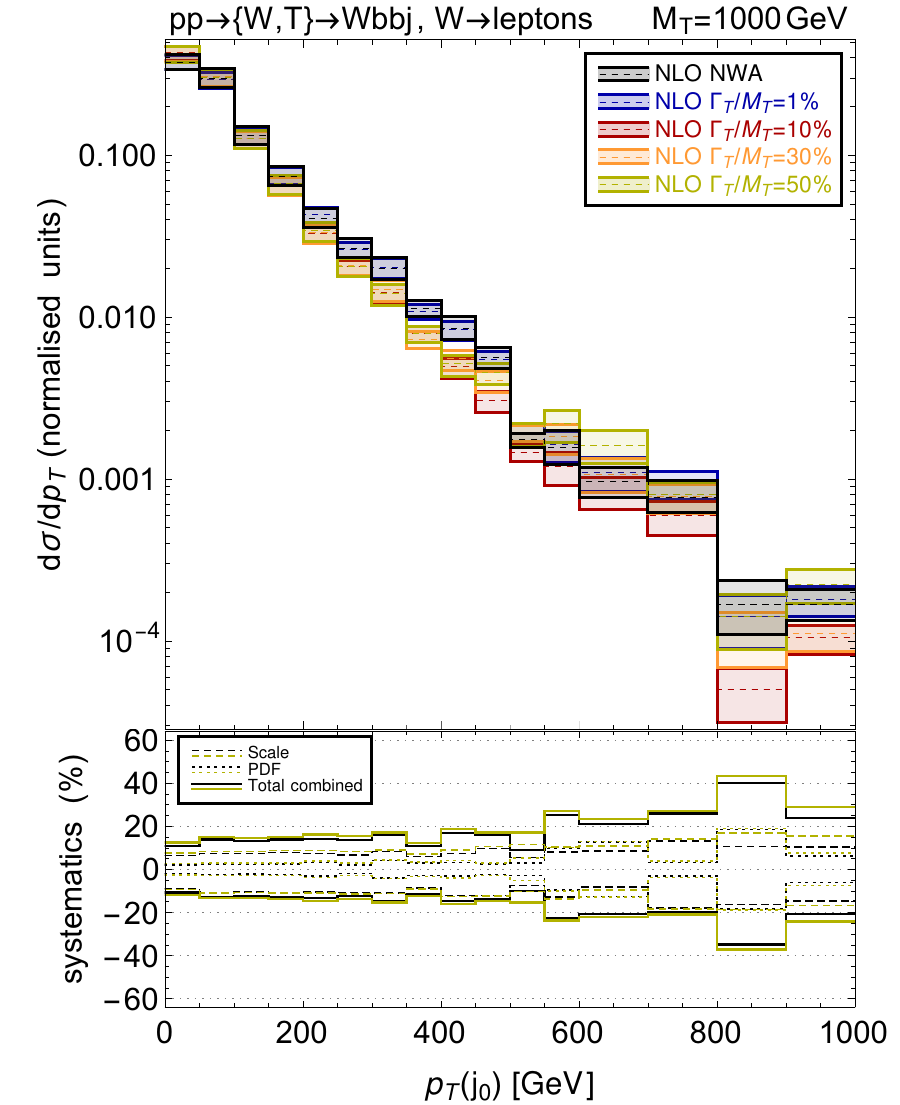}
\includegraphics[width=0.325\textwidth]{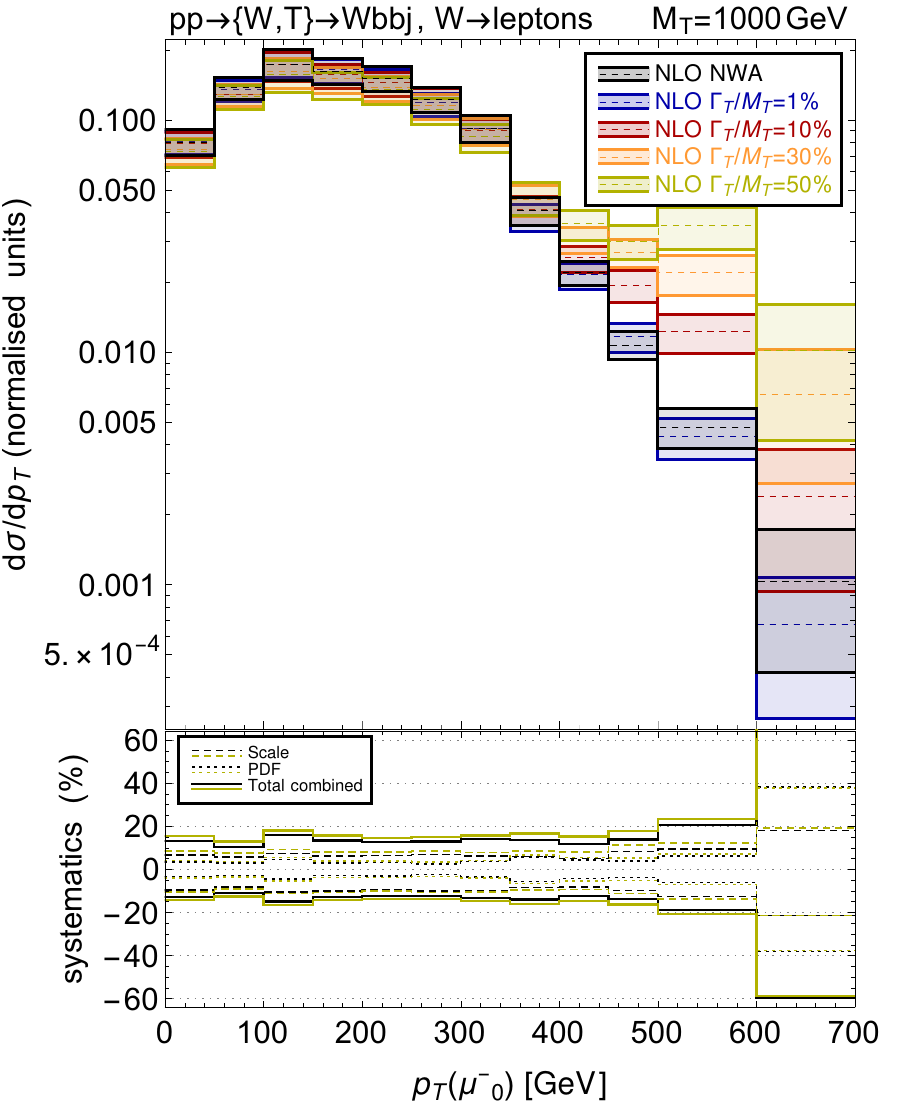}\vspace*{-0.2cm}
\caption{\label{fig:TWlep} Distributions of the transverse momentum of the leading jet (left) and of the leading muon (right) for single $T$ production in association with a jet.  We fix $m_T=1000$~ GeV and we consider the $pp\to\{W,T\}\to Wbbj$ channel when the $W$-boson decays leptonically. We refer to figure~\ref{fig:jet0pT} for the description of the presented curves.}
\end{figure}

To better isolate the role of the QCD corrections with respect to the one of the large width and to confirm the above findings, we now consider the distributions associated with the $Wbbj$ processes in a case where the $W$ boson decays leptonically. In figure~\ref{fig:TWlep} we compare the distributions of the transverse momentum of the leading jet ( $p_T(j_0)$) with the one of the leading muon ($p_T(\mu_0^-)$). The leading jet is here the spectator jet produced in association with the $T$ quark, as confirmed by the absence of a peak at $M_T/2$. As such a jet has no direct connection with the $T$ quark, the width of the latter should only mildly impact the jet kinematics. The role of the NLO corrections can therefore be fully extracted. In contrast, the properties of the muon do not receive meaningful contributions from NLO QCD corrections, but are affected in a sizeable way by the large width effects.

\subsubsection{Hadronic activity in the events}
\begin{figure}[t]
\centering
\includegraphics[width=0.325\textwidth]{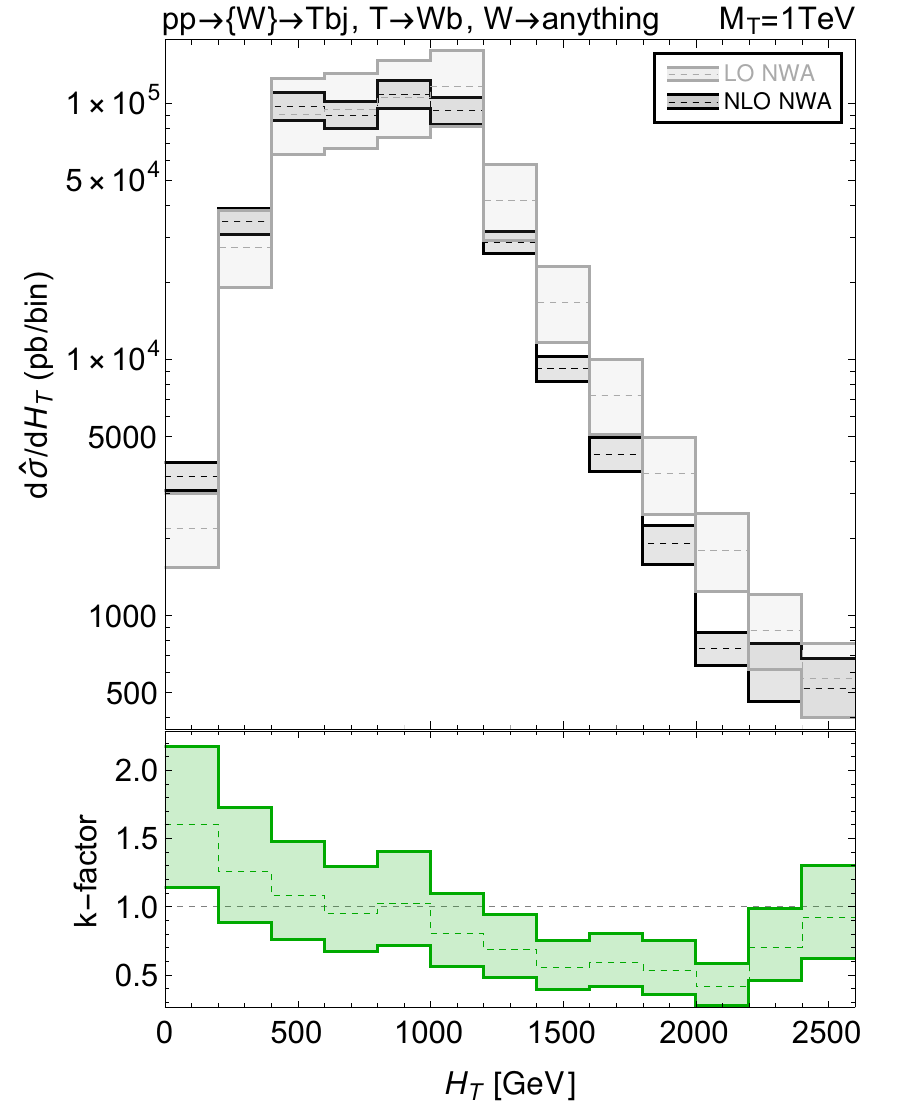}
\includegraphics[width=0.325\textwidth]{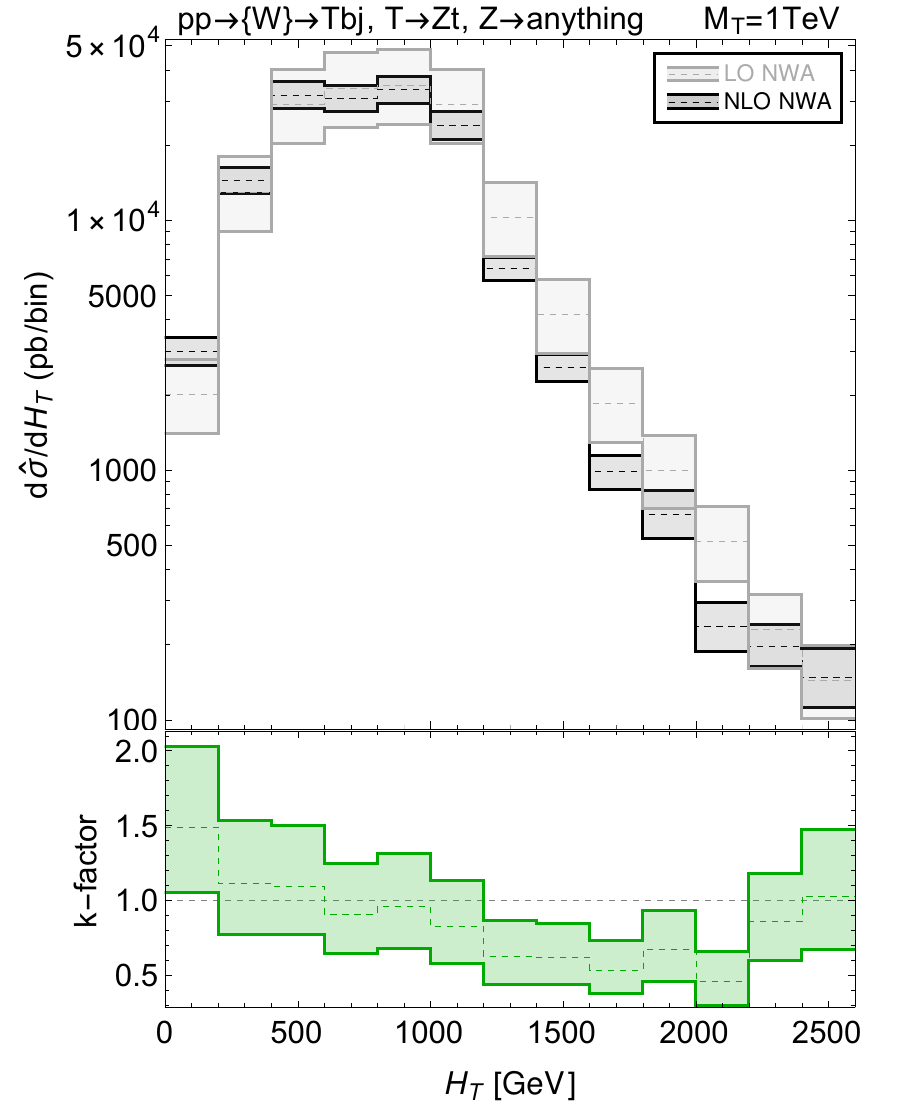}
\includegraphics[width=0.325\textwidth]{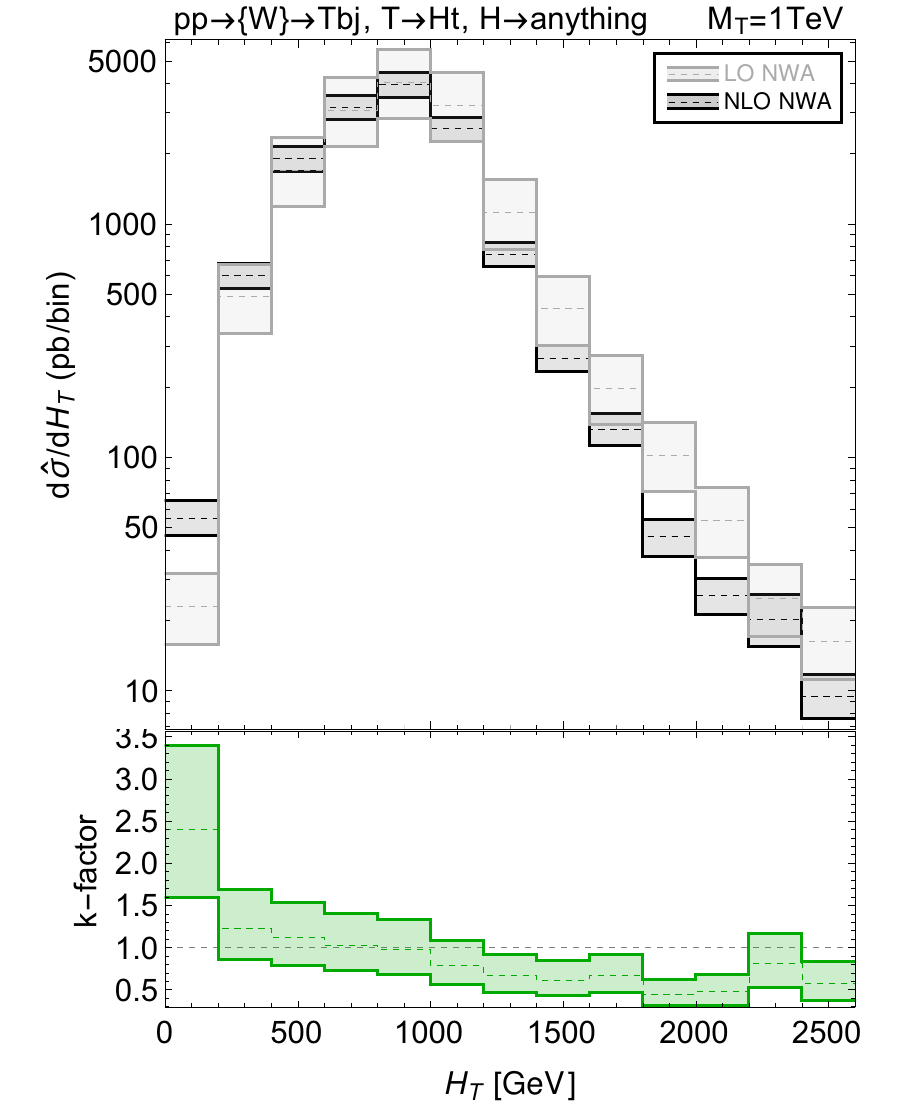}\\
\includegraphics[width=0.325\textwidth]{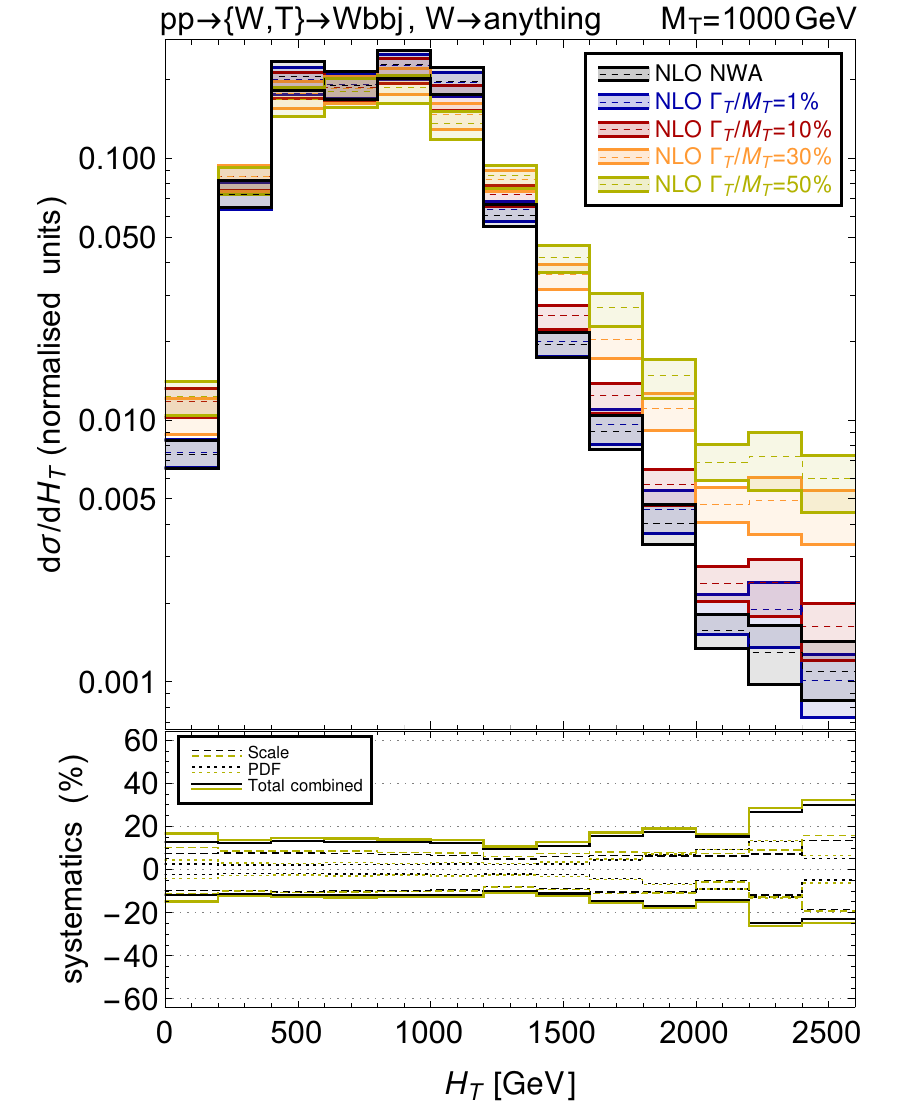}
\includegraphics[width=0.325\textwidth]{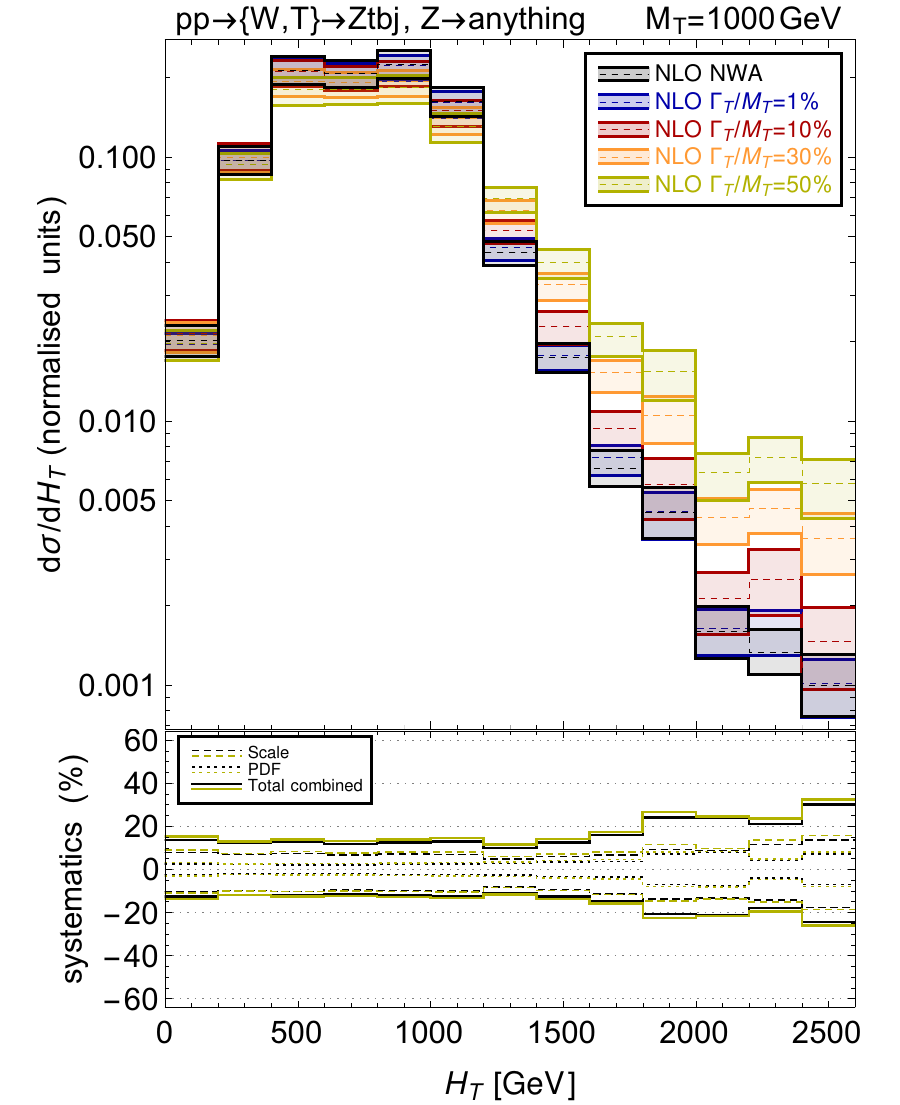}
\includegraphics[width=0.325\textwidth]{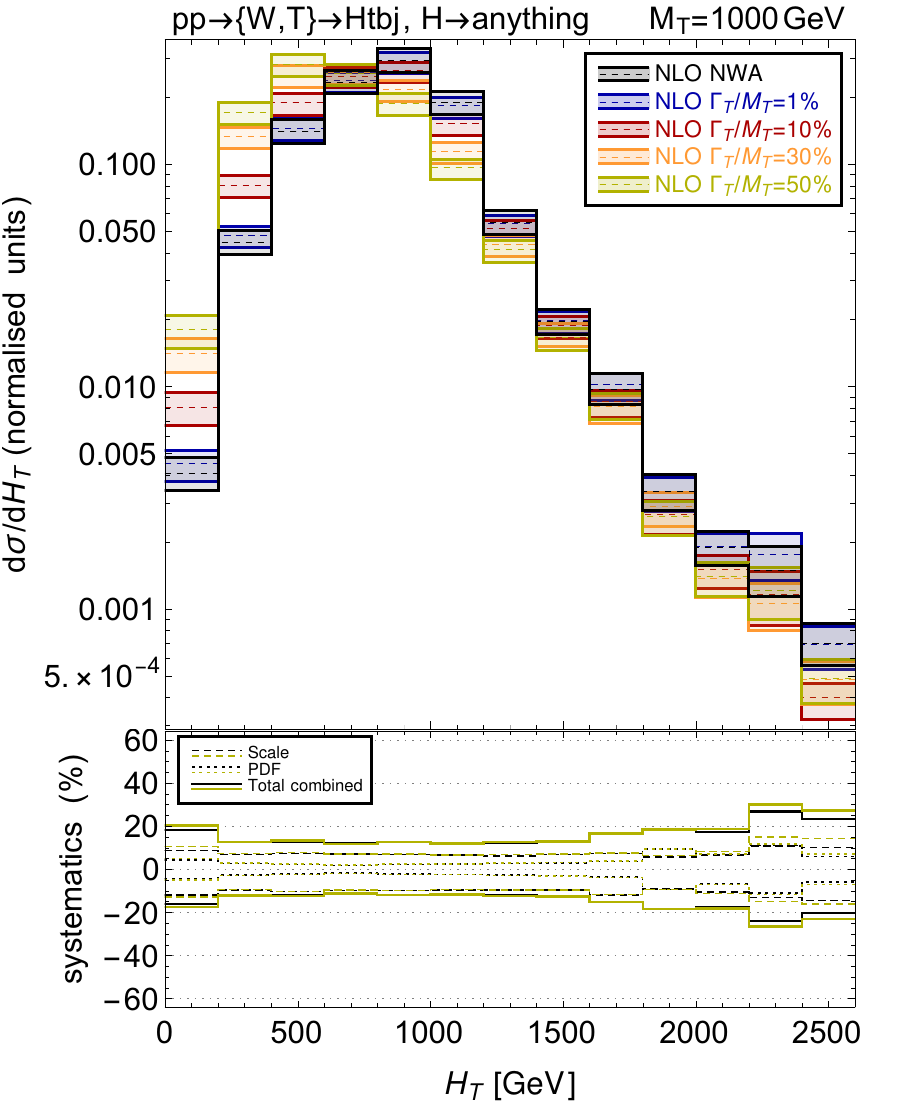}\vspace*{-0.2cm}
\caption{\label{fig:NLOHT} Same as in figure~\ref{fig:jet0pT} but for the hadronic activity defined as $H_T=\sum_{j,b} |\vec{p}_T|$.}
\end{figure}

In this section we consider the distribution in the scalar sum $H_T$ of the transverse momentum of all light jets and $b$-jets present in the final state. The predictions are shown in figure~\ref{fig:NLOHT}. They provide a clear illustrative example of the way the NLO QCD and large width effects can sum up or compensate each other in different regions of the phase space. In addition, the $H_T$ spectrum being an inclusive observable which tends to be sensitive to the difference between the SM (mostly populating the low-$H_T$ bins) and heavy new physics (usually populating the high-$H_T$ bins), the properties and features of this distribution could be useful for the design and optimisation of present and future experimental searches.

For all three considered processes, the \{NLO,NWA\} corrections strongly modify the shape of the distribution in roughly the same way: the bin-by-bin $K$-factor starts from values around 2 in the low-$H_T$ region and then decreases to values around 0.5 (barring fluctuations) in the high-$H_T$ regime. The region where the $K$-factor crosses 1 is in all cases for $H_T$ values around $M_T$. The $H_T$ distribution is therefore globally shifted towards lower $H_T$ values by NLO QCD corrections.

On the other hand, the impact of the width is different for each process, reflecting the different decay patterns of the SM boson originating from the $T$ decay. For the $Wbbj$ and $Ztbj$ processes, a large width-over-mass ratio implies that the  differential cross section increases with respect to the \{NLO,NWA\} reference case both in the low-$H_T$ and high-$H_T$ regimes, the increase being larger for larger $\Gamma_T/M_T$ values. The large-width spectra additionally remain lower than the \{NLO,NWA\} reference spectrum for $H_T$ values around $M_T$. The global large width effect is therefore to reduce the relative weight of the peak with respect to the more extreme regions of the distribution. The combination of the large width and QCD NLO effects therefore yields an enhancement of the positive differences with respect to the \{NLO,NWA\} distribution for low $H_T$ values, whereas they tend to balance each other in the high-$H_T$ region. For the $htbj$ process the situation is different. An increasing value of the $T$ width-over-mas ratio only affects the low-$H_T$ region, and such an increase further pushes the already-positive NLO QCD corrections with respect to the \{LO,NWA\} distribution towards positive values. For width values larger than 10\% of the mass, we hence get, for the first bins of the distribution, an enhancement of about one order of magnitude.

\subsubsection{Leading-jet pseudo-rapidity}
\begin{figure}[t]
\centering
\includegraphics[width=0.325\textwidth]{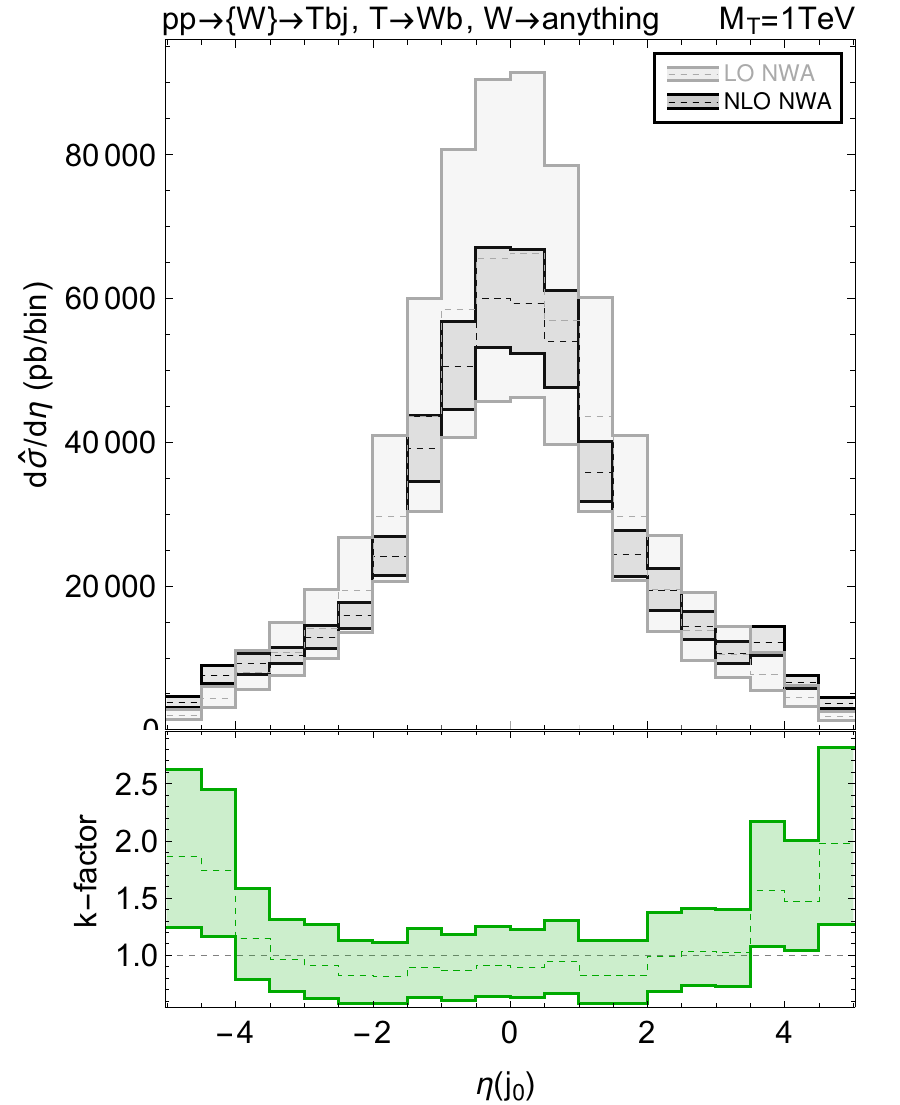}
\includegraphics[width=0.325\textwidth]{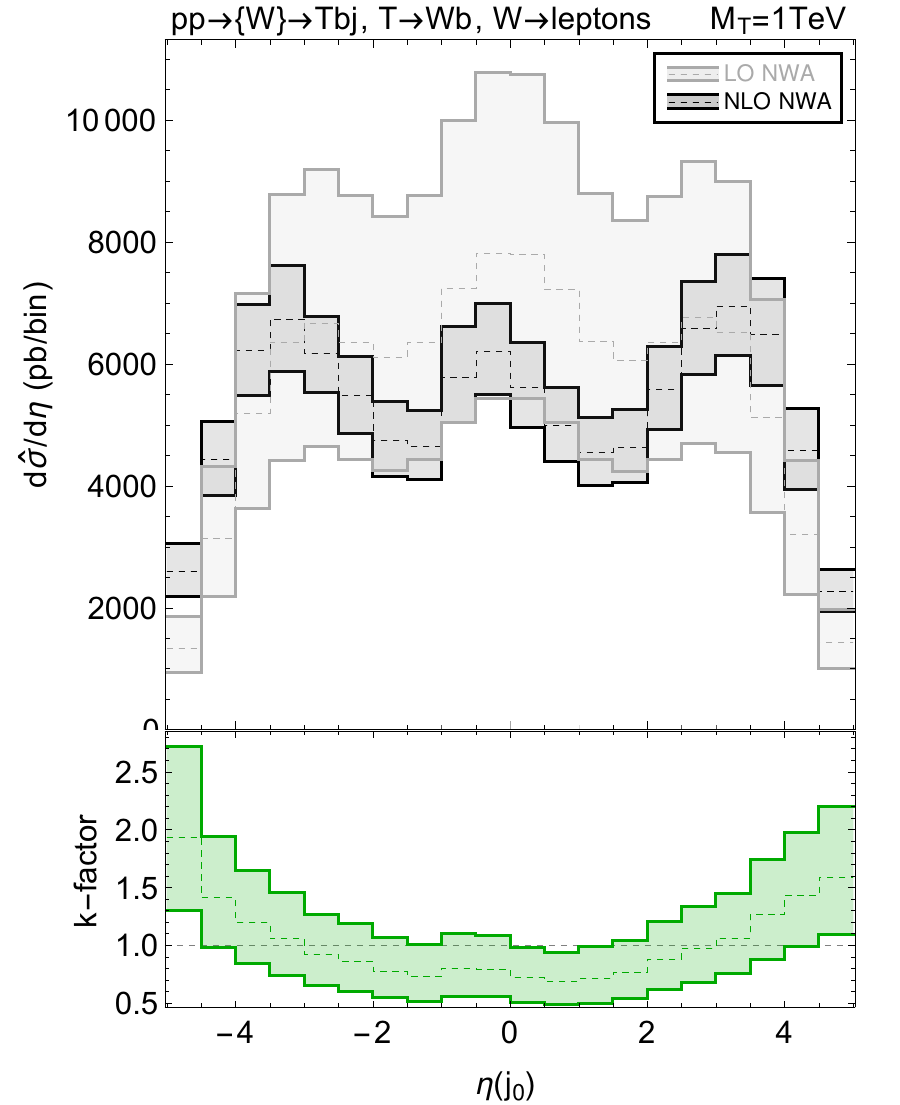}\\
\includegraphics[width=0.325\textwidth]{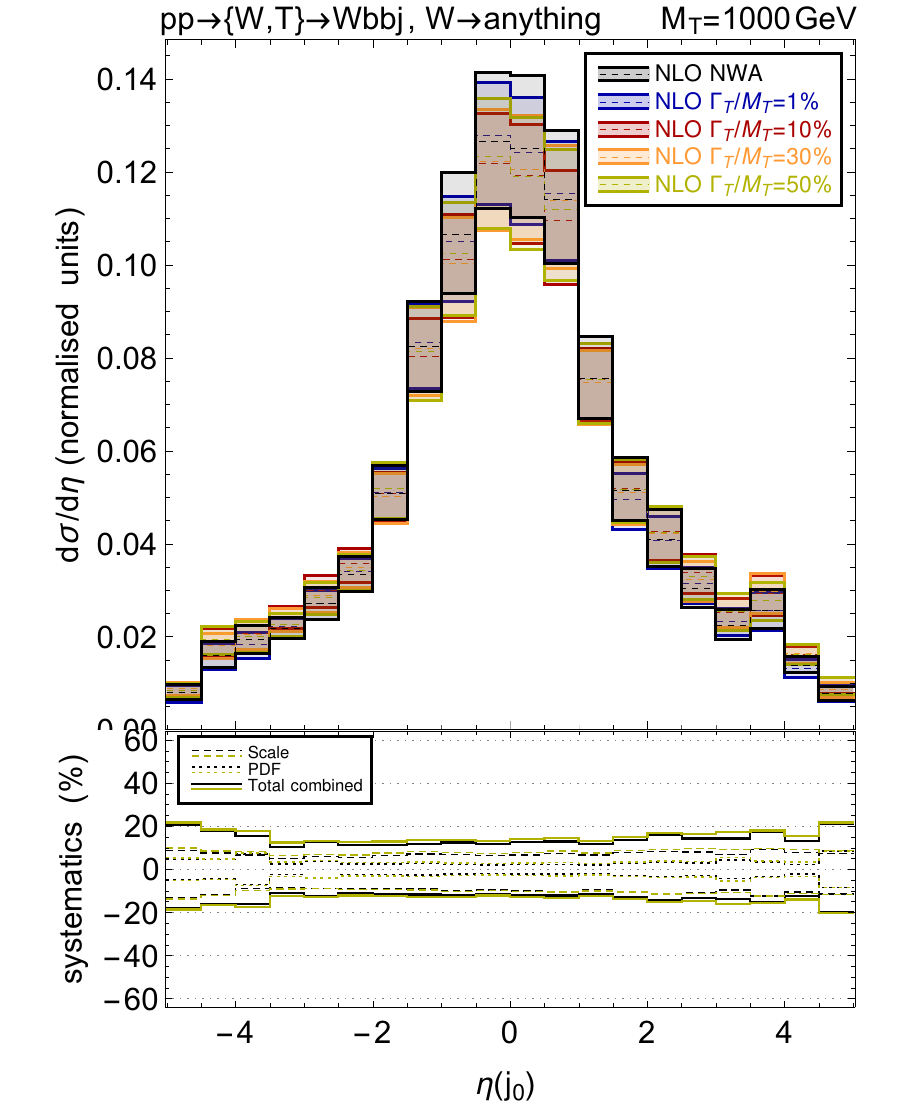}
\includegraphics[width=0.325\textwidth]{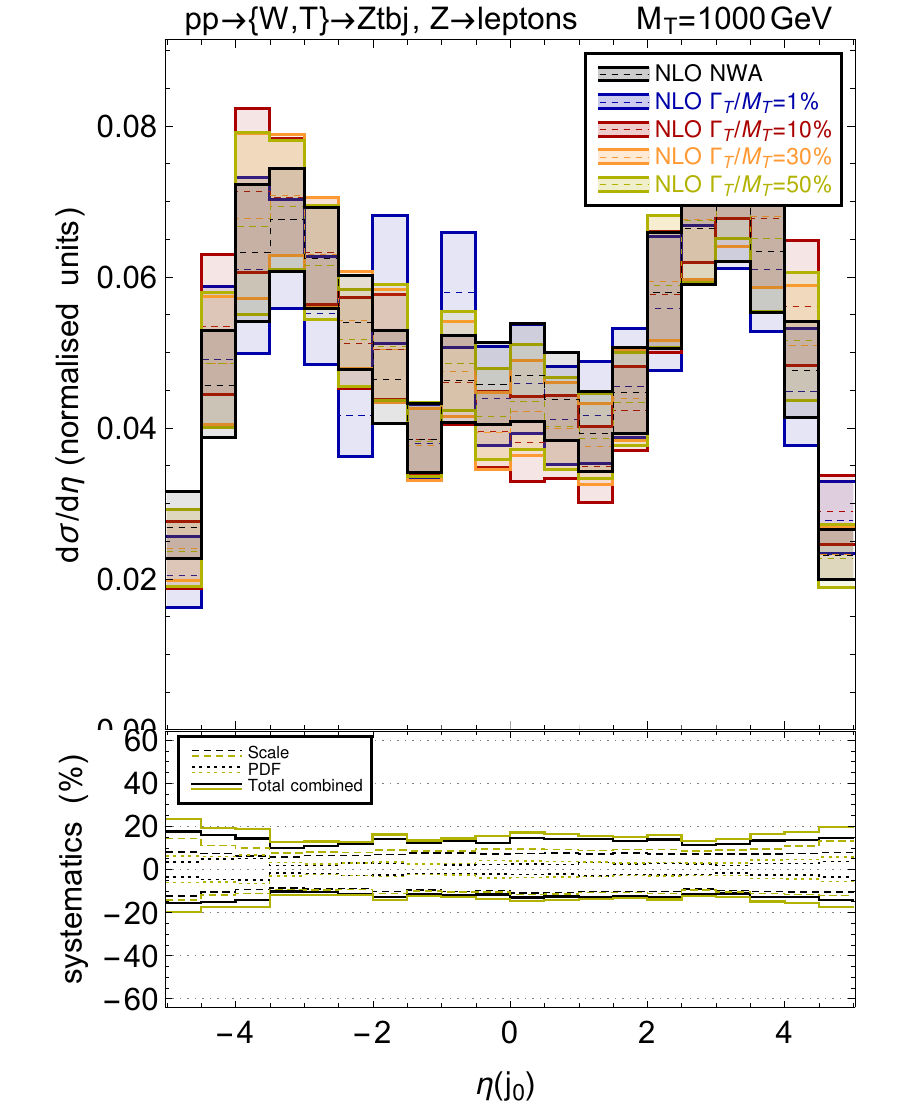}\vspace*{-0.2cm}
\caption{\label{fig:etaj0} Distributions of the pseudo-rapidity of the leading jet for single $T$ production in association with a jet.  We fix $m_T=1000$~ GeV and we consider the $pp\to\{W,T\}\to Wbbj$ channel when the $W$-boson decays inclusively (left) or leptonically (right). We refer to figure~\ref{fig:jet0pT} for the description of the presented curves.}
\end{figure}

As a last example, we focus in figure~\ref{fig:etaj0} on an angular variable, the pseudo-rapidity spectrum of the leading jet in the case of the $Wbbj$ channel. We consider both configurations in which the $W$-boson decays inclusively (left panels) or into leptons (right panels). Such a distinction allows us not only to emphasise the large impact of enforcing a selection on the presence of a single final-state lepton and of missing transverse energy, but also to highlight the role of the NLO corrections relatively to the finite width ones.

The NLO corrections have their largest impact in both cases when the jet has a high pseudo-rapidity and lies thus in the forward regime. We however recall that while in the inclusive case this jet can be either the spectator jet produced in association with the $T$ quark or the one emerging from the decay of the $W$ boson into a boosted and collimated di-jet system (thus reconstructed as a single jet), in the leptonic-decay case this jet can only be the spectator jet. The distributions are moreover mostly unaffected by the value of the $T$ width-over-mass ratio, at least when the NLO uncertainties are accounted for (that are nevertheless reduced with respect to those associated with the LO predictions). While such a  distribution is a powerful tool for the analysis of single $T$ production, it therefore cannot be used to further charaterise the VLQ width. 

Analogous results have been observed for the $Ztbj$ channel.

\subsubsection{Uncertainties}

From all figures, we can see that NLO predictions are associated with smaller uncertainties than the LO ones. The relative impact of the systematics due to the variation of the factorisation and renormalisation scales and to the fit of the parton densities are represented in the lower panels of the bottom rows of figures~\ref{fig:jet0pT}, \ref{fig:TWlep}, \ref{fig:NLOHT} and \ref{fig:etaj0}. The relative uncertainties for both contributions are rather constant in the low $p_T$ or $H_T$ range, and tend to increase for higher values. However, as the contribution of the systematics associated with the scales are around a few percents, those associated with the PDFs are larger and at least around 10\% for low $p_T$ and $H_T$ values. Above the resonant peak, fluctuations then become more important. This corresponds however to phase space regions more statistically limited, so that the interplay between the two contributions cannot be precisely evaluated. We now turn to the last dimension-less observable that has been considered, the pseudo-rapidity distribution of the leading jet. Here, the combined PDF and scale uncertainties are constant and reach roughly 10\%, with the exception of the forward regime where the PDF uncertainties are larger and dominate.

\subsubsection{Summary}
The few examples provided in this section  clearly show that including NLO corrections in predictions associated with the single production of a VLQ featuring a large width-over-mass ratio can be relevant as it can sizeably modify the kinematics of the final-state objects. Even with an approximate treatment that is valid only for processes for which the signal is dominated by resonant topologies and for which the interference with the SM background does not significantly contribute, the accuracy in the description of the distributions of the final state objects is definitely improved with respect to LO. Going beyond this approximation and reaching a full and consistent treatment of the interplay between the finite widths and NLO QCD contributions is however non-trivial. Embarking in such a task will be fully justified (and actually necessary) in the case where an excess would be observed in future analyses of single VLQ production processes.

%%%%%%%%%%%%%%%%%%%%%%%%%%%%%%%%%%%%%%%%%%%%%%%%%%%%%%%%%%%%%%%%%%%%%%%%%%%%%%

\section{Conclusions}\label{sec:conclusion}
Single production and decay of vector-like quarks is an important tool to explore a number of extensions of the Standard Model. As compared to VLQ pair-production, their single production provides a larger discovery potential in the high-mass regime, as it is less phase-space suppressed. It moreover offers sensitivity to the couplings of the VLQs with the SM particles. This however requires the understanding of important effects as QCD corrections, the dependence on the VLQ width and the interference between the VLQ signal and the SM background. 

In the present work we have provided a detailed study of such effects including guidelines for experimental searches in this field.  While it is nowadays not needed to stress the importance of NLO QCD effects at hadron colliders, given the large number of studies dedicated to this topic, those effects need to be carefully assessed in the specific case of broad vector-like quark production. Solely focusing on the NLO effects may indeed be misleading, as other effects can be of the same size or even larger. In the particular case of VLQ single production, a consistent treatment of the VLQ width is mandatory as the VLQ partial widths depend on the same couplings as those driving the production process. Given a fixed VLQ mass, a large production cross section indeed implies a large VLQ width, and such a large width then requires a specific treatment. Restraining ourselves to the standard narrow width approximation as usually implemented in the simulation tools may thus lead to inaccurate results.  This is especially important when exploring differential distributions associated with resonance searches, as the combined effect of large width, NLO corrections and interference can considerably modify the shape of the distributions. Invariant-mass distributions are not necessarily of a simple distorted Breit-Wigner type, and the modifications of various kinematical distributions are even process-dependent. The study that we have performed addresses these points in detail and quantifies the effects of finite width, signal-background interference and NLO QCD corrections (and their interplay) on the invariant mass and on other commonly considered distributions relevant for VLQ single production analyses.  

We have discussed in \cref{sec:model} the model description used in VLQ searches at colliders and introduced the formalism required to go beyond the narrow-width approximation. In our new physics parametrisation, the SM is extended by a single species of vector-like quarks $T$ that couples to the SM bosons and quarks. The higher-energy models leading to this simplified model typically contain a well-defined multiplet structure for the VLQs, and also predict the existence of exotic VLQ decay modes beyond the usual ones into $W$, $Z$ or Higgs boson. The $T$ quark can therefore become wider than the sum of partial widths $\Gamma^{th}_T + \Gamma^{tZ}_T + \Gamma^{Wb}_T$. Several experimental searches have explored this more general scenario with $T$ width-to-mass ratios ranging up to 30\%. In order to consistently discuss such a large width case, we have used the complex mass scheme and we have introduced an approximate method to obtain results that are accurate at the next-to-leading order in the strong coupling.

In \cref{sec:lo} we presented finite-width analyses of the processes $p p\rightarrow th+X$, $p p\rightarrow tZ+X$ and $p p\rightarrow tW+X$. Whereas we focused on $T$ production from $bW$ exchanges, we provide analogous key results for $T$ production from $tZ$ exchanges in \cref{app:TviaZ}. In  \cref{subsec:schemecomparison}  we compared large-width effects as estimated by using various schemes to treat the VLQ width, and we found that the predictions agree well. In \cref{sec:signal} we performed a LO parton-level analysis of the processes $p p\rightarrow \{W,T\}\rightarrow Wbbj$, $p\rightarrow \{W,T\}\rightarrow Ztbj$ and $p\rightarrow \{W,T\}\rightarrow htbj$. Our analysis included finite VLQ width effects, and it showed that in all cases, the signal is dominated by its resonant contributions. We determined the impact of a finite width on the total rates for several benchmark points (see \cref{tab:totalrates}). We then found a resulting Breit-Wigner-like invariant-mass distribution for the $Wbbj$ and $Ztbj$ final states, but \emph{not} for the $htbj$ final state (see figure~\ref{fig:res_tch_contributions}). Such a conclusion holds even in the narrow width approximation and can be understood analytically, as shown in \cref{subsec:parton_ME}. This difference also explains why ratios of total rates associated with different channels are \emph{not} proportional to the corresponding branching fractions for large-width scenarios. In \cref{subsec:plinterference} we investigated the impact of  signal-to-background interference when the VLQ width is large and we considered again the final-state invariant-mass distributions to this aim. In \cref{sec:lo_2d_scans} we provided cross section maps (see figure~\ref{fig:sigmahat_W_T_14TeV}) which allow us to calculate large width and interference effects on total rates for arbitrary VLQ couplings (beyond the benchmarks under consideration). We moreover demonstrated the impact on various distributions commonly used in searches for VLQ single production in section~\ref{sec:LOdifferential}. Those results have been promoted to the NLO QCD accuracy in section~\ref{sec:nlo}.

Most importantly, with the information provided in \cref{sec:model} and \cref{app:Tech}, we define a detailed prescription on how to use the publicly available {\sc Feynrules} implementation \cite{urlFR} to simulate VLQ single production at the LHC when the VLQ features a large width and when the signal contributions include the interference with the SM, as well as NLO QCD corrections. This prescription is applicable not only to the processes and distributions described in this article, but also for other distributions and other proceses involving VLQs with charge 5/3, 2/3, $-1/3$ and $-4/3$.

%%%%%%%%%%%%%%%%%%%%%%%%%%%%%%%%%%%%%%%%%%%%%%%%%%%%%%%%%%%%%%%%%%%%%%%%%%%%%%

\section*{Acknowledgements}
We wish to thank G.~Cacciapaglia for discussions in the early stage of this project, and O.~Mattelaer for his help with \mg. The authors are grateful to the Mainz Institute for Theoretical Physics (MITP) of the DFG Cluster of Excellence PRISMA$^+$ (Project ID 39083149), for its hospitality and partial support during the completion of this work.
AD is grateful to the LABEX Lyon Institute of Origins (ANR-10-LABX-0066) of the Université de Lyon for its financial support within the program ``Investissements d'Avenir'' (ANR-11-IDEX-0007) of the French government operated by the National Research Agency (ANR). 
TF’s work is supported by IBS under the project code IBS-R018-D1. LP's work is supported by the Knut and Alice Wallenberg foundation under the SHIFT project, grant KAW 2017.0100. LP acknowledges the use of the IRIDIS 4 HPC Facility at the University of Southampton. HSS is supported by the European Union’s Horizon 2020 research and innovation program under the grant agreement No.824093 in order to contribute to the EU Virtual Access ``NLOAccess'', the French ANR under the grant ANR-20-CE31-0015 (``PrecisOnium''), and the CNRS IEA under the grant agreement No.205210 (``GlueGraph'').

\clearpage

%%%%%%%%%%%%%%%%%%%%%%%%%%%%%%%%%%%%%%%%%%%%%%%%%%%%%%%%%%%%%%%%%%%%%%%%%%%%%%
\appendix

\section{Further results at leading order for single $T$ production via $W$-boson exchanges}\label{app:TviaW}
In this appendix we provide further results for single $T$ production when it is mediated by $W$-boson exchanges. Those predictions are included to provide a complete set of results, for allowing to be reproducible if needed and for further reference.
\begin{figure}[htbp]
  \centering
  \includegraphics[width=.325\textwidth]{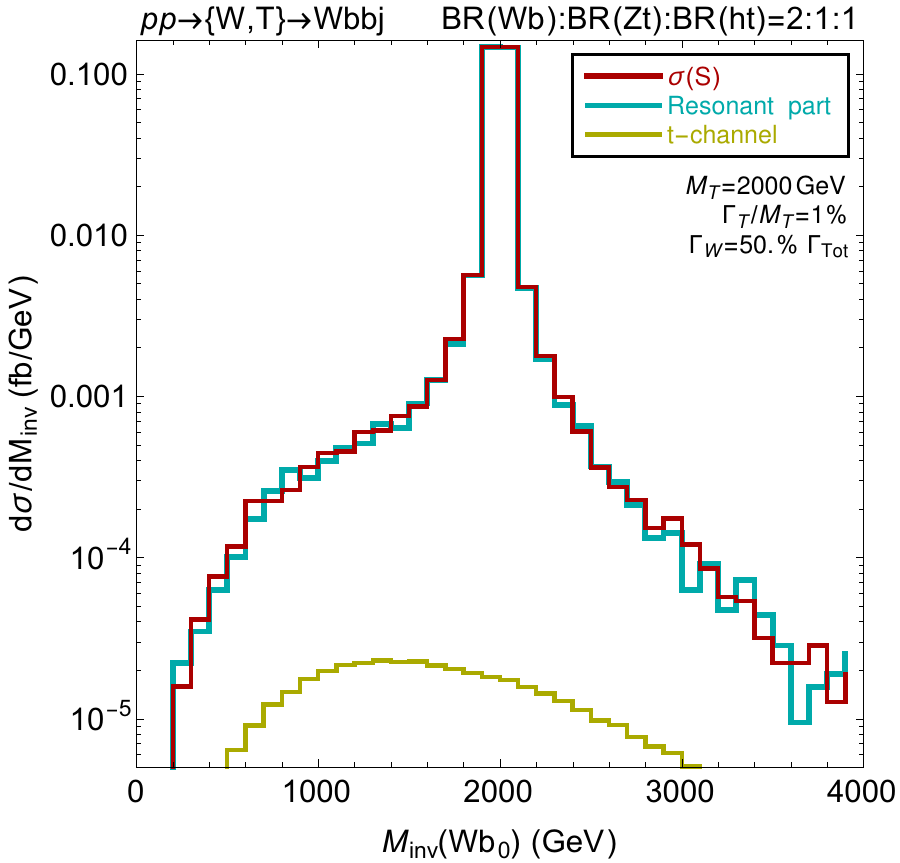}
  \includegraphics[width=.325\textwidth]{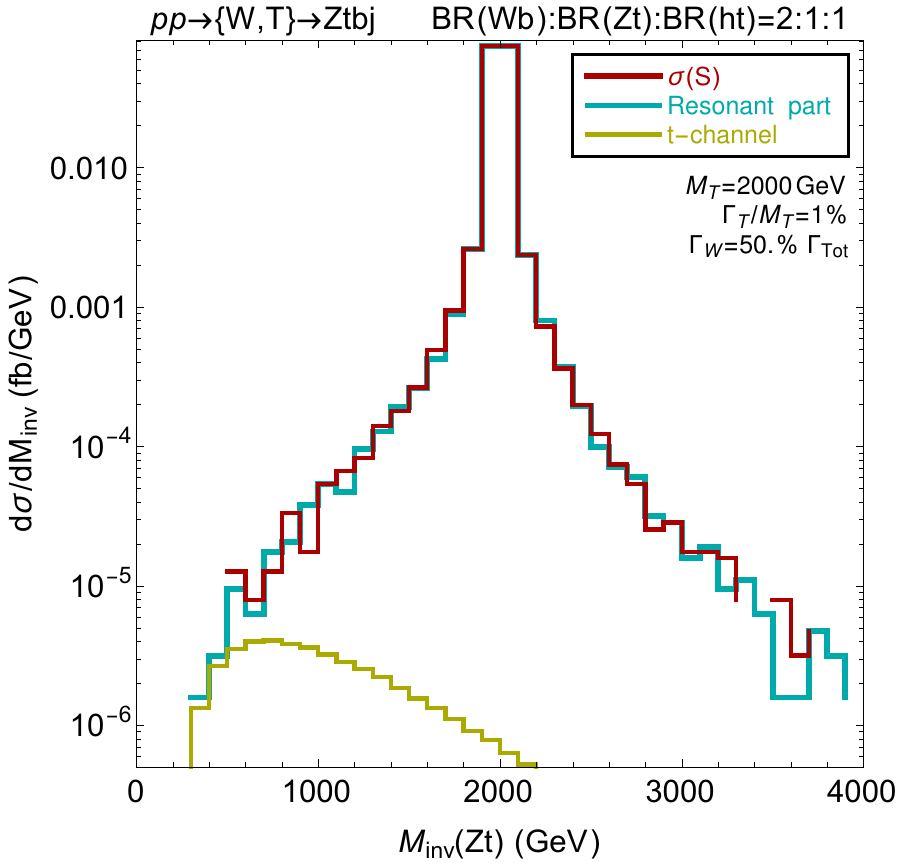}
  \includegraphics[width=.325\textwidth]{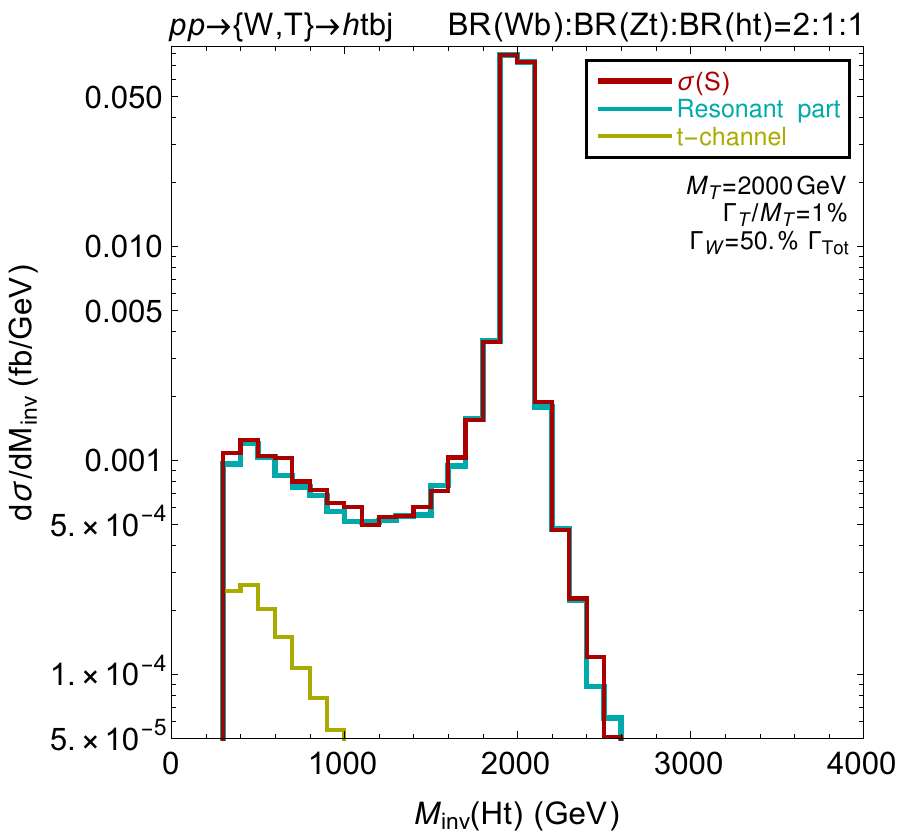}
  \\
  \includegraphics[width=.325\textwidth]{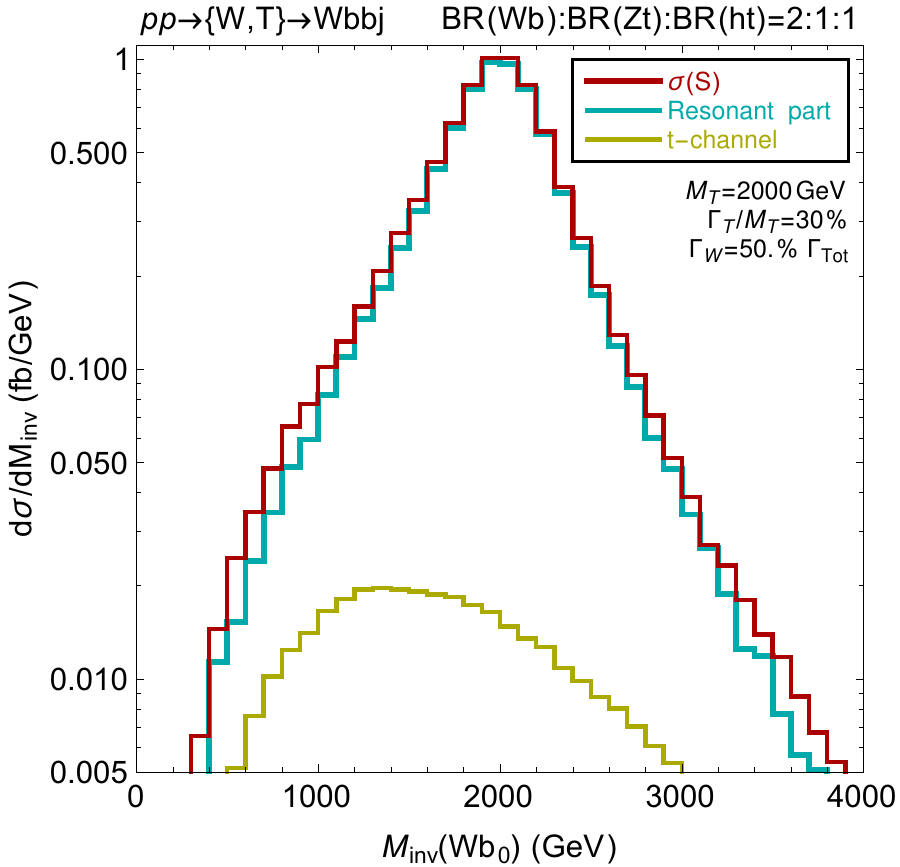}
  \includegraphics[width=.325\textwidth]{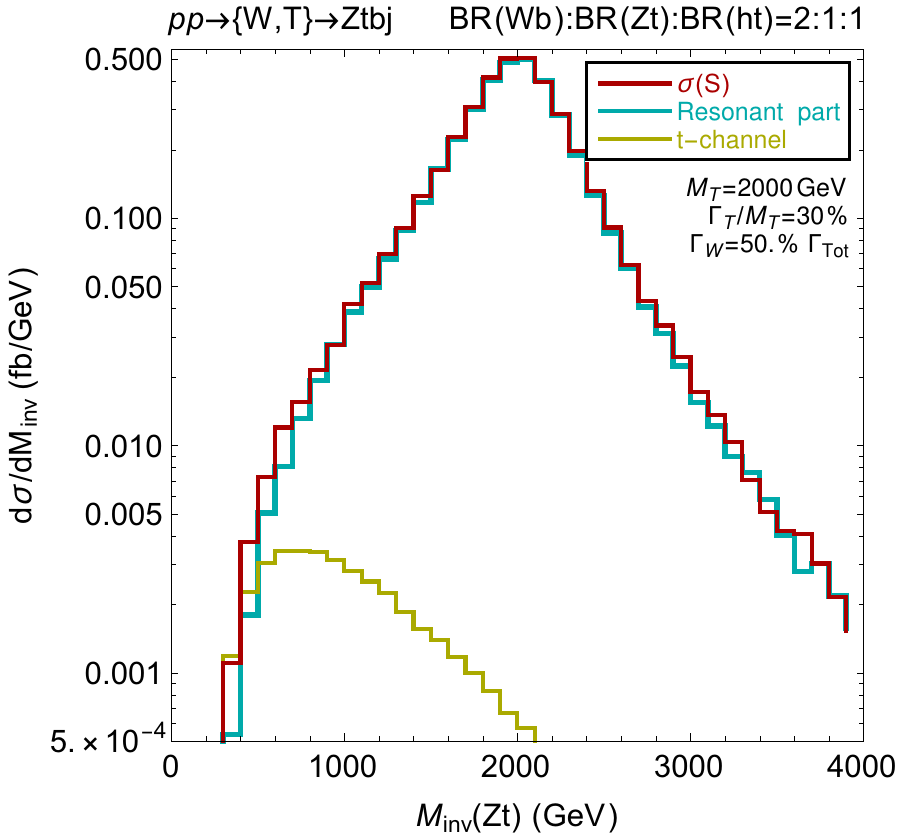}
  \includegraphics[width=.325\textwidth]{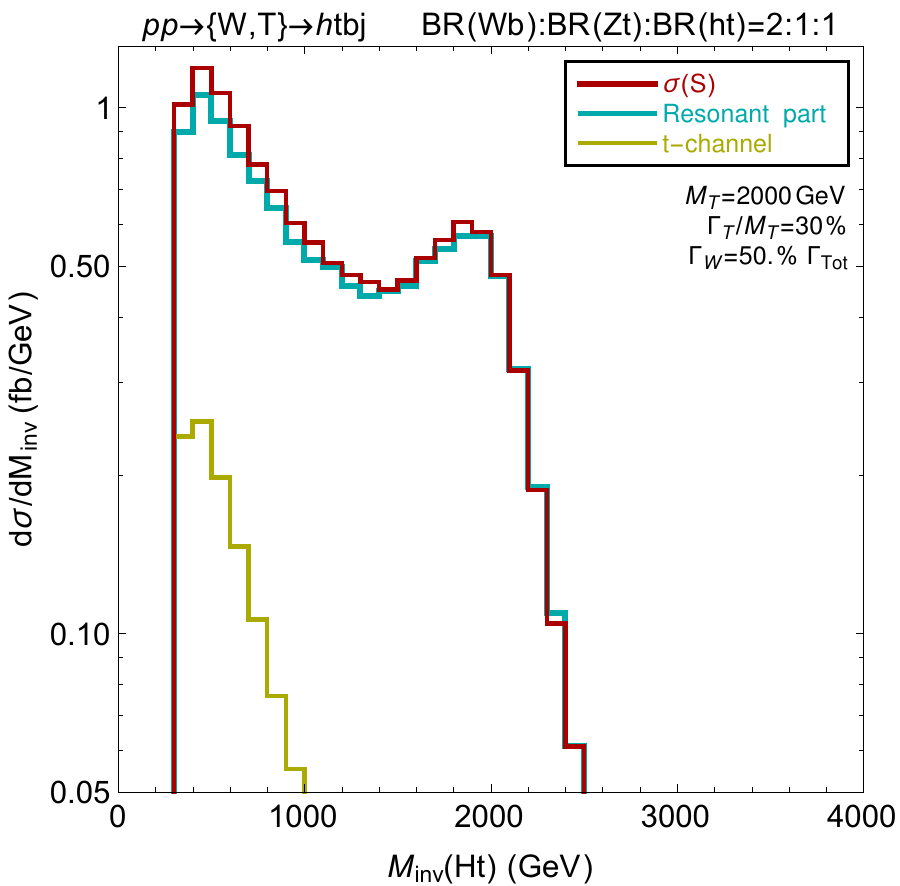}\vspace*{-0.2cm}
  \caption{\label{fig:res_tch_contributions_W_2000} Same as figure~\ref{fig:res_tch_contributions} for $M_T=2$~TeV.}
\end{figure}

\begin{figure}[!htbp]
  \centering
  \includegraphics[width=.325\textwidth]{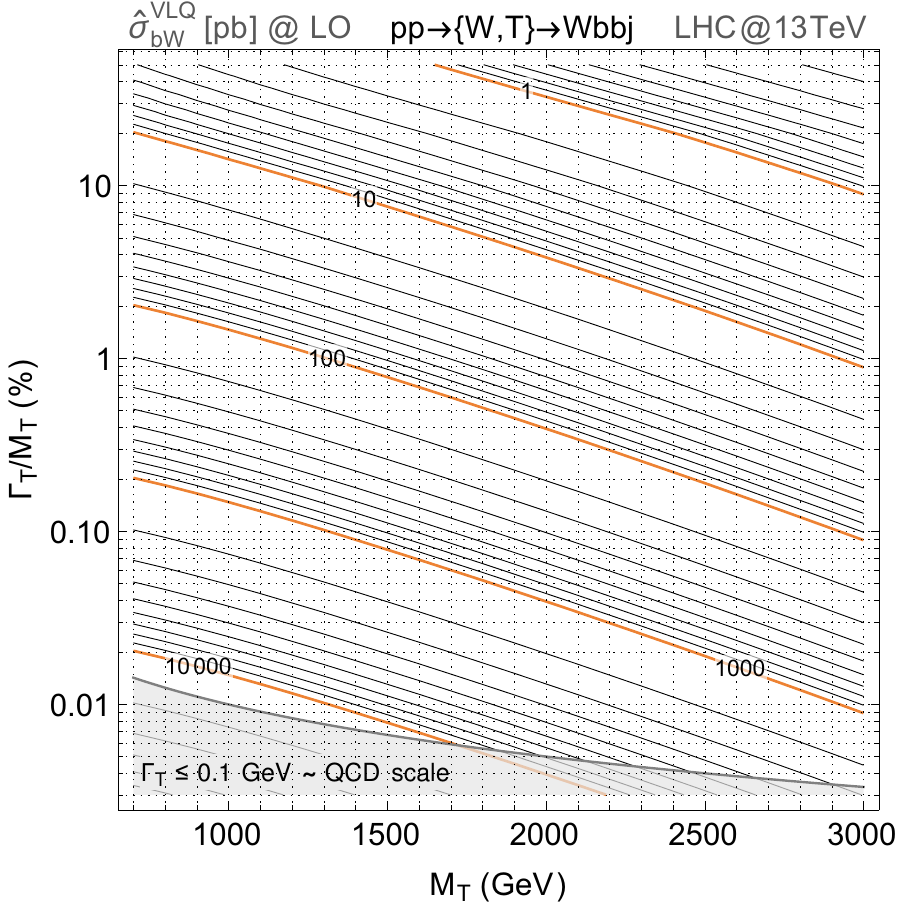}
  \includegraphics[width=.325\textwidth]{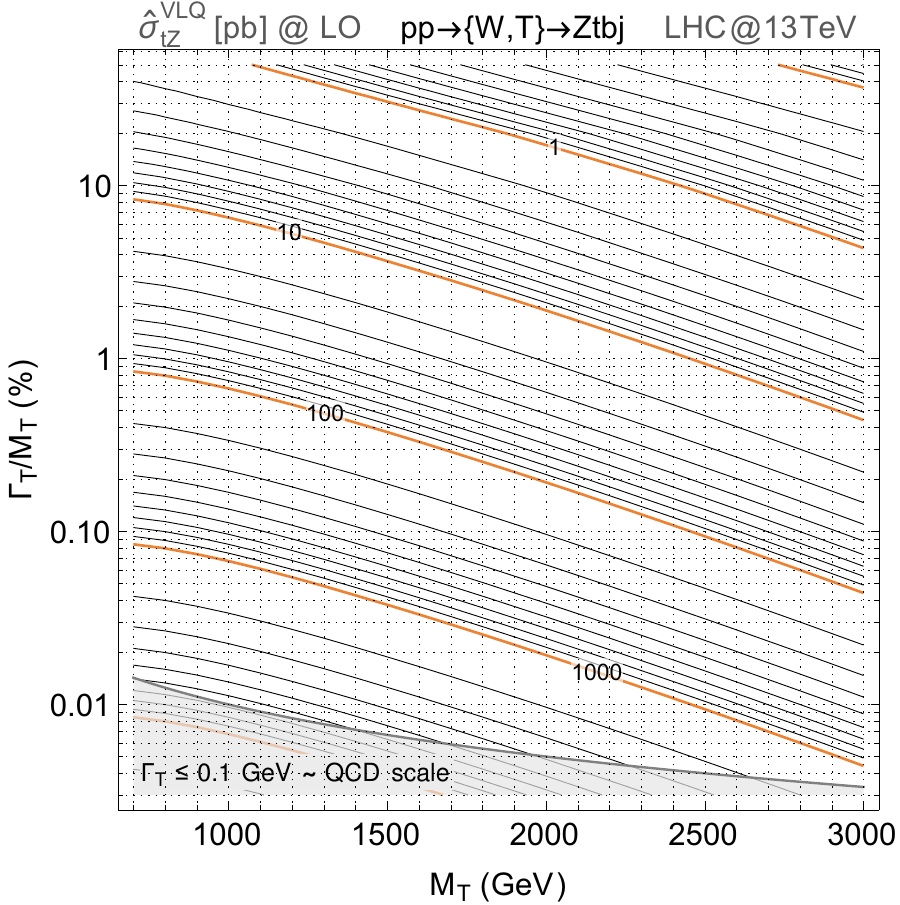}
  \includegraphics[width=.325\textwidth]{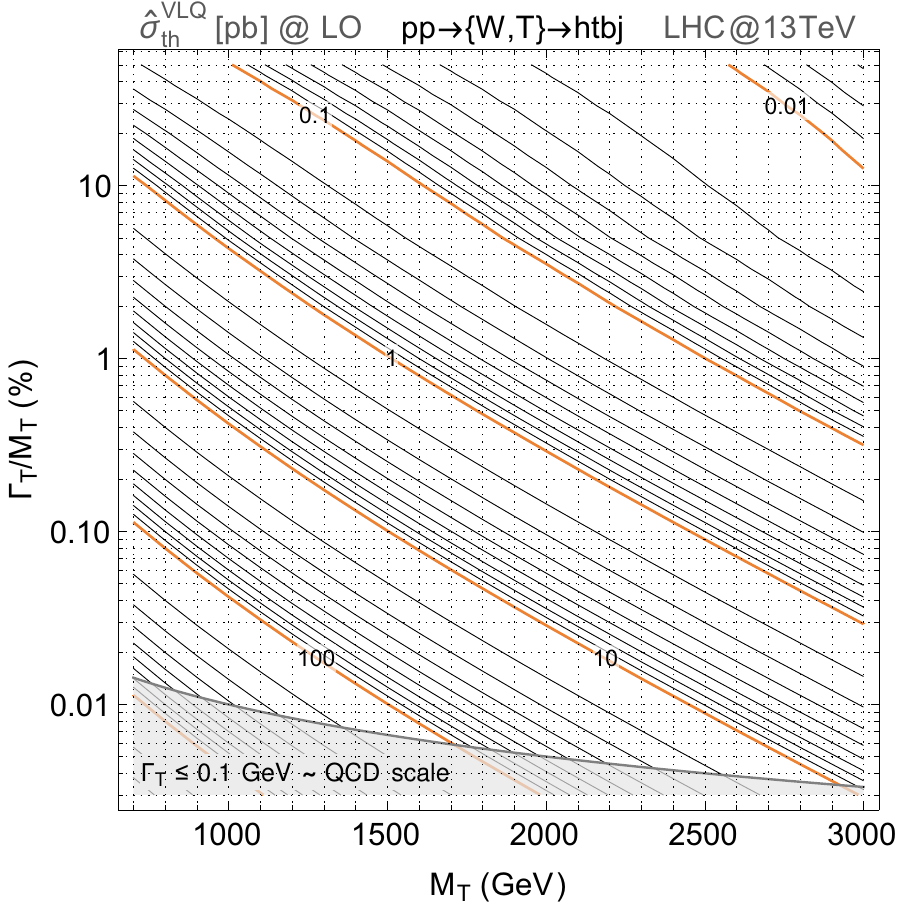}
  \\
  \includegraphics[width=.325\textwidth]{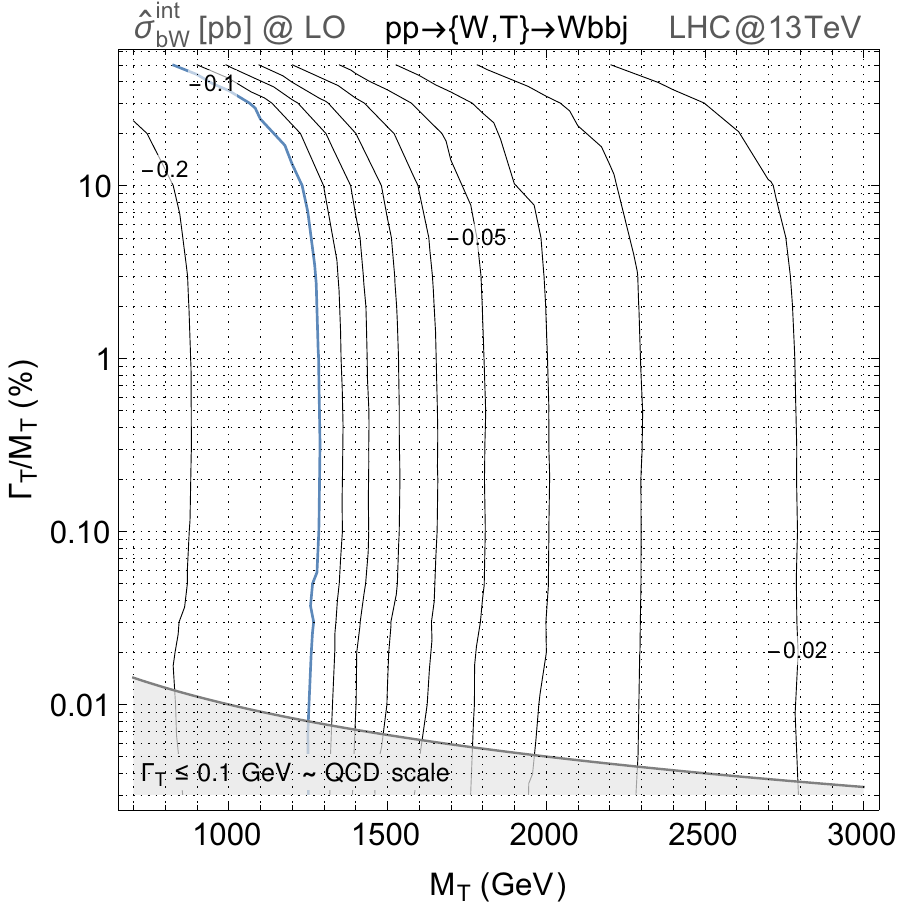}
  \includegraphics[width=.325\textwidth]{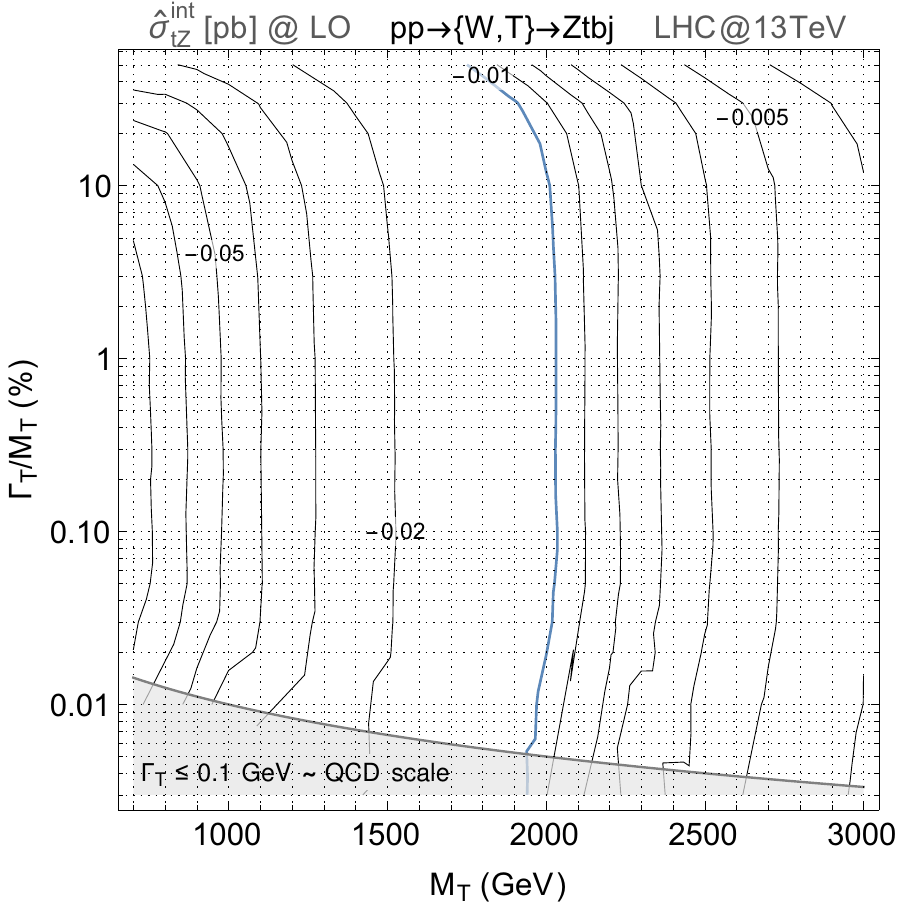}
  \includegraphics[width=.325\textwidth]{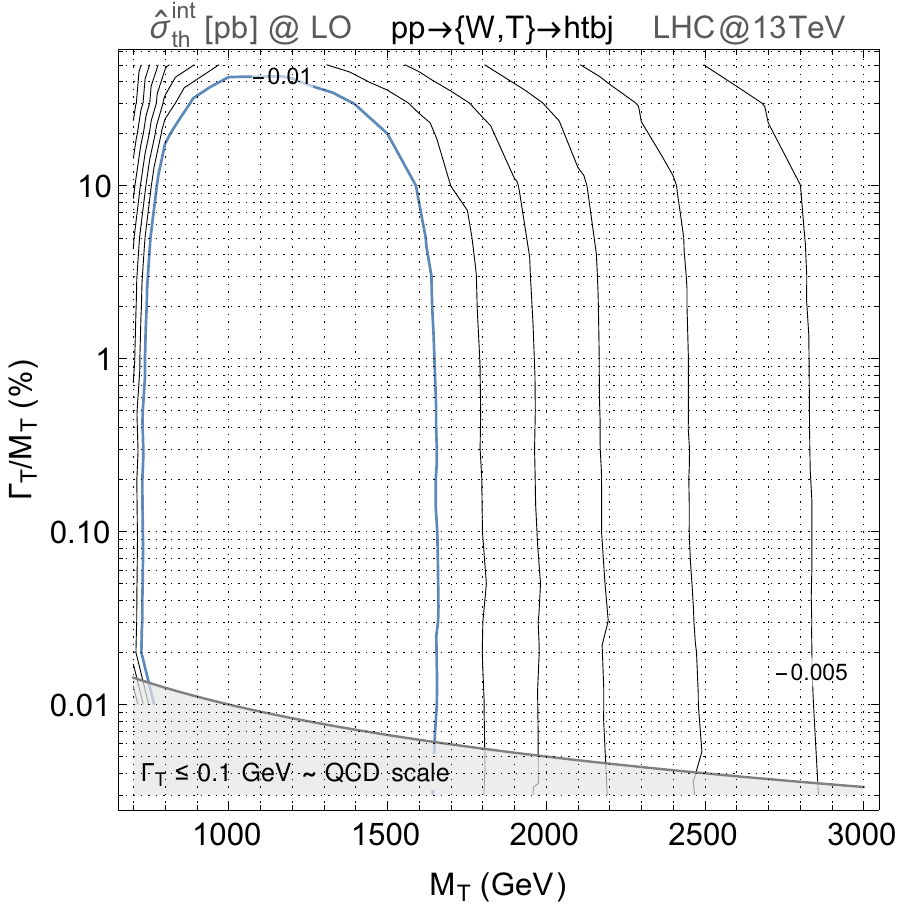}\vspace*{-0.2cm}
  \caption{\label{fig:sigmahat_W_T_13TeV}
  	 Values of the $\hat\sigma^{\rm VLQ}$ (top row) and $\hat\sigma^{\rm int}$ (bottom row) bare cross sections, in pb,
    for the $pp\to Wbbj$ (left), $Ztbj$ (centre) and $Htbj$ (right) processes. We consider $13\TeV$ LHC proton-proton collisions and present the results in the $(M_T, \Gamma_T/MT)$ plane.  }
\end{figure}

\section{Results at leading order for single $T$ production via $Z$-boson exchanges}\label{app:TviaZ}
We present here an extensive set of predictions for single $T$ production when it is mediated from $Z$-boson exchanges. Those predictions are in complete analogy to the results discussed in the main text for single $T$ production from $W$-boson exchanges. They are thus presented without any description, and we refer to the main body of this paper for more information.
\begin{figure}[htbp]
  \centering
  \begin{minipage}{.325\textwidth}
  \includegraphics[width=\textwidth]{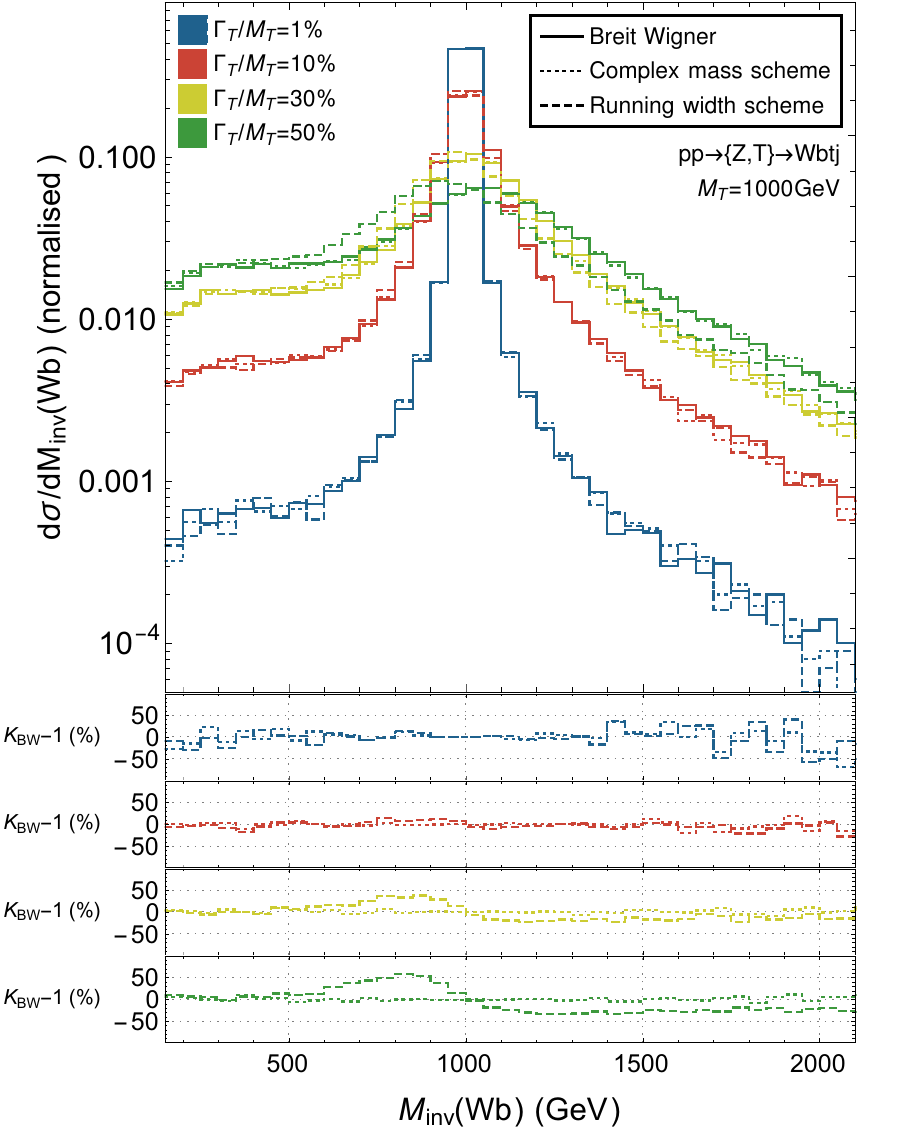}
  \end{minipage}
  \begin{minipage}{.325\textwidth}
  \includegraphics[width=\textwidth]{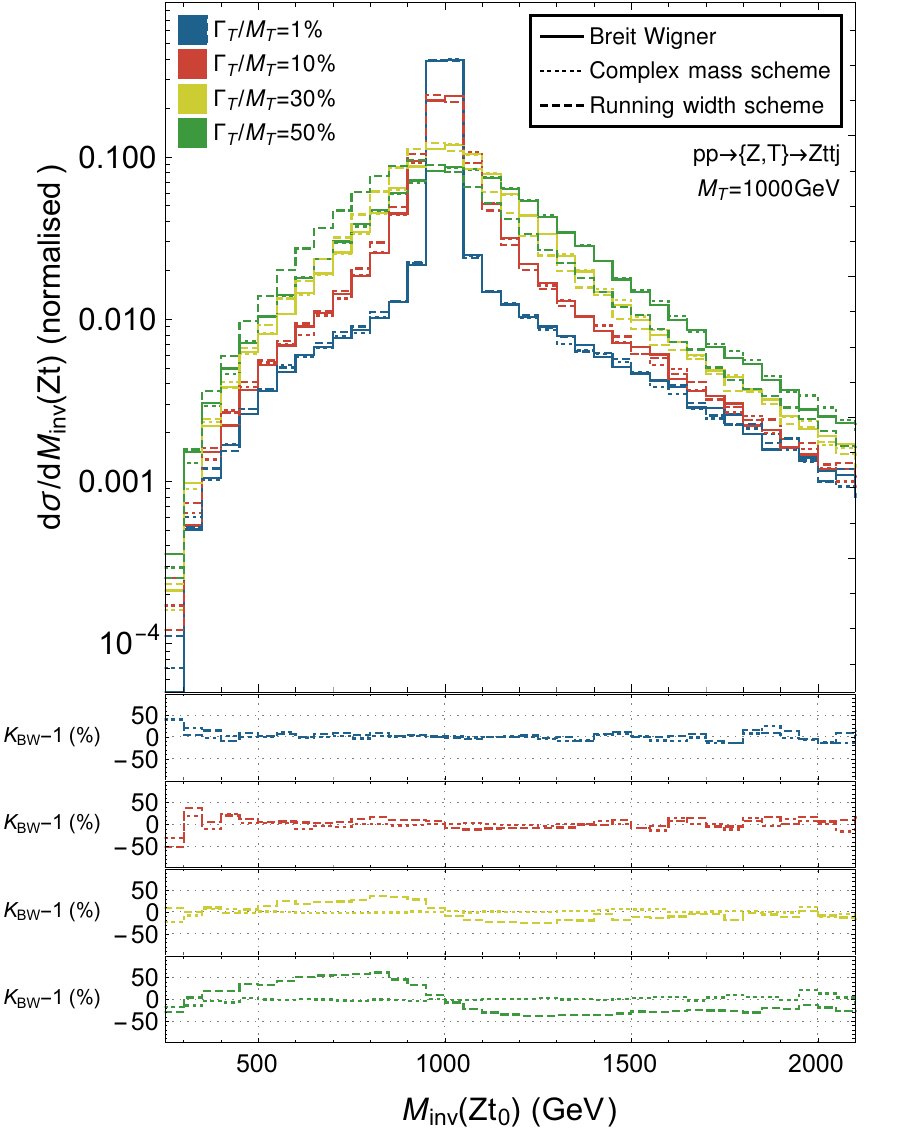}
  \includegraphics[width=\textwidth]{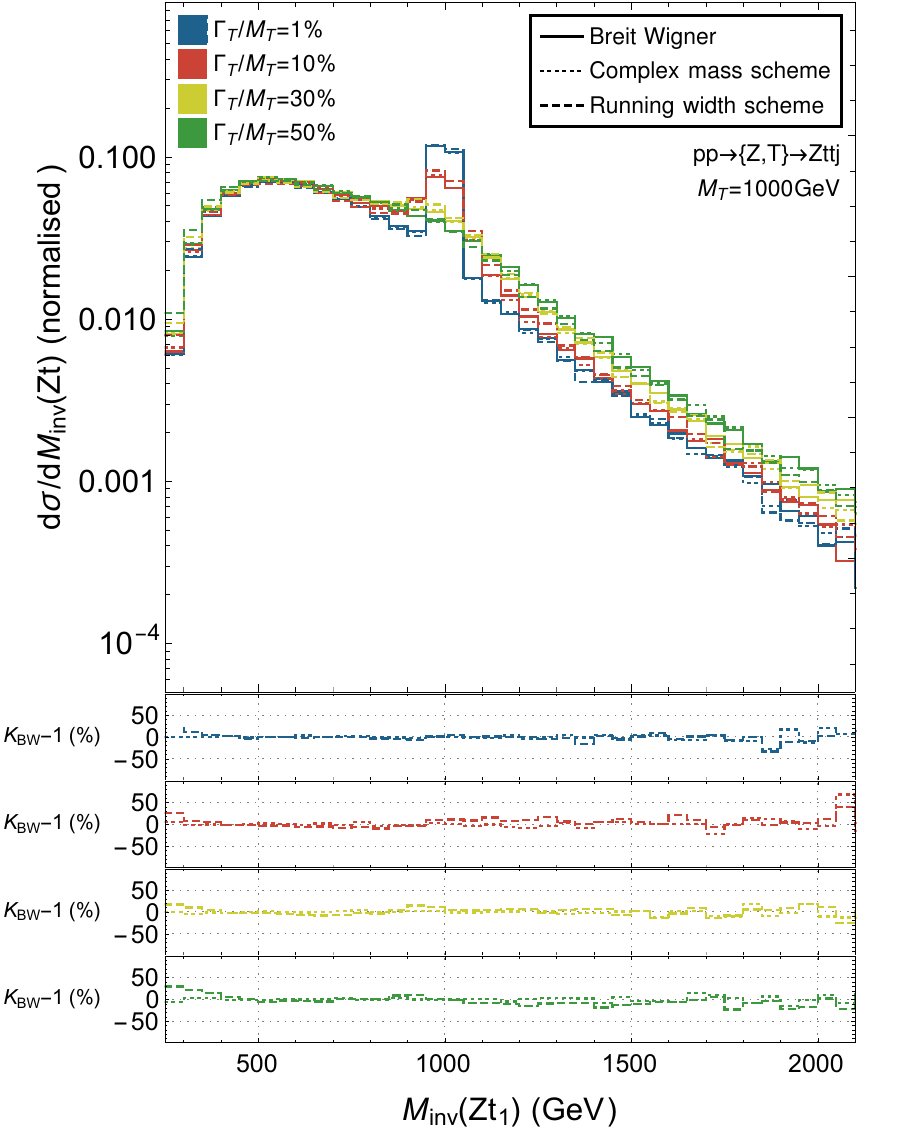}
  \end{minipage}
  \begin{minipage}{.325\textwidth}
  \includegraphics[width=\textwidth]{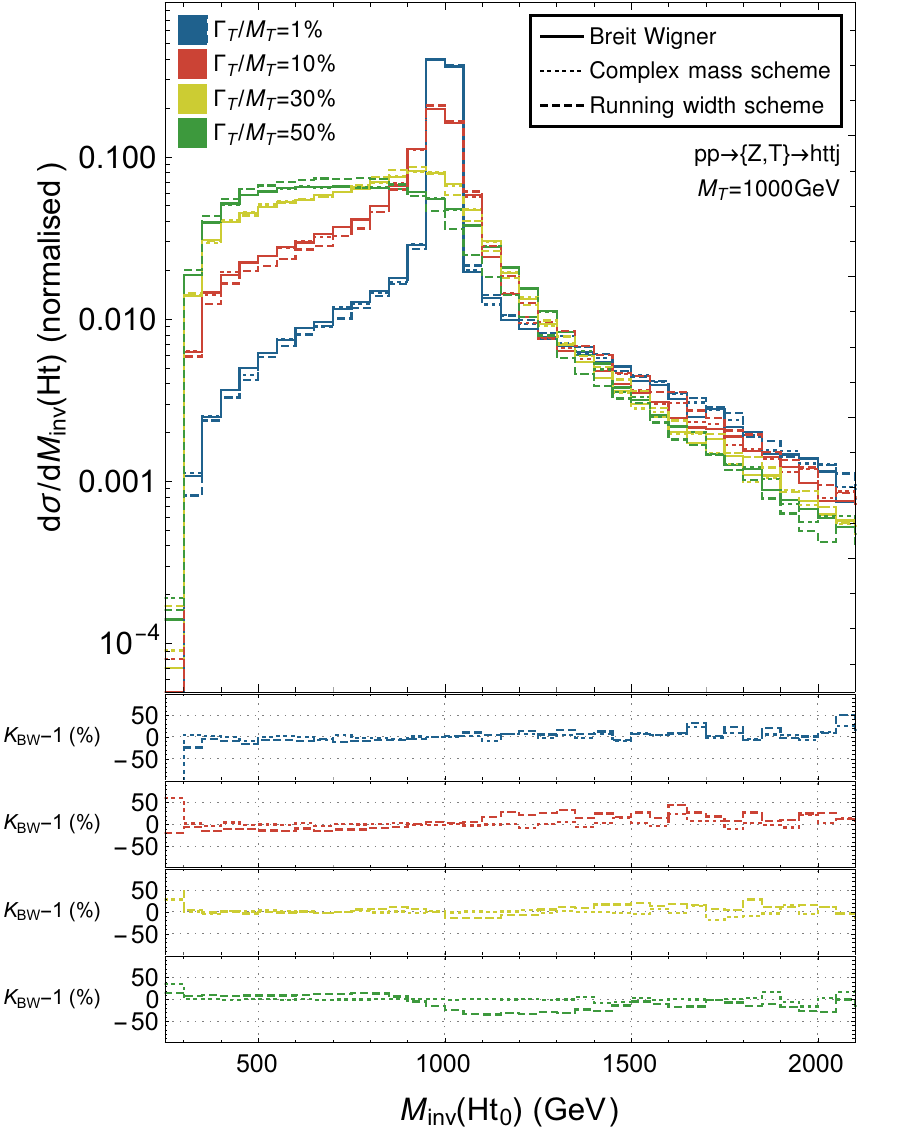}
  \includegraphics[width=\textwidth]{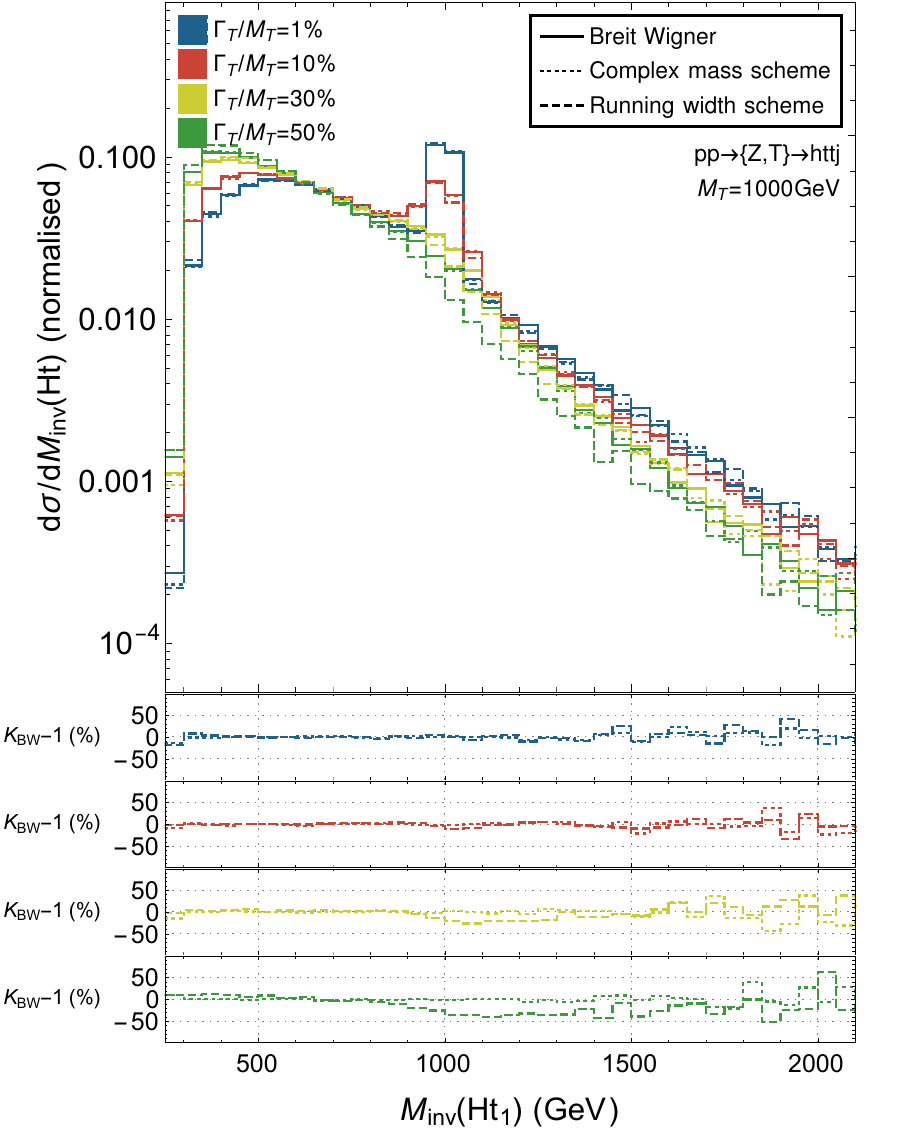}
  \end{minipage}
  \vspace*{-0.2cm}
  \caption{\label{fig:schemecomparisonZT} Comparison between predictions obtained in different schemes when treating the VLQ width. We consider $M_T=1000\GeV$ and different observables and channels: the invariant-mass spectrum of the $Wb$ system for the $pp\to\{Z,T\}\to Wbtj$ process (left), of the $Zt$ system (with the leading top quark) for $pp\to\{Z,T\}\to Zttj$ (centre top), of the $Zt$ system (with the sub-leading top quark) for $pp\to\{Z,T\}\to Zttj$ (centre bottom), of the $ht$ system (with the leading top quark) for $pp\to\{Z,T\}\to httj$ (right top) and of the $ht$ system (with the sub-leading top quark) for $pp\to\{Z,T\}\to httj$ (right bottom).}
\end{figure}

\begin{figure}[htbp]
  \centering
  \includegraphics[width=.325\textwidth]{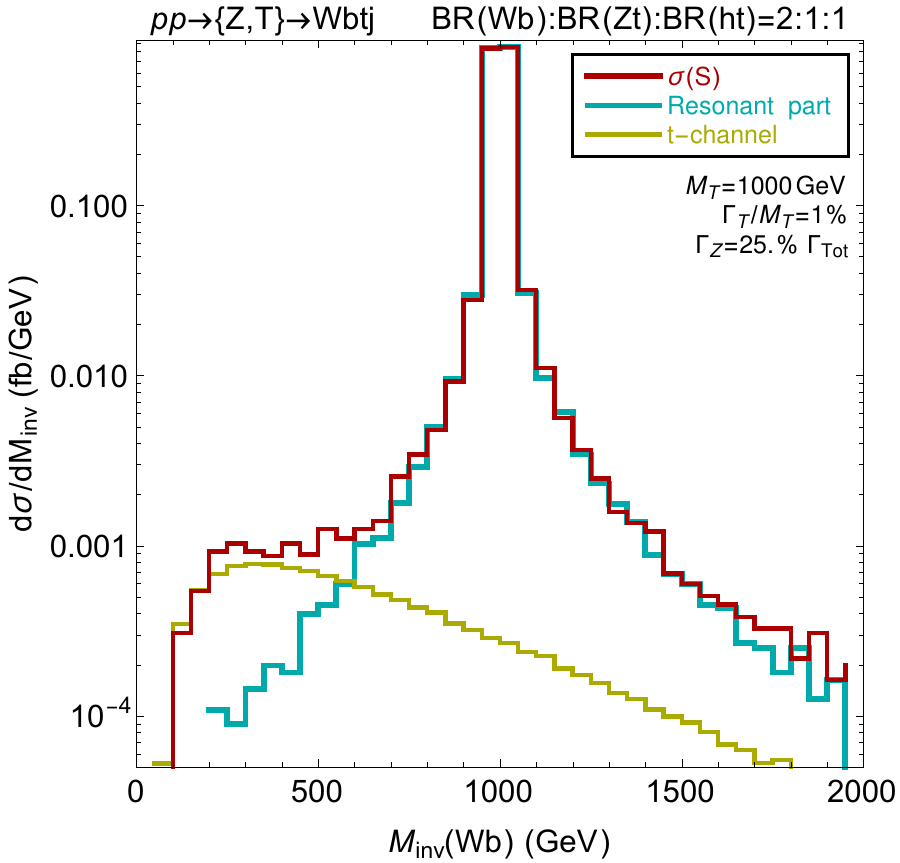}
  \includegraphics[width=.325\textwidth]{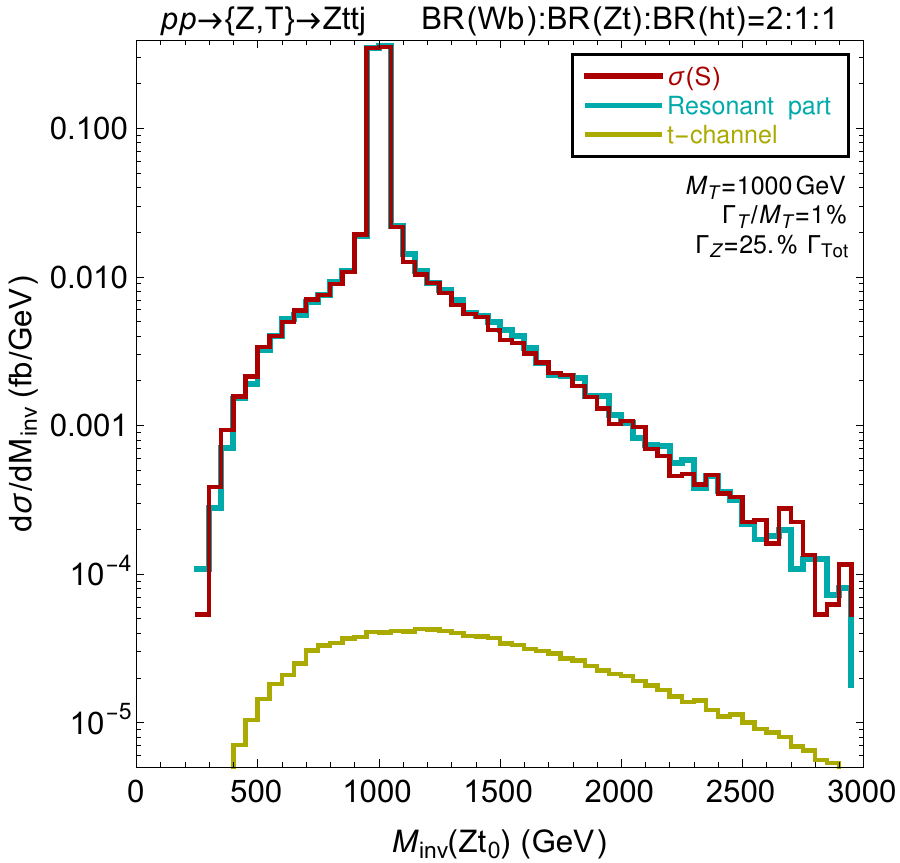}
  \includegraphics[width=.325\textwidth]{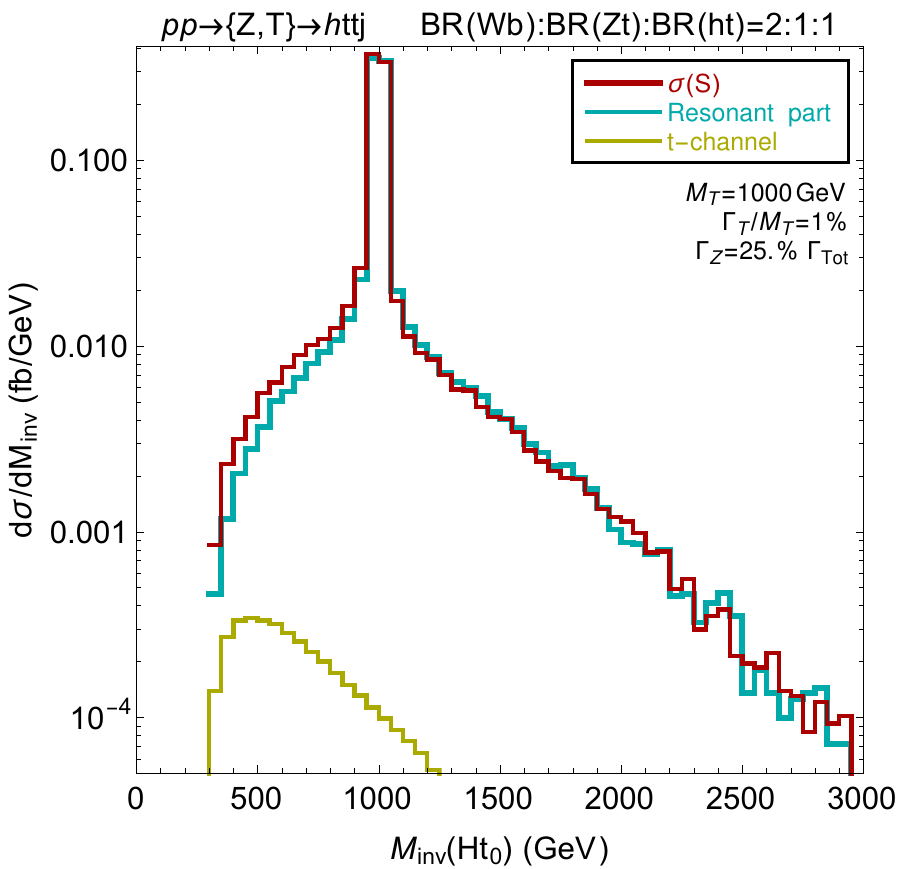}
  \\
  \includegraphics[width=.325\textwidth]{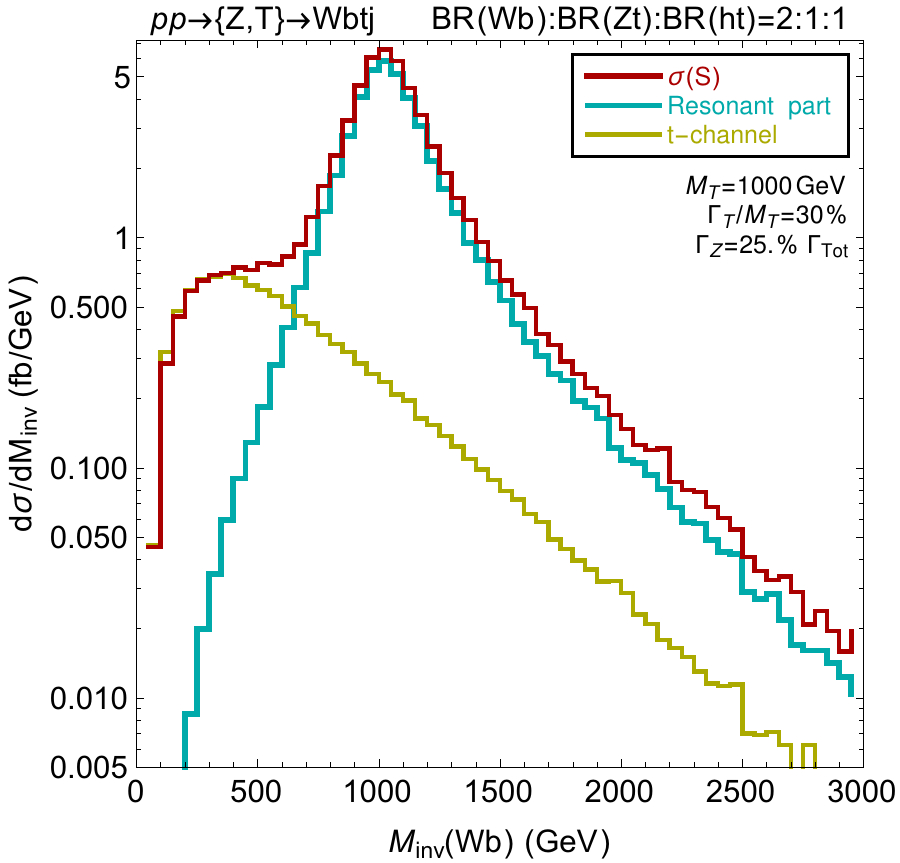}
  \includegraphics[width=.325\textwidth]{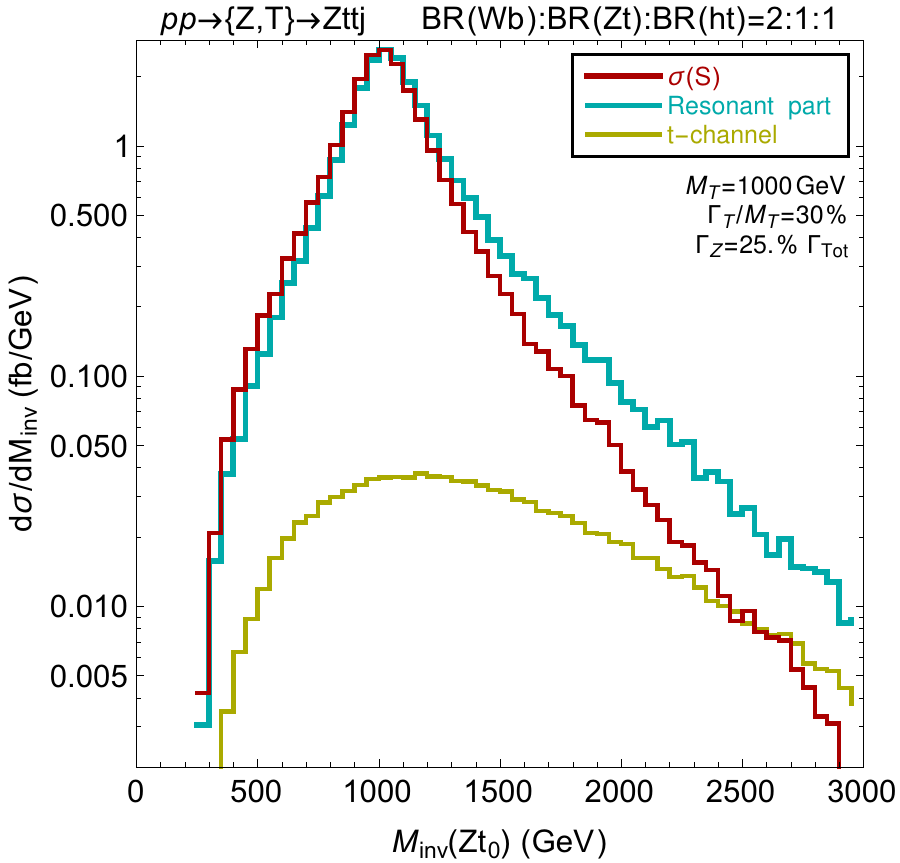}
  \includegraphics[width=.325\textwidth]{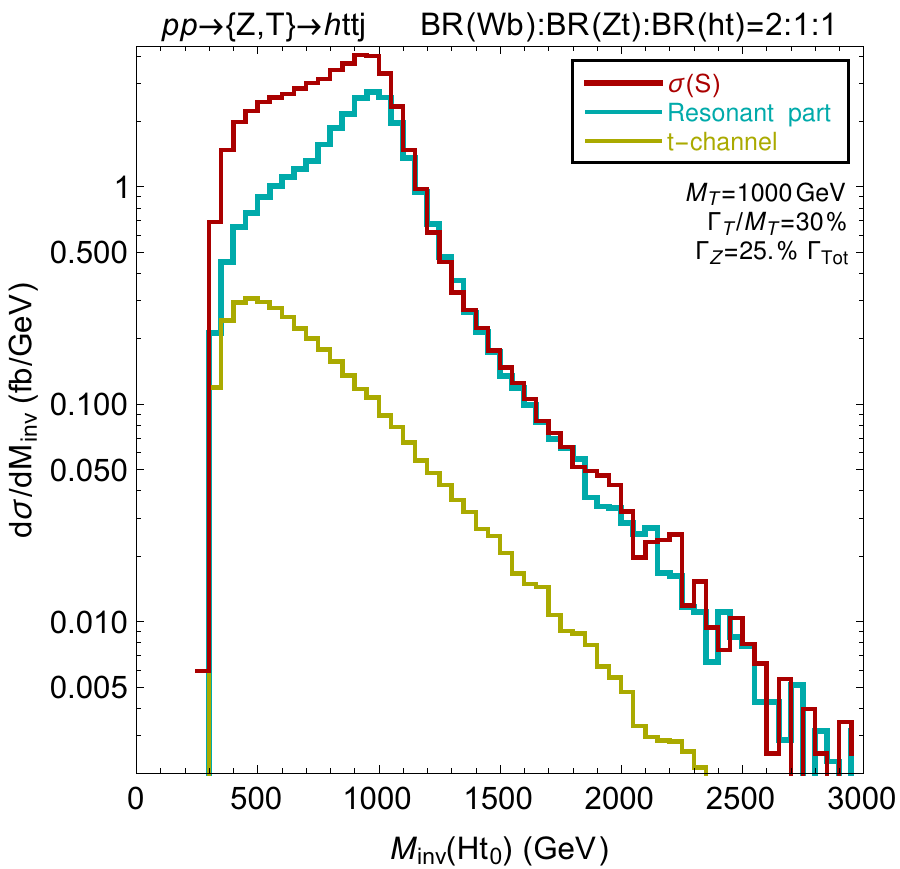}\vspace*{-0.2cm}
  \caption{\label{fig:res_tch_contributions_Z_1000} Same as figure~\ref{fig:res_tch_contributions} for the processes $pp\to Wbtj$ (left), $pp\to Zttj$ (centre) and $pp\to httj$ (right), with $m_T = 1$~TeV.}
\end{figure}

\begin{figure}[htbp]
  \centering
  \includegraphics[width=.325\textwidth]{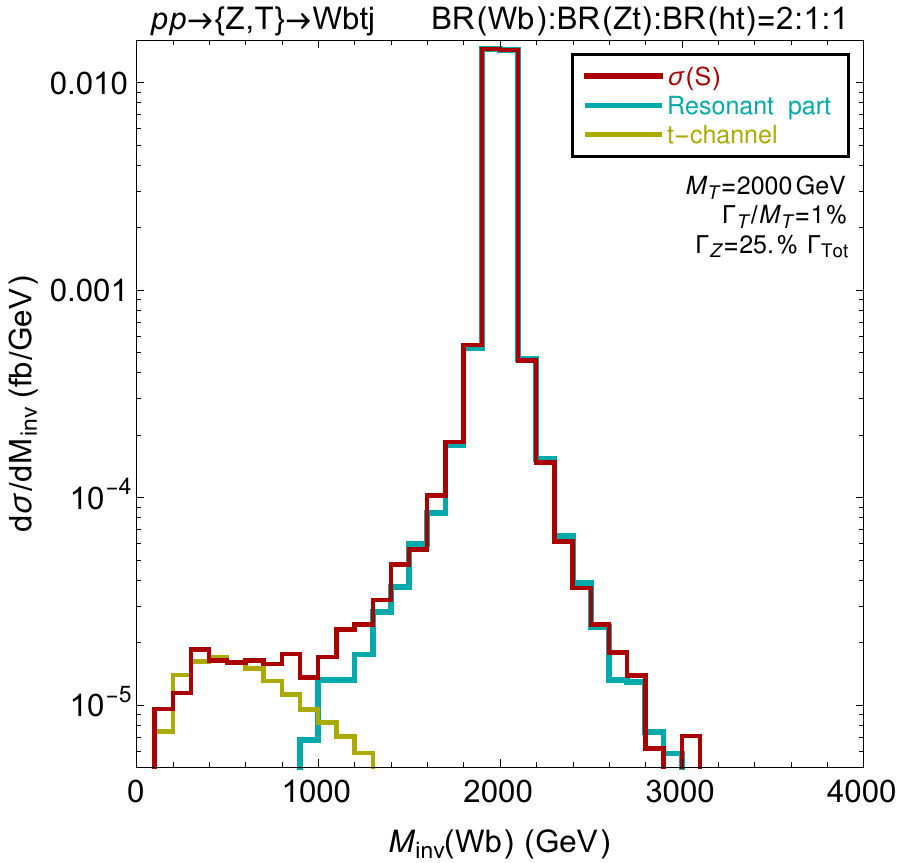}
  \includegraphics[width=.325\textwidth]{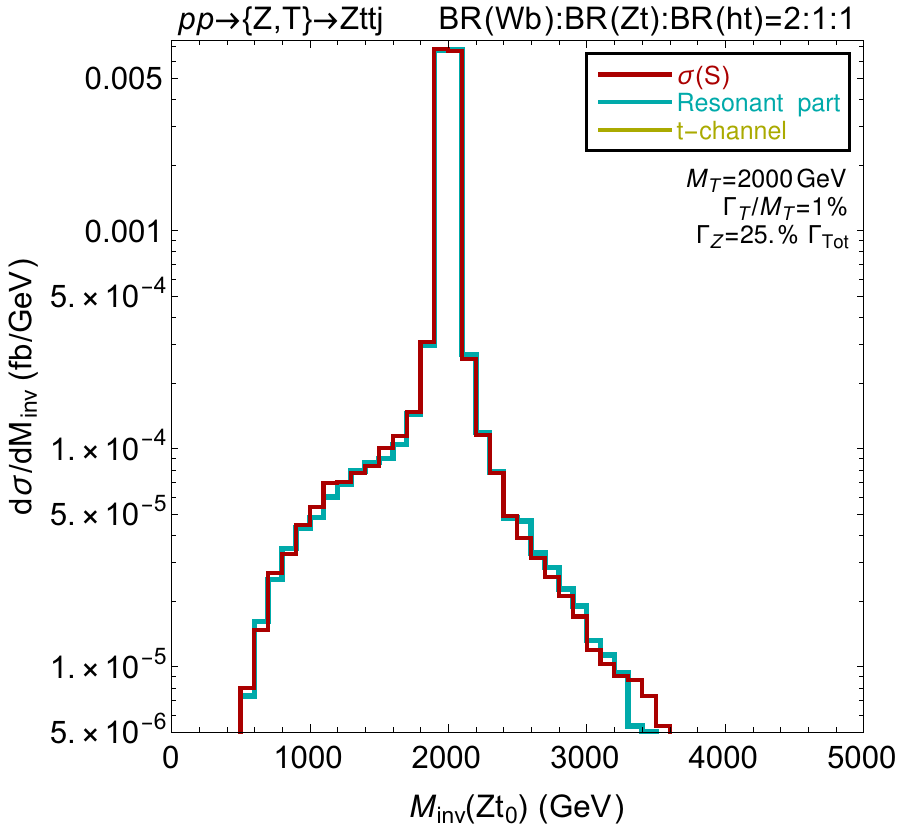}
  \includegraphics[width=.325\textwidth]{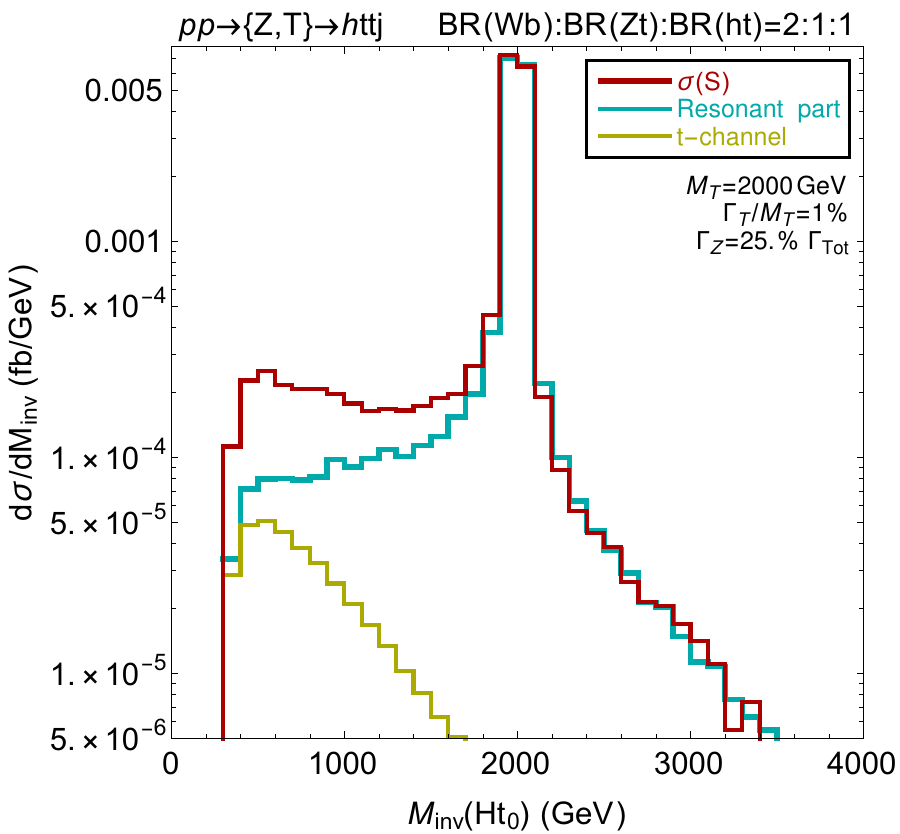}
  \\
  \includegraphics[width=.325\textwidth]{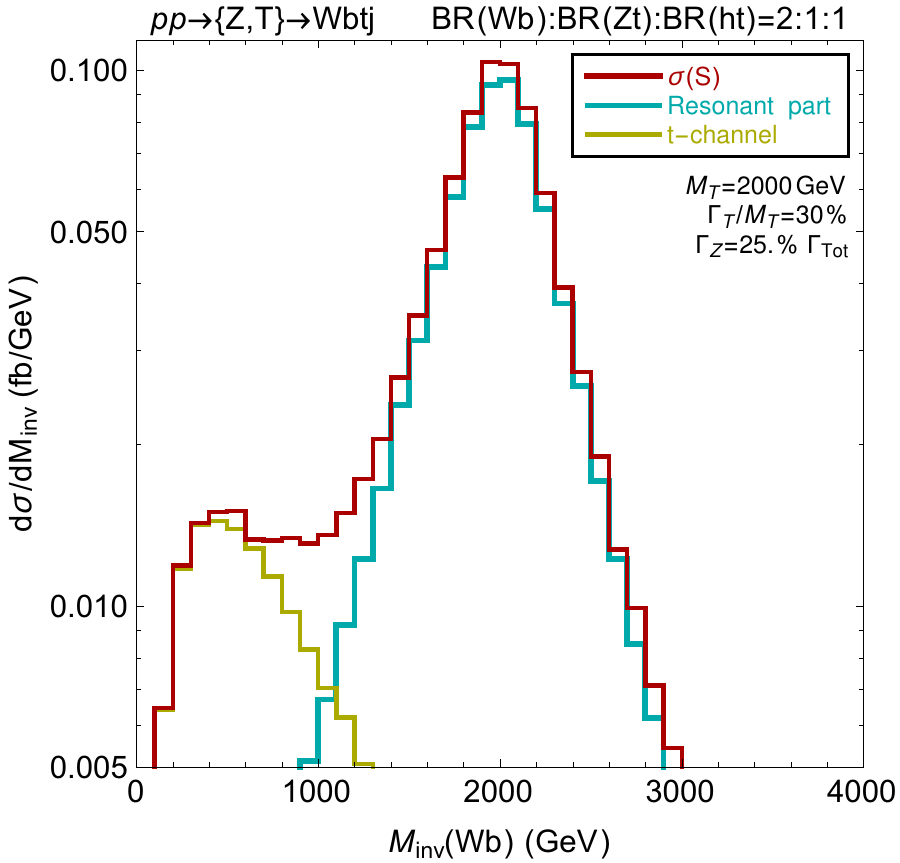}
  \includegraphics[width=.325\textwidth]{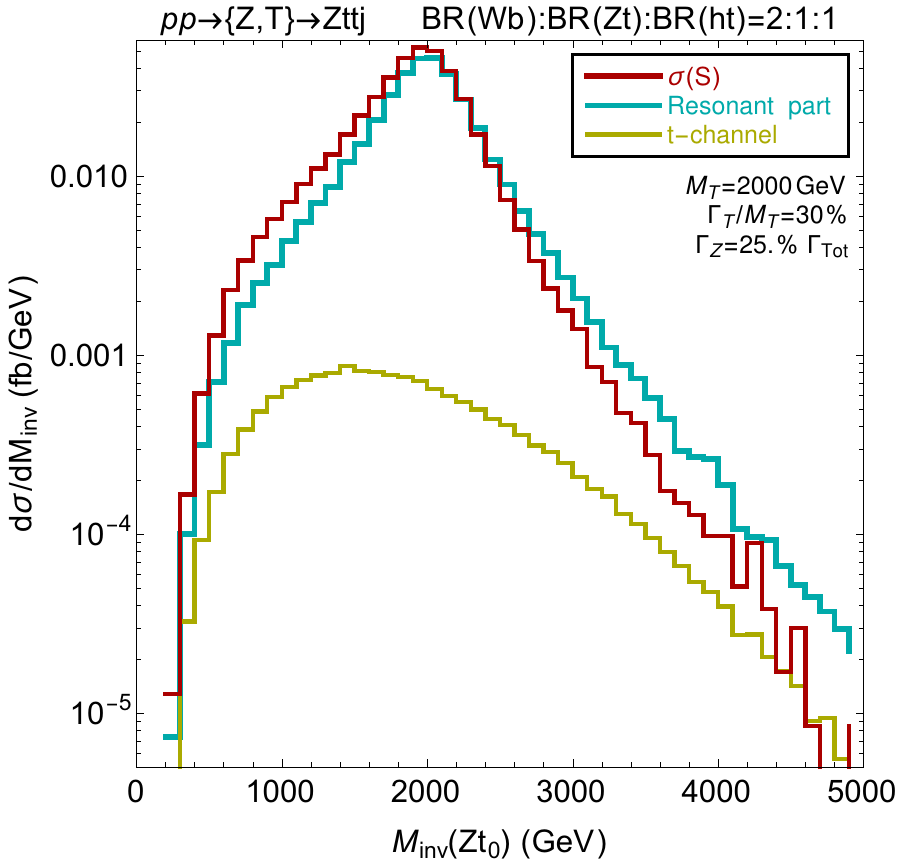}
  \includegraphics[width=.325\textwidth]{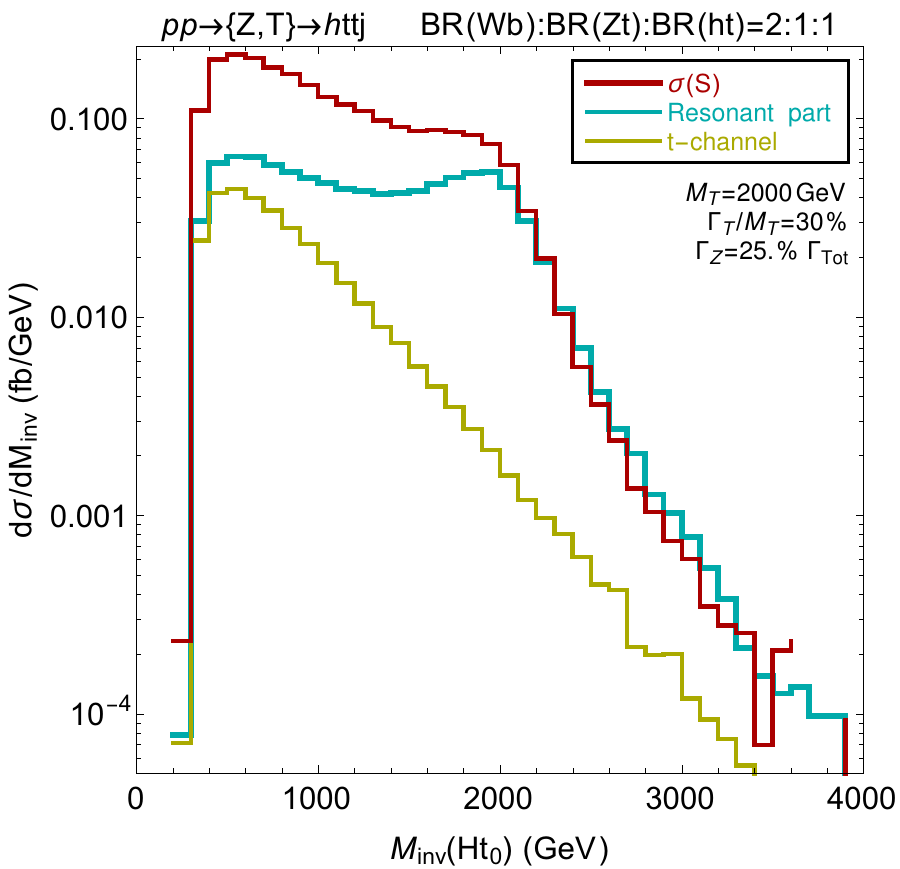}\vspace*{-0.2cm}
  \caption{\label{fig:res_tch_contributions_Z_2000} Same as figure~\ref{fig:res_tch_contributions_Z_1000} for $M_T=2$~TeV.}
\end{figure}

\begin{figure}[!htbp]
  \centering
  \includegraphics[width=.30\textwidth]{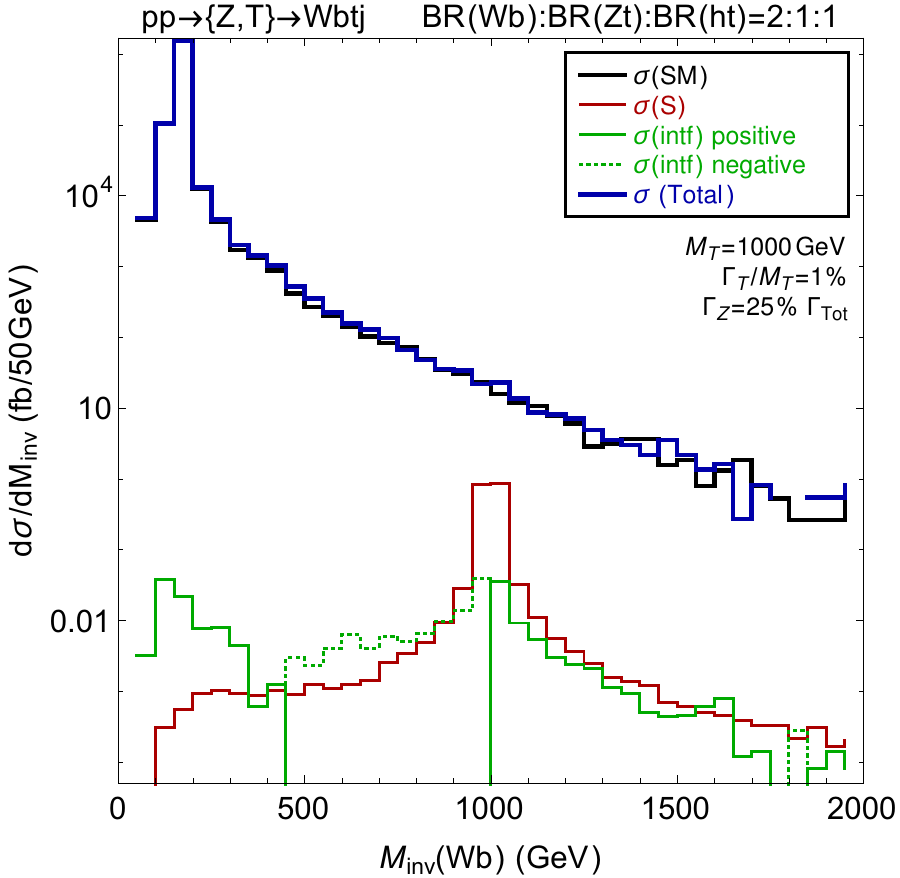}
  \includegraphics[width=.30\textwidth]{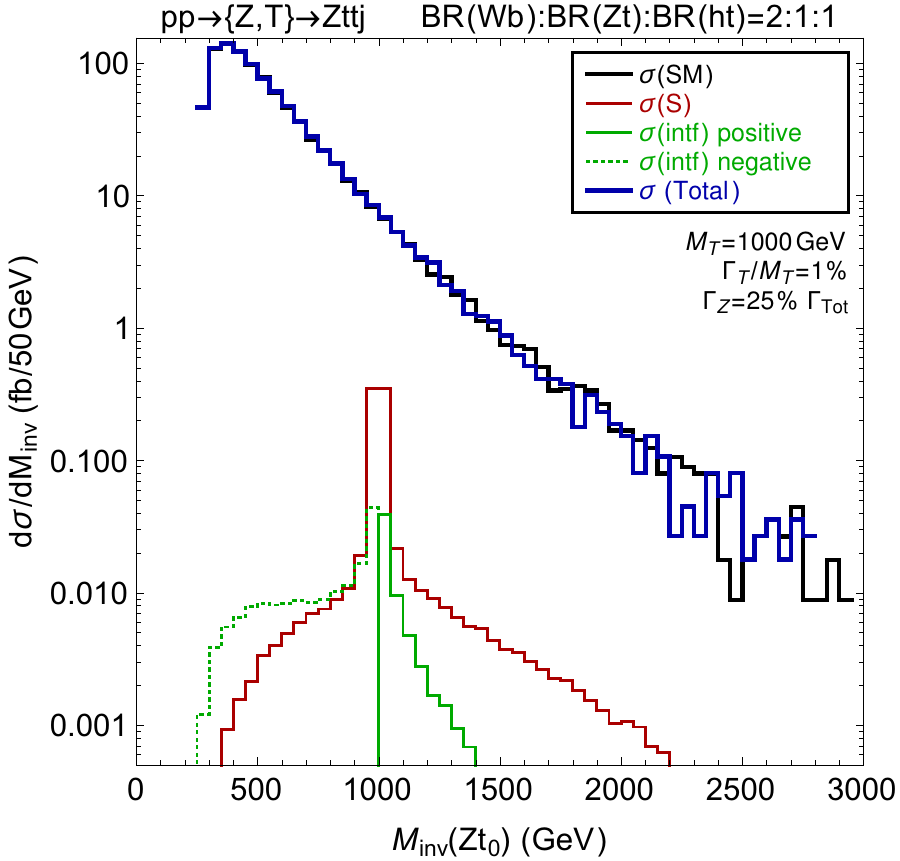}
  \includegraphics[width=.30\textwidth]{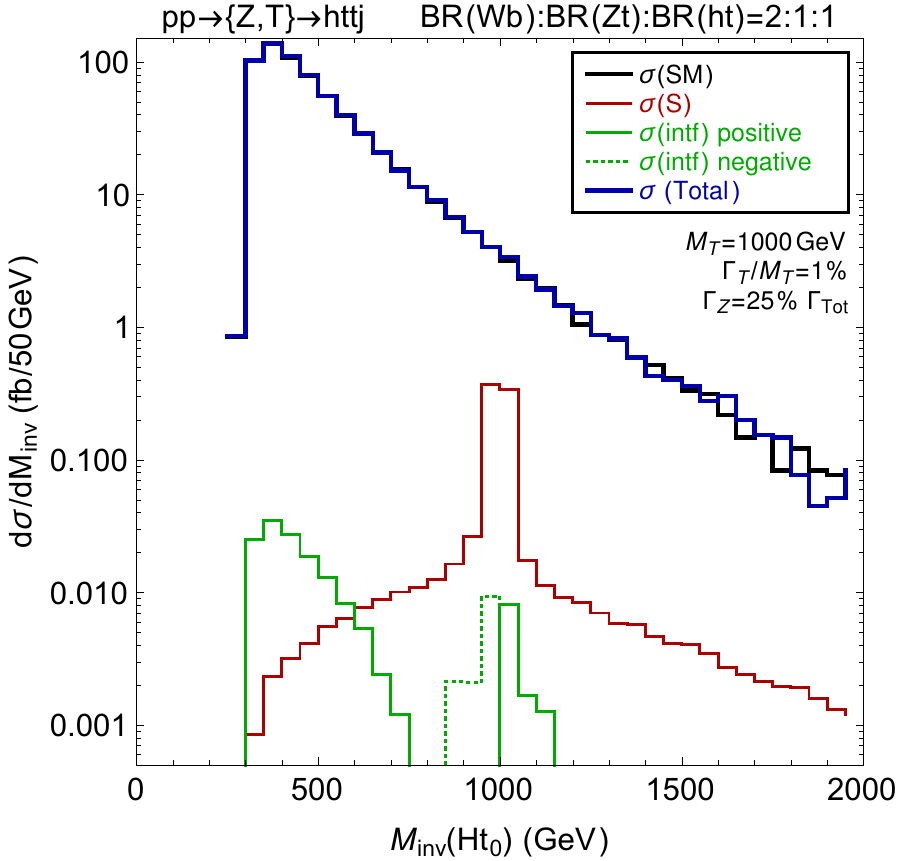}\\
  \includegraphics[width=.30\textwidth]{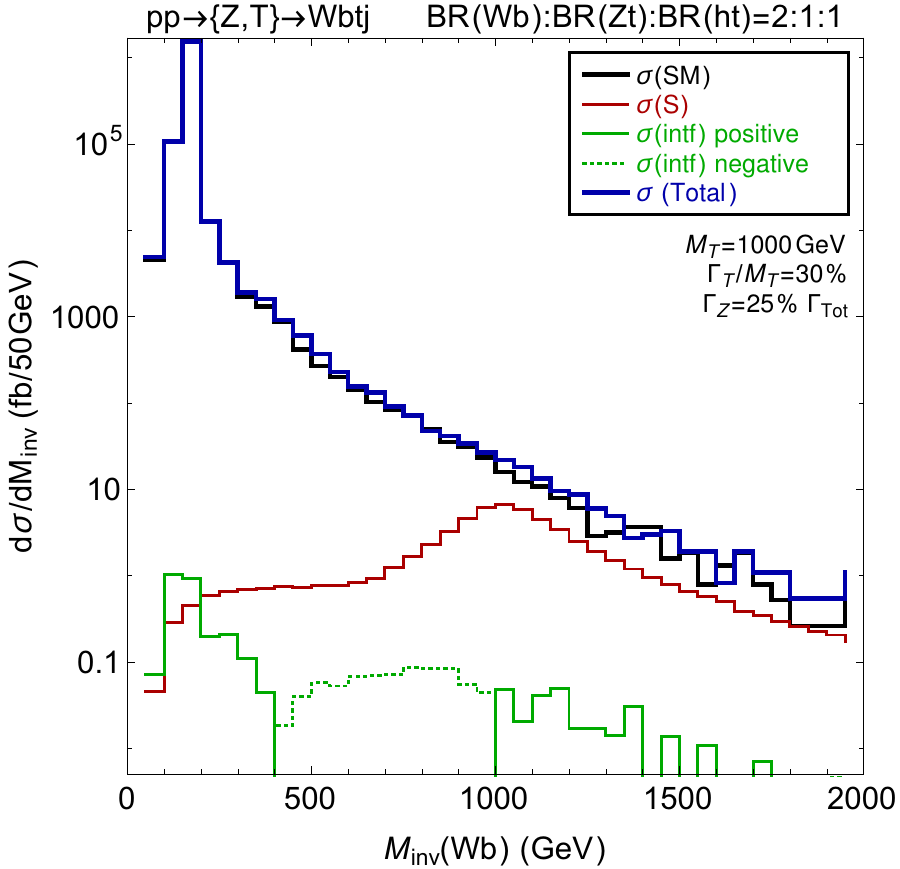}
  \includegraphics[width=.30\textwidth]{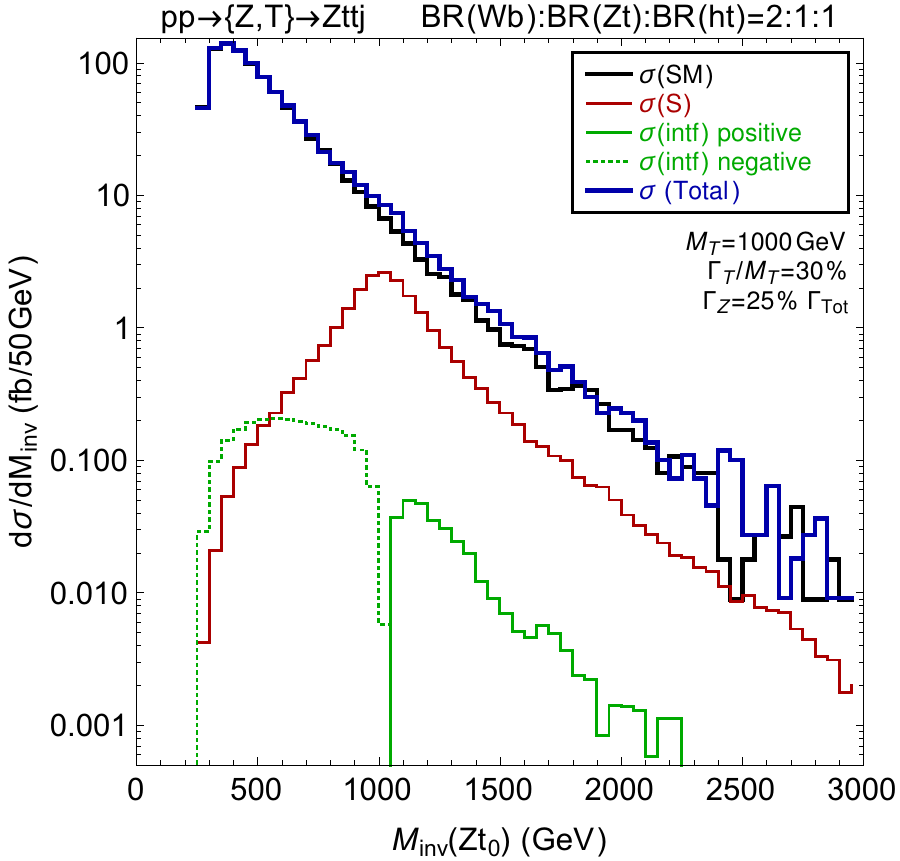}
  \includegraphics[width=.30\textwidth]{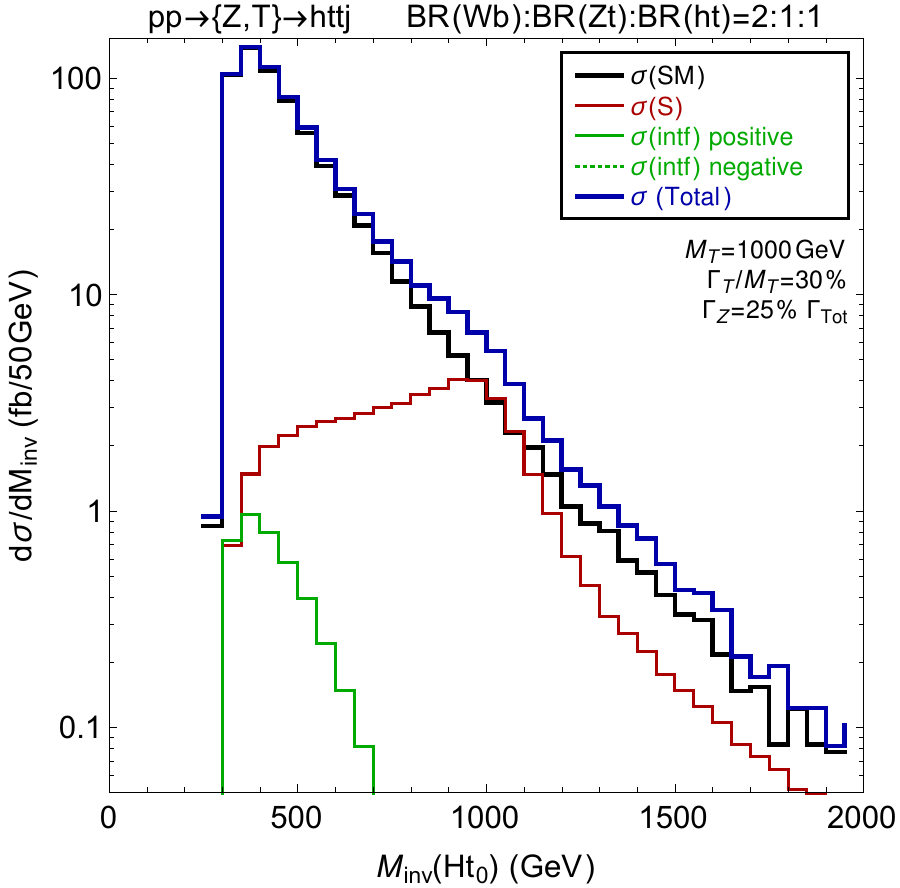}\vspace*{-.2cm}
  \caption{\label{fig:SBZ_int_sums_1000} Parton-level distributions in the
    invariant mass of the $Wb$ (left), $Zt$ (centre) and $ht$ (right) systems
    for the $pp\to Wbtj$, $pp\to Zttj$ and $pp\to httj$ processes respectively.
    We include the SM contribution (black), the new physics contributions
    stemming from a vector-like quark of mass $M_T=1\TeV$ (red) and the absolute
    value of their interference (green). We indicate by a dashed line the region
    in which the interference is negative. We consider a narrow
    width scenario ($\Gamma_T/M_T=1\%$, top row) and a large width one
    ($\Gamma_T/M_T=30\%$, bottom row), and the $T$ couplings are fixed as in
    eq.~\eqref{eq:benchmarks}.}\vspace*{.5cm}
  \includegraphics[width=.30\textwidth]{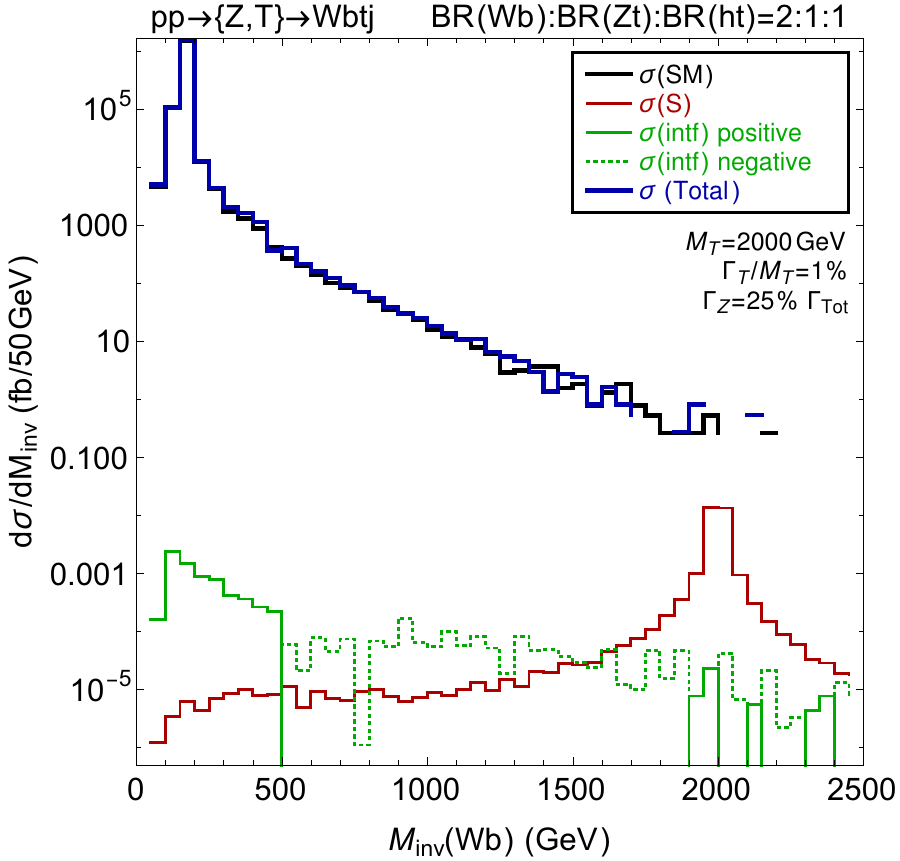}
  \includegraphics[width=.30\textwidth]{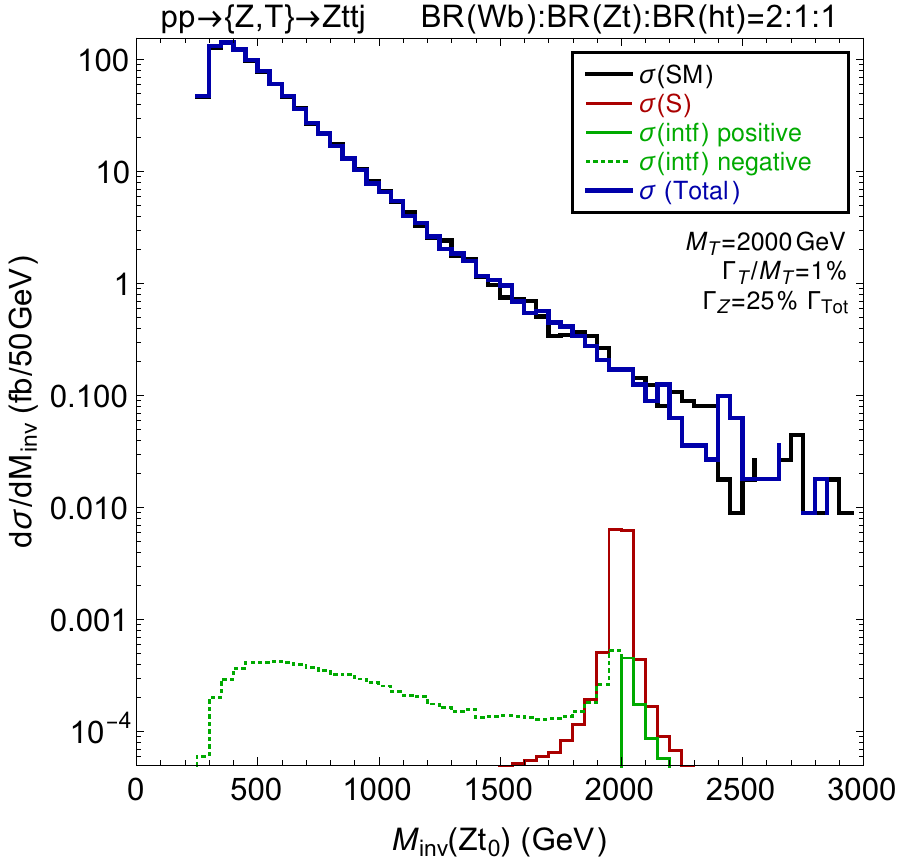}
  \includegraphics[width=.30\textwidth]{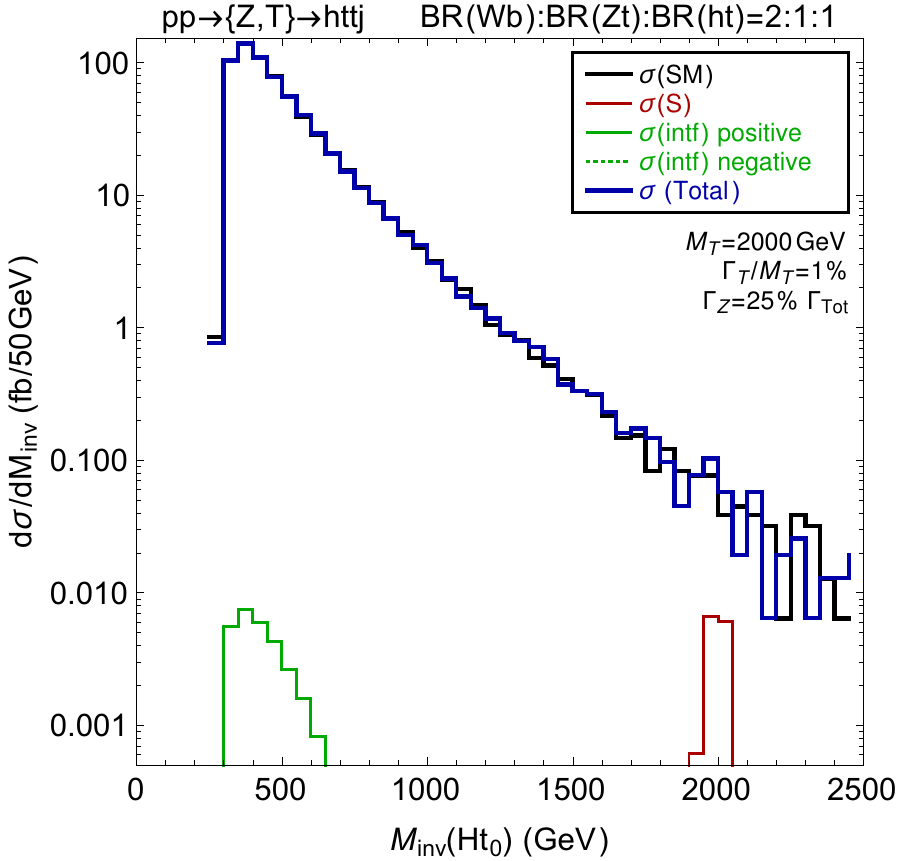}\\
  \includegraphics[width=.30\textwidth]{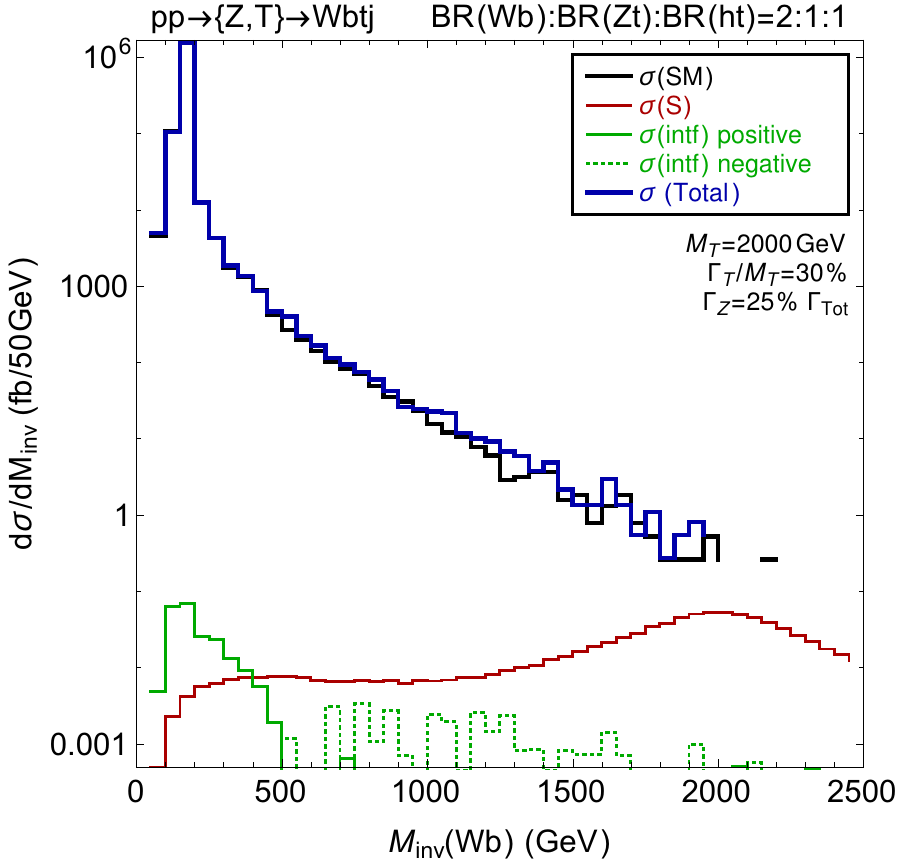}
  \includegraphics[width=.30\textwidth]{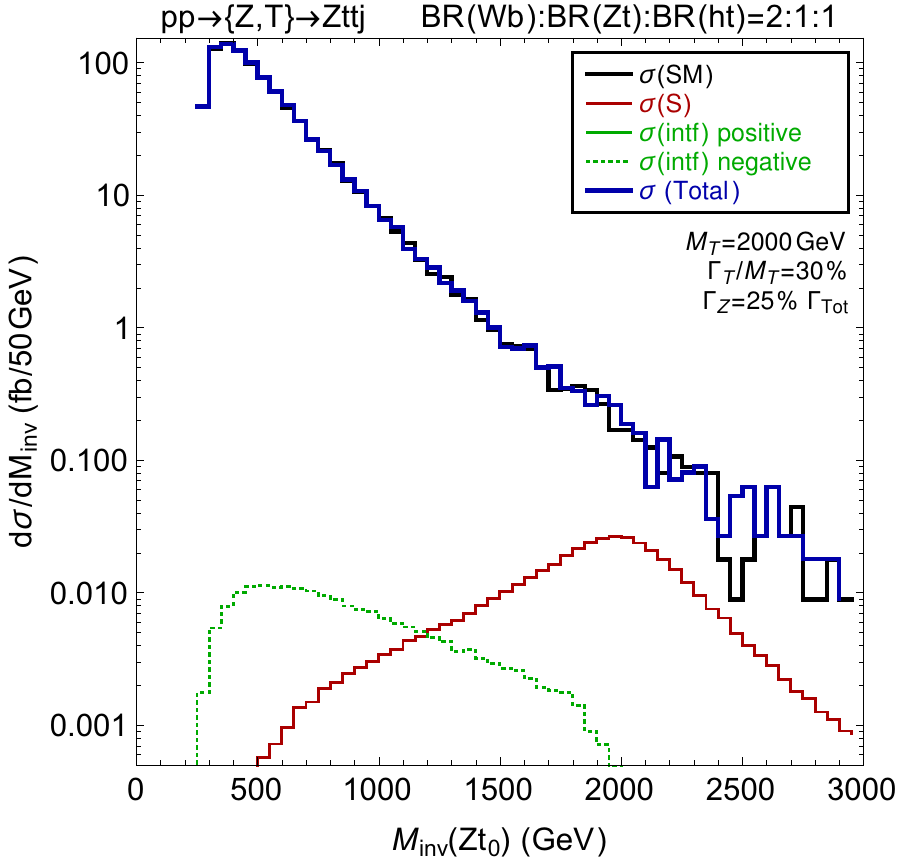}
  \includegraphics[width=.30\textwidth]{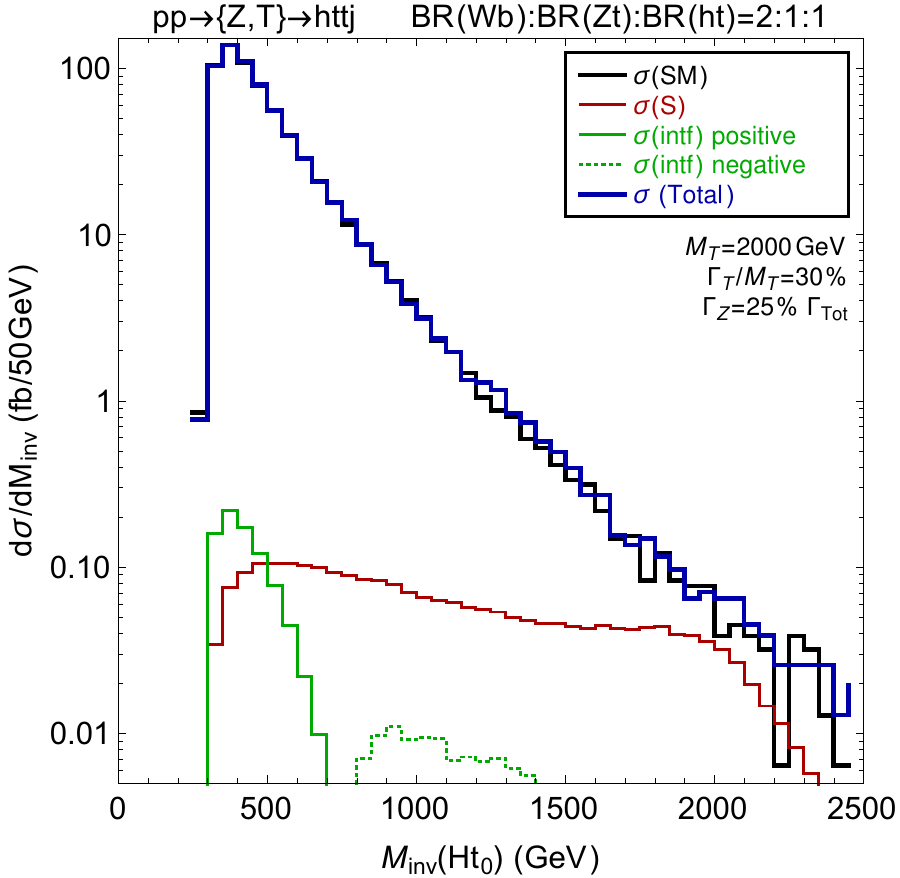}\vspace*{-.2cm}
  \caption{\label{fig:SBZ_int_sums_2000}
    Same as in figure~\ref{fig:SBZ_int_sums_1000} for $M_T=2\TeV$.}
\end{figure}
 
\begin{figure}
  \centering
  \vspace*{-0.4cm}
  \includegraphics[width=.325\textwidth]{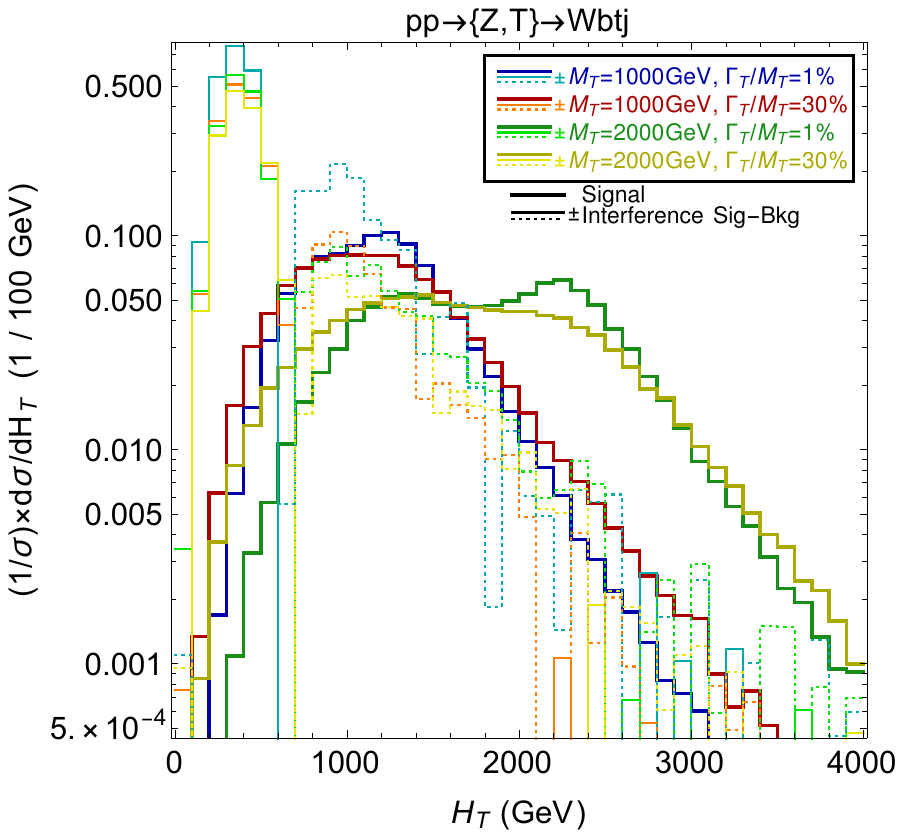}
  \includegraphics[width=.325\textwidth]{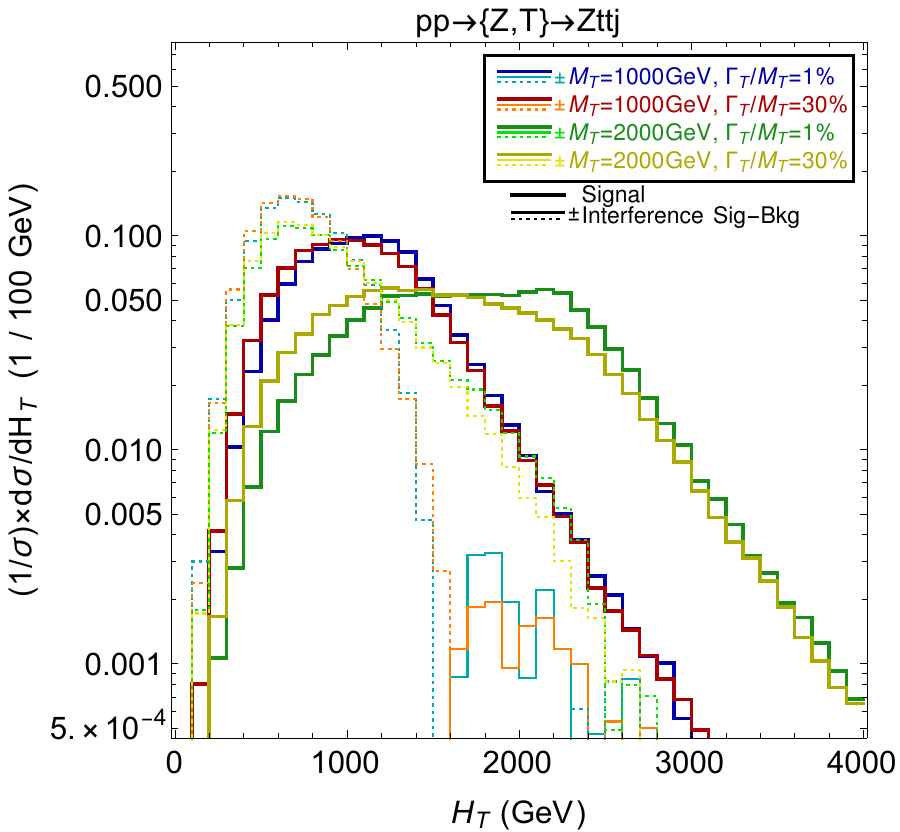}
  \includegraphics[width=.325\textwidth]{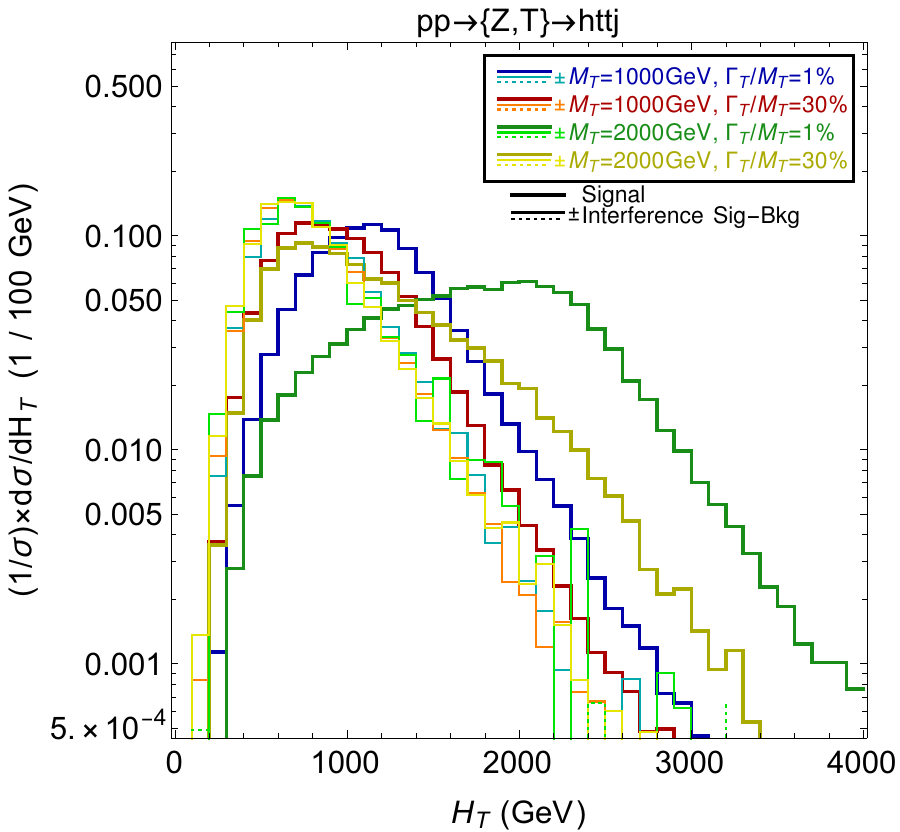}
  \\
  \includegraphics[width=.325\textwidth]{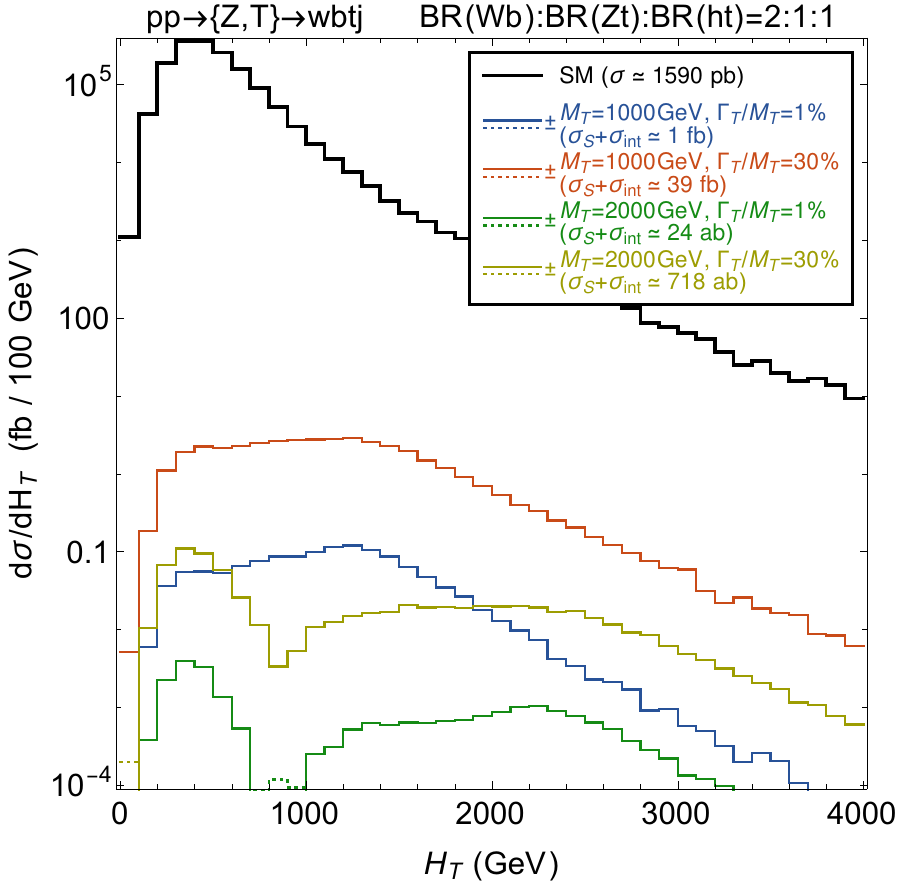}
  \includegraphics[width=.325\textwidth]{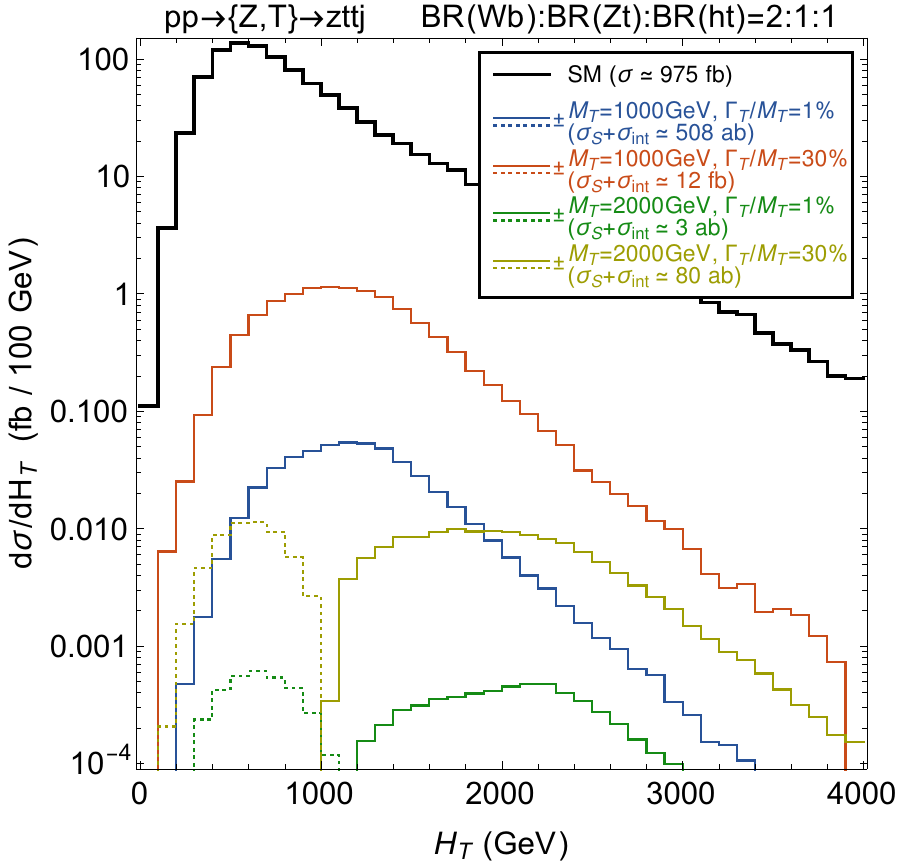}
  \includegraphics[width=.325\textwidth]{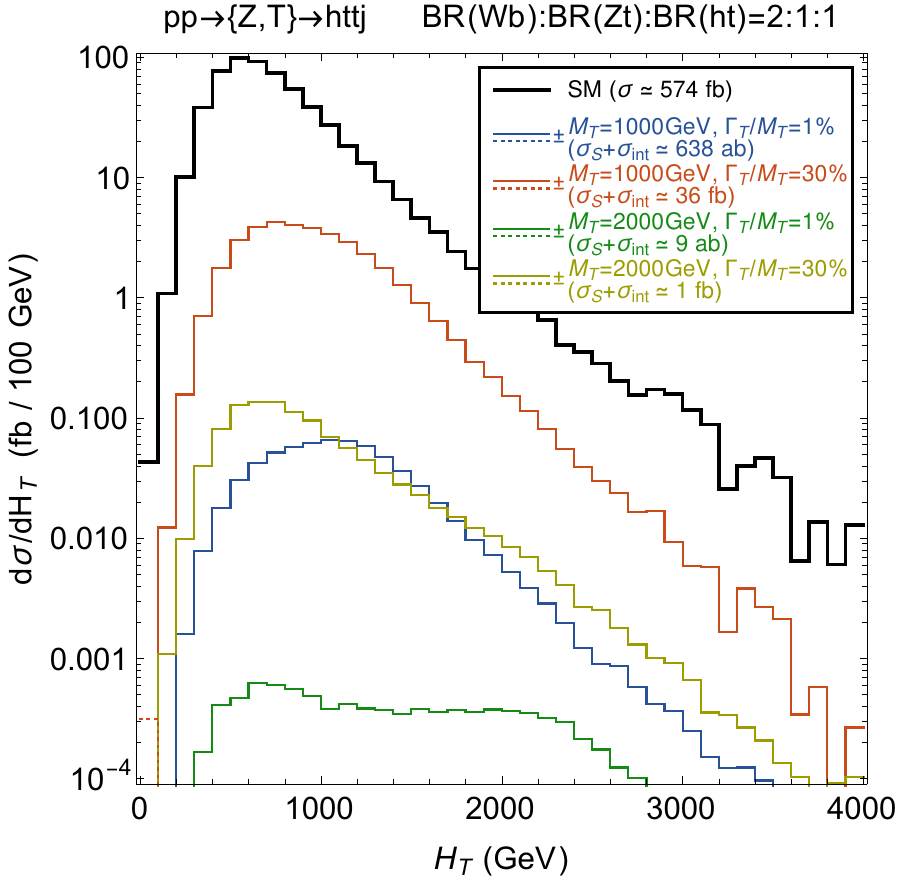}\vspace*{-0.2cm}
  \caption{\label{fig:recodistHT_Z} $H_T$ distribution at the reconstructed level for the $pp\to Wbtj$ (left), $Zttj$ (centre) and $httj$ (right) processes and for two values of the $T$ mass ($1000\GeV$ and $2000\GeV$) and two values of the $\Gamma_T/M_T$ ratio (1\% and 30\%). In the top panel we show the shapes of the $H_T$ distributions for the pure signal component, and for its interference with the SM. In the bottom panel the sums of signal and interference are displayed, as obtained following \cref{eq:xsec_param}. We consider a choice of couplings corresponding to the singlet case (see figure~\ref{fig:singletdoubletT}), and compared with the distribution of SM background. For the $Zttj$ case, the large particle multiplicities in the final state makes the Monte Carlo integration very resource-consuming. We have consequently simulated the decay of the $Z$-boson through {\sc Pythia 8}.}
\end{figure}

\begin{figure}
  \centering
  \includegraphics[width=.325\textwidth]{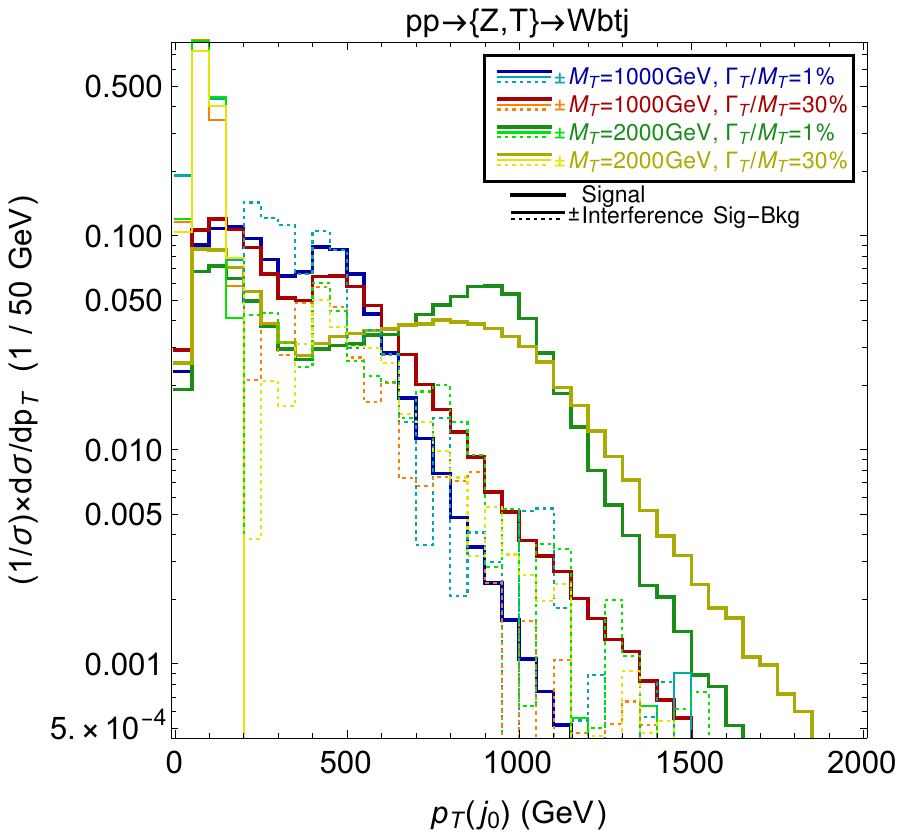}
  \includegraphics[width=.325\textwidth]{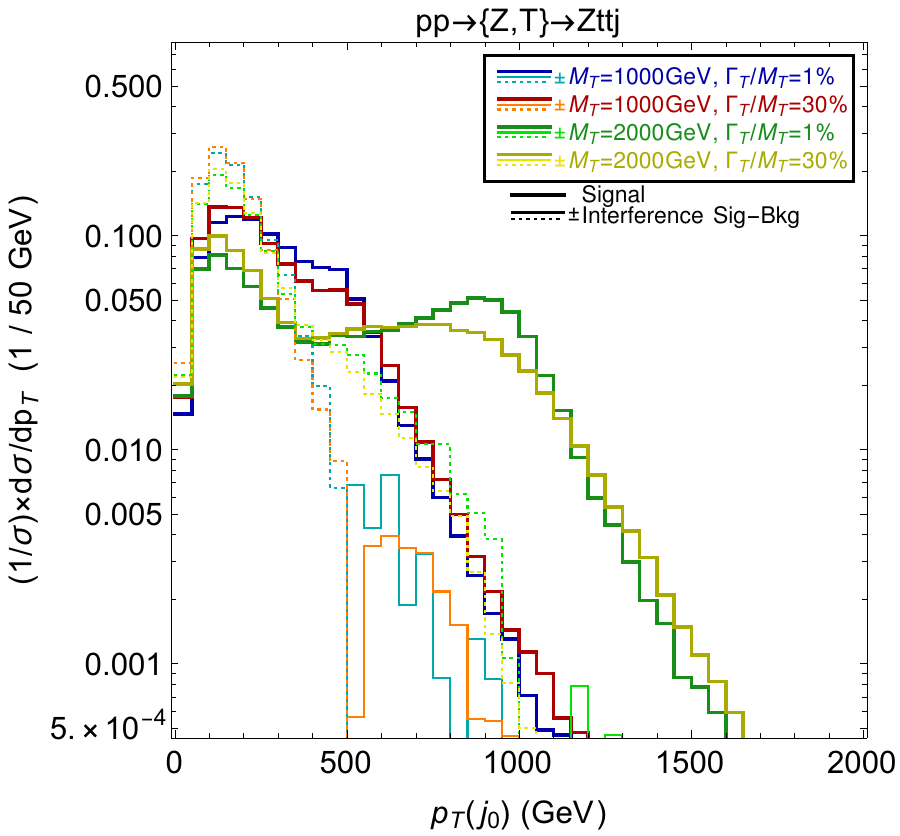}
  \includegraphics[width=.325\textwidth]{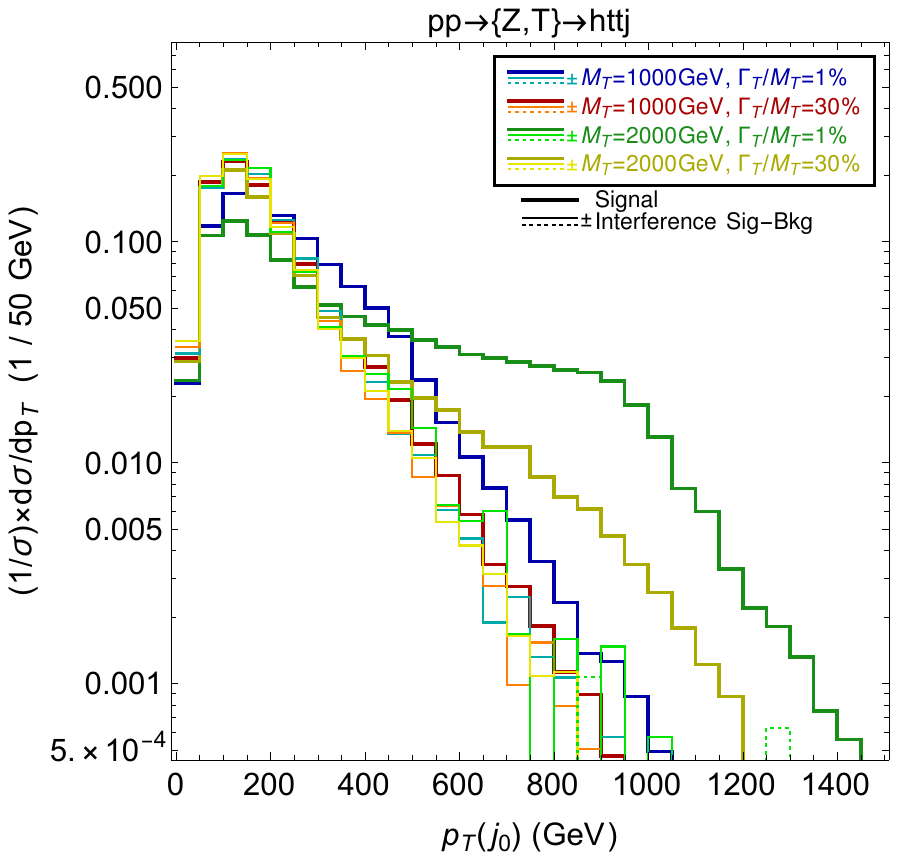}
  \\
  \includegraphics[width=.325\textwidth]{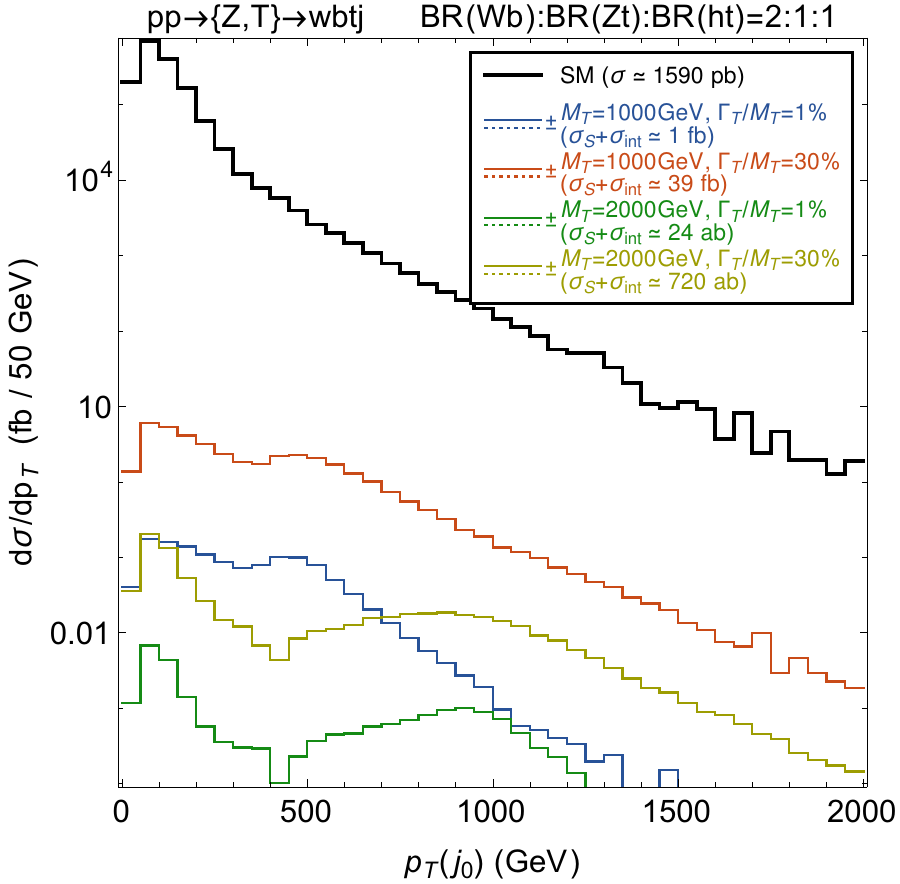}
  \includegraphics[width=.325\textwidth]{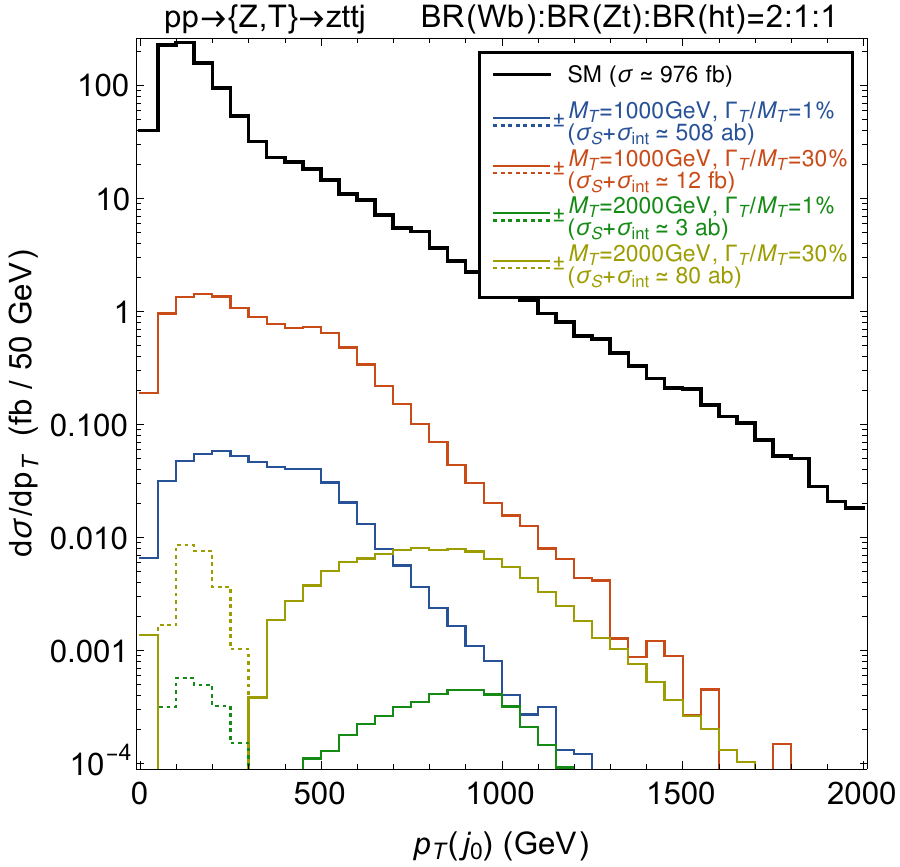}
  \includegraphics[width=.325\textwidth]{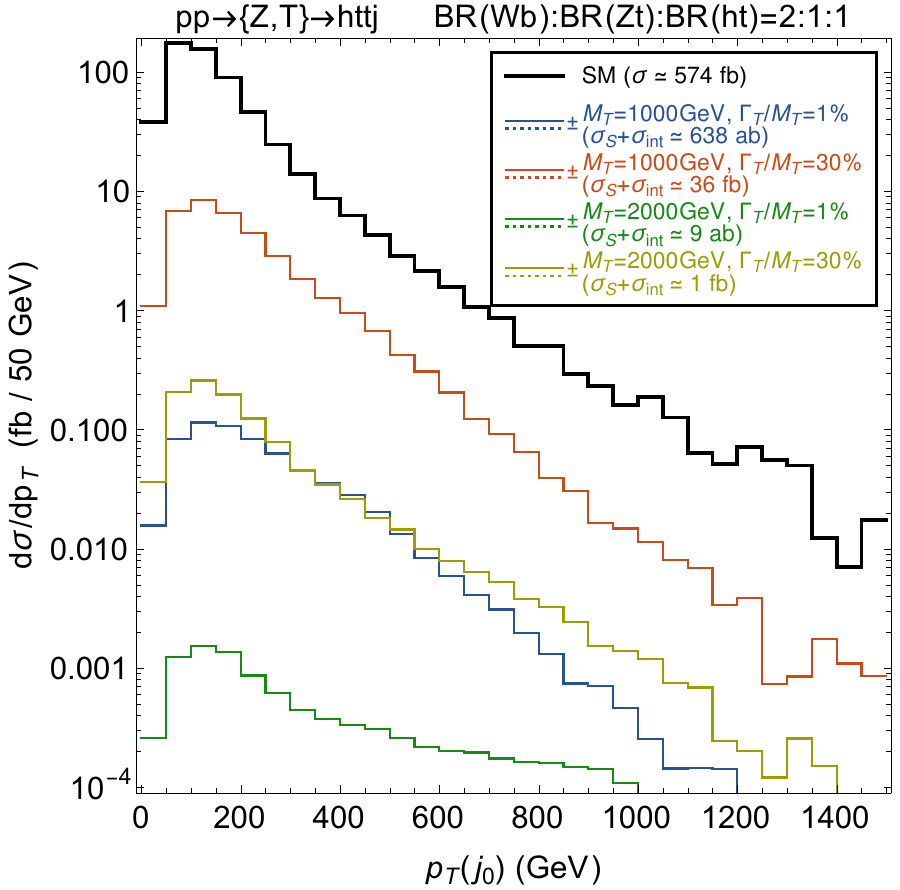}\vspace*{-0.2cm}
  \caption{\label{fig:recodistpTj0_Z} Same as figure~\ref{fig:recodistHT_Z} for the $p_T$ of the leading jet.}
\end{figure}

\begin{figure}
  \centering
  \includegraphics[width=.32\textwidth]{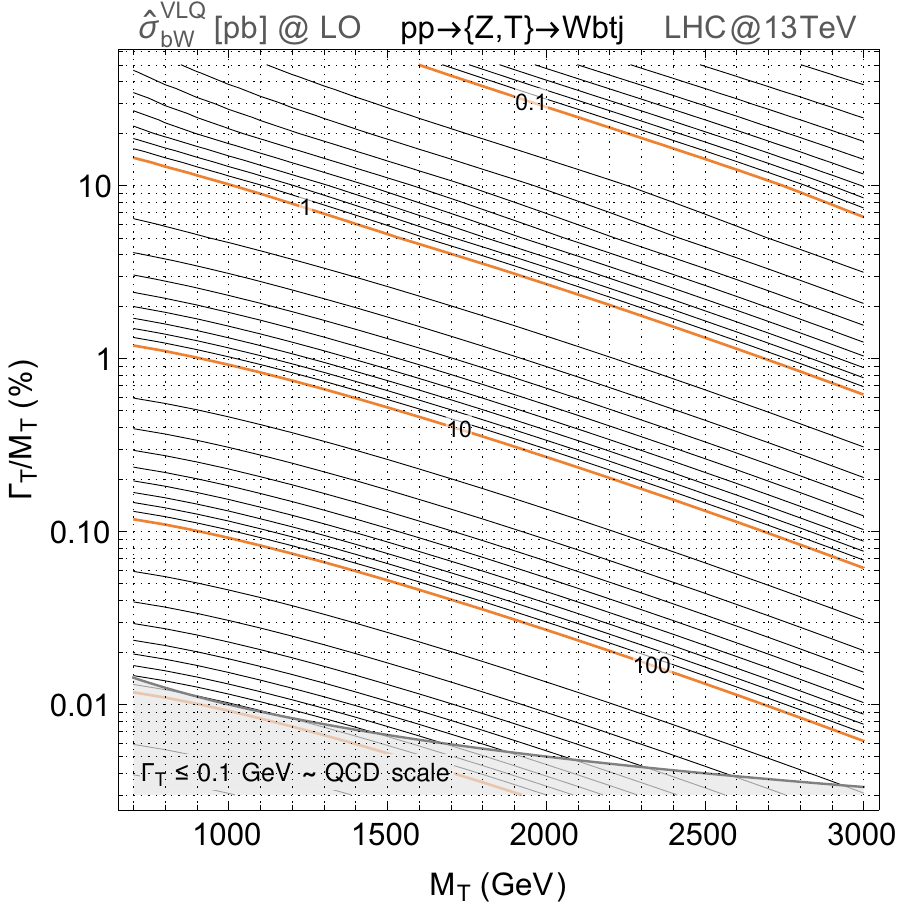}
  \includegraphics[width=.32\textwidth]{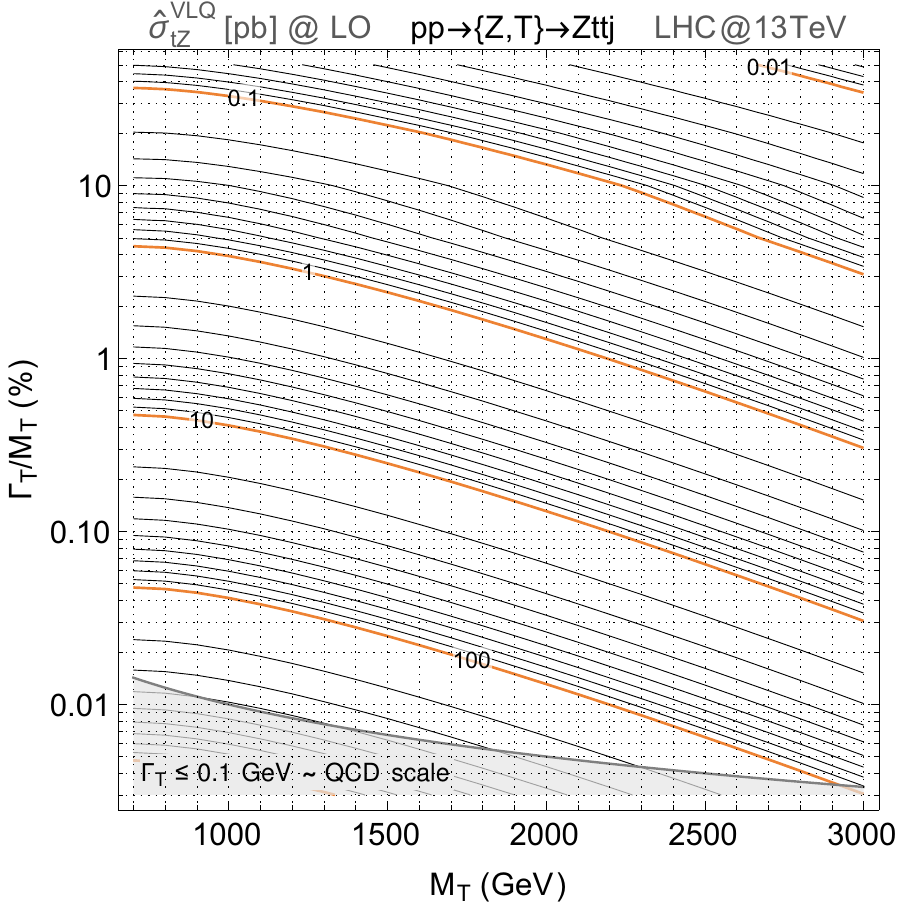}
  \includegraphics[width=.32\textwidth]{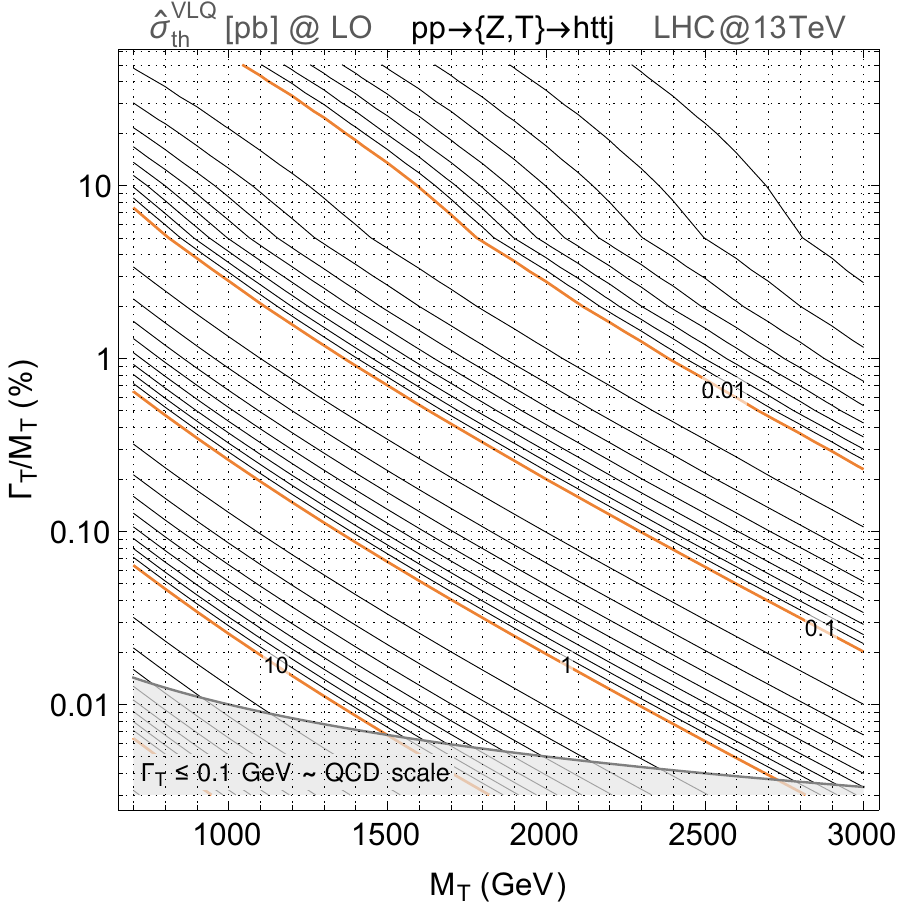}
  \\
  \includegraphics[width=.32\textwidth]{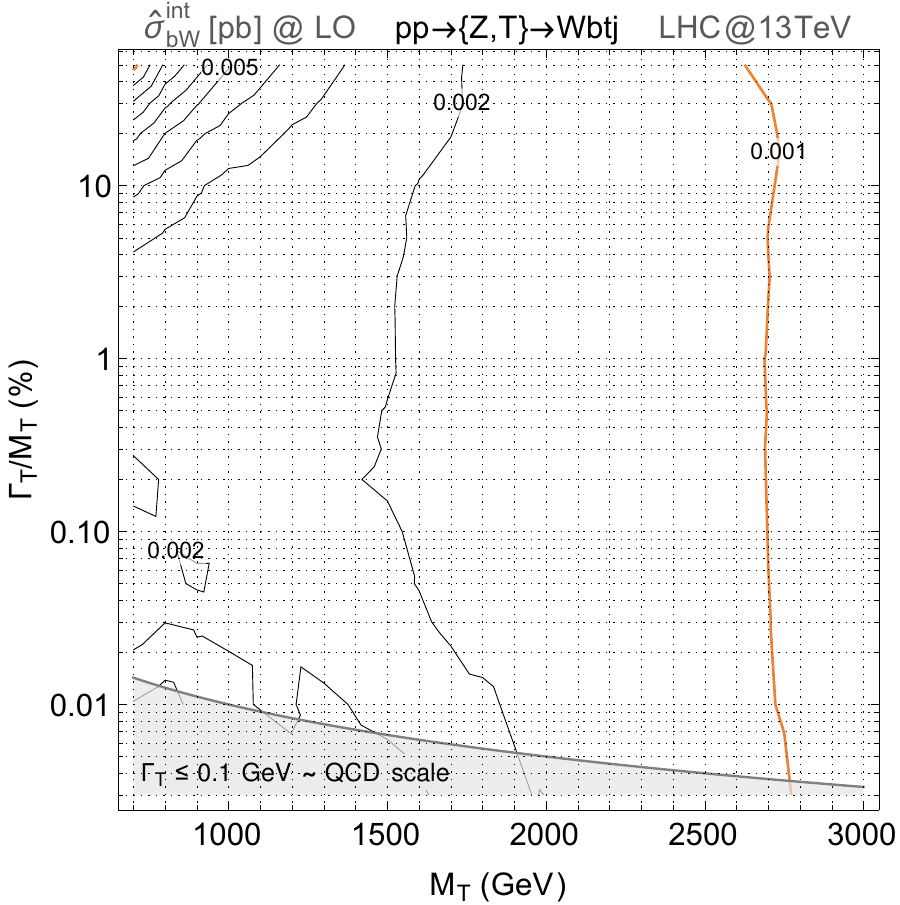}
  \includegraphics[width=.32\textwidth]{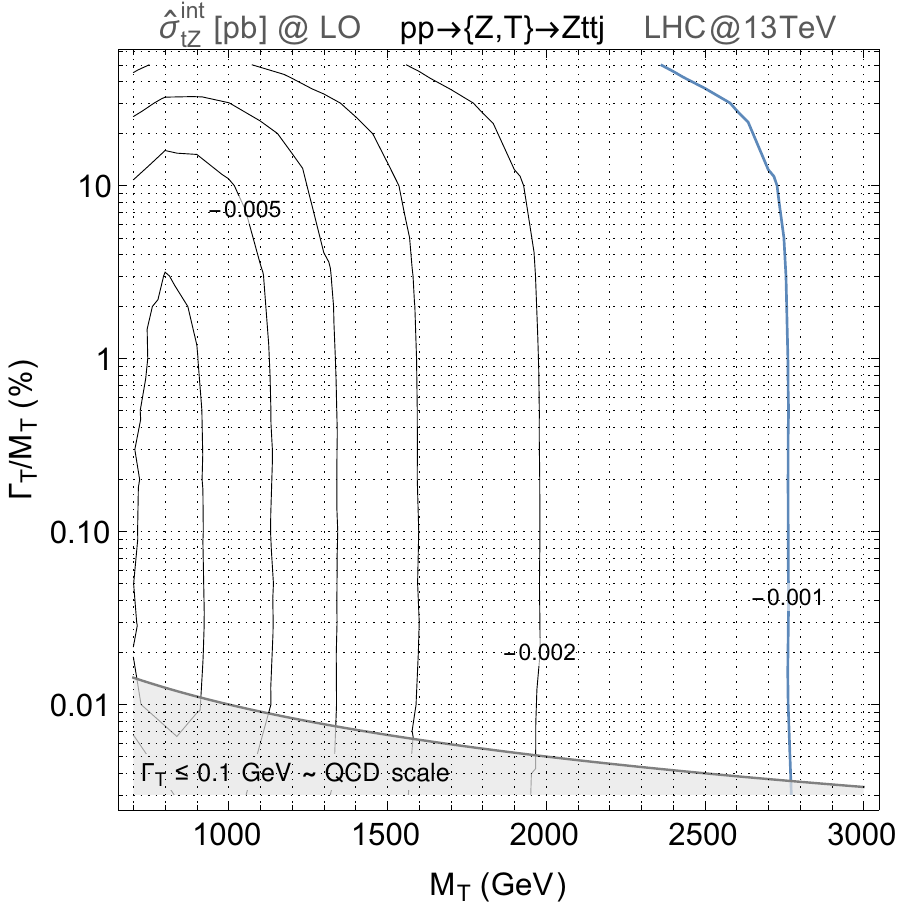}
  \includegraphics[width=.32\textwidth]{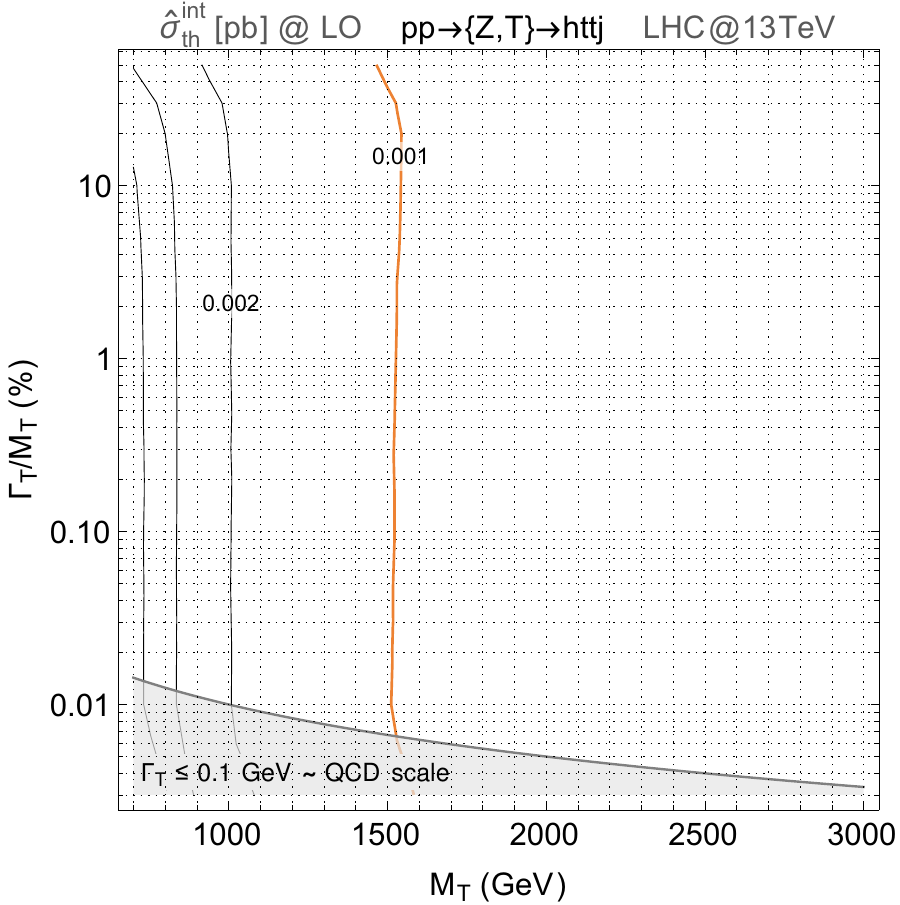}\vspace*{-0.3cm}
  \caption{\label{fig:sigmahat_Z_T_13TeV} Values of the $\hat\sigma^{\rm VLQ}$ (top row) and $\hat\sigma^{\rm int}$ (bottom row) bare cross sections, in pb,
    for the $pp\to Wbtj$ (left), $Zttj$ (centre) and $Httj$ (right) processes. We consider $13\TeV$ LHC proton-proton collisions and present the results in the $(M_T, \Gamma_T/MT)$ plane.}
\end{figure}

\begin{figure}
  \centering
  \includegraphics[width=.32\textwidth]{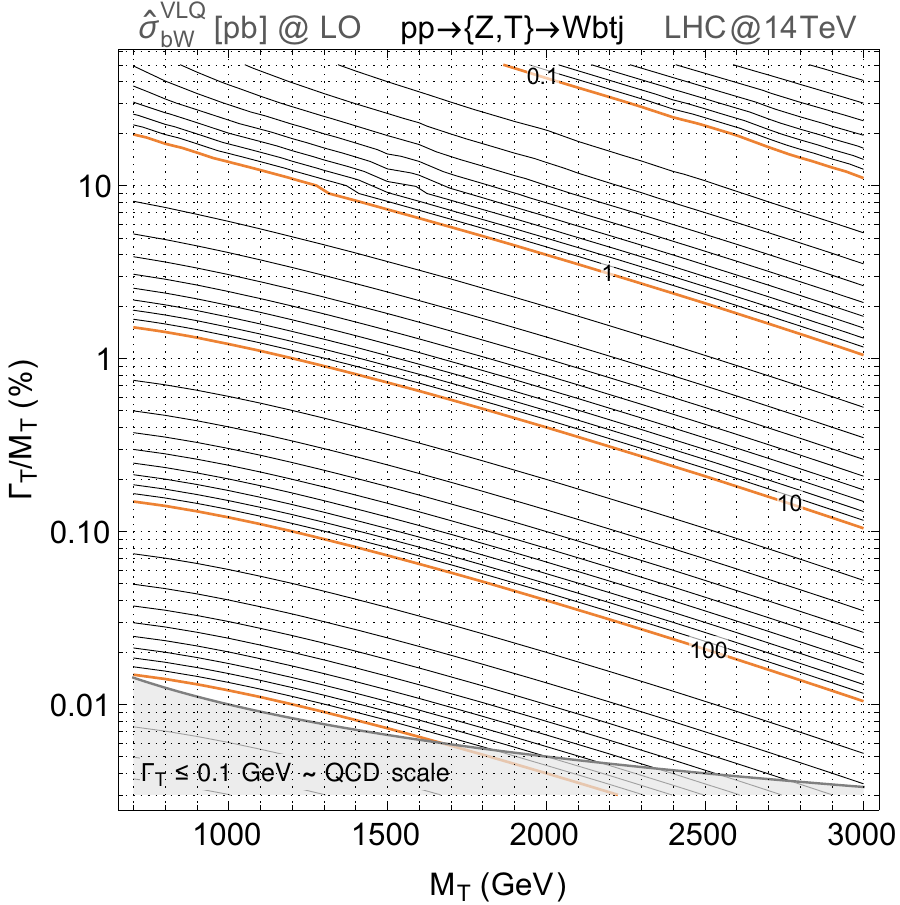}
  \includegraphics[width=.32\textwidth]{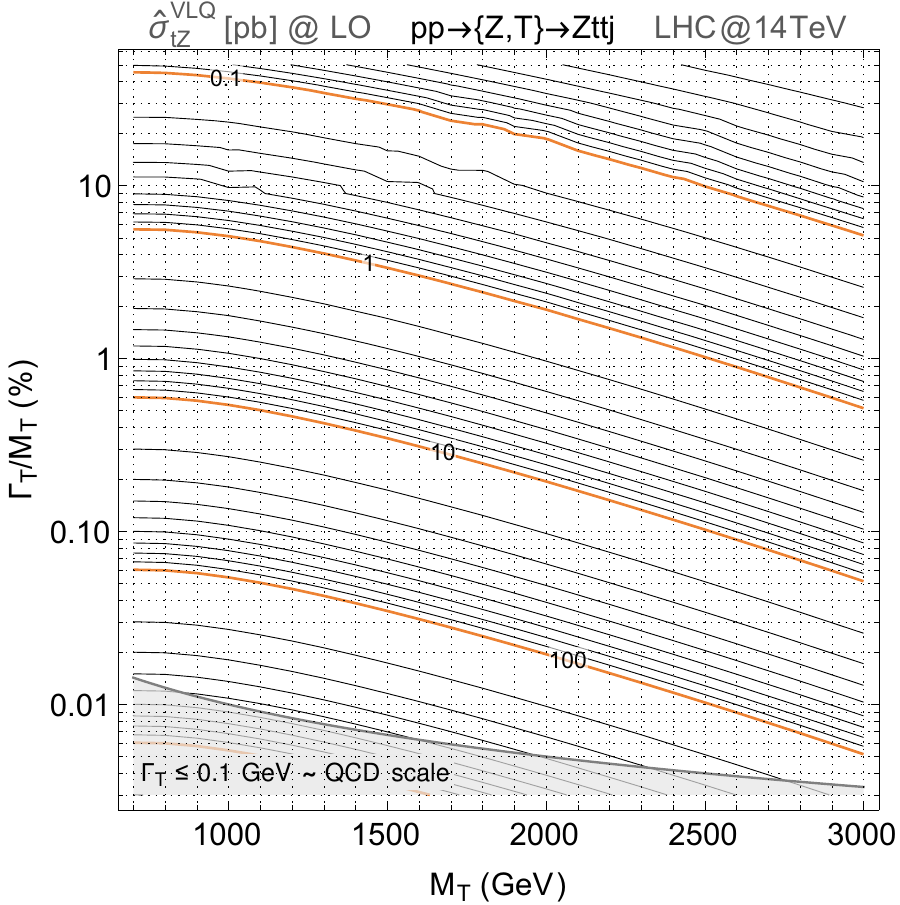}
  \includegraphics[width=.32\textwidth]{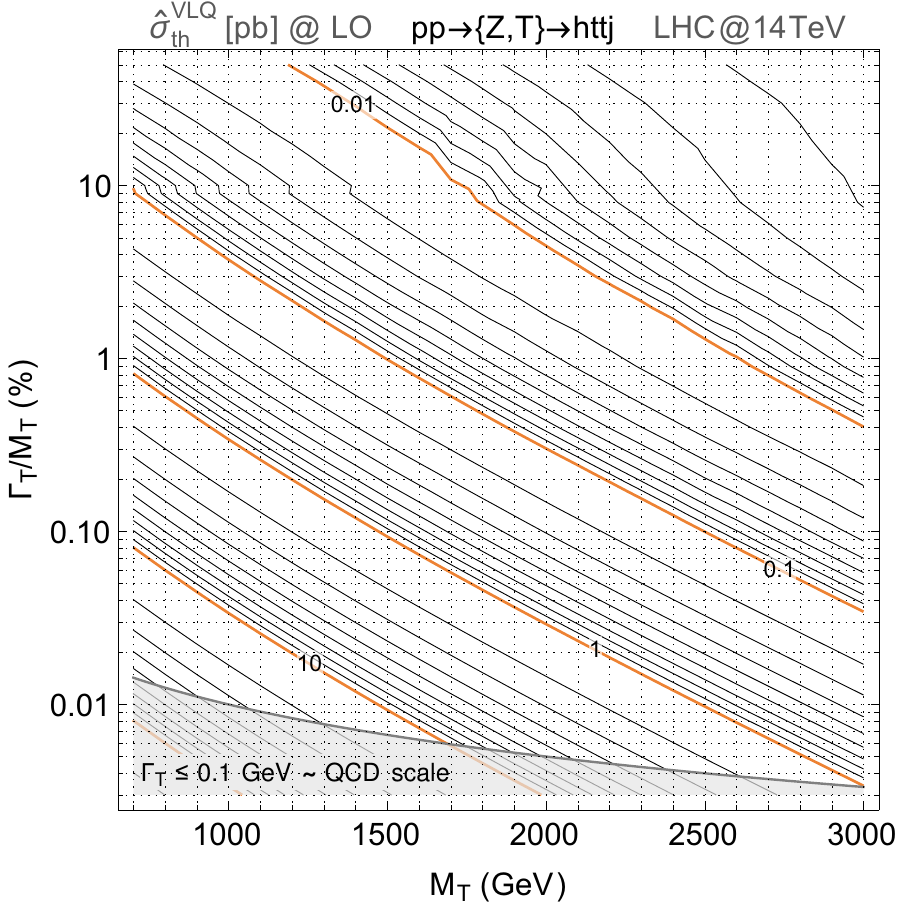}
  \\
  \includegraphics[width=.32\textwidth]{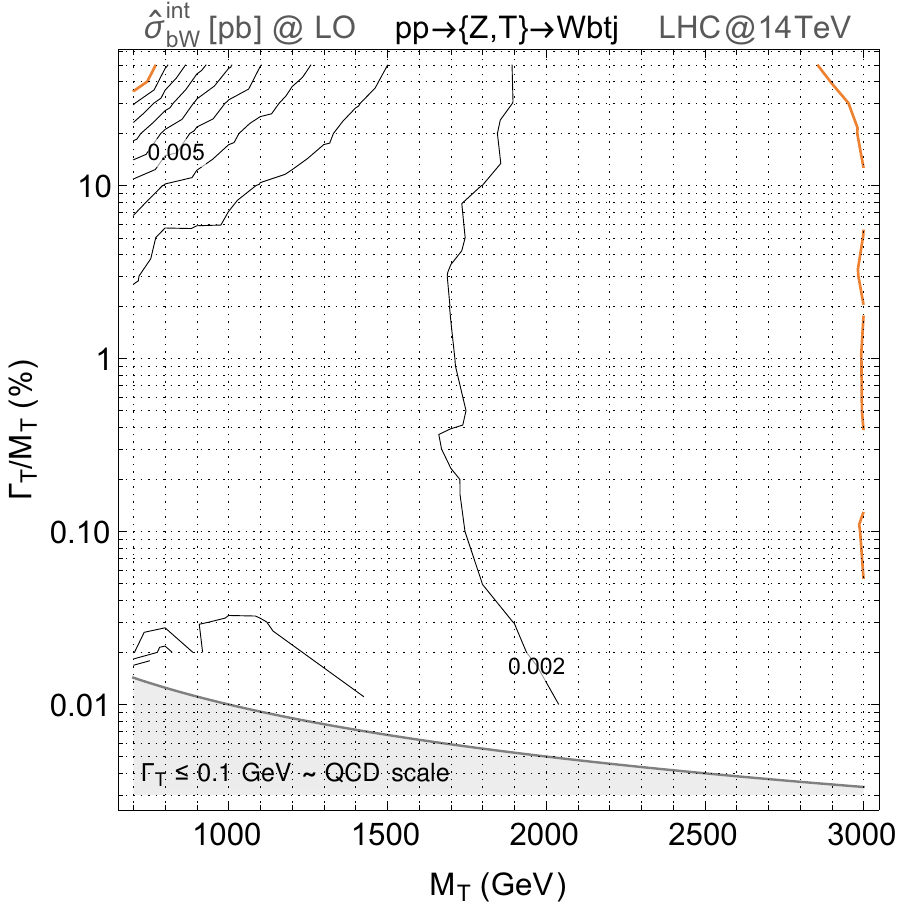}
  \includegraphics[width=.32\textwidth]{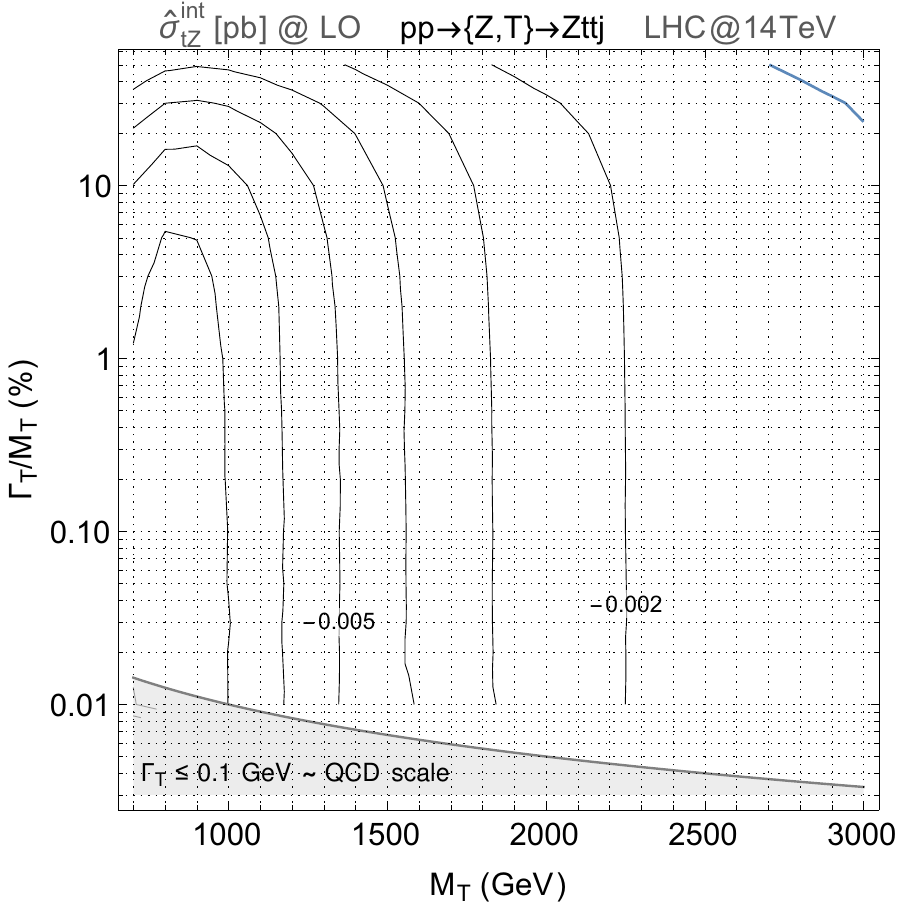}
  \includegraphics[width=.32\textwidth]{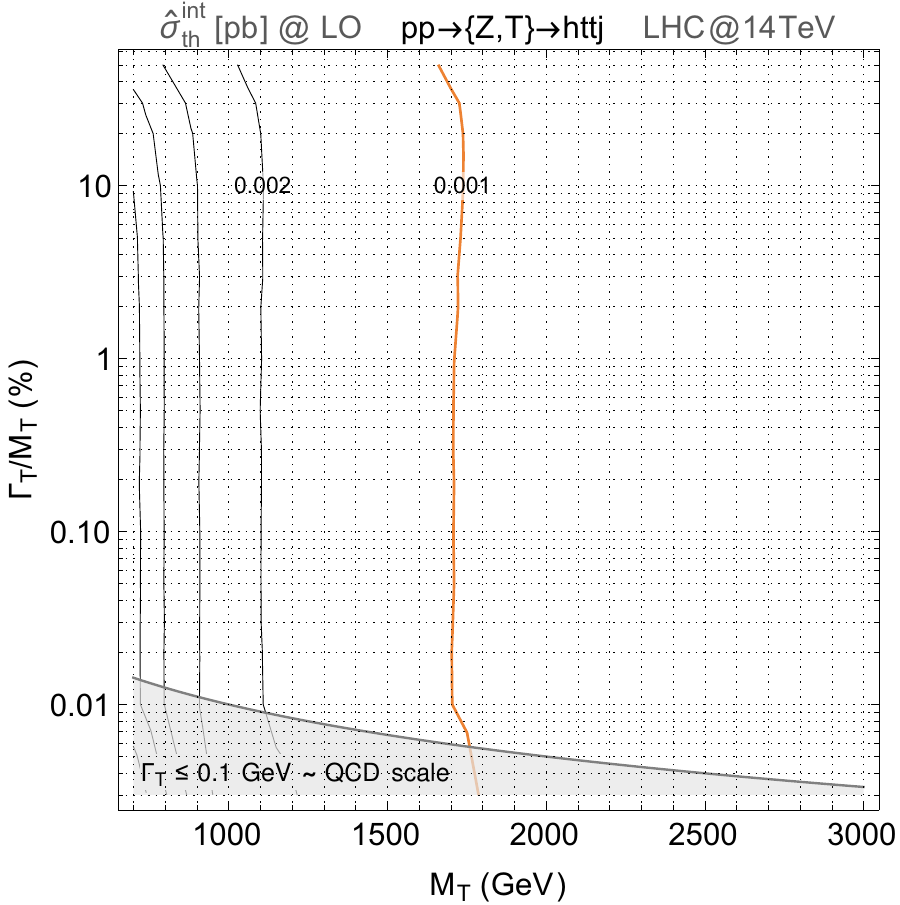}\vspace*{-0.2cm}
  \caption{\label{fig:sigmahat_Z_T_14TeV} Same as figure~\ref{fig:sigmahat_Z_T_13TeV}, but for proton-proton collisions at $14\TeV$.}
\end{figure}

\section{Simulation syntax with \mg~for single vector-like quark production}\label{app:Tech}
In this appendix we report and detail the \mg~syntax needed to simulate every process considered in this paper. The \mg~version used for the all our simulations is {\bf version 2.6.3.2}. The first step, common to every simulation, is to import the model\footnote{It can be downloaded from \url{http://feynrules.irmp.ucl.ac.be/wiki/NLOModels} in the ``Vector like quarks'' section.} with the restriction card that only turns on the VLQ interactions involving a third-generation SM quark via left-handed couplings and with a diagonal CKM matrix. This last simplification allows one to speed-up calculations when relevant. Then, the complex mass scheme has to be enabled. All of this is achieved by typing in the \mg\ command line interface:
\begin{tcolorbox}[top=-6pt,bottom=-6pt,left=-8pt,right=-8pt,width=\textwidth]
{\color{coolblack}
\begin{lstlisting}
import model VLQ_v4_4FNS_UFO-3rdL_noCKM
set complex_mass_scheme
\end{lstlisting}
}
\end{tcolorbox}
\noindent To consider a specific process and its charge-conjugate counterpart inclusively, the following labels are defined:
\begin{tcolorbox}[top=-6pt,bottom=-6pt,left=-8pt,right=-8pt,width=\textwidth]
{\color{coolblack}
\begin{lstlisting}
define tt = t t~
define bb = b b~
define ww = w+ w-
define tpp = tp tp~
\end{lstlisting}
}
\end{tcolorbox}
\noindent The generation of the {\bf LO SM irreducible background} (performed for consistency with the VLQ model) is achieved by forbidding VLQ propagation:\\[6pt]
\begin{minipage}{.3\textwidth}
\begin{itemize}[noitemsep,topsep=3pt,leftmargin=20pt]
 \item $pp\to Wbbj$
 \item $pp\to Ztbj$
 \item $pp\to htbj$
\end{itemize}
\end{minipage}\hfill
\begin{minipage}{.68\textwidth}
\begin{tcolorbox}[top=-6pt,bottom=-6pt,left=-8pt,right=-8pt,width=\textwidth]
{\color{coolblack}
\begin{lstlisting}
generate p p > ww bb bb j / tp bp x y 
generate p p > z tt bb j / tp bp x y 
generate p p > h tt bb j / tp bp x y 
\end{lstlisting}
}
\end{tcolorbox}
\end{minipage}\\[6pt]
\begin{minipage}{.3\textwidth}
\begin{itemize}[noitemsep,topsep=3pt,leftmargin=20pt]
 \item $pp\to Wbtj$
 \item $pp\to Zttj$
 \item $pp\to httj$
\end{itemize}
\end{minipage}\hfill
\begin{minipage}{.68\textwidth}
\begin{tcolorbox}[top=-6pt,bottom=-6pt,left=-8pt,right=-8pt,width=\textwidth]
{\color{coolblack}
\begin{lstlisting}
generate p p > ww bb tt j / tp bp x y 
generate p p > z tt ttj / tp bp x y 
generate p p > h tt tt j / tp bp x y 
\end{lstlisting}
}
\end{tcolorbox}
\end{minipage}\\[6pt]
\noindent The generation of the {\bf LO signal} contributions for processes mediated by $W$-boson or $Z$-boson exchanges is achieved through the following commands:\\[6pt]
\begin{minipage}{.3\textwidth}
\begin{itemize}[noitemsep,topsep=3pt,leftmargin=20pt]
 \item $pp\to\{W,T\}\to Wbbj$
 \item $pp\to\{W,T\}\to Ztbj$
 \item $pp\to\{W,T\}\to htbj$
\end{itemize}
\end{minipage}\hfill
\begin{minipage}{.68\textwidth}
\begin{tcolorbox}[top=-6pt,bottom=-6pt,left=-8pt,right=-8pt,width=\textwidth]
{\color{coolblack}
\begin{lstlisting}
generate p p > ww bb bb j / bp x y VLQ==2
generate p p > z tt bb j / bp x y VLQ==2
generate p p > h tt bb j / bp x y VLQ==2
\end{lstlisting}
}
\end{tcolorbox}
\end{minipage}\\[6pt]
\begin{minipage}{.3\textwidth}
\begin{itemize}[noitemsep,topsep=3pt,leftmargin=20pt]
 \item $pp\to\{Z,T\}\to Wbtj$
 \item $pp\to\{Z,T\}\to Zttj$
 \item $pp\to\{Z,T\}\to httj$
\end{itemize}
\end{minipage}\hfill
\begin{minipage}{.68\textwidth}
\begin{tcolorbox}[top=-6pt,bottom=-6pt,left=-8pt,right=-8pt,width=\textwidth]
{\color{coolblack}
\begin{lstlisting}
generate p p > ww bb tt j / bp x y VLQ==2
generate p p > z tt tt j / bp x y VLQ==2
generate p p > h tt tt j / bp x y VLQ==2
\end{lstlisting}
}
\end{tcolorbox}
\end{minipage}\\[6pt]
\noindent These instructions enforce the presence in the \textit{amplitude} of the process of \textit{exactly} two new couplings of the type \cbc{\tt VLQ}. This consists of a coupling order implemented in the {\tt UFO} model and that is associated with each coupling involving a VLQ, with the exception of the QCD and QED gauge interactions with gluons and photons. By doing so, the presence of a $T$ propagator, but not necessarily through a resonant configuration, is imposed. The results hence include both resonant and non-resonant topologies, as well as their interference. It is however not possible to include the subsequent decays of the SM particles in a way preserving the spin correlations, as the use of the \cbc{\verb~==~} syntax is incompatible with {\sc MadSpin} or with the addition of decays directly through in the \cbc{\verb~generate~} command (via, {\it e.g.}, \cbc{ \verb~generate ..., z > all all~}). Decays could only be handled through using {\sc Pythia}~8, albeit losing all spin correlations and describing the kinematics of the final state in an approximate way. 

An alternative syntax which allows us to include the decays of SM particles at the \mg\ level is:\\[6pt]
\begin{minipage}{.3\textwidth}
\begin{itemize}[noitemsep,topsep=3pt,leftmargin=20pt]
 \item $pp\to\{W,T\}\to Wbbj$
 \item $pp\to\{W,T\}\to Ztbj$
 \item $pp\to\{W,T\}\to htbj$
\end{itemize}
\end{minipage}\hfill
\begin{minipage}{.68\textwidth}
\begin{tcolorbox}[top=-6pt,bottom=-6pt,left=-8pt,right=-8pt,width=\textwidth]
{\color{coolblack}
\begin{lstlisting}
generate p p > ww bb bb j / bp x y QED=1 QCD=1
generate p p > z tt bb j / bp x y QED=1 QCD=1
generate p p > h tt bb j / bp x y QED=1 QCD=1
\end{lstlisting}
}
\end{tcolorbox}
\end{minipage}\\[6pt]
\begin{minipage}{.3\textwidth}
\begin{itemize}[noitemsep,topsep=3pt,leftmargin=20pt]
 \item $pp\to\{Z,T\}\to Wbtj$
 \item $pp\to\{Z,T\}\to Zttj$
 \item $pp\to\{Z,T\}\to httj$
\end{itemize}
\end{minipage}\hfill
\begin{minipage}{.68\textwidth}
\begin{tcolorbox}[top=-6pt,bottom=-6pt,left=-8pt,right=-8pt,width=\textwidth]
{\color{coolblack}
\begin{lstlisting}
generate p p > ww bb tt j / bp x y QED=1 QCD=1
generate p p > z tt tt j / bp x y QED=1 QCD=1
generate p p > h tt tt j / bp x y QED=1 QCD=1
\end{lstlisting}
}
\end{tcolorbox}
\end{minipage}\\[6pt]
\noindent With this syntax only the number of SM couplings is constrained to a specific value (as the \cbc{\tt =} syntax is interpreted by \mg~as \cbc{\tt <=}). No purely SM diagrams can be generated, leaving only topologies involving VLQ propagation. By decoupling all VLQs except the $T$ quark, the desired signal topologies (resonant and non-resonant, and their interference) are kept, and decays can be handled either directly through the \cbc{\verb~generate~} syntax, or via {\sc MadSpin}. This still relies (although not explicitly) on the \cbc{\tt VLQ} labelling of the couplings associated with the VLQs.\\

The generation of the processes describing the \textbf{LO interference between the signals and irreducible SM background} is done by enforcing the presence of a specific number of couplings in the {\it amplitude squared} through the coupling orders with the \cbc{\verb~\^{}2~} syntax of \mg. Unfortunately, decays which preserve spin correlations (in \mg~or {\sc MadSpin}) are not allowed once the \cbc{\verb~\^{}2~} syntax is used for process generation. For interference contributions there is thus no straightforward way to simulate the considered processes while keeping the spin correlations in the decay of final state particles. A workaround is however possible and described in \url{https://answers.launchpad.net/mg5amcnlo/+question/255413}, involving a manual modification of the information about the process generation in the {\tt LHE} event file, and the subsequent use of {\sc MadSpin}. The syntax for interference contributions used in this work, however, do not exploit this workaround. They are simply given by:\\[6pt]
\begin{minipage}{.325\textwidth}
\begin{itemize}[noitemsep,topsep=3pt,leftmargin=20pt]
 \item $pp\to\{W,T\},SM\to Wbbj$
 \item $pp\to\{W,T\},SM\to Ztbj$
 \item $pp\to\{W,T\},SM\to htbj$
\end{itemize}
\end{minipage}\hfill
\begin{minipage}{.66\textwidth}
\begin{tcolorbox}[top=-6pt,bottom=-6pt,left=-8pt,right=-8pt,width=\textwidth]
{\color{coolblack}
\begin{lstlisting}
generate p p > ww bb bb j / bp x y VLQ^2==2
generate p p > z tt bb j / bp x y VLQ^2==2
generate p p > h tt bb j / bp x y VLQ^2==2
\end{lstlisting}
}
\end{tcolorbox}
\end{minipage}\\[6pt]
\begin{minipage}{.325\textwidth}
\begin{itemize}[noitemsep,topsep=3pt,leftmargin=20pt]
 \item $pp\to\{Z,T\},SM\to Wbtj$
 \item $pp\to\{Z,T\},SM\to Zttj$
 \item $pp\to\{Z,T\},SM\to httj$
\end{itemize}
\end{minipage}\hfill
\begin{minipage}{.66\textwidth}
\begin{tcolorbox}[top=-6pt,bottom=-6pt,left=-8pt,right=-8pt,width=\textwidth]
{\color{coolblack}
\begin{lstlisting}
generate p p > ww bb tt j / bp x y VLQ^2==2
generate p p > z tt tt j / bp x y VLQ^2==2
generate p p > h tt tt j / bp x y VLQ^2==2
\end{lstlisting}
}
\end{tcolorbox}
\end{minipage}\\[6pt]
with the subsequent decays being done through {\sc Pythia~8}.

For what concerns the isolation of the \textbf{LO resonant and non-resonant signal contributions} in our study of the signal (which is relevant to understand which kind of accuracy is needed for the generation of the signal samples), the following syntax was used for the processes discussed in this paper. For the \textbf{LO resonant} contributions, we employed:\\[6pt]
\begin{minipage}{.31\textwidth}
\begin{itemize}[noitemsep,topsep=3pt,leftmargin=20pt]
 \item $pp\to\{W,T,\text{res}\}\to Wbbj$
\end{itemize}
\end{minipage}\hfill
\begin{minipage}{.66\textwidth}
\begin{tcolorbox}[top=-6pt,bottom=-6pt,left=-8pt,right=-8pt,width=1.09\textwidth]
{\color{coolblack}
\begin{lstlisting}
generate p p > tp > ww bb bb j / bp x y QED=1 QCD=1
add process p p > tp~ > ww bb bb j / bp x y QED=1 QCD=1
\end{lstlisting}
}
\end{tcolorbox}
\end{minipage}\\[6pt]
\begin{minipage}{.31\textwidth}
\begin{itemize}[noitemsep,topsep=3pt,leftmargin=20pt]
 \item $pp\to\{W,T,\text{res}\}\to Ztbj$
\end{itemize}
\end{minipage}\hfill
\begin{minipage}{.66\textwidth}
\begin{tcolorbox}[top=-6pt,bottom=-6pt,left=-8pt,right=-8pt,width=1.09\textwidth]
{\color{coolblack}
\begin{lstlisting}
generate p p > tp > z tt bb j / bp x y QED=1 QCD=1 
add process p p > tp~ > z tt bb j / bp x y QED=1 QCD=1 
\end{lstlisting}
}
\end{tcolorbox}
\end{minipage}\\[6pt]
\begin{minipage}{.31\textwidth}
\begin{itemize}[noitemsep,topsep=3pt,leftmargin=20pt]
 \item $pp\to\{W,T,\text{res}\}\to htbj$
\end{itemize}
\end{minipage}\hfill
\begin{minipage}{.66\textwidth}
\begin{tcolorbox}[top=-6pt,bottom=-6pt,left=-8pt,right=-8pt,width=1.09\textwidth]
{\color{coolblack}
\begin{lstlisting}
generate p p > tp > h tt bb j / bp x y QED=1 QCD=1 
add process p p > tp~ > h tt bb j / bp x y QED=1 QCD=1 
\end{lstlisting}
}
\end{tcolorbox}
\end{minipage}\\[6pt]
\begin{minipage}{.31\textwidth}
\begin{itemize}[noitemsep,topsep=3pt,leftmargin=20pt]
 \item $pp\to\{Z,T,\text{res}\}\to Wbtj$
\end{itemize}
\end{minipage}\hfill
\begin{minipage}{.66\textwidth}
\begin{tcolorbox}[top=-6pt,bottom=-6pt,left=-8pt,right=-8pt,width=1.09\textwidth]
{\color{coolblack}
\begin{lstlisting}
generate p p > tp > ww bb tt j / bp x y QED=1 QCD=1 
add process p p > tp~ > ww bb tt j / bp x y QED=1 QCD=1 
\end{lstlisting}
}
\end{tcolorbox}
\end{minipage}\\[6pt]
\begin{minipage}{.31\textwidth}
\begin{itemize}[noitemsep,topsep=3pt,leftmargin=20pt]
 \item $pp\to\{Z,T,\text{res}\}\to Zttj$
\end{itemize}
\end{minipage}\hfill
\begin{minipage}{.66\textwidth}
\begin{tcolorbox}[top=-6pt,bottom=-6pt,left=-8pt,right=-8pt,width=1.09\textwidth]
{\color{coolblack}
\begin{lstlisting}
generate p p > tp > z tt tt j / bp x y QED=1 QCD=1 
add process p p > tp~ > z tt tt j / bp x y QED=1 QCD=1 
\end{lstlisting}
}
\end{tcolorbox}
\end{minipage}\\[6pt]
\begin{minipage}{.31\textwidth}
\begin{itemize}[noitemsep,topsep=3pt,leftmargin=20pt]
 \item $pp\to\{Z,T,\text{res}\}\to httj$
\end{itemize}
\end{minipage}\hfill
\begin{minipage}{.66\textwidth}
\begin{tcolorbox}[top=-6pt,bottom=-6pt,left=-8pt,right=-8pt,width=1.09\textwidth]
{\color{coolblack}
\begin{lstlisting}
generate p p > tp > h tt tt j / bp x y QED=1 QCD=1 
add process p p > tp~ > h tt tt j / bp x y QED=1 QCD=1 
\end{lstlisting}
}
\end{tcolorbox}
\end{minipage}\\[6pt]

\noindent For the \textbf{LO non-resonant} ($t$-channel) contributions, the syntax reads:\\[6pt]
\begin{minipage}{.37\textwidth}
\begin{itemize}[noitemsep,topsep=3pt,leftmargin=20pt]
 \item $pp\to\{W,T,\text{non-res}\}\to Wbbj$
 \item $pp\to\{W,T,\text{non-res}\}\to Ztbj$
 \item $pp\to\{W,T,\text{non-res}\}\to htbj$
\end{itemize}
\end{minipage}\hfill
\begin{minipage}{.63\textwidth}
\begin{tcolorbox}[top=-6pt,bottom=-6pt,left=-8pt,right=-8pt,width=1.16\textwidth]
{\color{coolblack}
\begin{lstlisting}
generate p p > ww bb bb j $$ tp tp~ / bp x y QED=1 QCD=1 
generate p p > z tt bb j $$ tp tp~ / bp x y QED=1 QCD=1 
generate p p > h tt bb j $$ tp tp~ / bp x y QED=1 QCD=1 
\end{lstlisting}
}
\end{tcolorbox}
\end{minipage}\\[6pt]
\begin{minipage}{.37\textwidth}
\begin{itemize}[noitemsep,topsep=3pt,leftmargin=20pt]
 \item $pp\to\{Z,T,\text{non-res}\}\to Wbtj$
 \item $pp\to\{Z,T,\text{non-res}\}\to Zttj$
 \item $pp\to\{Z,T,\text{non-res}\}\to httj$
\end{itemize}
\end{minipage}\hfill
\begin{minipage}{.63\textwidth}
\begin{tcolorbox}[top=-6pt,bottom=-6pt,left=-8pt,right=-8pt,width=1.16\textwidth]
{\color{coolblack}
\begin{lstlisting}
generate p p > ww bb tt j $$ tp tp~ / bp x y QED=1 QCD=1 
generate p p > z tt tt j $$ tp tp~ / bp x y QED=1 QCD=1 
generate p p > h tt tt j $$ tp tp~ / bp x y QED=1 QCD=1 
\end{lstlisting}
}
\end{tcolorbox}
\end{minipage}\\[6pt]

The \textbf{LO simulations in the running width scheme} have been achieved by suitably modifying the VLQ propagator in the {\tt UFO} model to account for the modified width. As described in ref.~\cite{Christensen:2013aua}, we have modified the {\tt propagators.py} file by adding a specific propagator to be associated with the VLQ particle. This addition is given by:
\begin{tcolorbox}[top=-6pt,bottom=-6pt,left=-8pt,right=-8pt,width=\textwidth]
{\color{coolblack}
\begin{lstlisting}
VQ = Propagator(
 name = "VQ",
 numerator = "complex(0,1)*(Gamma('mu',1,2)*P('mu',id)+Mass(id)*Identity(1,2))",
 denominator = "P('mu',id)*P('mu',id)-Mass(id)*Mass(id)
                 +complex(0,1)*Mass(id)*Width(id)
                  *(P('mu',id)*P('mu',id)/(Mass(id)*Mass(id)))"
)
\end{lstlisting}
}
\end{tcolorbox}
\noindent where the running width parameter corresponding to eq.~\eqref{eq:runningwidth} is represented by the last term \cbc{\verb~complex(0,1)*Mass(id)*Width(id)*(P('mu',id)*P('mu',id)/(Mass(id)*Mass(id)))~} in the denominator. This running-width propagator has been assigned as a further attribute to the VLQ definition in {\tt particles.py}, as in the following example for the $T$ quark:
\begin{tcolorbox}[top=-6pt,bottom=-6pt,left=-8pt,right=-8pt,width=\textwidth]
{\color{coolblack}
\begin{lstlisting}
tp = Particle(pdg_code = 6000006,
              name = 'tp',
              antiname = 'tp~',
              spin = 2,
              color = 3,
              mass = Param.MTP,
              width = Param.WTP,
              propagator = Prop.VQ,
              texname = 'tp',
              antitexname = 'tp~',
              charge = 2/3,
              GhostNumber = 0,
              LeptonNumber = 0,
              Y = 0)
\end{lstlisting}
}
\end{tcolorbox}

\noindent Such a modification is not implemented by default in the public VLQ model. Moreover, to run simulations in this scheme, the complex mass scheme must not be enabled after importing the modified model.\\

Finally, \textbf{processes at NLO in QCD} are studied in the NWA approximation, and therefore the $T$ quark is produced only on-shell. This relies on the syntax:\\[6pt]
\begin{minipage}{.3\textwidth}
\begin{itemize}[noitemsep,topsep=3pt,leftmargin=20pt]
 \item $pp\to\{W,\text{NLO}\}\to Tbj$
 \item $pp\to\{Z,\text{NLO}\}\to Ttj$
\end{itemize}
\end{minipage}\hfill
\begin{minipage}{.68\textwidth}
\begin{tcolorbox}[top=-6pt,bottom=-6pt,left=-8pt,right=-8pt,width=\textwidth]
{\color{coolblack}
\begin{lstlisting}
generate p p > tpp bb j / bp x y [QCD] QED=1 QCD=1 
generate p p > tpp tt j / bp x y [QCD] QED=1 QCD=1 
\end{lstlisting}
}
\end{tcolorbox}
\end{minipage}\\[6pt]
The VLQ is subsequently decayed via {\sc MadSpin} to the final state under consideration. However, the NLO treatment of VLQ single production requires further modifications of the \mg~core files to account for the inclusion of all necessary loop topologies, including mixed electroweak and strong contributions. We refer to ref.~\cite{Fuks:2016ftf} for more information.

\newpage

%%%%%%%%%%%%%%%%%%%%%%%%%%%%%%%%%%%%%%%%%%%%%%%%%%%%%%%%%%%%%%%%%%%%%%%%%%%%%%
\bibliographystyle{JHEP}
\bibliography{bibliography}

\providecommand{\href}[2]{#2}\begingroup\raggedright\begin{thebibliography}{10}

\bibitem{Kaplan:1983fs}
D.~B. Kaplan and H.~Georgi, {\it {SU(2) x U(1) Breaking by Vacuum
  Misalignment}},  {\em Phys. Lett. B} {\bf 136} (1984) 183--186.

\bibitem{Kaplan:1991dc}
D.~B. Kaplan, {\it {Flavor at SSC energies: A New mechanism for dynamically
  generated fermion masses}},  {\em Nucl. Phys. B} {\bf 365} (1991) 259--278.

\bibitem{Agashe:2004rs}
K.~Agashe, R.~Contino, and A.~Pomarol, {\it {The Minimal composite Higgs
  model}},  {\em Nucl. Phys. B} {\bf 719} (2005) 165--187,
  [\href{http://arxiv.org/abs/hep-ph/0412089}{{\tt hep-ph/0412089}}].

\bibitem{Randall:1999ee}
L.~Randall and R.~Sundrum, {\it {A Large mass hierarchy from a small extra
  dimension}},  {\em Phys. Rev. Lett.} {\bf 83} (1999) 3370--3373,
  [\href{http://arxiv.org/abs/hep-ph/9905221}{{\tt hep-ph/9905221}}].

\bibitem{Chang:1999nh}
S.~Chang, J.~Hisano, H.~Nakano, N.~Okada, and M.~Yamaguchi, {\it {Bulk standard
  model in the Randall-Sundrum background}},  {\em Phys. Rev. D} {\bf 62}
  (2000) 084025, [\href{http://arxiv.org/abs/hep-ph/9912498}{{\tt
  hep-ph/9912498}}].

\bibitem{Gherghetta:2000qt}
T.~Gherghetta and A.~Pomarol, {\it {Bulk fields and supersymmetry in a slice of
  AdS}},  {\em Nucl. Phys. B} {\bf 586} (2000) 141--162,
  [\href{http://arxiv.org/abs/hep-ph/0003129}{{\tt hep-ph/0003129}}].

\bibitem{ArkaniHamed:2002qx}
N.~Arkani-Hamed, A.~G. Cohen, E.~Katz, A.~E. Nelson, T.~Gregoire, and J.~G.
  Wacker, {\it {The Minimal moose for a little Higgs}},  {\em JHEP} {\bf 08}
  (2002) 021, [\href{http://arxiv.org/abs/hep-ph/0206020}{{\tt
  hep-ph/0206020}}].

\bibitem{Perelstein:2003wd}
M.~Perelstein, M.~E. Peskin, and A.~Pierce, {\it {Top quarks and electroweak
  symmetry breaking in little Higgs models}},  {\em Phys. Rev. D} {\bf 69}
  (2004) 075002, [\href{http://arxiv.org/abs/hep-ph/0310039}{{\tt
  hep-ph/0310039}}].

\bibitem{Schmaltz:2005ky}
M.~Schmaltz and D.~Tucker-Smith, {\it {Little Higgs review}},  {\em Ann. Rev.
  Nucl. Part. Sci.} {\bf 55} (2005) 229--270,
  [\href{http://arxiv.org/abs/hep-ph/0502182}{{\tt hep-ph/0502182}}].

\bibitem{Bratchikov:2005vp}
D.~E. Lopez-Fogliani and C.~Munoz, {\it {Proposal for a Supersymmetric Standard
  Model}},  {\em Phys. Rev. Lett.} {\bf 97} (2006) 041801,
  [\href{http://arxiv.org/abs/hep-ph/0508297}{{\tt hep-ph/0508297}}].

\bibitem{Martin:2009bg}
S.~P. Martin, {\it {Extra vector-like matter and the lightest Higgs scalar
  boson mass in low-energy supersymmetry}},  {\em Phys. Rev. D} {\bf 81} (2010)
  035004, [\href{http://arxiv.org/abs/0910.2732}{{\tt arXiv:0910.2732}}].

\bibitem{Abdullah:2015zta}
M.~Abdullah and J.~L. Feng, {\it {Reviving bino dark matter with vectorlike
  fourth generation particles}},  {\em Phys. Rev. D} {\bf 93} (2016), no.~1
  015006, [\href{http://arxiv.org/abs/1510.06089}{{\tt arXiv:1510.06089}}].

\bibitem{Abdullah:2016avr}
M.~Abdullah, J.~L. Feng, S.~Iwamoto, and B.~Lillard, {\it {Heavy bino dark
  matter and collider signals in the MSSM with vectorlike fourth-generation
  particles}},  {\em Phys. Rev. D} {\bf 94} (2016), no.~9 095018,
  [\href{http://arxiv.org/abs/1608.00283}{{\tt arXiv:1608.00283}}].

\bibitem{Aguilar-Saavedra:2017giu}
J.~A. Aguilar-Saavedra, D.~E. L\'opez-Fogliani, and C.~Mu\~noz, {\it {Novel
  signatures for vector-like quarks}},  {\em JHEP} {\bf 06} (2017) 095,
  [\href{http://arxiv.org/abs/1705.02526}{{\tt arXiv:1705.02526}}].

\bibitem{Araz:2018uyi}
J.~Y. Araz, S.~Banerjee, M.~Frank, B.~Fuks, and A.~Goudelis, {\it {Dark matter
  and collider signals in an MSSM extension with vector-like multiplets}},
  {\em Phys. Rev. D} {\bf 98} (2018), no.~11 115009,
  [\href{http://arxiv.org/abs/1810.07224}{{\tt arXiv:1810.07224}}].

\bibitem{Zheng:2019kqu}
S.~Zheng, {\it {Minimal Vectorlike Model in Supersymmetric Unification}},  {\em
  Eur. Phys. J. C} {\bf 80} (2020), no.~3 273,
  [\href{http://arxiv.org/abs/1904.10145}{{\tt arXiv:1904.10145}}].

\bibitem{Barnard:2013zea}
J.~Barnard, T.~Gherghetta, and T.~S. Ray, {\it {UV descriptions of composite
  Higgs models without elementary scalars}},  {\em JHEP} {\bf 02} (2014) 002,
  [\href{http://arxiv.org/abs/1311.6562}{{\tt arXiv:1311.6562}}].

\bibitem{Ferretti:2013kya}
G.~Ferretti and D.~Karateev, {\it {Fermionic UV completions of Composite Higgs
  models}},  {\em JHEP} {\bf 03} (2014) 077,
  [\href{http://arxiv.org/abs/1312.5330}{{\tt arXiv:1312.5330}}].

\bibitem{Moretti:2016gkr}
S.~Moretti, D.~O'Brien, L.~Panizzi, and H.~Prager, {\it {Production of extra
  quarks at the Large Hadron Collider beyond the Narrow Width Approximation}},
  {\em Phys. Rev. D} {\bf 96} (2017), no.~7 075035,
  [\href{http://arxiv.org/abs/1603.09237}{{\tt arXiv:1603.09237}}].

\bibitem{Carvalho:2018jkq}
A.~Carvalho, S.~Moretti, D.~O'Brien, L.~Panizzi, and H.~Prager, {\it {Single
  production of vectorlike quarks with large width at the Large Hadron
  Collider}},  {\em Phys. Rev. D} {\bf 98} (2018), no.~1 015029,
  [\href{http://arxiv.org/abs/1805.06402}{{\tt arXiv:1805.06402}}].

\bibitem{Serra:2015xfa}
J.~Serra, {\it {Beyond the Minimal Top Partner Decay}},  {\em JHEP} {\bf 09}
  (2015) 176, [\href{http://arxiv.org/abs/1506.05110}{{\tt arXiv:1506.05110}}].

\bibitem{Chala:2017xgc}
M.~Chala, {\it {Direct bounds on heavy toplike quarks with standard and exotic
  decays}},  {\em Phys. Rev. D} {\bf 96} (2017), no.~1 015028,
  [\href{http://arxiv.org/abs/1705.03013}{{\tt arXiv:1705.03013}}].

\bibitem{Bizot:2018tds}
N.~Bizot, G.~Cacciapaglia, and T.~Flacke, {\it {Common exotic decays of top
  partners}},  {\em JHEP} {\bf 06} (2018) 065,
  [\href{http://arxiv.org/abs/1803.00021}{{\tt arXiv:1803.00021}}].

\bibitem{Han:2018hcu}
H.~Han, L.~Huang, T.~Ma, J.~Shu, T.~M. Tait, and Y.~Wu, {\it {Six Top Messages
  of New Physics at the LHC}},  {\em JHEP} {\bf 10} (2019) 008,
  [\href{http://arxiv.org/abs/1812.11286}{{\tt arXiv:1812.11286}}].

\bibitem{Xie:2019gya}
K.-P. Xie, G.~Cacciapaglia, and T.~Flacke, {\it {Exotic decays of top partners
  with charge 5/3: bounds and opportunities}},  {\em JHEP} {\bf 10} (2019) 134,
  [\href{http://arxiv.org/abs/1907.05894}{{\tt arXiv:1907.05894}}].

\bibitem{Benbrik:2019zdp}
R.~Benbrik et~al., {\it {Signatures of vector-like top partners decaying into
  new neutral scalar or pseudoscalar bosons}},  {\em JHEP} {\bf 05} (2020) 028,
  [\href{http://arxiv.org/abs/1907.05929}{{\tt arXiv:1907.05929}}].

\bibitem{Cacciapaglia:2019zmj}
G.~Cacciapaglia, T.~Flacke, M.~Park, and M.~Zhang, {\it {Exotic decays of top
  partners: mind the search gap}},  {\em Phys. Lett. B} {\bf 798} (2019)
  135015, [\href{http://arxiv.org/abs/1908.07524}{{\tt arXiv:1908.07524}}].

\bibitem{Aguilar-Saavedra:2019ghg}
J.~Aguilar-Saavedra, J.~Alonso-Gonz\'alez, L.~Merlo, and J.~No, {\it {Exotic
  vectorlike quark phenomenology in the minimal linear \ensuremath{\sigma}
  model}},  {\em Phys. Rev. D} {\bf 101} (2020), no.~3 035015,
  [\href{http://arxiv.org/abs/1911.10202}{{\tt arXiv:1911.10202}}].

\bibitem{Aaboud:2017zfn}
{\bf ATLAS} Collaboration, M.~Aaboud et~al., {\it {Search for pair production
  of heavy vector-like quarks decaying to high-p$_{T}$ W bosons and b quarks in
  the lepton-plus-jets final state in pp collisions at $ \sqrt{s}=13 $ TeV with
  the ATLAS detector}},  {\em JHEP} {\bf 10} (2017) 141,
  [\href{http://arxiv.org/abs/1707.03347}{{\tt arXiv:1707.03347}}].

\bibitem{Aaboud:2017qpr}
{\bf ATLAS} Collaboration, M.~Aaboud et~al., {\it {Search for pair production
  of vector-like top quarks in events with one lepton, jets, and missing
  transverse momentum in $ \sqrt{s}=13 $ TeV $pp$ collisions with the ATLAS
  detector}},  {\em JHEP} {\bf 08} (2017) 052,
  [\href{http://arxiv.org/abs/1705.10751}{{\tt arXiv:1705.10751}}].

\bibitem{Aaboud:2018xuw}
{\bf ATLAS} Collaboration, M.~Aaboud et~al., {\it {Search for pair production
  of up-type vector-like quarks and for four-top-quark events in final states
  with multiple $b$-jets with the ATLAS detector}},  {\em JHEP} {\bf 07} (2018)
  089, [\href{http://arxiv.org/abs/1803.09678}{{\tt arXiv:1803.09678}}].

\bibitem{Aaboud:2018saj}
{\bf ATLAS} Collaboration, M.~Aaboud et~al., {\it {Search for pair- and
  single-production of vector-like quarks in final states with at least one $Z$
  boson decaying into a pair of electrons or muons in $pp$ collision data
  collected with the ATLAS detector at $\sqrt{s} = 13$ TeV}},  {\em Phys. Rev.
  D} {\bf 98} (2018), no.~11 112010,
  [\href{http://arxiv.org/abs/1806.10555}{{\tt arXiv:1806.10555}}].

\bibitem{Aaboud:2018xpj}
{\bf ATLAS} Collaboration, M.~Aaboud et~al., {\it {Search for new phenomena in
  events with same-charge leptons and $b$-jets in $pp$ collisions at $\sqrt{s}=
  13$ TeV with the ATLAS detector}},  {\em JHEP} {\bf 12} (2018) 039,
  [\href{http://arxiv.org/abs/1807.11883}{{\tt arXiv:1807.11883}}].

\bibitem{Aaboud:2018wxv}
{\bf ATLAS} Collaboration, M.~Aaboud et~al., {\it {Search for pair production
  of heavy vector-like quarks decaying into hadronic final states in $pp$
  collisions at $\sqrt{s} = 13$ TeV with the ATLAS detector}},  {\em Phys. Rev.
  D} {\bf 98} (2018), no.~9 092005,
  [\href{http://arxiv.org/abs/1808.01771}{{\tt arXiv:1808.01771}}].

\bibitem{Aaboud:2018pii}
{\bf ATLAS} Collaboration, M.~Aaboud et~al., {\it {Combination of the searches
  for pair-produced vector-like partners of the third-generation quarks at
  $\sqrt{s} =$ 13 TeV with the ATLAS detector}},  {\em Phys. Rev. Lett.} {\bf
  121} (2018), no.~21 211801, [\href{http://arxiv.org/abs/1808.02343}{{\tt
  arXiv:1808.02343}}].

\bibitem{Sirunyan:2017pks}
{\bf CMS} Collaboration, A.~M. Sirunyan et~al., {\it {Search for pair
  production of vector-like quarks in the bW$\overline{\mathrm{b}}$W channel
  from proton-proton collisions at $\sqrt{s} =$ 13 TeV}},  {\em Phys. Lett. B}
  {\bf 779} (2018) 82--106, [\href{http://arxiv.org/abs/1710.01539}{{\tt
  arXiv:1710.01539}}].

\bibitem{Sirunyan:2018qau}
{\bf CMS} Collaboration, A.~M. Sirunyan et~al., {\it {Search for vector-like
  quarks in events with two oppositely charged leptons and jets in
  proton-proton collisions at $\sqrt{s} =$ 13 TeV}},  {\em Eur. Phys. J. C}
  {\bf 79} (2019), no.~4 364, [\href{http://arxiv.org/abs/1812.09768}{{\tt
  arXiv:1812.09768}}].

\bibitem{Sirunyan:2018omb}
{\bf CMS} Collaboration, A.~M. Sirunyan et~al., {\it {Search for vector-like T
  and B quark pairs in final states with leptons at $\sqrt{s} =$ 13 TeV}},
  {\em JHEP} {\bf 08} (2018) 177, [\href{http://arxiv.org/abs/1805.04758}{{\tt
  arXiv:1805.04758}}].

\bibitem{Sirunyan:2019sza}
{\bf CMS} Collaboration, A.~M. Sirunyan et~al., {\it {Search for pair
  production of vectorlike quarks in the fully hadronic final state}},  {\em
  Phys. Rev. D} {\bf 100} (2019), no.~7 072001,
  [\href{http://arxiv.org/abs/1906.11903}{{\tt arXiv:1906.11903}}].

\bibitem{Sirunyan:2018yun}
{\bf CMS} Collaboration, A.~M. Sirunyan et~al., {\it {Search for top quark
  partners with charge 5/3 in the same-sign dilepton and single-lepton final
  states in proton-proton collisions at $ \sqrt{s}=13 $ TeV}},  {\em JHEP} {\bf
  03} (2019) 082, [\href{http://arxiv.org/abs/1810.03188}{{\tt
  arXiv:1810.03188}}].

\bibitem{Sirunyan:2020qvb}
{\bf CMS} Collaboration, A.~M. Sirunyan et~al., {\it {A search for bottom-type,
  vector-like quark pair production in a fully hadronic final state in
  proton-proton collisions at $\sqrt{s} =$ 13 TeV}},  {\em Phys. Rev. D} {\bf
  102} (2020) 112004, [\href{http://arxiv.org/abs/2008.09835}{{\tt
  arXiv:2008.09835}}].

\bibitem{Aaboud:2018ifs}
{\bf ATLAS} Collaboration, M.~Aaboud et~al., {\it {Search for single production
  of vector-like quarks decaying into $Wb$ in $pp$ collisions at $\sqrt{s} =
  13$ TeV with the ATLAS detector}},  {\em JHEP} {\bf 05} (2019) 164,
  [\href{http://arxiv.org/abs/1812.07343}{{\tt arXiv:1812.07343}}].

\bibitem{Sirunyan:2017ynj}
{\bf CMS} Collaboration, A.~Sirunyan et~al., {\it {Search for single production
  of a vector-like T quark decaying to a Z boson and a top quark in
  proton-proton collisions at $\sqrt s$ = 13 TeV}},  {\em Phys. Lett. B} {\bf
  781} (2018) 574--600, [\href{http://arxiv.org/abs/1708.01062}{{\tt
  arXiv:1708.01062}}].

\bibitem{Sirunyan:2018fjh}
{\bf CMS} Collaboration, A.~M. Sirunyan et~al., {\it {Search for single
  production of vector-like quarks decaying to a b quark and a Higgs boson}},
  {\em JHEP} {\bf 06} (2018) 031, [\href{http://arxiv.org/abs/1802.01486}{{\tt
  arXiv:1802.01486}}].

\bibitem{Sirunyan:2018ncp}
{\bf CMS} Collaboration, A.~M. Sirunyan et~al., {\it {Search for single
  production of vector-like quarks decaying to a top quark and a W boson in
  proton-proton collisions at $\sqrt{s} =$ 13 TeV}},  {\em Eur. Phys. J. C}
  {\bf 79} (2019) 90, [\href{http://arxiv.org/abs/1809.08597}{{\tt
  arXiv:1809.08597}}].

\bibitem{Sirunyan:2019xeh}
{\bf CMS} Collaboration, A.~M. Sirunyan et~al., {\it {Search for electroweak
  production of a vector-like T quark using fully hadronic final states}},
  {\em JHEP} {\bf 01} (2020) 036, [\href{http://arxiv.org/abs/1909.04721}{{\tt
  arXiv:1909.04721}}].

\bibitem{Matsedonskyi:2014mna}
O.~Matsedonskyi, G.~Panico, and A.~Wulzer, {\it {On the Interpretation of Top
  Partners Searches}},  {\em JHEP} {\bf 12} (2014) 097,
  [\href{http://arxiv.org/abs/1409.0100}{{\tt arXiv:1409.0100}}].

\bibitem{CMS:2013xfa}
{\bf CMS} Collaboration, {\it {Projected Performance of an Upgraded CMS
  Detector at the LHC and HL-LHC: Contribution to the Snowmass Process}},  in
  {\em {Community Summer Study 2013}: {Snowmass on the Mississippi}}, 7, 2013.
\newblock \href{http://arxiv.org/abs/1307.7135}{{\tt arXiv:1307.7135}}.

\bibitem{Barducci:2017xtw}
D.~Barducci and L.~Panizzi, {\it {Vector-like quarks coupling discrimination at
  the LHC and future hadron colliders}},  {\em JHEP} {\bf 12} (2017) 057,
  [\href{http://arxiv.org/abs/1710.02325}{{\tt arXiv:1710.02325}}].

\bibitem{CidVidal:2018eel}
X.~Cid~Vidal et~al., {\it {Report from Working Group 3}: {Beyond the Standard
  Model physics at the HL-LHC and HE-LHC}},  {\em CERN Yellow Rep. Monogr.}
  {\bf 7} (2019) 585--865, [\href{http://arxiv.org/abs/1812.07831}{{\tt
  arXiv:1812.07831}}].

\bibitem{Fuks:2016ftf}
B.~Fuks and H.-S. Shao, {\it {QCD next-to-leading-order predictions matched to
  parton showers for vector-like quark models}},  {\em Eur. Phys. J.} {\bf C77}
  (2017), no.~2 135, [\href{http://arxiv.org/abs/1610.04622}{{\tt
  arXiv:1610.04622}}].

\bibitem{Cacciapaglia:2018qep}
G.~Cacciapaglia, A.~Carvalho, A.~Deandrea, T.~Flacke, B.~Fuks, D.~Majumder,
  L.~Panizzi, and H.-S. Shao, {\it {Next-to-leading-order predictions for
  single vector-like quark production at the LHC}},  {\em Phys. Lett. B} {\bf
  793} (2019) 206--211, [\href{http://arxiv.org/abs/1811.05055}{{\tt
  arXiv:1811.05055}}].

\bibitem{Alwall:2014hca}
J.~Alwall, R.~Frederix, S.~Frixione, V.~Hirschi, F.~Maltoni, O.~Mattelaer,
  H.~S. Shao, T.~Stelzer, P.~Torrielli, and M.~Zaro, {\it {The automated
  computation of tree-level and next-to-leading order differential cross
  sections, and their matching to parton shower simulations}},  {\em JHEP} {\bf
  07} (2014) 079, [\href{http://arxiv.org/abs/1405.0301}{{\tt
  arXiv:1405.0301}}].

\bibitem{Alloul:2013bka}
A.~Alloul, N.~D. Christensen, C.~Degrande, C.~Duhr, and B.~Fuks, {\it
  {FeynRules 2.0 - A complete toolbox for tree-level phenomenology}},  {\em
  Comput. Phys. Commun.} {\bf 185} (2014) 2250--2300,
  [\href{http://arxiv.org/abs/1310.1921}{{\tt arXiv:1310.1921}}].

\bibitem{Christensen:2009jx}
N.~D. Christensen, P.~de~Aquino, C.~Degrande, C.~Duhr, B.~Fuks, M.~Herquet,
  F.~Maltoni, and S.~Schumann, {\it {A Comprehensive approach to new physics
  simulations}},  {\em Eur. Phys. J. C} {\bf 71} (2011) 1541,
  [\href{http://arxiv.org/abs/0906.2474}{{\tt arXiv:0906.2474}}].

\bibitem{Degrande:2011ua}
C.~Degrande, C.~Duhr, B.~Fuks, D.~Grellscheid, O.~Mattelaer, and T.~Reiter,
  {\it {UFO - The Universal FeynRules Output}},  {\em Comput. Phys. Commun.}
  {\bf 183} (2012) 1201--1214, [\href{http://arxiv.org/abs/1108.2040}{{\tt
  arXiv:1108.2040}}].

\bibitem{Degrande:2014vpa}
C.~Degrande, {\it {Automatic evaluation of UV and R2 terms for beyond the
  Standard Model Lagrangians: a proof-of-principle}},  {\em Comput. Phys.
  Commun.} {\bf 197} (2015) 239--262,
  [\href{http://arxiv.org/abs/1406.3030}{{\tt arXiv:1406.3030}}].

\bibitem{Denner:1999gp}
A.~Denner, S.~Dittmaier, M.~Roth, and D.~Wackeroth, {\it {Predictions for all
  processes e+ e- ---\ensuremath{>} 4 fermions + gamma}},  {\em Nucl. Phys. B}
  {\bf 560} (1999) 33--65, [\href{http://arxiv.org/abs/hep-ph/9904472}{{\tt
  hep-ph/9904472}}].

\bibitem{Denner:2005fg}
A.~Denner, S.~Dittmaier, M.~Roth, and L.~Wieders, {\it {Electroweak corrections
  to charged-current e+ e- ---\ensuremath{>} 4 fermion processes: Technical
  details and further results}},  {\em Nucl. Phys. B} {\bf 724} (2005)
  247--294, [\href{http://arxiv.org/abs/hep-ph/0505042}{{\tt hep-ph/0505042}}].
  [Erratum: Nucl.Phys.B 854, 504--507 (2012)].

\bibitem{Frederix:2018nkq}
R.~Frederix, S.~Frixione, V.~Hirschi, D.~Pagani, H.~S. Shao, and M.~Zaro, {\it
  {The automation of next-to-leading order electroweak calculations}},  {\em
  JHEP} {\bf 07} (2018) 185, [\href{http://arxiv.org/abs/1804.10017}{{\tt
  arXiv:1804.10017}}].

\bibitem{Berdine:2007uv}
D.~Berdine, N.~Kauer, and D.~Rainwater, {\it {Breakdown of the Narrow Width
  Approximation for New Physics}},  {\em Phys. Rev. Lett.} {\bf 99} (2007)
  111601, [\href{http://arxiv.org/abs/hep-ph/0703058}{{\tt hep-ph/0703058}}].

\bibitem{Bardin:1988xt}
D.~{\relax Yu}. Bardin, A.~Leike, T.~Riemann, and M.~Sachwitz, {\it {Energy
  Dependent Width Effects in e+ e- Annihilation Near the Z Boson Pole}},  {\em
  Phys. Lett.} {\bf B206} (1988) 539--542.

\bibitem{Chen:2017hak}
C.-Y. Chen, S.~Dawson, and E.~Furlan, {\it {Vectorlike fermions and Higgs
  effective field theory revisited}},  {\em Phys. Rev. D} {\bf 96} (2017),
  no.~1 015006, [\href{http://arxiv.org/abs/1703.06134}{{\tt
  arXiv:1703.06134}}].

\bibitem{Cacciapaglia:2015ixa}
G.~Cacciapaglia, A.~Deandrea, N.~Gaur, D.~Harada, Y.~Okada, and L.~Panizzi,
  {\it {Interplay of vector-like top partner multiplets in a realistic mixing
  set-up}},  {\em JHEP} {\bf 09} (2015) 012,
  [\href{http://arxiv.org/abs/1502.00370}{{\tt arXiv:1502.00370}}].

\bibitem{Cacciapaglia:2018lld}
G.~Cacciapaglia, A.~Deandrea, N.~Gaur, D.~Harada, Y.~Okada, and L.~Panizzi,
  {\it {The LHC potential of Vector-like quark doublets}},  {\em JHEP} {\bf 11}
  (2018) 055, [\href{http://arxiv.org/abs/1806.01024}{{\tt arXiv:1806.01024}}].

\bibitem{Christensen:2013aua}
N.~D. Christensen, P.~de~Aquino, N.~Deutschmann, C.~Duhr, B.~Fuks,
  C.~Garcia-Cely, O.~Mattelaer, K.~Mawatari, B.~Oexl, and Y.~Takaesu, {\it
  {Simulating spin-$ \frac{3}{2}$ particles at colliders}},  {\em Eur. Phys. J.
  C} {\bf 73} (2013), no.~10 2580, [\href{http://arxiv.org/abs/1308.1668}{{\tt
  arXiv:1308.1668}}].

\bibitem{Ball:2014uwa}
{\bf NNPDF} Collaboration, R.~D. Ball et~al., {\it {Parton distributions for
  the LHC Run II}},  {\em JHEP} {\bf 04} (2015) 040,
  [\href{http://arxiv.org/abs/1410.8849}{{\tt arXiv:1410.8849}}].

\bibitem{Buckley:2014ana}
A.~Buckley, J.~Ferrando, S.~Lloyd, K.~Nordstr\"om, B.~Page, M.~R\"ufenacht,
  M.~Sch\"onherr, and G.~Watt, {\it {LHAPDF6: parton density access in the LHC
  precision era}},  {\em Eur. Phys. J. C} {\bf 75} (2015) 132,
  [\href{http://arxiv.org/abs/1412.7420}{{\tt arXiv:1412.7420}}].

\bibitem{Artoisenet:2012st}
P.~Artoisenet, R.~Frederix, O.~Mattelaer, and R.~Rietkerk, {\it {Automatic
  spin-entangled decays of heavy resonances in Monte Carlo simulations}},  {\em
  JHEP} {\bf 03} (2013) 015, [\href{http://arxiv.org/abs/1212.3460}{{\tt
  arXiv:1212.3460}}].

\bibitem{Alwall:2014bza}
J.~Alwall, C.~Duhr, B.~Fuks, O.~Mattelaer, D.~G. \"Ozt\"urk, and C.-H. Shen,
  {\it {Computing decay rates for new physics theories with FeynRules and
  MadGraph 5 \_aMC@NLO}},  {\em Comput. Phys. Commun.} {\bf 197} (2015)
  312--323, [\href{http://arxiv.org/abs/1402.1178}{{\tt arXiv:1402.1178}}].

\bibitem{Dawson:1984gx}
S.~Dawson, {\it {The Effective W Approximation}},  {\em Nucl. Phys. B} {\bf
  249} (1985) 42--60.

\bibitem{Kane:1984bb}
G.~L. Kane, W.~Repko, and W.~Rolnick, {\it {The Effective W+-, Z0 Approximation
  for High-Energy Collisions}},  {\em Phys. Lett. B} {\bf 148} (1984) 367--372.

\bibitem{Kunszt:1987tk}
Z.~Kunszt and D.~E. Soper, {\it {On the Validity of the Effective $W$
  Approximation}},  {\em Nucl. Phys. B} {\bf 296} (1988) 253--289.

\bibitem{delAguila:2000rc}
F.~del Aguila, M.~Perez-Victoria, and J.~Santiago, {\it {Observable
  contributions of new exotic quarks to quark mixing}},  {\em JHEP} {\bf 09}
  (2000) 011, [\href{http://arxiv.org/abs/hep-ph/0007316}{{\tt
  hep-ph/0007316}}].

\bibitem{Buchkremer:2013bha}
M.~Buchkremer, G.~Cacciapaglia, A.~Deandrea, and L.~Panizzi, {\it {Model
  Independent Framework for Searches of Top Partners}},  {\em Nucl. Phys.} {\bf
  B876} (2013) 376--417, [\href{http://arxiv.org/abs/1305.4172}{{\tt
  arXiv:1305.4172}}].

\bibitem{Sjostrand:2014zea}
T.~Sj\"ostrand, S.~Ask, J.~R. Christiansen, R.~Corke, N.~Desai, P.~Ilten,
  S.~Mrenna, S.~Prestel, C.~O. Rasmussen, and P.~Z. Skands, {\it {An
  introduction to PYTHIA 8.2}},  {\em Comput. Phys. Commun.} {\bf 191} (2015)
  159--177, [\href{http://arxiv.org/abs/1410.3012}{{\tt arXiv:1410.3012}}].

\bibitem{Conte:2012fm}
E.~Conte, B.~Fuks, and G.~Serret, {\it {MadAnalysis 5, A User-Friendly
  Framework for Collider Phenomenology}},  {\em Comput. Phys. Commun.} {\bf
  184} (2013) 222--256, [\href{http://arxiv.org/abs/1206.1599}{{\tt
  arXiv:1206.1599}}].

\bibitem{Conte:2014zja}
E.~Conte, B.~Dumont, B.~Fuks, and C.~Wymant, {\it {Designing and recasting LHC
  analyses with MadAnalysis 5}},  {\em Eur. Phys. J. C} {\bf 74} (2014), no.~10
  3103, [\href{http://arxiv.org/abs/1405.3982}{{\tt arXiv:1405.3982}}].

\bibitem{Araz:2020lnp}
J.~Y. Araz, B.~Fuks, and G.~Polykratis, {\it {Simplified fast detector
  simulation in MADANALYSIS~5}},  {\em Eur. Phys. J. C} {\bf 81} (2021), no.~4
  329, [\href{http://arxiv.org/abs/2006.09387}{{\tt arXiv:2006.09387}}].

\bibitem{Cacciari:2008gp}
M.~Cacciari, G.~P. Salam, and G.~Soyez, {\it {The anti-$k_t$ jet clustering
  algorithm}},  {\em JHEP} {\bf 04} (2008) 063,
  [\href{http://arxiv.org/abs/0802.1189}{{\tt arXiv:0802.1189}}].

\bibitem{Cacciari:2011ma}
M.~Cacciari, G.~P. Salam, and G.~Soyez, {\it {FastJet User Manual}},  {\em Eur.
  Phys. J. C} {\bf 72} (2012) 1896, [\href{http://arxiv.org/abs/1111.6097}{{\tt
  arXiv:1111.6097}}].

\bibitem{Cacciapaglia:2009pa}
G.~Cacciapaglia, A.~Deandrea, and J.~Llodra-Perez, {\it {A Dark Matter
  candidate from Lorentz Invariance in 6D}},  {\em JHEP} {\bf 03} (2010) 083,
  [\href{http://arxiv.org/abs/0907.4993}{{\tt arXiv:0907.4993}}].

\bibitem{Cacciapaglia:2013wha}
G.~Cacciapaglia, A.~Deandrea, J.~Ellis, J.~Marrouche, and L.~Panizzi, {\it {LHC
  Missing-Transverse-Energy Constraints on Models with Universal Extra
  Dimensions}},  {\em Phys. Rev. D} {\bf 87} (2013), no.~7 075006,
  [\href{http://arxiv.org/abs/1302.4750}{{\tt arXiv:1302.4750}}].

\bibitem{Butterworth:2015oua}
J.~Butterworth et~al., {\it {PDF4LHC recommendations for LHC Run II}},  {\em J.
  Phys.} {\bf G43} (2016) 023001, [\href{http://arxiv.org/abs/1510.03865}{{\tt
  arXiv:1510.03865}}].

\bibitem{urlFR}
``{FeynRules NLO models, Vector like quarks section}.''
  \url{http://feynrules.irmp.ucl.ac.be/wiki/NLOModels}.

\end{thebibliography}\endgroup

\end{document}